\documentclass[cernpreprint,backref=false,texlive=2020,UKenglish,dvipsnames,block=none,texmf]{atlasdoc}
 
\usepackage[subfigure]{atlaspackage}
\usepackage{atlasbiblatex}

\usepackage[BSM,jetetmiss,hepparticle]{atlasphysics}
\usepackage{atlasbsm}
\usepackage{atlasmisc}
\clearpage
\def\nnpdftwo{\texttt{NNPDF2.3lo}\xspace}

\def\evtgen{\textsc{EvtGen}\xspace}
\def\a14{\texttt{A14}\xspace}

\def\hdamp{\ensuremath{h_{\mathrm{damp}}}\xspace}
\def\ttbar{\ensuremath{t\bar{t}}\xspace}

\def\hdamp{\ensuremath{h_{\mathrm{damp}}}\xspace}

\clearpage
\usepackage{adjustbox}
\usepackage{array,makecell}
\usepackage{colortbl}
\newcolumntype{P}[1]{>{\centering\arraybackslash}p{#1}}
\newcolumntype{L}[1]{>{\raggedright\let\newline\\\arraybackslash\hspace{0pt}}m{#1}}
\newcolumntype{R}[1]{>{\raggedleft\let\newline\\\arraybackslash\hspace{0pt}}m{#1}}
\usepackage{diagbox}
\usepackage{fancyvrb}
\usepackage{longtable}
\usepackage{multirow}
\usepackage{rotating}
\usepackage{slashed}
\usepackage{soul}
\usepackage{xcolor}
\usepackage{pdflscape}
 
\graphicspath{{logos/}{figures/}}
 
\usepackage{ANA-SUSY-2019-09-PAPER-defs}
\usepackage{moremacros}

\AtlasTitle{Search for chargino--neutralino pair production in final states with three leptons and missing transverse momentum
in $\sqrt{s} = 13$~TeV $pp$ collisions with the ATLAS detector}
 
\PreprintIdNumber{CERN-EP-2021-059}
 
\AtlasAbstract{
A search for chargino--neutralino pair production in three-lepton final states with missing transverse momentum is presented.
The study is based on a dataset of $\sqrt{s} = 13$~TeV $pp$ collisions recorded with the ATLAS detector at
the LHC, corresponding to an integrated luminosity of 139~fb$^{-1}$.
No significant excess relative to the Standard Model predictions is found in data.
The results are interpreted in simplified models of supersymmetry, and statistically combined with results from a previous ATLAS search for compressed spectra in two-lepton final states.
Various scenarios
for the production and decay of charginos ($\tilde\chi^\pm_1$) and neutralinos ($\tilde\chi^0_2$) are considered.
For pure higgsino $\tilde\chi^\pm_1\tilde\chi^0_2$ pair-production scenarios,
exclusion limits at 95\% confidence level are set on $\tilde\chi^0_2$ masses up to 210~GeV.
Limits are also set for pure wino $\tilde\chi^\pm_1\tilde\chi^0_2$ pair production, on $\tilde\chi^0_2$ masses up to 640~GeV~for decays via on-shell $W$ and $Z$ bosons,
up to 300~GeV~for decays via off-shell $W$ and $Z$ bosons, and up to 190~GeV~for decays via $W$ and Standard Model Higgs bosons.
}
 
\author{The ATLAS Collaboration}
 
\AtlasRefCode{SUSY-2019-09}
\AtlasJournal{EPJC}
 
\hypersetup{pdftitle={ATLAS document},pdfauthor={The ATLAS Collaboration}}
 
\makeatletter
\newcommand{\thesize}{\f@size pt}
\makeatother
\begin{document}
 
\maketitle
 
% The next lines are included from the .//sections/introduction.tex input file
\section{Introduction}
\label{sec:intro}
 
{
\begin{DIFnomarkup}\enlargethispage*{1\baselineskip}\end{DIFnomarkup}
Supersymmetry (SUSY)~\cite{Golfand:1971iw,Volkov:1973ix,Wess:1974tw,Wess:1974jb,Ferrara:1974pu,Salam:1974ig}
postulates a symmetry between bosons and fermions, and predicts the existence of new partners for each Standard Model (SM) particle.
This extension offers a solution to the hierarchy problem~\cite{Girardello:1981wz,Sakai:1981gr,Dimopoulos:1981yj,Ibanez:1981yh,Dimopoulos:1981zb}
and provides a candidate for dark matter as the lightest supersymmetric particle (LSP),
which will be stable in the case of conserved $R$-parity~\cite{Farrar:1978xj}.

This paper describes a search for direct production of charginos and neutralinos, mixtures of
the SUSY partners of the electroweak gauge and Higgs ($h$) bosons, decaying to three charged leptons, and significant missing transverse momentum (\ptmissvec, of magnitude \met).
The search uses the full \RunTwo~dataset of proton--proton collisions recorded between 2015 and 2018 with the ATLAS detector at the CERN Large Hadron Collider (LHC).
Protons were collided at a centre-of-mass energy $\sqrt{s}$ of $13~\TeV$ and the dataset corresponds to an integrated luminosity of 139~\ifb~\cite{ATLAS-CONF-2019-021}.
Similar searches at the LHC have been reported by the ATLAS~\cite{SUSY-2016-24,SUSY-2017-03,SUSY-2018-06,SUSY-2013-12,SUSY-2018-16,SUSY-2017-01,SUSY-2018-23}
and CMS collaborations~\cite{CMS-SUS-17-004,CMS-SUS-16-034,CMS-SUS-17-007,CMS-SUS-16-048,CMS-SUS-16-039,CMS-SUS-16-043,CMS-SUS-16-045}.

Previous results are extended by analysing the full ATLAS \RunTwo~dataset,
improving the signal selection strategies --~particularly for intermediately compressed mass spectra,
and exploiting improved particle reconstruction performance.
Significant gains in lepton identification and isolation performance follow from
updates in the electron reconstruction as well as from the use of a novel multivariate discriminant~\cite{HIGG-2017-02}.
Furthermore, the new results are statistically combined with a previous ATLAS search~\cite{SUSY-2018-16}
targeting compressed mass spectra and two-lepton final states.
Finally, the paper reports updated results for a previous ATLAS search which observed excesses of three-lepton events in the partial, 36~\ifb, \RunTwo~dataset~\cite{SUSY-2017-03}.
The original analysis using the \RJR~(RJR) technique~\cite{Jackson:2016mfb,Jackson:2017gcy}
is repeated using the full \RunTwo~dataset, and no significant excesses relative to the SM expectation are observed.
A related follow-up search emulating the RJR technique with conventional laboratory-frame variables, also using the full \RunTwo~dataset,
was published in Ref.~\cite{SUSY-2018-06}.
The updated RJR results are not included in the combination with the new results,
as they are not statistically independent and not competitive with the results of the new search optimised for the full \RunTwo dataset.

Section~\ref{sec:scenario} introduces the target SUSY scenarios,
while a brief overview of the ATLAS detector is presented in Section~\ref{sec:detector},
followed by a description of the dataset and Monte Carlo simulation in Section~\ref{sec:datamc}.
After a discussion of the event reconstruction and physics objects used in the analysis in Section~\ref{sec:objdef},
Section~\ref{sec:analysisstrategy} covers the general analysis strategy,
including the definition of signal regions, background estimation techniques, and systematic uncertainties.
This is followed by Section~\ref{sec:onshellAndWh}, with details specific to the \onShell selection and the \Wh selection, and Section~\ref{sec:offshell}, with details specific to the \offShell selection.
Results are presented in Section~\ref{sec:results}, together with the interpretation in the context of relevant SUSY scenarios.
Section~\ref{sec:rj} reports the follow-up RJR analysis,
and finally Section~\ref{sec:conclusion} summarises the main conclusions.

\begin{DIFnomarkup}
\vskip 1em\section{Target scenarios}
\label{sec:scenario}
\end{DIFnomarkup}
The bino, the winos, and the higgsinos are respectively the superpartners of the $U(1)_Y$ and $SU(2)_L$ gauge fields, and the Higgs field.
In the minimal supersymmetric extension of the SM (MSSM)~\cite{Fayet:1976et,Fayet:1977yc},
$M_1$, $M_2$, and $\mu$ are the mass parameters for the bino, wino, and higgsino states, respectively.
Through mixing of the superpartners, chargino ($\widetilde{\chi}^\pm_{1,2}$) and neutralino ($\widetilde{\chi}^{0}_{1,2,3,4}$) mass eigenstates are formed.
These are collectively referred to as electroweakinos, and the subscripts indicate increasing electroweakino mass.
If the \textninoone is stable, e.g.\ as the lightest supersymmetric particle (LSP) and with $R$-parity conservation assumed,
it is a viable dark-matter candidate~\cite{Goldberg:1983nd,Ellis:1983ew}.
}
 
\begin{figure}[t!]
\centering
\begin{DIFnomarkup}
\includegraphics[width=0.28\columnwidth]{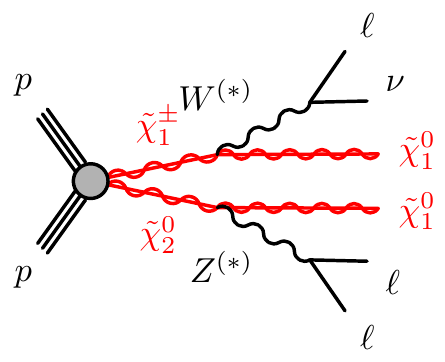}\hspace{0.5cm} 
\includegraphics[width=0.28\columnwidth]{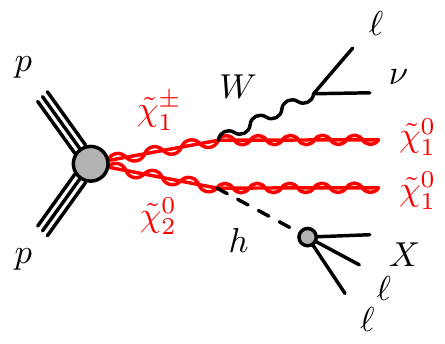}\vskip -0.3em
\end{DIFnomarkup}
\caption{Diagrams of the targeted simplified models: \chinoonepmninotwo pair production with subsequent decays into two \ninoone,
via leptonically decaying $W$, $Z$ and SM Higgs bosons, three leptons and a neutrino.
Diagrams are shown for (left) intermediate \WZ(\WZStar)
as well as (right) intermediate \Wh,
with the Higgs boson decaying indirectly into leptons+$X$ (where $X$ denotes additional decay products) via \WW, \ZZ, or $\tau\tau$.
}
\label{fig:intro:feynman}
\end{figure}

{
\begin{DIFnomarkup}\pagebreak\enlargethispage*{2\baselineskip}\end{DIFnomarkup}
Two physics scenarios are considered in this search.
In the first scenario, referred to as the `wino/bino scenario',
mass parameters $|M_1| < |M_2| \ll |\mu|$ are assumed such that the produced electroweakinos have a wino and/or bino nature,
with the \textchinoonepm and \textninotwo being wino dominated,
and the \textninoone LSP being bino dominated.
Such a hierarchy is typically predicted by
either a class of models in the framework of gaugino mass unification at the GUT scale (including mSUGRA~\cite{Chamseddine:1982jx,Barbieri:1982eh} and cMSSM~\cite{Kane:1993td}),
or a MSSM parameter space where the discrepancy between
the measured muon anomalous magnetic moment~\cite{Albahri:2021ixb},
and its SM predictions~\cite{Aoyama:2020ynm}
can be explained~\cite{moroi1996muon,feng2000supernatural,endo2020muon}.
When the mass-splitting between \textchinoonepm and \textninoone is 15--30~\GeV,
this hierarchy is also motivated by the fact that the LSP can naturally be
a thermal-relic dark-matter candidate that was depleted in the early universe
through co-annihilation processes to match the observed dark-matter density~\cite{Griest:1990kh,Edsjo:1997bg,duan2018probing}.
These models are poorly constrained by dark-matter direct-detection experiments, and collider searches constitute the only direct probe for $|\mu| > 800~\GeV$~\cite{Profumo:2017ntc}.
 
The second scenario, referred to as the `higgsino scenario', considers a triplet of higgsino-like states (\textchinoonepm, \textninotwo, \textninoone) to be the lightest SUSY particles.
This type of scenario is motivated by naturalness arguments~\cite{Barbieri:1987fn,deCarlos:1993yy},
which suggest that $|\mu|$ should be near the weak scale~\cite{Barbieri:2009ev,Baer:2011ec,Papucci:2011wy,Baer:2012up},
while $M_1$ and/or $M_2$ can be larger.
The mass-splittings between the light higgsino states are determined by the magnitude of $M_1$ or $M_2$ relative to $|\mu|$.
For the higgsino scenario this paper considers the regime where
the mass-splitting between \textninotwo and \textninoone is about 5--60~\GeV,
corresponding to cases where the wino and bino states are moderately decoupled ($M_1,M_2 > 0.5~\TeV$).
 
Simplified SUSY models~\cite{Alwall:2008ve,Alwall:2008ag,Alves:2011wf} for the two scenarios are considered for optimisation of the selections and interpretation of the results.
For the wino/bino scenario, the \textchinoonepm and \textninotwo are assumed to be mass degenerate and purely wino, while the \textninoone is purely bino.
The product of the two signed neutralino eigenmass parameters \mNN can be either positive or negative
\footnote{
The mixing matrix used to diagonalise the neutral electroweakino states can be complex,
even in the absence of CP violation,
but can be made real at the cost of introducing negative mass eigenstates.
The sign will affect the couplings and thus the distributions in the decay under consideration.
For additional discussion of this, see Ref.~\cite{Fuks:2017rio} and Appendix A of Ref.~\cite{Gunion:1984yn}.
},
and the two cases are referred to as the wino/bino `{(+)}' or `{($-$)}' scenario, respectively.
For the higgsino scenario, the \textchinoonepm, \textninotwo and \textninoone are purely higgsino states,
and the mass of the \textchinoonepm is assumed to be exactly the mean of the \textninoone and \textninotwo masses.
In both scenarios, all other SUSY particles are assumed to be heavier, such that they do not affect the production and decay of the \textchinoonepm and \textninotwo.

The search targets direct pair production of the lightest chargino and the next-to-lightest neutralino, \textchinoonepmninotwo,
decaying into a pair of \textninoone LSPs
via an intermediate state with a $W$ boson and a $Z$ boson (\WZ mediated), or a $W$ boson and a SM Higgs boson (\Wh mediated).
Final states with three light-flavour leptons (electrons or muons, referred to as `leptons' in the rest of this paper) are explored.
One lepton originates from a leptonic decay of a $W$ boson, and two leptons come from the direct decay of a $Z$ boson or the indirect decay of a Higgs boson.
The signatures are also characterised by the presence of \met originating from the LSPs,
and this \met component is enhanced when hadronic initial-state radiation (ISR) is present,
due to recoil between the \textchinoonepmninotwo system and the jets.
 
The following three simplified model scenarios of \textchinoonepmninotwo pair production, as illustrated in Figure~\ref{fig:intro:feynman}, are considered with dedicated selections:
\begin{DIFnomarkup}
\begin{itemize}\setlength{\itemindent}{-0.2cm}
\item \parbox{4.15cm}{{\bfseries \OnShell} selection:} \parbox[t]{10.2cm}{$\ninotwo \rightarrow Z\ninoone$ with 100\% branching ratio, where $\dmNN \gtrsim \mZ$, for the wino/bino {(+)} scenario.}
\item \parbox{4.15cm}{{\bfseries \OffShell} selection:} \parbox[t]{10.2cm}{$\ninotwo \rightarrow \ZBstar\ninoone$ with 100\% branching ratio, where $\dmNN < \mZ$, for the wino/bino {(+)}, the wino/bino {($-$)}, and the higgsino scenarios.}
\item \parbox{4.15cm}{{\bfseries \Wh} selection:} \parbox[t]{10.2cm}{$\ninotwo \rightarrow h\ninoone$ with 100\% branching ratio, where $\dmNN > \mh$, for the wino/bino {(+)} scenario.}
\end{itemize}
\end{DIFnomarkup}
}
 
A 100\% branching ratio is assumed for $\chinoonepm \rightarrow W^{(*)}\ninoone$ for all models.
Unless otherwise indicated, mass splitting \dm refers to \dmNN~in the rest of this paper.
For the considered \Wh-mediated scenarios,
the Higgs boson has SM properties and branching fractions;
and three-lepton final states are expected when the Higgs boson
decays into \WW, \ZZ, or $\tau\tau$, and each $W$ boson, $Z$ boson or $\tau$-lepton decays leptonically.
 
\begin{DIFnomarkup}
\end{DIFnomarkup}

For \textchinoonepmninotwo pair production with decays via \WZ to $3\ell$ final states,
in the wino/bino (+) scenario,
limits were previously set at the LHC for \textchinoonepm/\textninotwo masses
up to 500~\GeV\ for massless \textninoone,
up to 200~\GeV\ for $\Dm \sim \mZ$,
and up to 240~\GeV\ for $50~\GeV < \Dm < \mZ$~\cite{CMS-SUS-17-004}.
Limits for mass splittings $\Dm < 50~\GeV$ were set in $2\ell$ final states
for \textchinoonepm/\textninotwo masses
up to 250~\GeV~\cite{SUSY-2018-16}.
For decays via \Wh to $3\ell$ final states (including hadronically decaying $\tau$-leptons),
limits reached 150~\GeV\ for massless \textninoone,
and as high as 145~\GeV\ for a \textninoone mass of 20~\GeV~\cite{SUSY-2013-12}.

For the higgsino scenario, the most stringent limits for $5~\GeV < \dm < 55~\GeV$
were set by ATLAS using $2\ell$ final states~\cite{SUSY-2018-16}
where \textninotwo masses up to 130--190~\GeV\ are excluded depending on $\dm$.
For $\dm > 55~\GeV$ the best limits were reported by LEP~\cite{LEPlimits,Heister:2001nk,Heister:2002mn,Heister:2002jca,Heister:2003zk,Abdallah:2003xe,Abdallah:2003xe,Acciarri:2000wy,Achard:2003ge,Abbiendi:2003ji,Abbiendi:2002vz},
excluding \textchinoonepm masses up to $103.5~\GeV$.
 
% End of text imported from the .//sections/introduction.tex input file
% The next lines are included from the .//sections/detector.tex input file
\section{ATLAS detector}
\label{sec:detector}
{
\newcommand{\AtlasCoordFootnote}{
ATLAS uses a right-handed coordinate system with its origin at the nominal interaction point (IP)
in the centre of the detector and the $z$-axis along the beam pipe.
The $x$-axis points from the IP to the centre of the LHC ring,
and the $y$-axis points upwards.
Cylindrical coordinates $(r,\phi)$ are used in the transverse plane,
$\phi$ being the azimuthal angle around the $z$-axis.
Pseudorapidity is defined in terms of the polar angle $\theta$ as $\eta = -\ln \tan(\theta/2)$.
Angular distance is measured in units of $\smash{\Delta R \equiv \sqrt{(\Delta\eta)\raisebox{-0.2em}{\ensuremath{^{2}}} + (\Delta\phi)\raisebox{-0.2em}{\ensuremath{^{2}}}}}$.
Rapidity is defined by $y = \smash{\frac{1}{2}}\ln[(E+p_z)/(E-p_z)]$,
where $E$ is the energy and $p_z$ is the longitudinal component of the momentum
along the beam direction. }
 
The ATLAS detector~\cite{PERF-2007-01} is a general-purpose particle detector with almost $4\pi$
solid angle coverage around the interaction point.\footnote{\AtlasCoordFootnote}
It consists of an inner tracking system surrounded by
a superconducting solenoid, sampling electromagnetic and hadronic calorimeters, 
and a muon spectrometer encompassing superconducting toroidal magnets. 
 
The inner detector (ID) reconstructs charged-particle tracks in the pseudorapidity range $|\eta| < 2.5$,
using silicon pixel and microstrip subsystems followed by a transition radiation tracker.
For $\sqrts = 13$~\TeV\ data-taking an additional innermost layer, the insertable B-layer~\cite{ATLAS-TDR-2010-19,PIX-2018-001},
was added to the pixel tracker to improve tracking performance
and flavour identification of quark-initiated jets. The ID is immersed in a \SI{2}{T} axial
magnetic field provided by the surrounding thin, superconducting solenoid.
 
Beyond the ID a high-granularity lead/liquid-argon (LAr) electromagnetic sampling calorimeter (ECAL) and
a steel/scintillator-tile hadronic sampling calorimeter cover $|\eta| < 3.2$ and $|\eta| < 1.7$ respectively.
In the forward regions a copper/LAr endcap calorimeter extends the hadronic coverage to $1.7 < |\eta| < 3.2$,
while copper/LAr and tungsten/LAr forward calorimeters are used for
electromagnetic and hadronic measurements in the  $3.1 < |\eta| < 4.9$ region.
The muon spectrometer (MS) surrounds the calorimeters and comprises three layers of trigger and
high-precision tracking chambers spanning $|\eta| < 2.4$ and $|\eta| < 2.7$, respectively.
A magnetic field is provided by a system of three superconducting air-core toroidal magnets with eight coils each.
 
Events of interest are selected using a two-level trigger system~\cite{TRIG-2016-01} consisting of
a custom hardware-based first-level (L1) trigger followed by a software-based high-level trigger (HLT).
The L1 trigger accepts events from the \SI{40}{\MHz} bunch crossings at a rate below \SI{100}{\kHz},
which the high-level trigger reduces in order to record events to disk at about \SI{1}{\kHz}.
% End of text imported from the .//sections/detector.tex input file
% The next lines are included from the .//sections/datamc.tex input file
\begin{DIFnomarkup}
\section{Data and Monte Carlo simulated event samples}
\label{sec:datamc}
\end{DIFnomarkup}
{
This analysis exploits the full \RunTwo~$\sqrt{s}=13~\TeV$ $pp$ dataset recorded by the ATLAS experiment
during stable beam conditions between 2015 and 2018.
The LHC collided protons with bunch-crossing intervals of \SI{25}{ns}, and
the average number of interactions per crossing in data was $\langle\mu\rangle = 34$.
After applying beam, detector and data-quality requirements~\cite{DAPR-2018-01}, the dataset corresponds to a total integrated luminosity of \totallumi~\ifb~\cite{ATLAS-CONF-2019-021},
with an uncertainty in the integrated luminosity of 1.7\%,
obtained using the LUCID-2 detector~\cite{LUCID2} for the primary luminosity measurements.

The expected contributions of SM processes and $\chinoonepmninotwo$ SUSY signals are estimated using Monte Carlo (MC) simulation.
The MC samples are used in the optimisation of event selection criteria,
as well as for yield prediction and the estimation of systematic uncertainties in the yield prediction.
The yield prediction for the dominant \WZ background is improved by
extracting normalisation factors from data in dedicated control regions,
as discussed in Section~\ref{ssec:analysisstrategy:bkgs}.
The background contribution from events with one or more misidentified or non-prompt leptons
is estimated using a data-driven method also outlined in Section~\ref{ssec:analysisstrategy:bkgs}.
For all other processes, the MC-predicted yields are used directly.
The samples are produced including an ATLAS detector simulation~\cite{SOFT-2010-01} based on \textsc{Geant4}~\cite{Agostinelli:2002hh},
or a faster simulation using a parameterised calorimeter response~\cite{ATL-PHYS-PUB-2010-013} and \textsc{Geant4} for all other detector systems.
Simulated events are reconstructed in the same way as data events.
Details of the MC simulation,
including the generators used for the matrix element (ME) calculation and the parton shower (PS), hadronisation and underlying event (UE) modelling,
the parton distribution function (PDF) sets used in the ME and PS, the set of tuned parameter values used as the UE tune,
and the order of the cross-section calculations used for yield normalisation
are given in Table~\ref{tab:mcgen} and briefly discussed below.
 
% The next lines are included from the .//tables/mclongsplit.tex input file
\begin{table}[t!]
\begin{minipage}{\textwidth}
\caption{
Monte Carlo simulation details by physics process.
The table lists the event generators used for ME and PS calculations,
the accuracy of the ME calculation, the PDF sets and UE parameter tunes used,
and the order in $\alpha_\text{s}$ of cross-section calculations used for yield normalisation (`-' if the cross section is taken directly from MC simulation).
}
\vskip 0.2em
\label{tab:mcgen}
\centering
\adjustbox{max width=0.98\textwidth}{
\begin{tabular}{L{6.5cm}*{4}{l}}
\hline
\hline
\textbf{Process}
& \textbf{Event generator}
& \textbf{ME accuracy}
& \textbf{ME PDF set}
& \textbf{Cross-section}
\\
&
&
&
& \textbf{normalisation}
\\
\hline 
\tabvspace
\smash{\chinoonepmninotwo}      & \MADGRAPH[2.6]~\cite{Alwall:2014hca}      & 0,1,2j@LO\ \ \footnotemark[3]     & \newnnpdftwo~\cite{Ball:2012cx} &  NLO+NLL~\cite{Beenakker:1999xh,Debove:2010kf,Fuks:2012qx,Fuks:2013vua,Fiaschi:2018hgm,Borschensky:2014cia}\ \footnotemark[3] \\
Diboson~\cite{ATL-PHYS-PUB-2017-005}             & \SHERPA[2.2.2]~\cite{Bothmann:2019yzt}  & 0, 1j@NLO + 2,3j@LO   & \newnnpdfnlo~\cite{Ball:2014uwa}  & - \\
Triboson~\cite{ATL-PHYS-PUB-2017-005}            & \SHERPA[2.2.2]          & 0j@NLO + 1,2j@LO       & \newnnpdfnlo & - \\
Triboson (alternative)~\cite{ATL-PHYS-PUB-2017-005}      & \SHERPA[2.2.1]          & 0,1j@LO                & \newnnpdftwo & - \\
$Z$+jets~\cite{ATL-PHYS-PUB-2017-006}              & \SHERPA[2.2.1]          & 0,1,2j@NLO + 3,4j@LO   & \newnnpdfnlo & NNLO~\cite{Anastasiou:2003ds}\ \footnotemark[3] \\
\ttbar~\cite{Frixione:2007nw}                                & \POWHEGBOX[2]~\cite{Nason:2004rx, Frixione:2007vw,Alioli:2010xd}       & NLO   & \newnnpdfnlo & NNLO+NNLL~\cite{Beneke:2011mq,Cacciari:2011hy,Baernreuther:2012ws,Czakon:2012zr,Czakon:2012pz,Czakon:2013goa,Czakon:2011xx}\ \footnotemark[3] \\
\Pqt{}\PW{}~\cite{Re:2010bp}                           & \POWHEGBOX[2]       & NLO   & \newnnpdfnlo & NLO+NNLL~\cite{Kidonakis:2010ux,Kidonakis:2013zqa} \\
single-\Pqt{} (t-channel~\cite{Frederix:2012dh}, s-channel~\cite{Alioli:2009je})  & \POWHEGBOX[2]       & NLO  & \newnnpdfnlo & NLO~\cite{Aliev:2010zk,Kant:2014oha} \\
\ttbar{}$h$~\cite{Hartanto:2015uka}                           & \POWHEGBOX[2]       & NLO   & \newnnpdfnlo  & NLO~\cite{deFlorian:2016spz} \\ 
\ttV, \Pqt{}\PZ{}, \Pqt{}\WZ                                  & \MGNLO[2.3]         & NLO  & \newnnpdfnlo  & - \\
$\ttbar\ell\ell$ ($t\ra\PW{}\Pqb{}+(\gamma^{*}\!/Z\ra\ell\ell)$)~\cite{Quintero:2014lqa}
& \MGNLO[2.3]         & LO   & \newnnpdftwo  & - \\
\ttbar$VV$, 3-top, 4-top                                      & \MGNLO[2.2]         & LO   & \newnnpdftwo  & - \\
Higgs (ggF)                   & \POWHEGBOX[2]       & NNLO+NNLL   & \newnnpdfnlo   & NNNLO+NLO(EWK)~\cite{deFlorian:2016spz,Anastasiou:2016cez,Anastasiou:2015ema,Dulat:2018rbf,Aglietti:2004nj,Actis:2008ug,Bonetti:2018ukf}\ \footnotemark[3] \\ 
Higgs (VBF)                   & \POWHEGBOX[2]       & NLO+NNLL    & \newnnpdfnlo   & NNLO+NLO(EWK)~\cite{deFlorian:2016spz,Ciccolini:2007jr,Ciccolini:2007ec,Bolzoni:2010xr} \\ 
Higgs ($Vh$)                    & \POWHEGBOX[2]       & NLO         & \newnnpdfnlo   & NNLO+NLO(EWK)~\cite{deFlorian:2016spz} \\
\hline\hline
\textbf{Process}
& \textbf{PS and}
& \textbf{PS PDF set}
& \textbf{UE tune}
&
\\
& \textbf{hadronisation}
&
&
&
\\
\hline 
\tabvspace
\smash{\chinoonepmninotwo}      & \PYTHIA[8.2]~\cite{Sjostrand:2014zea}     & \newnnpdftwo & A14~\cite{ATL-PHYS-PUB-2014-021}  & \\
Diboson, triboson, $Z$+jets      & \SHERPA[2.2.2]   & default \SHERPA~\cite{Schumann:2007mg} & default \SHERPA & \\
Triboson (alternative)      & \SHERPA[2.2.1]   & default \SHERPA & default \SHERPA & \\
\ttbar, \Pqt{}\PW{}, single-\Pqt{},  \ttbar{}$h$
& \PYTHIA[8.2]     & \newnnpdftwo & A14  & \\
\ttV, \Pqt{}\PZ{}, \Pqt{}\WZ, $\ttbar\ell\ell$
& \PYTHIA[8.2]    & \newnnpdftwo    & A14 \\
\ttbar$VV$, 3-top, 4-top                & \PYTHIA[8.1]    & \newnnpdftwo    & A14 \\
Higgs (ggF, VBF, $Vh$)          & \PYTHIA[8.2]   & \newcteq~\cite{Pumplin:2002vw}   & AZNLO~\cite{STDM-2012-23}  & \\
\hline
\hline 
\end{tabular}
}
\renewcommand{\thempfootnote}{\arabic{mpfootnote}}
\footnotetext[3]{{\fontsize{8.8pt}{\baselineskip}\selectfont Abbreviations used: jet (j), leading order (LO), next-to-leading order (NLO), next-to-next-to-leading order (NNLO), next-to-\linebreak
next-to-next-to-leading order (NNNLO), next-to-leading-log (NLL), next-to-next-to-leading-log (NNLL), electroweak (EWK).}}
\addtocounter{footnote}{1}
\end{minipage}
\end{table}

% End of text imported from the .//tables/mclongsplit.tex input file

The SUSY $\chinoonepmninotwo\rightarrow \WZ/\Wh\rightarrow 3\ell$ signal samples were generated
from leading-order (LO) matrix elements with up to two additional partons
using \MADGRAPH[2.6] and \PYTHIA[8.2],
for both the wino/bino and the higgsino scenarios.
\MADSPIN~\cite{Artoisenet:2012st} was used to model off-shell $\WZ$ decays.
The ME--PS matching was done using the CKKW-L prescription~\cite{Lonnblad:2001iq,Lonnblad:2011xx},
with the matching scale set to one quarter of the \textchinoonepm/\textninotwo mass.
Samples were generated
for \textchinoonepm/\textninotwo masses between 100~\GeV\ and 850~\GeV, and mass splittings $\Dm$ between 5~\GeV\ and 850~\GeV.
Only
\textchinoonepm/\textninotwo decays via bosons, which in turn decay leptonically via SM branching fractions, are considered.
For the \Wh samples, only Higgs boson decays via $WW$, $ZZ$ and $\tau\tau$ were generated,
with cross section times branching fractions corrected to match the SM Higgs branching fractions~\cite{deFlorian:2016spz}. 
The generated signal events are required to have at least two leptons for the on-shell $\WZ$ samples,
and at least three leptons for the off-shell $\WZ$ samples and the $\Wh$ samples;
hadronically decaying $\tau$-leptons are not considered in the requirement.
 
The only difference between the two wino/bino scenarios (positive or negative \mNN)
is the mass lineshape of the $Z$ boson from the \textninotwo decay,
particularly when $\dm<\mZ$ and the $Z$ boson is off-shell.\footnote{See also Section~3 of Ref.~\cite{SUSY-2018-16}}
The samples were generated for the (+) scenario and a reweighting in $m_{Z^{(*)}}$, based on an analytic function presented in Ref.~\cite{DeSanctis:2007yoa}, was used to simulate the ($-$) scenario.
 
Inclusive production cross sections are computed at next-to-leading order (NLO) plus next-to-leading-log (NLL) precision~\cite{Beenakker:1999xh,Debove:2010kf,Fuks:2012qx,Fuks:2013vua,Fiaschi:2018hgm,Borschensky:2014cia}.
For wino production the computation is performed
in the limit of mass-degenerate \textchinoonepm and \textninotwo, and with light \textninoone,
while for higgsino production a partially degenerate case is considered,
with the \textchinoonepm mass equal to the mean of the \textninoone and \textninotwo masses;
all the other supersymmetric particles (sparticles) are assumed to be heavy and decoupled.
For production at a centre-of-mass energy of $\sqrt{s} = 13~\TeV$, the wino (higgsino) \textchinoonepmninotwo cross section ranges between
$22.67 \pm 0.97$~pb
($12.22 \pm 0.26$~pb, $\dm=80$~\GeV)
and
$3.42 \pm 0.41$~fb
($87.2 \pm 3.2$~fb, $\dm=20$~\GeV)
for \textninotwo masses between 100~\GeV\ and 850~(320)~\GeV,
with the higgsino cross section depending additionally on \dm.
 
Diboson, triboson and $Z$+jets processes were simulated with the \SHERPA[2.2] generator.
ME--PS matching and merging is based on Catani--Seymour dipole factorisation~\cite{Gleisberg:2008fv,Schumann:2007mg,Hoeche:2011fd},
using improved CKKW matching~\cite{Catani:2001cc,Hoeche:2009rj} extended to NLO accuracy using the MEPS@NLO prescription~\cite{Hoeche:2011fd,Hoeche:2012yf,Catani:2001cc,Hoeche:2009rj},
and including NLO virtual QCD corrections for the ME~\cite{Cascioli:2011va,Denner:2016kdg}. 
The diboson samples cover dilepton masses down to \begin{DIFnomarkup}\smash{$4~\GeV$ for $\ptl{1},\ptl{2} > 5~\GeV$}\end{DIFnomarkup},
and down to $\mll > 2 m_{\ell}+250~\MeV$ if \begin{DIFnomarkup}\smash{$\ptl{1} > 5~\GeV$}\end{DIFnomarkup}
and any of \begin{DIFnomarkup}\smash{$\mll>4~\GeV$}\end{DIFnomarkup}, \begin{DIFnomarkup}\smash{$\ptl{1}>20~\GeV$}\end{DIFnomarkup}, or \begin{DIFnomarkup}\smash{$\met>50~\GeV$}\end{DIFnomarkup} are satisfied.
The standard multiboson samples do not include Higgs boson production.
An alternative triboson sample
including \ofs contributions and leptonically decaying $h\rightarrow VV$ (with $V$ = $W$ or $Z$) contributions
is used in the \ofs \WZ selection,
where \WZStar decays are targeted and \ofs triboson processes are non-negligible in the estimation of the SM background;
dilepton masses down to 4~\GeV\ are considered in the sample.

The \ttbar, single-top $\Pqt{}W$, t-channel, s-channel and \tth processes were modelled using \POWHEGBOX[2] + \PYTHIA[8].
The \hdamp\ parameter\footnote{The \hdamp\ parameter is a resummation damping factor and one of the
parameters that controls the matching of \POWHEG matrix elements to the parton shower and thus effectively regulates the
high-\pt\ radiation against which the \ttbar\ system recoils.}
was set to 1.5 times the top-quark mass~\cite{ATL-PHYS-PUB-2016-020}.
The samples were generated employing the five-flavour scheme (four-flavour in case of single-top t-channel),
and a diagram removal scheme~\cite{Frixione:2008yi} was used in the case of $\Pqt{}W$
to remove interference and overlap with \ttbar\ production.
Other top-quark processes ($\ttbar V$, \Pqt{}$Z$, \Pqt{}\WZ, $\ttbar VV$, $\ttbar\ell\ell$ ($t\ra\PW{}\Pqb{}+(\gamma^{*}\!/Z\ra\ell\ell))$, 3-top and 4-top) were modelled using \MGNLO[2] + \PYTHIA[8].
Samples of Higgs boson production via gluon fusion, vector-boson fusion and associated production were generated using \POWHEGBOX[2] + \PYTHIA[8].

All background and signal samples make use of \evtgen 1.6.0 and 1.2.0~\cite{EvtGen} for the modelling of $b$- and $c$-hadrons,
except those generated using \SHERPA.
The effect of additional interactions in the same and neighbouring bunch crossings (pile-up) was included
by overlaying simulated minimum-bias interactions onto each hard-scatter process.
The simulation was done using \PYTHIA[8.2] with the A3 tune~\cite{ATL-PHYS-PUB-2016-017} and the \nnpdftwo set of PDFs,
and the samples were reweighted such that the pile-up distribution matches the one in data.

}
% End of text imported from the .//sections/datamc.tex input file
% The next lines are included from the .//sections/objectreco.tex input file
\begin{DIFnomarkup}
\vskip 1em\section{Event reconstruction and preselection}\label{sec:objdef}
\end{DIFnomarkup}
 
The strategy for event reconstruction and preselection is defined here, where a common approach has been adopted for all regions in the analysis, unless specified otherwise.
Further selection specific to individual regions is discussed in Sections~\ref{sec:analysisstrategy} to~\ref{sec:offshell}.
 
Events are chosen for the \Wh and \ons \WZ selections using dilepton triggers
and for the \ofs \WZ selection using single-lepton, dilepton and trilepton triggers~\cite{TRIG-2018-05,TRIG-2018-01}.
The \ofs \WZ selection is complemented at high \met with softer-lepton events selected using \met triggers~\cite{TRIG-2019-01}.
The lepton triggers use various \pT thresholds, depending on the lepton type, quality and multiplicity.
To ensure trigger efficiencies are well understood in the analysis phase space,
tighter quality and \pT requirements are applied to fully reconstructed signal leptons, as defined below. 
Single-electron triggers are not used, to facilitate looser signal-lepton identification criteria.
The number of leptons in the event that activate the trigger must be at least as many as the number of leptons required in the trigger,
and electrons (muons) activating the trigger must have a fully calibrated \pT above 18~\GeV\ (27.3, 14.7 or 6.5~\GeV, for increasing trigger-lepton multiplicity).
For events selected by a \met trigger, an offline requirement of $\met > 200~\GeV$ is imposed to similarly
ensure well-understood trigger efficiencies in the analysis phase space.
 
Events are required to have at least one reconstructed $pp$ interaction vertex~\cite{ATL-PHYS-PUB-2015-026,PERF-2015-01} with a minimum of two associated tracks with $\pt > 500~\MeV$.
In events with multiple vertices, the primary vertex is defined as the one with the highest $\sum\pt^2$ of associated tracks.
 
The primary objects used in this analysis are electrons, muons and jets.
To be considered, reconstructed objects must satisfy `baseline' loose identification criteria;
to be selected for the analysis regions, they must also survive a second, tighter set of `signal' identification requirements.
Additionally, a lepton `\antiID' requirement is defined, corresponding to leptons that satisfy the baseline criteria but not the signal criteria. These \antiID leptons are used in the \ZjetZgam background estimation in Section~\ref{ssec:analysisstrategy:bkgs}.
Hadronically decaying $\tau$-leptons are not considered in the analysis, and the term `lepton' always refers to electrons or muons in this document.

Electron candidates are reconstructed from three-dimensional clustered energy deposits in the electromagnetic calorimeter (ECAL),
matched to an ID track~\cite{EGAM-2018-01}.
Muon candidates are reconstructed by matching MS tracks or track segments to ID tracks~\cite{MUON-2018-03}.
Electron and muon candidates are calibrated \textit{in situ}~\cite{EGAM-2018-01,MUON-2018-03},
using $Z\ra ee$, $J/\psi\ra ee$, $Z\ra\mu\mu$ and $J/\psi\ra \mu\mu$ decays.
Baseline electrons are required to have $\pt > 4.5~\GeV$ and fall within the acceptance of the ID ($|\eta| < 2.47$).
They are further required to satisfy the calorimeter- and tracking-based `\emph{Loose and B-layer} likelihood'
identification~\cite{EGAM-2018-01}.
Baseline muons must have $\pt > 3~\GeV$ and $|\eta|<2.5$,
and satisfy \emph{Medium} identification criteria~\cite{MUON-2018-03}.
To suppress pile-up, both the baseline electrons and baseline muons are required to have a trajectory consistent with the primary vertex,
i.e.\ $|z_0\sin\theta|<0.5$~mm.\footnote{The transverse impact parameter, $d_0$, is defined as the distance of closest approach in the transverse plane between a track and the beam-line. The longitudinal impact parameter, $z_0$, corresponds to the $z$-coordinate distance between the point along the track at which the transverse impact parameter is defined and the primary vertex.}
 
Jet candidates are reconstructed from topological energy clusters in the electromagnetic and hadronic calorimeters~\cite{PERF-2014-07},
grouped using the anti-$k_t$ algorithm~\cite{Cacciari:2008gp,Fastjet} with radius parameter $R = 0.4$.
After subtracting the expected energy contribution from pile-up following the jet area technique~\cite{PERF-2014-03},
the jet energy scale (JES) and resolution (JER) are corrected to particle level using MC simulation,
and then calibrated \textit{in situ} using $Z$+jets, $\gamma$+jets and multijet events~\cite{PERF-2016-04,JETM-2018-05}.
Baseline jets must then have $\pt > 20~\GeV$, and fall within the full calorimeter acceptance ($|\eta|<4.5$).
 
Photon candidates are reconstructed from energy clusters in the ECAL
provided they have no matched track, or have one or more matched tracks consistent with photon conversion origin.
Baseline photons, while not used in the signal regions,
are included in the calculation of missing transverse momentum,
and used in SM background estimation validation.
They are required to have $\pt>25~\GeV$, fall inside the ECAL strip detector acceptance ($|\eta|<2.37$), but outside the ECAL transition region ($|\eta|\in [1.37,1.52]$).
Candidates must also satisfy \emph{Tight} identification criteria~\cite{EGAM-2018-01}.
 
Ambiguities may exist between reconstructed objects.
To prevent single detector signatures from being identified as multiple objects,
the following overlap removal procedure is applied to baseline leptons and jets.
First, all electrons sharing an ID track with a muon are discarded to remove bremsstrahlung from muons that is followed by a photon conversion.
Second, all jets separated from remaining electrons by less than $\DeltaR = 0.2$ are removed.
Also, all jets within $\DeltaR = 0.4$ of a muon
and associated with fewer than three tracks with $\pt \geq 500~\MeV$ are removed.
Finally, electrons or muons separated from surviving jets by less than $\DeltaR = 0.4$
are discarded to reject non-prompt leptons from decays of $b$- and $c$-hadrons.
 
The missing transverse momentum is defined as the negative vector sum of the transverse momenta
of all baseline objects (electrons, muons, jets, and photons) and an additional soft term~\cite{ATLAS-CONF-2018-023}.
The soft term is constructed from all tracks that pass basic quality requirements and are associated with the primary vertex,
but are not associated with any baseline object.
In this way, the \ptmissvec is adjusted for the calibration of the contributing objects,
while maintaining robustness against pile-up~\cite{PERF-2016-07}.
Additionally, an `object-based \met significance'~\cite{ATLAS-CONF-2018-038} is defined as
\begin{DIFnomarkup}$\sqrt{\scalebox{0.85}{$|\ptmissvec|^{2} / (\sigma_{\text{L}}^2 (1-\rho_{\text{LT}}^{2}))$}}$\end{DIFnomarkup}.
The \pt resolution of the contributing objects, at a given \pt and $|\eta|$,
is determined from parameterised Monte Carlo simulation which well reproduces the resolution measured in data.
The quantity $\sigma_{\textup{L}}$ denotes the \pt resolution of the system, and $\rho_{\textup{LT}}$ is a correlation factor
between the resolutions of the \pt components parallel (L) and perpendicular (T) to \ptmissvec.
The \met significance is used to discriminate events where the \met arises from undetected particles in the final state
or from events where the \met arises from poorly measured particles (and jets).
It is also useful in discriminating between signal events with large \met and e.g.\ \Zjet events with medium-to-low \met.

{\begin{DIFnomarkup}\enlargethispage*{1.2\baselineskip}\end{DIFnomarkup}
To ensure high-quality object measurement and selection purity for the analysis regions,
leptons and jets must satisfy additional tighter `signal' criteria and isolation requirements to be selected.
Signal jets are selected within $|\eta|<2.8$, and must satisfy \emph{Loose} quality criteria
to reject contamination from non-collision backgrounds or noise bursts~\cite{ATLAS-CONF-2015-029}.
In order to suppress jets originating from pile-up, signal jet candidates with $\pt < 120~\GeV$~and $|\eta| < 2.5$ (within the ID acceptance)
are further required to satisfy the \emph{Medium} working point of the track-based jet vertex tagger (JVT)~\cite{PERF-2014-03,ATLAS-CONF-2014-018}.
For jets with $|\eta|<2.5$ a multivariate discriminant
--~constructed using track impact parameters, information about displaced secondary vertices, and trajectories of $b$- and $c$-hadrons inside the jet~\cite{FTAG-2018-01}~--
is used for the identification of $b$-hadron decays, referred to as $b$-jets.
The $b$-tagging algorithm working point is chosen such that $b$-jets from simulated \ttbar events are identified
with 85\% efficiency, with rejection factors of 2.7 for charm-quark jets and 25 for light-quark and gluon jets~\cite{FTAG-2018-01}.
Signal electrons must satisfy \emph{Medium} identification criteria~\cite{EGAM-2018-01}.
All signal leptons are then required to be compatible with originating from the primary vertex;
the significance of the transverse impact parameter 
must satisfy $|d_0/\sigma(d_0)| < 5~(3)$ for electrons (muons),
where $\sigma(d_0)$ is the track-by-track estimated impact parameter resolution.
 
Isolation requirements are applied to suppress contributions
from conversions, semileptonic decays of heavy-flavour hadrons,
or hadrons and jets wrongly identified as leptons,
collectively referred as fake or non-prompt (FNP) leptons.
The criteria rely on isolation energy variables calculated as $\sum\pt$ of tracks or calo-clusters within a certain size of cone around the lepton candidate;
the energy of the lepton candidate itself is not considered in this calculation.
The isolation working points used in this analysis are based on those described in Refs.~\cite{EGAM-2018-01}~and~\cite{MUON-2018-03},
including updates to improve the performance under the increased pile-up conditions encountered during 2017 and 2018 data-taking.
The choice of isolation working points is optimised per selection region and per lepton-flavour to account for different levels of contribution from the FNP lepton background.
The \emph{Tight} 
working point is used for both electrons and muons in the \ons \WZ and \Wh selections,
while the looser working point \emph{Gradient} (\emph{Loose}) is employed for electrons (muons) in the \offShell selection to maintain a reasonable efficiency down to low \pt.
 
To further suppress FNP lepton backgrounds in the \offShell selection, a dedicated multivariate discriminant `non-prompt lepton BDT'~\cite{HIGG-2017-02}
is used to tighten the requirements on the lepton with the lowest $\pt$
(which is commonly also the most FNP-like lepton of the three),
after selecting exactly three baseline leptons in the event.
The discriminant uses eight input variables including the isolation information, combined lepton and track quantities, and
the $b$-jet likeliness calculated from the energy deposits and tracks in a cone around the lepton using the \texttt{DL1mu} or \texttt{RNNIP} algorithms~\cite{ATL-PHYS-PUB-2017-013}.
The non-prompt lepton BDT selection is designed to maintain 70\%--90\% efficiency for real leptons, for lepton \pT below 20~\GeV, with a rejection factor of 2--3 for FNP leptons passing the isolation selection.
Figure~\ref{fig:obj:lepEff} shows the combined signal lepton selection efficiency (including the reconstruction, identification, isolation, vertex association and non-prompt BDT selection) for the leptons from \textchinoonepmninotwo~signal events,
as well as the differential probability for a \ZjetZgam event to be accompanied by a FNP lepton satisfying the signal lepton selection criteria.
 
\begin{DIFnomarkup}
\begin{figure}[t!]
\centering
\includegraphics[width=0.485\textwidth]{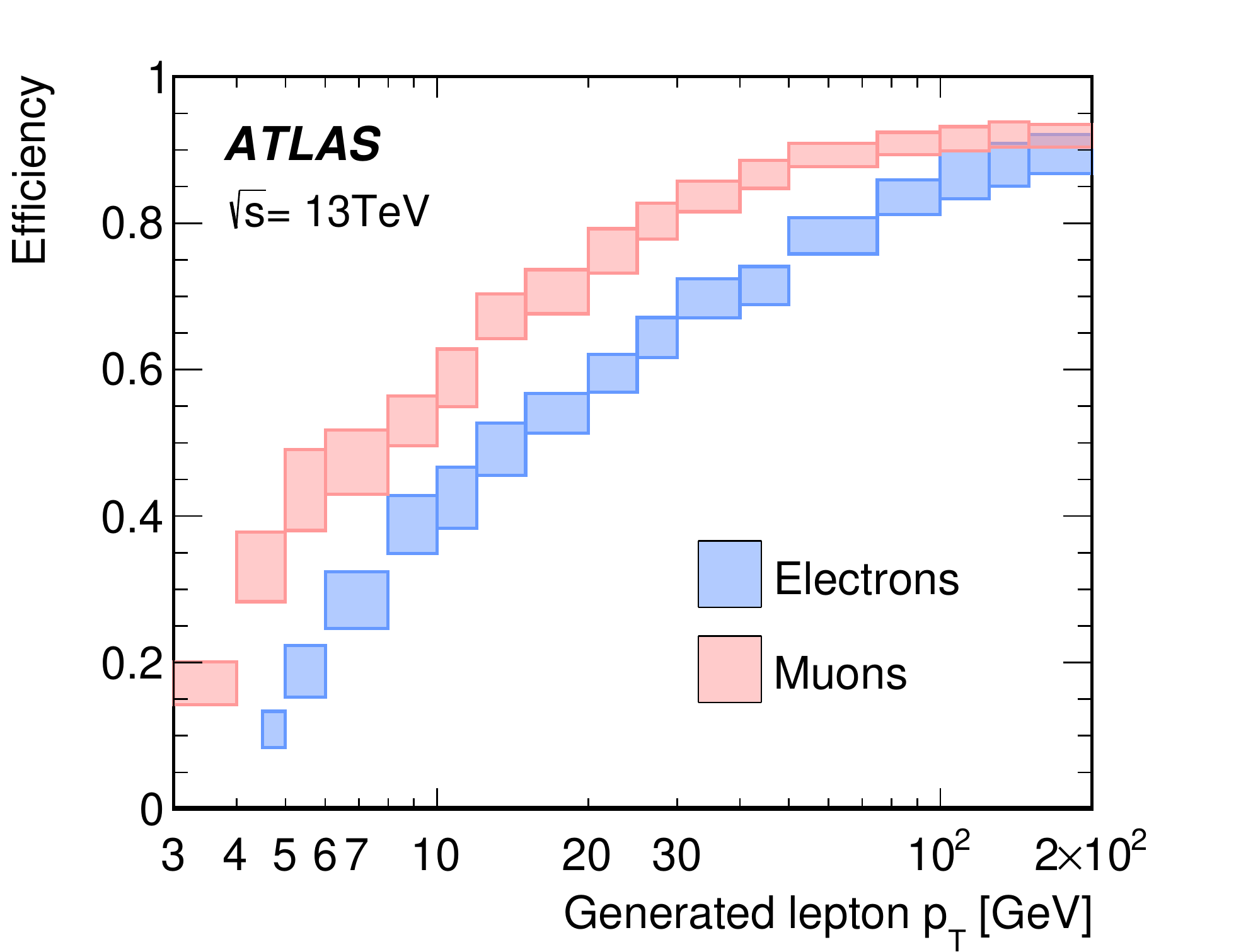}
\includegraphics[width=0.485\textwidth]{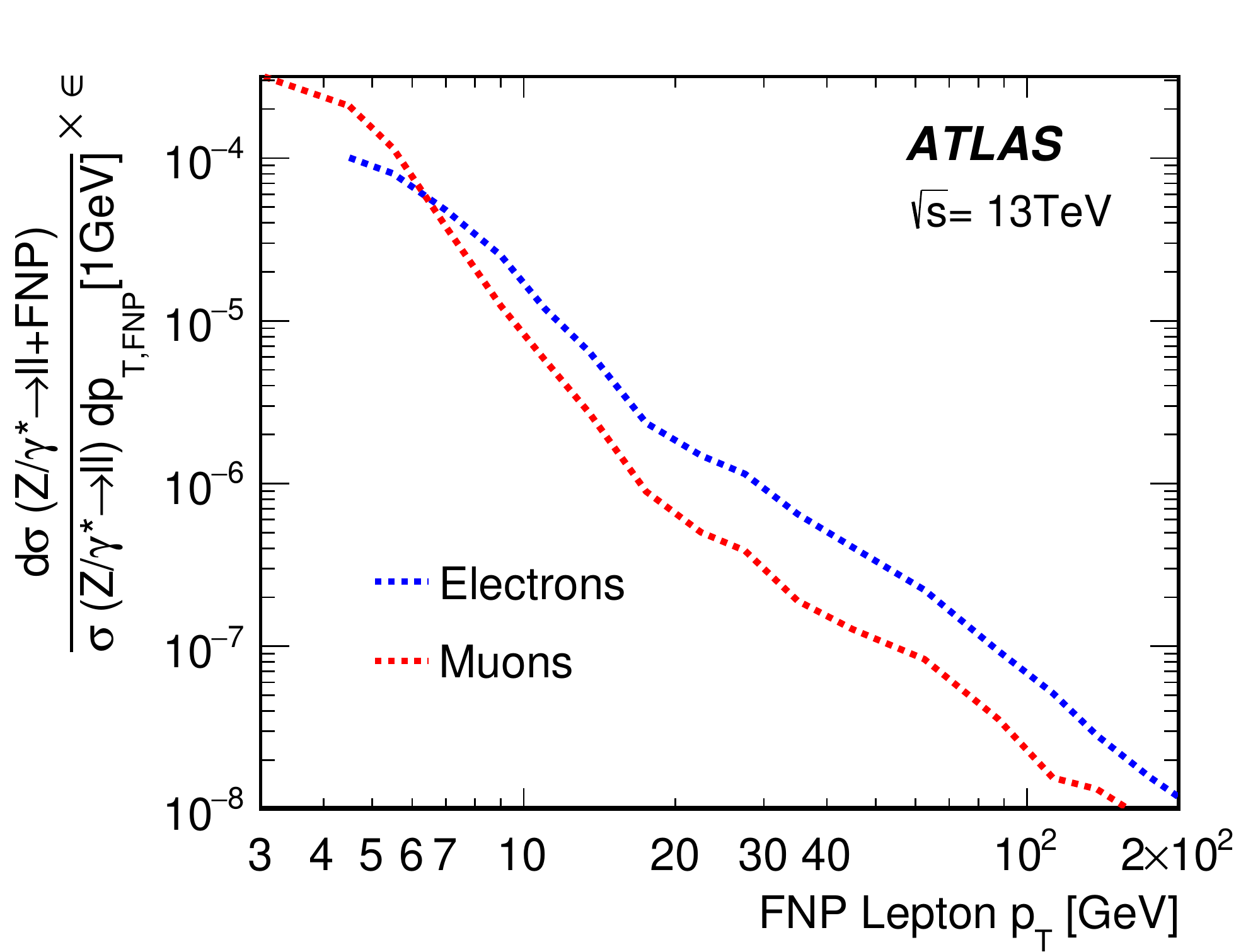}
\caption{The left panel shows the combined lepton selection efficiencies for the signal electron/muon requirements
applied to the lowest-\pt lepton after selecting three baseline leptons in the \offShell selection.
The efficiencies
are calculated using simulated samples of wino/bino (+) $\textchinoonepm/\textninotwo$ decays and shown as a function of the generated lepton \pt.
The associated uncertainties represent the range of efficiencies observed across all signal samples used for the given \pt bin.
The right panel illustrates the differential probability for a \ZjetZgam event to be accompanied by a FNP lepton satisfying the signal lepton criteria, as a function of the FNP lepton \pt.
This probability is measured using data events in a region with at least two signal leptons, with the other processes subtracted using the MC samples.
}
\label{fig:obj:lepEff}
\end{figure}
\end{DIFnomarkup}

To account for small efficiency differences between simulation and data,
simulated events are corrected with scale factors
covering lepton reconstruction, identification, isolation and trigger efficiencies,
as well as jet pile-up rejection and flavour-tagging efficiencies.
}
 
A common preselection is applied for all search regions requiring exactly three signal leptons.
Events are also required to have exactly three baseline leptons. 
This additional baseline requirement ensures orthogonality with other ATLAS SUSY analyses~\cite{SUSY-2018-16,SUSY-2018-32,SUSY-2019-08}
and facilitates statistical combinations; 
it also simplifies the FNP lepton background estimation.
Muons in the region $2.5 < |\eta| < 2.7$ are exceptionally included in this count if they satisfy all other baseline muon criteria,
in order to harmonise with the definition applied in the other analyses.

% End of text imported from the .//sections/objectreco.tex input file
% The next lines are included from the .//sections/analysisstrategy.tex input file
\newcommand{\uncs}{uncertainties\xspace}
\newcommand{\unc}{uncertainty\xspace}
\newcommand{\statuncs}{statistical uncertainties\xspace}
\newcommand{\sysuncs}{systematic uncertainties\xspace}
\newcommand{\sysunc}{systematic uncertainty\xspace}
\newcommand{\theo}{theoretical\xspace}

{\begin{DIFnomarkup}\enlargethispage*{1.5\baselineskip}\end{DIFnomarkup}
\begin{DIFnomarkup}
\vskip 1em\section{Analysis strategy}
\label{sec:analysisstrategy}
\end{DIFnomarkup}
The selections in this paper
-- while targeting different simplified model scenarios --
all consider final states with exactly three leptons, possible ISR jets, and \met.
Therefore, a common approach is used throughout most steps of the analyses.
The \ons \WZ, \ofs \WZ, and \Wh selections are optimised independently.
 
This section describes the general analysis strategy, introducing the common parts of the
search region definitions (Section~\ref{ssec:analysisstrategy:sr}),
the background estimation (Section~\ref{ssec:analysisstrategy:bkgs}),
and the uncertainty treatment (Section~\ref{ssec:analysisstrategy:systematics}).
The statistical methods used are outlined in Section~\ref{ssec:analysisstrategy:fit}.
Further details specific to either the \ons \WZ selection and the \Wh selection, or the \ofs \WZ selection,
are then discussed in dedicated Sections~\ref{sec:onshellAndWh} and \ref{sec:offshell}.

\begin{DIFnomarkup}
\vskip 1em\subsection{Search regions}
\label{ssec:analysisstrategy:sr}
\end{DIFnomarkup}
 
Event selections enriched in signal (signal regions or SRs)
are designed independently for the three targeted models,
i.e.\ for the \ons \WZ, \ofs \WZ or \Wh selections.
All the SRs are optimised to the wino/bino (+) scenario.
The SRs of the \ons \WZ selection, \SRonWZ, are optimised for \textchinoonepmninotwo signals
with \WZ-mediated decays and mass splittings near or above the $Z$-boson mass, $\dm \gtrsim \mZ$,
while the SRs of the \ofs \WZ selection, \SRoffWZ, target \WZStar-mediated decays and mass splittings $\dm < \mZ$.
The SRs of the \Wh selection, \SRWh, are optimised for \Wh-mediated decays and veto $Z$-boson candidates.
 
}

{\begin{DIFnomarkup}\enlargethispage*{1.5\baselineskip}\end{DIFnomarkup}
For SRs targeting \WZBStar-mediated scenarios, 
two leptons are assigned to the \ZBstar-boson candidate by selecting a same-flavour opposite-charge-sign (SFOS) lepton pair in the event,
and the remaining lepton is assigned to the \WBstar boson (labelled $W$ lepton or \lW).
If more than one SFOS lepton pair is present in the event, the invariant mass \mll of the SFOS lepton pairs is used to select
which pair is assigned to the \ZBstar-boson candidate.
The \ons \WZ selection selects the SFOS lepton pair with \mll nearest the $Z$-boson mass, \mllmZ,
while the \ofs \WZ selection selects the SFOS lepton pair with the smallest \mll, \minmll.
In the rest of this document, these two types of lepton assignment are referred to as
\mllmZ-based and \minmll-based lepton assignment,
and \mll refers to \mllmZ unless otherwise indicated.
\linebreak In \Wh-mediated scenarios, the opposite-sign leptons are the indirect product of the Higgs boson decay and can be of either the same or different flavour.
Two subsets of SRs are defined depending on lepton flavour composition: the \SRWhSF target events with at least one SFOS pair
(using \mllmZ-based lepton assignment), and the \SRWhDF target complementary events without a SFOS lepton pair.

For events with at least one SFOS lepton pair the transverse mass, \mt, is constructed using the $W$ lepton and the \met,
and assuming the SM \WZ event hypothesis:
\begin{DIFnomarkup}
$\mt = \scalebox{0.9}{$\sqrt{2\smash{\pt^{\raisebox{-4pt}{\scalebox{0.75}{$\ell_{\W}$}}}}\met(1-\cos{(\Delta\phi)})}$}$,
where $\Delta\phi$ is the separation in the transverse plane between the lepton and the \met.
\end{DIFnomarkup}
This exploits the difference between SM \WZ,
which has a Jacobian peak with a sharp cut-off at $\mt \sim m_W$ (the $W$-boson mass),
and the targeted signals, which have relatively flat distributions.

For the initial SR segmentation, events with at least one SFOS lepton pair are divided into three \mll slices:
below, in, and above the $Z$-boson mass window,
defined as $\mll \in [75,105]~\GeV$.
The \SRoffWZ and \SRonWZ use the first and second slice, respectively,
while the \SRWhSF use the first and third slice.
The \SRonWZ are orthogonal to the \SRoffWZ and the \SRWhSF through the \mll selection.
The \SRoffWZ and the \SRWhSF can overlap, but are never used in the same interpretation.
The \SRWhDF are orthogonal to all other SRs through lepton flavour composition.

For the final selection, a few key discriminating variables are used to further segment and refine the SRs.
The \SRonWZ and \SRWh have a shared binning strategy aside from the \mll range,
while \SRoffWZ binning focuses on \mllmin and properties of more compressed \textchinoonepmninotwo~signals.
Ultimately, 20, 31, 19, and 2 SR bins are defined for the \SRonWZ, \SRoffWZ, \SRWhSF, and \SRWhDF, respectively.
The complete definitions of these nominal SRs are further detailed per selection
in subsequent Sections~\ref{ssec:onshellAndWh:reg} (\SRonWZ and \SRWh) and~\ref{ssec:offshell:reg} (\SRoffWZ).
The bins within each subset are explicitly disjoint,
and are statistically combined when calculating the constraints on the target models.
A more detailed overview of the fit configuration is given in Section~\ref{ssec:analysisstrategy:fit}.
Additionally, discovery-oriented inclusive SRs are designed by grouping sets of adjoining nominal-SR bins
in order to facilitate quantifying the size of data excesses in a model-independent manner.
The inclusive-SR definitions are discussed in Section~\ref{ssec:results:incl}.
 
}

\begin{DIFnomarkup}
\vskip 1em\subsection{Background estimation}
\label{ssec:analysisstrategy:bkgs}
\end{DIFnomarkup}
 
The dominant SM background in most of the SRs in this analysis
is from SM \WZ events with only leptonic decays,
followed in importance by \ttbar and \ZjetZgam processes associated with at least one FNP lepton.
In \SRWhDF, SM Higgs, triboson and \ttbar production are the dominant processes.
 
A partially data-driven method is used for the estimation of the \WZ background,
which produces three real and prompt leptons.
The background is predicted using MC simulation samples and
normalised to data in dedicated control regions (CRs).
This normalisation improves the estimation in the phase space of the selections,
and constrains the systematic uncertainties.
The CRs are designed to be both orthogonal and similar to the SRs,
whilst also having little signal contamination;
this is achieved by taking the SR definitions and inverting some of the selection criteria.
Dedicated validation regions (VRs) are defined kinematically in between the CRs and SRs,
and are used to assess the quality of the background estimation and its extrapolation to the SRs.
The final estimation of the yields and uncertainties is performed with a simultaneous fit
to the CRs and SRs, as discussed in Section~\ref{ssec:analysisstrategy:fit}.
 
{\begin{DIFnomarkup}\enlargethispage*{1.5\baselineskip}\end{DIFnomarkup}
The \ttbar background is predicted using MC simulation samples and validated in VRs.
It is dominated by decays with a dileptonic final state and an additional lepton from a $b$- or $c$-hadron decay.
As the MC modelling is found to be of good quality, no additional corrections are applied to the MC events.
Rare SM processes, including multiboson and Higgs boson production, top-pair production in association with a boson, and single-top production, are estimated from MC simulation in all analysis regions.
 
The \Zjetcomplex background has two prompt leptons and one FNP lepton from jets or photons.
In the rest of this document, `\ZjetZgam' is used to refer to this set of processes.
As there are no invisible particles in these processes at tree level,
the observed \met is mostly due to mismeasured leptons and/or jets, or due to the \met soft term.
The FNP leptons originate from a mix of sources,
including light-flavour jets faking leptons,
electrons from photon conversion,
and non-prompt leptons from $b$- or $c$-hadron decays.
Such FNP leptons often arise from instrumental effects, hadronisation, and the underlying event, all of which are challenging to model reliably in simulation.
Therefore a data-driven method, referred to as the `fake-factor method'~\cite{STDM-2011-24,ATL-PHYS-PUB-2010-005}, is used to estimate the \ZjetZgam background.
The fake factor (FF) is defined as the ratio of the probability for a given lepton candidate to pass the signal lepton requirements to that to fulfil the \antiID requirements.
This is measured using data in a control region, \CRFF, designed to target \ZjetZgam events with FNP leptons whose sources are representative of those expected in the SRs.
Exactly three baseline leptons and at least one SFOS lepton pair are required in \CRFF.
The $Z$-boson candidate in the event is identified as the SFOS pair yielding the invariant mass closest to the $Z$-boson mass, and the remaining lepton is tagged as the FNP lepton candidate.
The two leptons from the $Z$-boson candidate must activate the dilepton trigger to ensure there is no selection bias from FNP leptons.
The \ZjetZgam prediction in a given region is obtained by applying the FFs to the events in its corresponding `\antiID region'.
This region is defined by the same selection criteria as used for the nominal region with three signal leptons,
except that at least one of the leptons is \antiID instead of signal.
Each event in the \antiID region is scaled by a weight based on the FF assigned to each \antiID lepton in the region.
The FFs are derived separately per lepton flavour
and are parameterised as a function of lepton $\pt$ and lepton $\eta$ or $\met$ in the event, depending on the analysis selection.
In both the FF measurement and the FF application procedure,
contributions from processes other than \ZjetZgam are subtracted using MC simulation samples.
 
While sharing a common approach,
the estimation and validation procedures for the main SM backgrounds were optimised independently for the different selections,
which each target a different primary phase-space region with different relative background composition and importance.
Details are given in Section~\ref{ssec:onshellAndWh:bkg} (\CRWZ/\VRWZ) and Section~\ref{ssec:offshell:bkg} (\CRoffWZ/\VRoffWZ).\begin{DIFnomarkup}\vskip 1em\end{DIFnomarkup}
 
\begin{DIFnomarkup}
\subsection{Systematic uncertainties}
\label{ssec:analysisstrategy:systematics}
\end{DIFnomarkup}
The analysis considers uncertainties in the predicted yields of signal or background processes due to
instrumental \sysuncs as well as \statuncs and \theo \sysuncs of the MC simulated samples.
Uncertainties are assigned to the yield in each region, except for $WZ$ processes constrained in CRs,
in which case they are assigned to the acceptance in each SR relative to that in the CR.
The uncertainty treatment is largely common to the \onShell, \Wh and \offShell selections; exceptions are discussed in Sections~\ref{ssec:onshellAndWh:bkg} (\SRonWZ and \SRWh) and \ref{ssec:offshell:bkg} (\SRoffWZ).
Relative uncertainties are illustrated in a breakdown per SR in the same sections.
}
 
{\begin{DIFnomarkup}\pagebreak\enlargethispage*{0.5\baselineskip}\end{DIFnomarkup}
The dominant instrumental uncertainties are the jet energy scale (JES) and resolution (JER).
The jet uncertainties are derived as a function of \pt\ and $\eta$ of the jet,
as well as of the pile-up conditions and the jet flavour composition of the selected jet sample.
They are determined using a combination of simulated samples and studies in data, such as measurements of the jet \pt\ balance in dijet, $Z$+jet and $\gamma$+jet events~\cite{PERF-2016-04,JETM-2018-05,PERF-2014-02}.
Another significant instrumental uncertainty is that in the modelling of \met,
evaluated by propagating the uncertainties in the energy and momentum scale of each of the objects entering the calculation,
as well as the uncertainties in the \met soft-term resolution and scale~\cite{ATLAS-CONF-2018-023}.
Other instrumental uncertainties concerning the efficiency of the trigger selection, flavour-tagging and JVT,
as well as reconstruction, identification, impact parameter selection and isolation for leptons, are found to have minor impact.
Each experimental \unc is treated as fully correlated across the analysis regions and physics processes considered.
 
For the processes estimated using the MC simulation,
the predicted yield is also affected by different sources of theoretical modelling uncertainty.
All theoretical uncertainties are treated as fully correlated across analysis regions, except those related to MC statistics.
The \uncs for the dominant background processes, \WZ, \ZZ, and \ttbar, are derived using MC simulation samples.
For the \WZ background, which is normalised to data in CRs, these uncertainties are implemented as transfer factor uncertainties
that reflect differences in the SR-to-CR or VR-to-CR ratio of yields, and therefore provide an uncertainty in the assumed shape of MC distributions across analysis regions.
The \uncs related to the choice of QCD renormalisation and factorisation scales are represented by three Gaussian nuisance parameters in the fit (see Section~\ref{ssec:analysisstrategy:fit}):
the first varies the renormalisation scale up and down,
where a one-sigma deviation represents varying that scale up or down by a factor of two,
while the factorisation scale is fixed to its nominal value;
the second varies the factorisation scale in the same way while fixing the renormalisation scale;
and the third nuisance coherently varies both the renormalisation and factorisation scales.
There is no nuisance parameter to account for anti-correlated configurations of the renormalisation and factorisation scales, as these are deemed unphysical.
For the \WZ and \ZZ samples, the \uncs due to the resummation and matching scales between ME and PS as well as the PS recoil scheme are evaluated by varying the corresponding parameters in \textsc{Sherpa}.
For \ttbar, modelling uncertainties at ME and PS level are determined by comparing the predictions of nominal and alternative generators, considering \POWHEGBOX versus \MGNLO and \PYTHIA[8] versus \HERWIG[7]~\cite{Bahr:2008pv,Bellm:2015jjp}, respectively.
Uncertainties in the \ttbar prediction due to ISR and final-state radiation (FSR) uncertainties are evaluated by varying the relevant generator parameters.
The uncertainties associated with the choice of PDF set (NNPDF~\cite{Ball:2012cx,Ball:2014uwa}) and the uncertainty in the strong coupling constant, $\alpha_\mathrm{s}$, are also considered for the major backgrounds.
Uncertainties in the cross section of 13\%, 12\%, 10\% and 20\% are applied for minor backgrounds $\ttbar W$, $\ttbar Z$, $\ttbar h$, and triboson, respectively~\cite{deFlorian:2016spz};
for all other rare top processes a conservative uncertainty of 50\% is applied.
 
The data-driven \ZjetZgam estimation is subject to the statistical uncertainty due to the limited data sample size in \CRFF or
in the anti-ID regions used when applying the FF method,
the uncertainty due to varying choice of parameterisation, and the uncertainty in the subtraction of non-\ZjetZgam processes.
The uncertainties are evaluated by considering the variations in the FF and propagating the effects to the estimated yields.
The prescription applied for the estimation in the \ofs \WZ selection is different from that in the \ons \WZ and \Wh selections,
reflecting the higher presence of \ZjetZgam in \SRoffWZ.
Details are included in Sections~\ref{ssec:onshellAndWh:bkg} and~\ref{ssec:offshell:bkg}.
 
Uncertainties in the expected yields for SUSY signals are estimated by varying by a factor of two the \MGNLO parameters
corresponding to the renormalisation, factorisation and CKKW-L matching scales,
as well as the \textsc{Pythia8} shower tune parameters.
\begin{DIFnomarkup}
The overall uncertainties in the signal acceptance range from 5\% to 20\% depending on the analysis region.
Uncertainties are smallest in jet-veto regions and slightly larger for higher \met and jet-inclusive regions.
This uncertainty estimates match the results of a dedicated study using data and MC $Z\ra \mu\mu$ events in Ref.~\cite{SUSY-2018-16}.
\end{DIFnomarkup}
 
In the following results,
the uncertainties related to experimental effects are grouped and shown as `Experimental' uncertainty.
This uncertainty is applied for all processes whose yield is estimated from simulation.
The `Modelling' uncertainty groups the uncertainties due to the theoretical uncertainties, including the \WZ transfer factor uncertainties.
The `Fakes' group represents the uncertainties for FNP background processes whose yield is estimated from data.
`MC stat' stands for the statistical uncertainties of the simulated event samples.
Finally, the `Normalisation' group describes the uncertainties related to the normalisation factors derived from the CRs.
}
 
\begin{DIFnomarkup}\vskip 1em
\subsection{Statistical analysis}
\label{ssec:analysisstrategy:fit}
\end{DIFnomarkup}
{\begin{DIFnomarkup}\enlargethispage*{1.8\baselineskip}\end{DIFnomarkup}
Final background estimates are obtained by performing a profile log-likelihood fit~\cite{Cowan:2010js},
implemented in the \textsc{HistFitter}~\cite{Baak:2014wma} framework,
simultaneously on all CRs and SRs relevant to a given interpretation.
The statistical and systematic uncertainties are implemented as nuisance parameters in the likelihood;
Poisson constraints are used to estimate the uncertainties arising from limited numbers of events in the MC samples or in the data-driven \ZjetZgam estimation,
whilst Gaussian constraints are used for experimental and theoretical systematic uncertainties.
Neither the VRs, which solely serve to validate the background estimation in the SRs,
nor the CRs used for data-driven \ZjetZgam estimation,
are included in any of the fits.
 
Three types of fit configuration are used to derive the results.
\begin{itemize}
\item A `\textit{background-only fit}' is performed considering only the CRs and assuming no signal presence.
The normalisation of the \WZ background is allowed to float and is constrained by the \WZ CRs.
The normalisation factors and nuisance parameters are adjusted by maximising the likelihood.
The background prediction as obtained from this fit is compared with data in the VRs to assess the quality of the background modelling,
as well as in the SRs.
The significance of the difference between the observed and expected yields is calculated with the profile likelihood method from Ref.~\cite{Cousins:2007bmb}, adding a minus sign if the yield is
below the prediction.
\item A `\textit{discovery fit}' is performed to derive model-independent constraints, setting upper limits on the new-physics cross section.
The fit considers the target single-bin SR and the associated CRs, constraining the backgrounds by following the same method as in the background-only fit.
Considering only one SR at a time avoids introducing a dependence on the signal model, which may arise from correlations across multiple SR bins.
A signal contribution is allowed only in the SR, and a non-negative signal-strength parameter assuming generic beyond-the-SM (BSM) signals is derived.
\item An `\textit{exclusion fit}' is performed to set exclusion limits on the target models.
The backgrounds are again constrained by following the same method as in the background-only fit, considering the CRs and the SRs,
and the signal contribution to each region participating in the fit is taken into account according to the model predictions.
\end{itemize}
For each discovery or exclusion fit, the compatibility of the observed data with the signal-plus-background hypotheses
is checked using the \CLs prescription~\cite{Read:2002hq},
and limits on the cross section are set at 95\% confidence level (CL).
 
Following the independent optimisation of the CRs and SRs, the simultaneous fits are performed separately for the different selections:
once for the \ons \WZ and \Wh selections combined, and once for the \ofs \WZ selection.
The results are presented in Section~\ref{sec:results}.
}
 
{\begin{DIFnomarkup}\pagebreak\end{DIFnomarkup}
The new results of the \ons and \ofs \WZ searches,
as well as the results of a previous ATLAS search for electroweak SUSY with compressed mass spectra~\cite{SUSY-2018-16},
are statistically combined and interpreted in the simplified models discussed in Section~\ref{sec:intro}.
Exclusion limits are calculated by statistically combining the results from
the signal regions of the contributing searches, which are designed to be orthogonal.
The combination is implemented in the \texttt{pyhf} framework~\cite{Heinrich:2021gyp,pyhfzenodo},
which was validated against the \textsc{HistFitter} framework~\cite{ATL-PHYS-PUB-2019-029}.
The results are presented in Section~\ref{ssec:results:excl}.
}
 
% End of text imported from the .//sections/analysisstrategy.tex input file
% The next lines are included from the .//sections/onshellAndWh.tex input file
\begin{DIFnomarkup}
\vskip 1em\section{\OnShell and \Wh selections}\label{sec:onshellAndWh}
\end{DIFnomarkup}
 
The following subsections discuss the implementation specific to the \ons \WZ selection and the \Wh selection,
expanding on the general strategy outlined in Section~\ref{sec:analysisstrategy}. 
The selection is applied on top of the common preselection as defined in Section~\ref{sec:objdef},
and the SRs are optimised to the wino/bino (+) scenario.

\begin{DIFnomarkup}
\vskip 1em\subsection{Search regions} \label{ssec:onshellAndWh:reg}
\end{DIFnomarkup}
The \SRonWZ and \SRWh selections as introduced in Section~\ref{ssec:analysisstrategy:sr} are further refined, 
taking into consideration differences in signal and background kinematics and composition.
Driven by the \pt thresholds of the dilepton triggers used in this selection,
the leading and sub-leading leptons in the event must satisfy $\pt >$ 25, 20~\GeV, while the third lepton  must satisfy $\pt > 10$~\GeV.
To reduce SM backgrounds with little to no real \met, events are required to have $\met > 50$~\GeV. 
To suppress the contribution of \ttbar events and single-boson production in association with a \ttbar pair, events with at least one \bjet are rejected.
 
To reduce the contribution from processes with low-mass dilepton resonances, events are vetoed if they contain a SFOS lepton pair with an invariant mass below 12~\GeV.
Additionally, in events with a SFOS pair, the three-lepton invariant mass \mtl is required to be inconsistent with the mass of a $Z$ boson, $|\mtl-\mZ| > 15$~\GeV,
in order to suppress contributions from asymmetric photon conversions from the \ZjetZgam process with $Z\rightarrow \ell\ell\gamma^{(*)}$and $\gamma^{(*)}\rightarrow \ell\ell$,
where one of the leptons is out of acceptance.
A summary of the preselection criteria is presented in Table~\ref{tab:presel}.
The \SRWZ and \SRWh regions are further segmented as discussed below, and indexed with \SRj{-i}.

\begin{DIFnomarkup}
\begin{table}[bth!]
\centering
\caption{
Summary of the preselection criteria applied in the SRs of the \ons \WZ and \Wh selections.
In rows where only one value is given it applies to all regions. `-' indicates no requirement is applied for a given variable/region.
}
 
\label{tab:presel}
\adjustbox{max width=0.80\textwidth}{
\begin{tabular}{ l *{3}{P{2.4cm}}}
\hline\hline
& \multicolumn{3}{c}{Preselection requirements} \\\cline{2-4}
\rule{0pt}{\dimexpr.7\normalbaselineskip+1mm}
Variable     & \SRonWZ & \SRWhSF & \SRWhDF \\ \hline
\nlbl, \nlsig & \multicolumn{3}{c}{= 3}                               \\
Trigger & \multicolumn{3}{c}{dilepton}                                \\
\ptl{1}, \ptl{2}, \ptl{3} [\GeV] & \multicolumn{3}{c}{$>$ 25, 20, 10} \\
\MET [\GeV] & \multicolumn{3}{c}{$>$ 50}                               \\
\nbjets & \multicolumn{3}{c}{= 0}                                          \\\hline
Resonance veto \mll [\GeV] & $>$ 12 & $>$ 12 & -                \\
\nSFOS & $\geq 1$ & $\geq 1$ & = 0                                      \\
\mll [\GeV] & $\in [75,105]$ & $\notin [75,105]$ & -    \\
$|\mtl-\mZ|$ [\GeV] & $>15$ & $>15$ & -                           \\
\hline\hline
\end{tabular}
}
\end{table}
\end{DIFnomarkup}

{\begin{DIFnomarkup}\pagebreak\enlargethispage*{1.2\baselineskip}\end{DIFnomarkup}
Events with at least one SFOS lepton pair are divided into three \mll bins,
in order to separate processes that include a $Z$ boson in the decay chain from processes where a SM Higgs boson is involved.
The first bin is defined as the $Z$-boson mass window ($\mll \in [75,105]~\GeV$), and is used for the \SRonWZ selection.
The second and third bins are defined below and above the $Z$-boson mass ($\mll \leq 75$~\GeV~and $\mll \geq 105$~\GeV),
and are used for the \SRWhSF selection.
The $Z$-boson mass window bin is expected to contain a larger irreducible SM background contribution than the other bins.

Each \mll bin is further divided into \mt and \met bins, which enhances the sensitivity to various \dm\linebreak scenarios.
The \mt distribution falls steeply in the region around the $W$-boson mass,
and facilitates discrimination against the background from SM \WZ production.
Three \mt bins, $\mt < 100$, $100 \leq \mt \leq 160$, and $\mt > 160$~\GeV, are defined to separate processes with and without a leptonic $W$-boson decay.
The lower and upper bounds on the \met bins vary with the \mll and \mt thresholds.
The SM background contribution is expected to be higher in low \mt and \met bins,
while the signal populates different \mt and \met bins, depending on the mass splitting.
Signals with smaller \dm tend to have more events in the lower \met and \mt range,
shifting to higher \met and \mt bins as the mass difference increases.
 
Furthermore, events are separated by jet multiplicity,
with jet-veto (\smash{$\nJ = 0$; \SRWZi{1} to \SRk{8}, \SRWhSFi{1} to \SRk{7}} and \smash{\SRk{17} to \SRk{19}})
and jet-inclusive (\smash{$\nJ > 0$; \SRWZi{9} to \SRk{20}, \SRWhSFi{8} to \SRk{16}})
SRs.
The ISR topology is exploited further in the jet-inclusive regions of \SRonWZ and \SRWhSF
by categorising the events with at least one jet according to their \HT, the scalar \pT sum of the jets with $\pt > 20$~\GeV.
At higher \HT, signals with mass splitting $\dm \approx \mz$
tend to have more events at high values of \MET and \mt than the SM background, due to the recoil against ISR jets.
In the high \HT ($\HT > 200~\GeV$)  regions, softer lepton-\pt spectra are expected for the signal because of the presence of a massive \textninoone,
which carries most of the transverse momenta of the boosted \textchinoonepmninotwo system.
Therefore \htl, the scalar \pT sum of the three selected leptons, is required to be less than 350~\GeV.
The \HT categorisation is applied in regions with $\mll < 105$~\GeV. 
Finally, in the high-mass off-peak region ($\mll \geq 105$~\GeV), only jet-veto events are considered.
The full set of 20 \SRonWZ and 19 \SRWhSF signal regions is summarised in Tables~\ref{tab:SR_onshell} and~\ref{tab:SR_mll}.

\begin{table}[p!]
\begin{center}
\caption{
Summary of the selection criteria for the SRs targeting events with at least one SFOS lepton pair and $\mll \in [75,105]~\GeV$,
for the \ons \WZ search regions.
Region selections are binned by \mt (rows) and \met for the two sets of regions, where each set has different \nJ and \HT requirements.
\SRonWZ preselection criteria are applied (Table~\ref{tab:presel}).
}
\label{tab:SR_onshell}
\begin{tabular}{l| c c c  c}
\hline\hline
\multicolumn{5}{c}{Selection requirements} \\ \cline{1-5}
\multicolumn{1}{l}{} & \multicolumn{4}{c}{$\nJ = 0$} \\ \cline{1-5}
\mt [\GeV] & \multicolumn{4}{c}{\MET [\GeV]} \\ \hline\tabvthinspace
[100, 160]   & \SRWZi{1}: [50, 100]& \SRWZi{2}: [100, 150] & \SRWZi{3}: [150, 200]& \SRWZi{4}: $> 200$ \\ \hline\tabvthinspace
$ > $ 160  &  \SRWZi{5}: [50, 150]& \SRWZi{6}: [150, 200] & \SRWZi{7}: [200, 350]& \SRWZi{8}: $>$ 350 \\ \hline \hline
\multicolumn{1}{l}{}       & \multicolumn{4}{c}{$\nJ > 0 $, $\HT < 200$~\GeV} \\ \cline{1-5}
\mt [\GeV]  & \multicolumn{4}{c}{\MET [\GeV]} \\ \hline\tabvthinspace
[100, 160]   & \SRWZi{9}: [100, 150] & \SRWZi{10}: [150, 250]& \SRWZi{11}: [250, 300] & \SRWZi{12}: $> 300$ \\ \hline\tabvthinspace
$ > $ 160   & \SRWZi{13}: [50, 150]& \SRWZi{14}: [150, 250] & \SRWZi{15}: [250, 400]& \SRWZi{16}: $> 400$ \\ \hline \hline
\multicolumn{1}{l}{}   & \multicolumn{4}{c}{$\nJ > 0 $, $\HT > 200$~\GeV, $\htl < 350$~\GeV } \\ \cline{1-5}
\mt [\GeV]& \multicolumn{4}{c}{\MET [\GeV]} \\ \hline\tabvthinspace
$ > $ 100  &  \SRWZi{17}: [150, 200]& \SRWZi{18}: [200, 300] & \SRWZi{19}: [300, 400]& \SRWZi{20}: $> 400$ \\ \hline \hline
\end{tabular}
\end{center}
\end{table}

\begin{table}[p!]
\begin{center}
\caption{
Summary of the selection criteria for the SRs targeting events with at least one SFOS lepton pair and $\mll \notin [75,105]~\GeV$, for the \Wh search regions.
Region selections are binned by \mt (rows) and \met for the three sets of regions, where each set has different \mll, \nJ, and \HT requirements.
\SRWhSF preselection criteria are applied (Table~\ref{tab:presel}).
}
\label{tab:SR_mll}
\adjustbox{max width=0.95\textwidth}{
\begin{tabular}{l| c c c }
\hline\hline
\multicolumn{4}{c}{Selection requirements} \\ \cline{1-4}
\multicolumn{1}{l}{} & \multicolumn{3}{c}{$\mll \leq 75~\GeV$, $\nJ = 0$} \\ \cline{1-4}
\mt [\GeV] & \multicolumn{3}{c}{\MET [\GeV]} \\ \hline\tabvthinspace
[0, 100]  &  \SRWhSFi{1}: [50, 100] & \SRWhSFi{2}: [100, 150]& \SRWhSFi{3} $>$ 150   \\ \hline\tabvthinspace
[100, 160]  & \SRWhSFi{4}: [50, 100]&  \multicolumn{2}{c}{\SRWhSFi{5}: $> 100$} \\ \hline\tabvthinspace
$ > $ 160  &  \SRWhSFi{6}: [50, 100]&  \multicolumn{2}{c}{\SRWhSFi{7}: $> 100$}\\ \hline\hline
\multicolumn{1}{l}{}         & \multicolumn{3}{c}{$\mll \leq 75~\GeV$, $\nJ > 0$, $\HT < 200~\GeV$} \\ \cline{1-4}
\mt [\GeV] & \multicolumn{3}{c}{\MET [\GeV]} \\ \hline\tabvthinspace
[0, 50] &  \multicolumn{3}{c}{\SRWhSFi{8}: [50, 100]}    \\ \hline\tabvthinspace
[50, 100] & \multicolumn{3}{c}{\SRWhSFi{9}: [50, 100]}   \\ \hline\tabvthinspace
[0, 100]  &  \multicolumn{2}{c}{\SRWhSFi{10}: [100, 150]}& \SRWhSFi{11}: $>$ 150   \\ \hline\tabvthinspace
[100, 160]  & \SRWhSFi{12}: [50, 100] & \SRWhSFi{13}: [100, 150]& \SRWhSFi{14}: $>$ 150  \\ \hline\tabvthinspace
$ > $ 160  &  \SRWhSFi{15}: [50, 150]&  \multicolumn{2}{c}{\SRWhSFi{16}: $> 150$}\\ \hline \hline
\multicolumn{1}{l}{}  & \multicolumn{3}{c}{$\mll \geq 105~\GeV$, $\nJ = 0$} \\ \cline{1-4}
\mt [\GeV] & \multicolumn{3}{c}{\MET [\GeV]} \\ \hline\tabvthinspace
$ > $ 100  &  \SRWhSFi{17}: [50, 100]& \SRWhSFi{18}: [100, 200] & \SRWhSFi{19}: $> 200$ \\ \hline \hline
\end{tabular}
}
\end{center}
\end{table}

In the \SRWhDF regions, 
events are required to have one same-flavour same-charge-sign (SFSS) lepton pair
as well as a third lepton which has a different flavour and opposite sign to the SFSS pair, and is referred to as the DFOS lepton.
After this selection, \ttbar production dominates the SM background and is minimised by keeping events with low jet multiplicity (\smash{$\nJ < 3$}).
These are then further split into two SR bins, one with \smash{$\nJ = 0$} \smash{(\SRWhDFi{1})} and the other satisfying \smash{$\nJ \in [1, 2]$} \smash{(\SRWhDFi{2})}.
Due to the presence of the \textninoone, signals tend to have higher \metsig than the SM background, and therefore the events are required to have $\metsig > 8$.
The third lepton in \ttbar production usually arises from a heavy flavour quark decay and is typically lower in \pT than the third lepton in the SUSY signal scenarios.
To reduce this contribution
the lower bound on the third lepton's \pt is increased to 15 and 20~\GeV\ in the \smash{\SRWhDFi{1} and \SRWhDFi{2}} regions, respectively.
Angular proximity between leptons coming from a Higgs-boson decay is used for further event separation,
using the variable $\dRnear$, defined as the $\Delta R$ between the DFOS lepton and the SFSS lepton nearest in $\phi$.
The signal is expected to populate the lower range in \dRnear, while the SM background tends to have a flatter distribution.
Events in \SRWhDFi{1} are required to satisfy $\dRnear < 1.2$. To suppress the higher \ttbar contribution in the \SRWhDFi{2}, a tighter selection on $\dRnear$ is imposed.
A complete summary of the selection criteria in \SRWhDF is presented in Table~\ref{tab:DFOSSR}.
 
For the \WZ-mediated \textchinoonepmninotwo signal sample with NLSP mass of 600~\GeV\ and massless \textninoone, the \smash{\SRonWZzj} and \smash{\SRonWZnj} regions
have selection acceptance times efficiency values of $2.0\times 10^{-3}$ and $3.0\times 10^{-3}$, respectively.
For the \Wh-mediated \textchinoonepmninotwo signal sample with NLSP mass of 200~\GeV\ and massless~\textninoone, the \smash{\SRWhlmzj}, \smash{\SRWhlmnj}, and \smash{\SRWhDF} regions
have selection acceptance times efficiency values of $9.1\times 10^{-5}$, $1.0\times 10^{-4}$, and $3.7\times 10^{-5}$, respectively.
}

\begin{DIFnomarkup}
\begin{table}[hbtp!]
\centering
\caption{
Summary of the selection criteria for the SRs targeting events with a DFOS lepton pair, for the \Wh selection.
\SRWhDF preselection criteria are applied (Table~\ref{tab:presel}).}
\label{tab:DFOSSR}
\begin{tabular}{r*{2}{P{2cm}}}
\hline\hline
& \multicolumn{2}{c}{Selection requirements} \\ \cline{2-3}\tabvspace
Variable & \SRWhDFi{1} & \SRWhDFi{2} \\ \hline
\nJ         & $ = 0$ & $ \in{[1,2]}$ \\
\metsig & $>$ 8 & $>$ 8 \\ \hline
$\ptl{3}$ [\GeV] & $>$ 15 & $>$ 20\\
\dRnear & $<$ 1.2 & $<$ 1.0 \\
\hline\hline
\end{tabular}
\end{table}
\end{DIFnomarkup}
 
\FloatBarrier

\FloatBarrier
{
\begin{DIFnomarkup}
\vskip 1em\subsection{Background estimation} \label{ssec:onshellAndWh:bkg}
\end{DIFnomarkup}
The normalisation of the \WZ background is measured in CRs characterised by moderate values of the \met and \mt variables.
The CRs contain only events with at least one SFOS pair with an invariant mass of $75 < \mll < 105$~\GeV, targeting on-shell decays.
Additional requirements of $50 < \met < 100~\GeV$~and $20 < \mt < 100~\GeV$~improve the \WZ purity, the upper bound on \mt at $100$~\GeV\ also ensures orthogonality between the \WZ CRs and \SRonWZ.
To address the possible mis-modelling of the jet multiplicity in the \WZ simulated samples,
the cross-section normalisation factor is extracted separately in each jet multiplicity and \HT category, using \CRonWZzj, \CRonWZnjl, and \CRonWZnjh.
The estimation is cross-checked in kinematically similar, orthogonal VRs: \VRonWZzj, \VRonWZnjl, and \VRonWZnjh.
A summary of the selection criteria defining the \WZ CRs and VRs is presented in Table~\ref{tab:CR_sel}.
The \WZ purity is about 80\%  in all CRs and VRs.
The signal contamination is almost negligible in the CRs and increases to 10\% in the VRs.

\begin{table}[b!]
\centering
\caption{
Summary of the selection criteria for the CRs and VRs for \WZ, for the \ons \WZ and \Wh selections.
In rows where only one value is given it applies to all regions.
`-' indicates no requirement is applied for a given variable/region.
}
\label{tab:CR_sel}
\adjustbox{max width=\textwidth}{
\begin{tabular}{ l *{3}{P{1.8cm}} | *{3}{P{1.8cm}} }
\hline\hline
Variable      & \multicolumn{3}{c|}{\CRonWZ} & \multicolumn{3}{c}{\VRonWZ} \\
& \rzj & \rlo-\rHT & \rhi-\rHT & \rzj & \rlo-\rHT & \rhi-\rHT \\\hline
\nlbl, \nlsig & \multicolumn{3}{c|}{= 3} & \multicolumn{3}{c}{= 3}                               \\
Trigger & \multicolumn{3}{c|}{dilepton} & \multicolumn{3}{c}{dilepton}                          \\
\ptl{1}, \ptl{2}, \ptl{3} [\GeV] & \multicolumn{3}{c|}{$>$ 25, 20, 10} & \multicolumn{3}{c}{$>$ 25, 20, 10} \\
\nbjets & \multicolumn{3}{c|}{= 0}    & \multicolumn{3}{c}{= 0}                                       \\
\nSFOS & \multicolumn{3}{c|}{$\geq 1$} & \multicolumn{3}{c}{$\geq 1$} \\
\mt [\GeV] & \multicolumn{3}{c|}{$\in [20,100]$} & \multicolumn{3}{c}{$\in [20,100]$} \\
\mll [\GeV] & \multicolumn{3}{c|}{$\in [75,105]$} & \multicolumn{3}{c}{$\in [75,105]$} \\
$|\mtl-\mZ|$ [\GeV] & \multicolumn{3}{c|}{$>15$} & \multicolumn{3}{c}{$>15$} \\
\met [\GeV] & \multicolumn{3}{c|}{$\in [50,100]$} &  \multicolumn{3}{c}{$>100$} \\ \hline
\nJ & =0 & $\geq 1$ & $\geq 1$ & =0 & $\geq 1$ & $\geq 1$ \\
\HT [\GeV] & - & $<200$ & $>200$ & - &  $<200$ & $>200$ \\
\hline\hline
\end{tabular}
}
\end{table}
 
Performing the simultaneous background-only fit for the \onShell and \Wh selections,
normalisation factors for \WZ of
$1.07\pm0.02$ (\CRonWZzj),
$0.94\pm 0.03 $ (\CRonWZnjl) and
$0.85\pm 0.05$ (\CRonWZnjh)
are found.
}

\begin{DIFnomarkup}\pagebreak\end{DIFnomarkup}
A good description of the \mt and \met distributions in the \WZ simulation is crucial in this analysis,
especially in the high-\mt and high-\met tails where new physics may appear.
The tail of the \mt distribution is a result of, in decreasing order of importance: the use of a wrong pair of leptons
to compute the mass of the $Z$-boson candidate and the \mt of the $W$-boson candidate (`mis-pairing' of the leptons),
the \met resolution, and the $W$-boson width.
The prediction of lepton mis-pairing in simulation is validated in a control sample in data similar to the one used to calculate the cross-section normalisation factor,
but only allowing events with a SFOS pair of different flavour than the $W$ lepton.
The $Z$-boson candidate can then be identified unambiguously,
and a mis-paired control sample is obtained using the DFOS pair in the \mll computation and using the third lepton to calculate \mt.
Finally, the modelling of the \mt and \met distributions is validated in a \wg control sample.
The \wg and \WZ processes have very similar \mt shapes because their production mechanisms are similar,
with the exception that the FSR production diagram of \wg is much more common than the corresponding diagram in \WZ,
which is doubly suppressed due to the mass of the $Z$ boson and its weak coupling to leptons.
Furthermore, a photon is a good proxy for a leptonically decaying $Z$ boson since photons and leptons are reconstructed with comparable resolutions,
and no large extra mismeasurements are expected.
The enhancement of the FSR diagram in the \wg process leads to differences in the \mt distribution shapes between \WZ and \wg.
When a photon is radiated, leptons lose energy, resulting in a lower \mt.
In order to use the \wg \mt shape to validate the \WZ MC prediction, the FSR contribution in the \wg control region has to be suppressed.
This is done by placing threshold requirements on the \pt of the photon, $\pt^{\gamma} > 50$~\GeV,
and the separation between the lepton and the photon, $\dR(\ell,\gamma) > 0.4$, in \wg events,
as FSR photons are expected to be close to the lepton radiating them and also tend to have low \pt.
The distribution shapes of \mt and \met, as well as other kinematic variables, are compared in data and MC events in the \wg region.
The \mt distribution in the validation region with mis-paired leptons and the \wg validation region are shown in Figure~\ref{fig:bg:onShell:CRWZ}.
Good agreement in both control samples is observed and no extra corrections or scale factors are applied to correct the \mt distribution for the \WZ background.

{
The \ttbar MC modelling is validated in VRs, enhancing the \ttbar contribution by requiring a DFOS lepton pair and using a moderate $\met > 50~\GeV$~selection.
The main VR, \VRontt, requires the presence of one or two $b$-jets, further increasing the \ttbar contribution.
To validate the modelling in the $\nJ = 0$ region as well,
an additional VR inclusive in $b$-jets, \VRonttinc, is considered,
with a $\metsig < 8$ requirement to ensure orthogonality with the \SRWhDF regions. The \ttbar purity is about 80\% in the \VRontt and 72\% in the \VRonttinc.
The selection requirements for the \ttbar VRs are summarised in Table~\ref{tab:ons:ttbar}.
 
\begin{table}[p!]
\centering
\caption{
Summary of the selection criteria for the CRs and VRs for \ttbar and \ZjetZgam, for the \ons \WZ and \Wh selections.
The corresponding anti-ID regions used for the \ZjetZgam prediction follow the same selection criteria,
except that at least one of the leptons is anti-ID instead of signal.
`-' indicates no requirement is applied for a given variable/region.
}
\label{tab:ons:ttbar}
\begin{tabular}{ l *{2}{P{2.3cm}}|*{2}{P{2.3cm}}}
\hline\hline
Variable    & \VRontt & \VRonttinc & \CRonFF & \VRonFF\\ \hline
\nlbl, \nlsig & = 3 & = 3 & = 3 & = 3 \\
\nSFOS & = 0 & = 0 & $\geq 1$ & $\geq 1$ \\
Trigger & dilepton & dilepton & dilepton & dilepton \\
\nbjets     & $\in [1,2]$ & - & =0 & =0 \\
$|\mll-\mZ|$ [\GeV] & - & - & $<15$ & $<15$ \\
\ptl{Z1}, \ptl{Z2} [\GeV] & - & - & $>25$, $>20$ & - \\
\met [\GeV] & $>50$ & $>50$ & $\in [20,50]$ & $\in [50,100]$ \\
\metsig     & - & $<8$ & - & - \\
\mT [\GeV]  & - & - & $<20$ & $<20$ \\
\mtl [\GeV] & - & - & - & $\in [105,160]$ \\
\hline\hline
\end{tabular}
\end{table}
}
 
The \ZjetZgam estimation uses the FF method as described in Section~\ref{ssec:analysisstrategy:bkgs}.
For measurement region \CRonFF, the $Z$-boson candidate mass must be compatible with the $Z$-boson mass within 15~\GeV,
and low \met and \mt are required to minimise \WZ contributions.
The typical value of FFs varies from 0.2 to 0.4, depending on the lepton \pt and $\eta$.
The \ZjetZgam estimation is then validated in \VRonFF,
considering the intermediate \met range closer to, but orthogonal to, the SRs,
and adding a \mtl lower bound to reduce \WZ contamination.
The selection criteria for \CRonFF as well as those of \VRonFF are summarised in Table~\ref{tab:ons:ttbar}.
 
\begin{DIFnomarkup}
Figure~\ref{fig:bg:onShell:CRWZ}\ presents the \mt distribution in \VRonWZnjh, and the \met distribution in \VRontt,
showing good agreement between the observed data and the estimated background.
The comparisons between the expected and observed yields in the \CRonWZ{} and all \VR{}{\rWZon}{} are given in Figure~\ref{fig:results:WZ_pull}.
\end{DIFnomarkup}

\begin{figure}[p!]
\centering
\begin{DIFnomarkup}
\vskip 1.2em
\includegraphics[width=0.49\columnwidth]{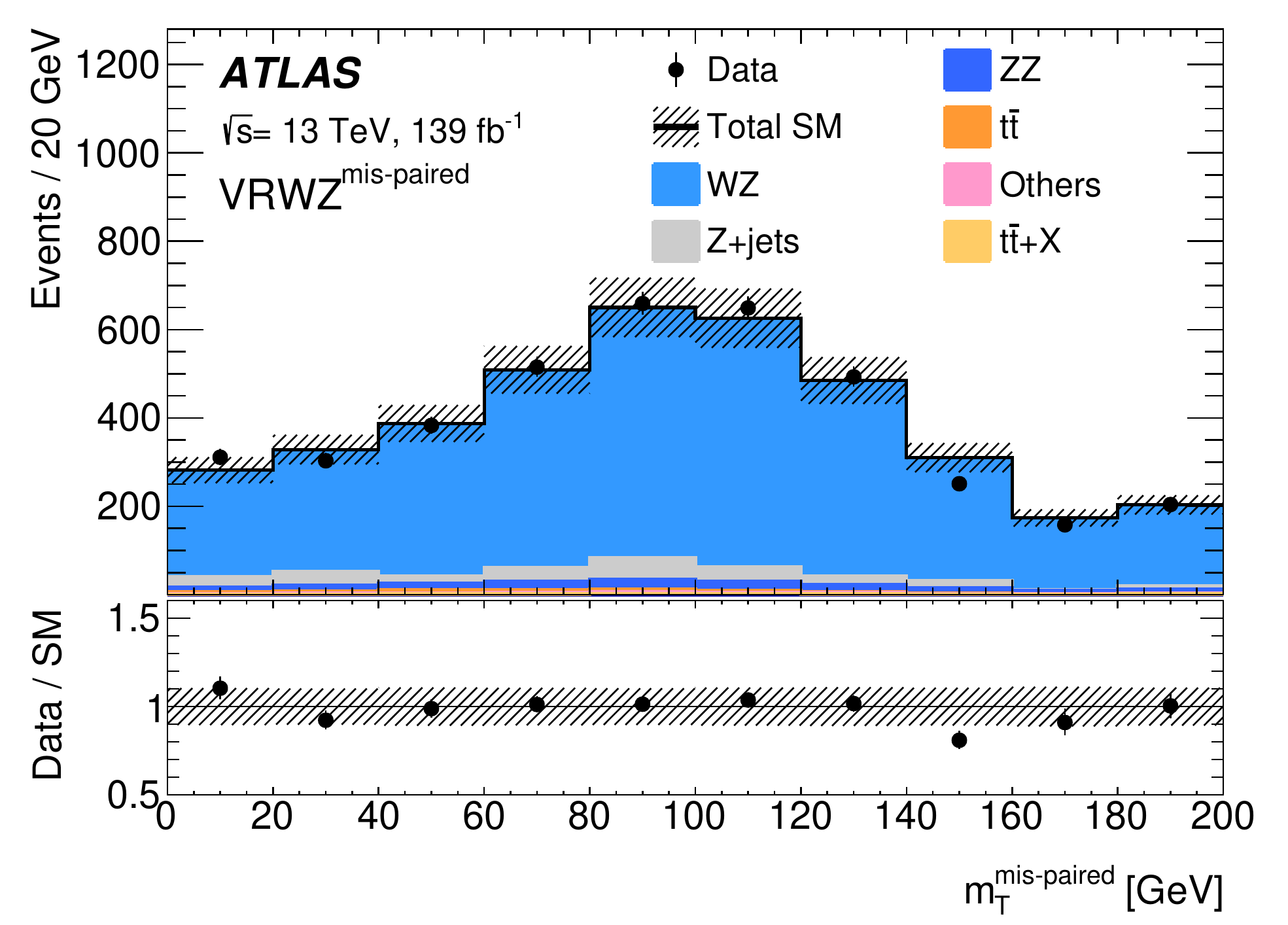}
\includegraphics[width=0.49\columnwidth]{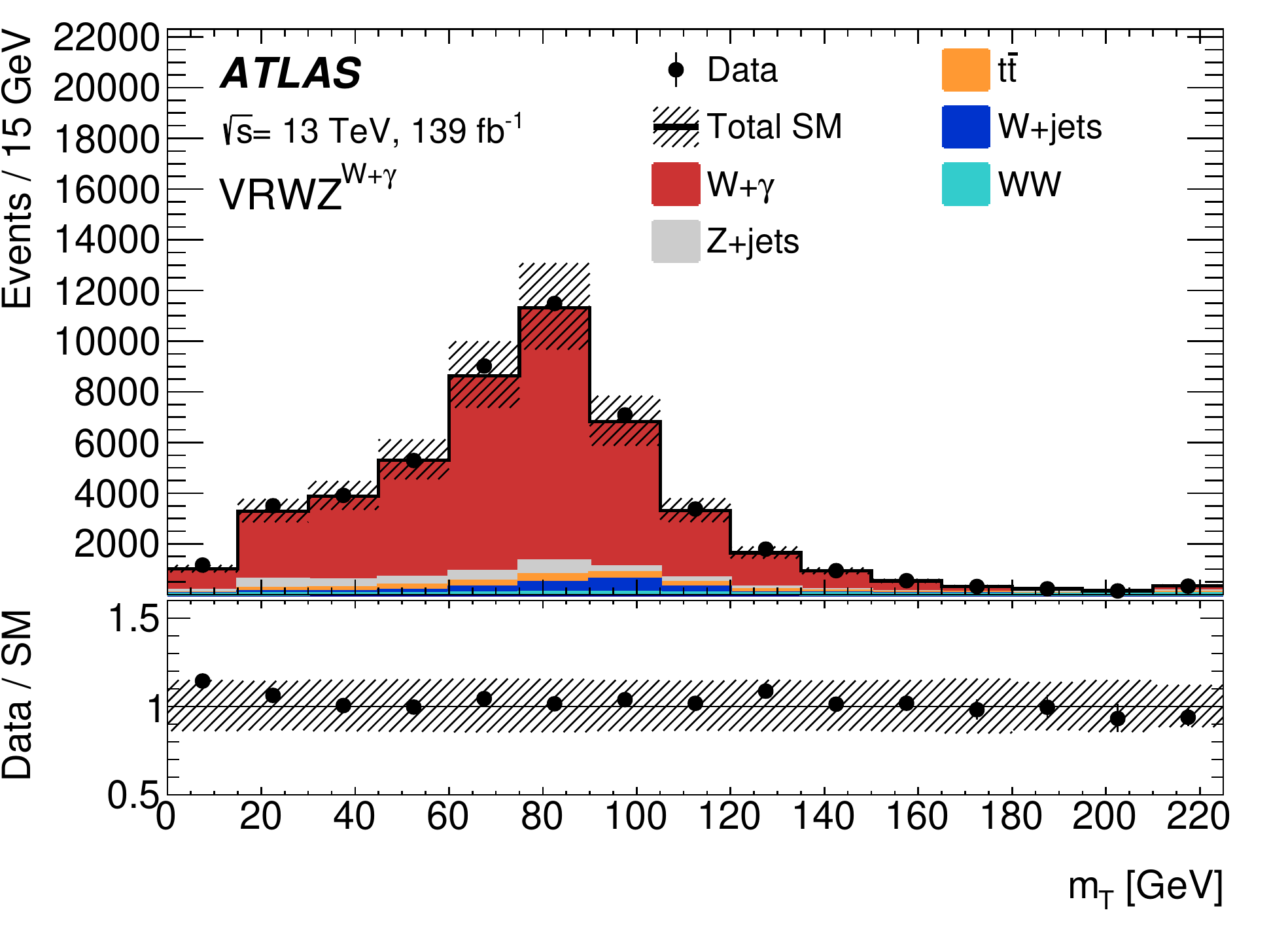}
\includegraphics[width=0.49\columnwidth]{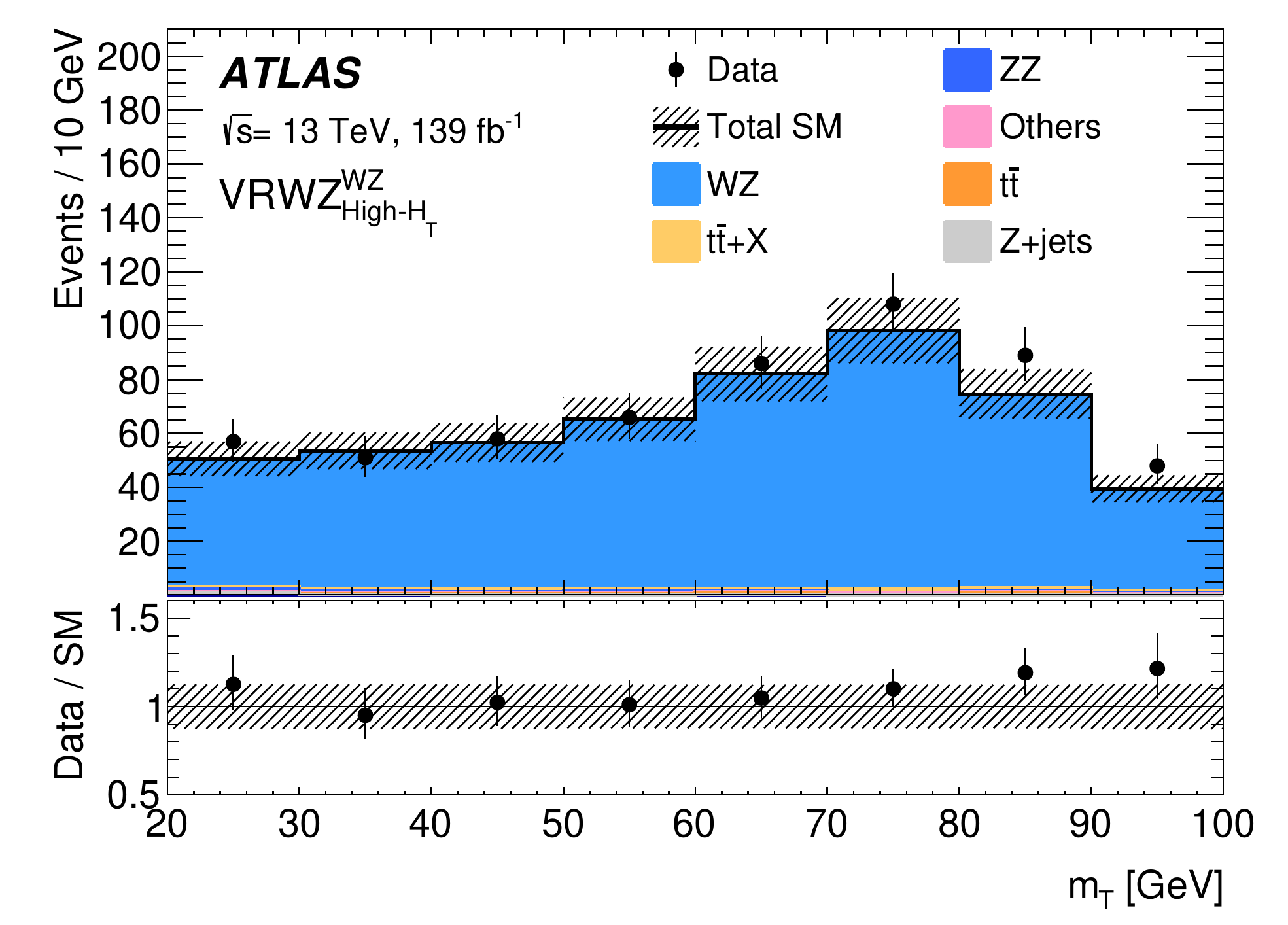}
\includegraphics[width=0.49\columnwidth]{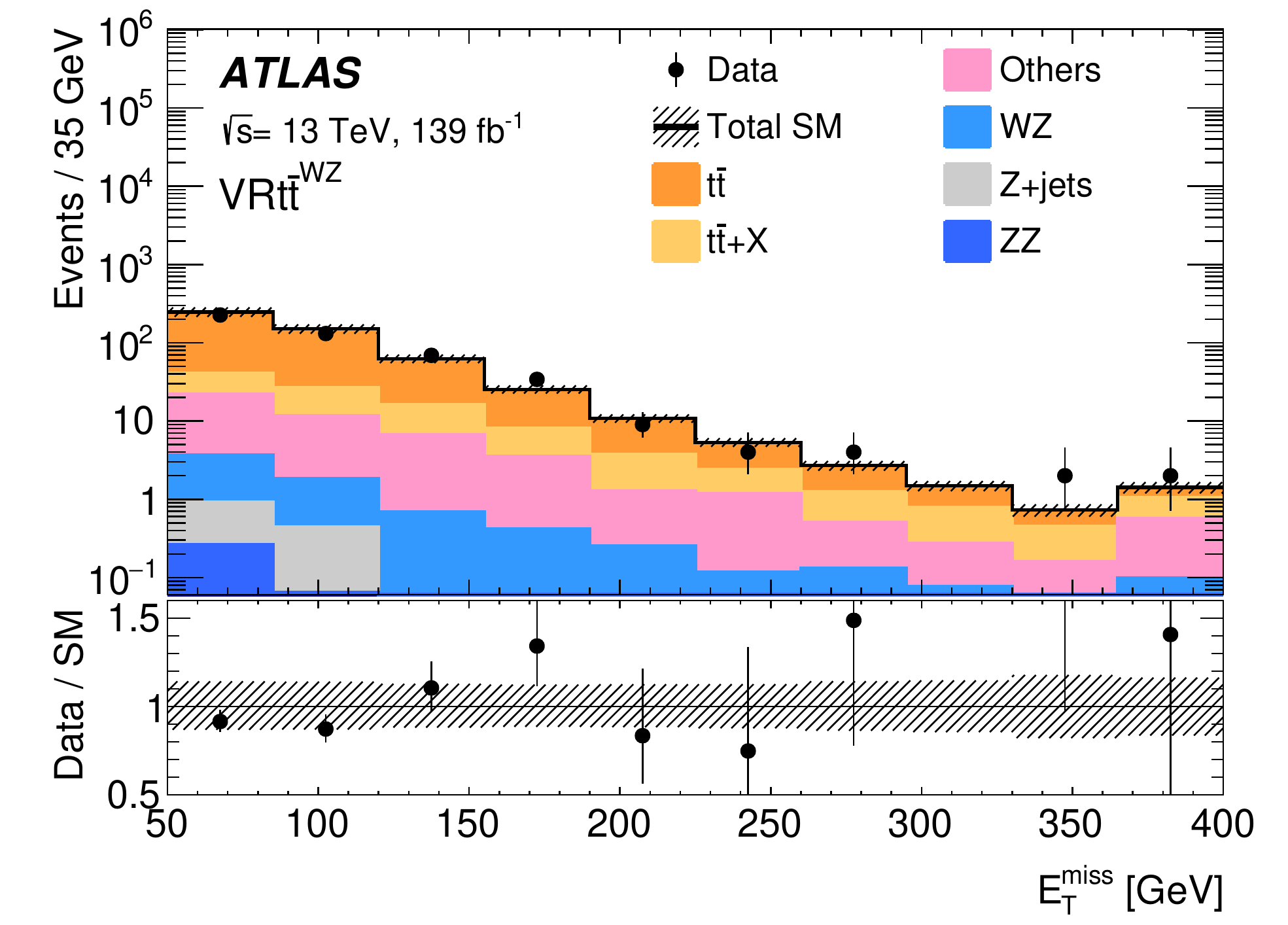}
\end{DIFnomarkup}
\caption{
Distributions of \mT showing the data and the pre-fit expected background in
(top left) the mis-paired lepton validation region and
(top right) the \wg validation region, used to validate the \WZ background.
Distributions of
(bottom left) \mt in \VRonWZnjh and
(bottom right) \met in \VRontt,
showing the data and the post-fit expected background in each region.
The last bin includes overflow.
The `Others' category contains backgrounds from single-top, \WW, triboson, Higgs and rare top processes.
The bottom panel shows the ratio of the observed data to the predicted yields.
The hatched bands indicate the combined theoretical, experimental, and MC statistical uncertainties.
}
\begin{DIFnomarkup}
\vskip -1.2em
\end{DIFnomarkup}
\label{fig:bg:onShell:CRWZ}
\end{figure}

\FloatBarrier
\pagebreak
\begin{figure}[t!]
\centering
{\includegraphics[width=0.90\textwidth]{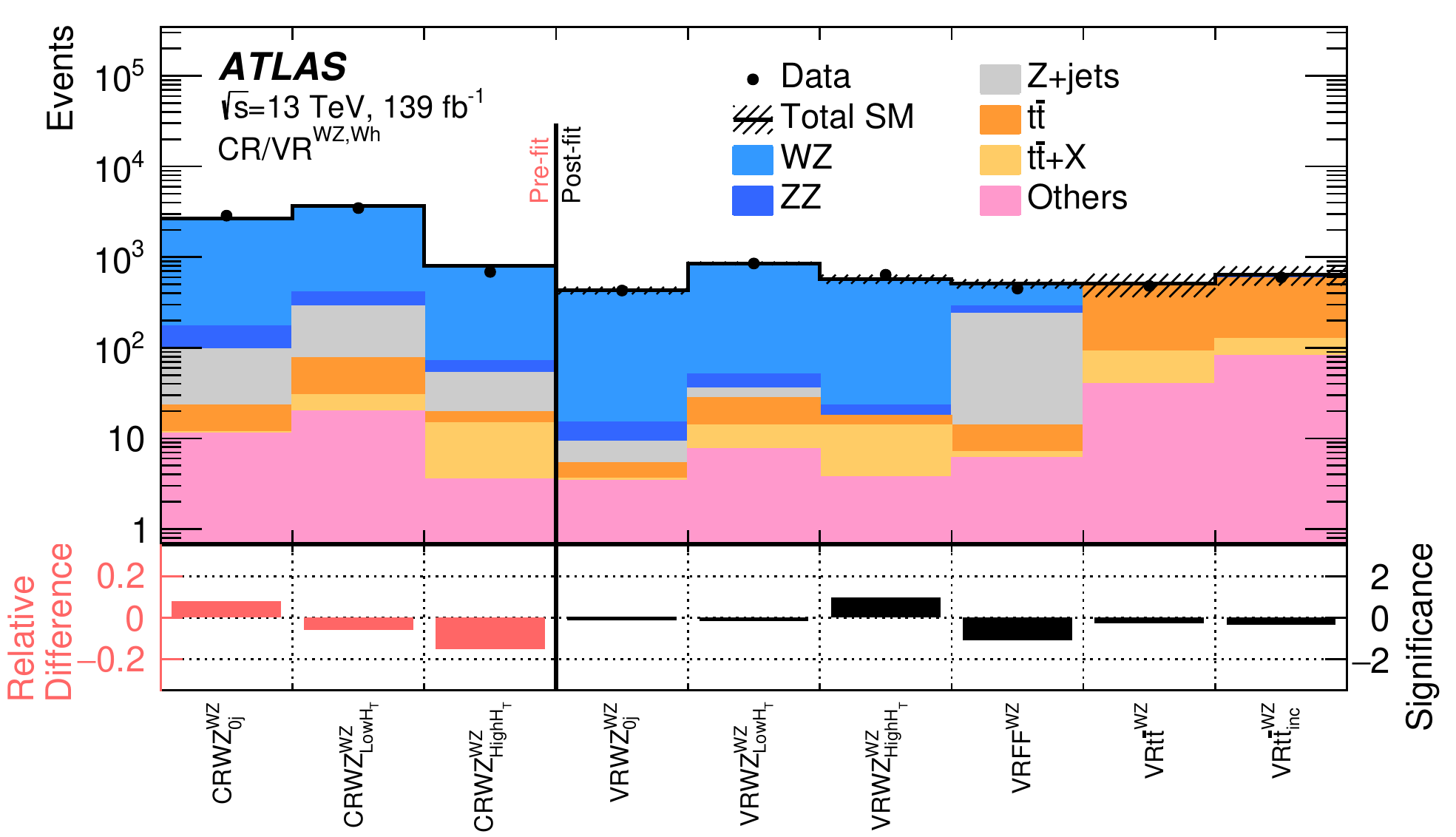}}
\caption{
Comparison of the observed data and expected SM background yields in the CRs (pre-fit) and VRs (post-fit) of the \ons \WZ and \Wh selections.
The `Others' category contains the single-top, \WW, triboson, Higgs and rare top processes.
The hatched band indicates the combined theoretical, experimental, and MC statistical uncertainties.
The bottom panel shows the relative difference between the observed data and expected yields for the CRs and the significance of the difference for the VRs,
calculated with the profile likelihood method from Ref.~\cite{Cousins:2007bmb}, adding a minus sign if the yield is below the prediction.
}
\label{fig:results:WZ_pull}
\end{figure}

The systematic uncertainties considered in the \onShell and the \Wh SRs follow the approach discussed in Section~\ref{ssec:analysisstrategy:systematics}.
The relative composition of FNP muons is similar between the \CRFF and \SRWZ,
whereas for FNP electrons the main source in the \SRWZ is photon conversions,
while in the \CRFF the heavy-flavour decay contribution dominates.
An additional source of uncertainty that is considered accounts for the different FNP lepton compositions in the \CRFF and \SRWZ.
This uncertainty arises from the method's performance in the simulation (closure) in various regions of parameter space
and is given by the differences between the estimated and simulated yields of events in the given region.
In the DFOS region where the triboson contribution becomes dominant,
the uncertainties related to the QCD renormalisation and factorisation scales are also evaluated
for this background component, in the same way as previously described for diboson and \ttbar.
A summary of the considered systematic uncertainties is presented in Figure~\ref{fig:sysWZ},
with uncertainties grouped as discussed in Section~\ref{ssec:analysisstrategy:systematics}.
 
Bin-to-bin fluctuations in the statistical uncertainty as well as the experimental uncertainty reflect the difference in expected yields in the various search regions, which varies by an order of magnitude.
These uncertainties become the dominant ones in \SRWZi{3--4, 6--8, 11--12} and \texttt{15--16} of the \ons \WZ selection,
and \SRWhSFi{5}, \SRWhSFi{14}, and \SRWhSFi{19} of the \Wh selection, due to limited number of MC events at high \met and \mt.
Although the FNP lepton uncertainty is negligible in the majority of the search bins, its relative size reaches 30\% in \SRWhDFi{2},
due to the small number of events in the corresponding anti-ID sample.

\begin{figure}[bt!]
\centering
\includegraphics[width=0.46\columnwidth]{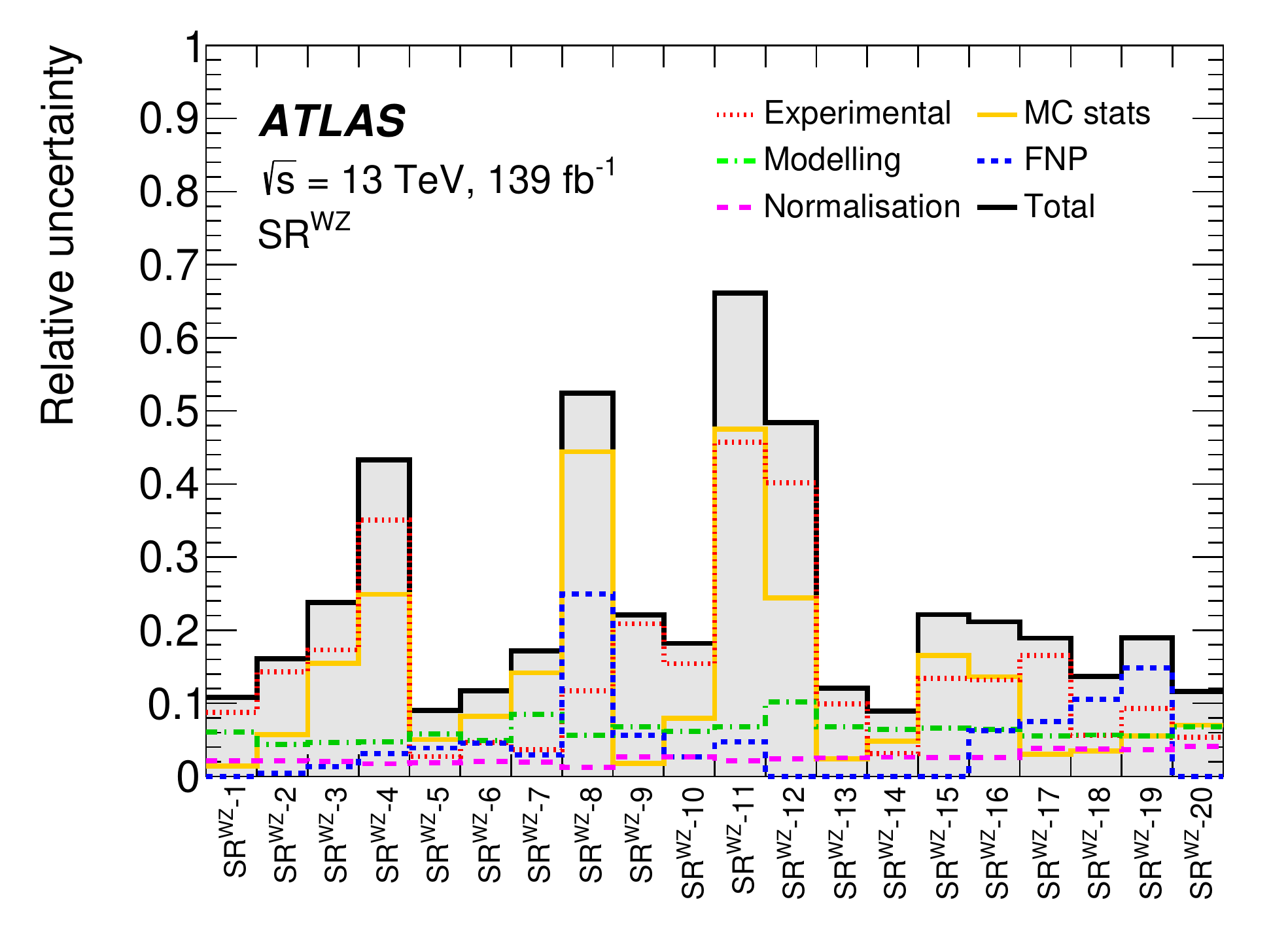}\hspace{0.1cm}
\includegraphics[width=0.46\columnwidth]{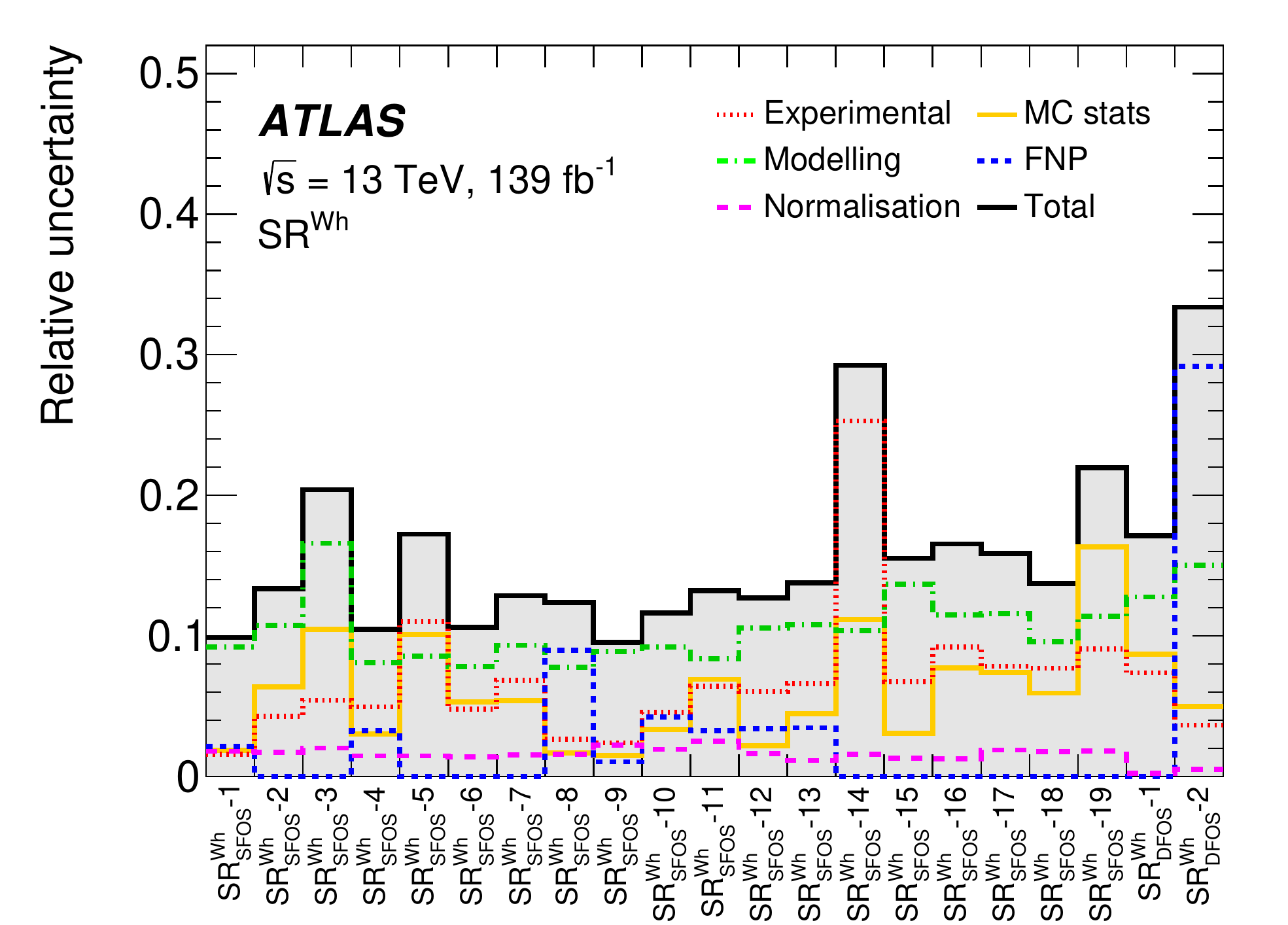}
\caption{
Breakdown of the total systematic uncertainties in the background prediction for the SRs of (left) the \ons \WZ selection and (right) the \Wh selection.
}
\label{fig:sysWZ}
\end{figure}
% End of text imported from the .//sections/onshellAndWh.tex input file
% The next lines are included from the .//sections/offshell.tex input file
\FloatBarrier
{
\begin{DIFnomarkup}
\vskip 1em\section{\OffShell selection}\label{sec:offshell}
\enlargethispage*{5\baselineskip}
\end{DIFnomarkup}
 
The following subsections discuss the implementation specific to the \ofs \WZ selection,
expanding on the general strategy outlined in Section~\ref{sec:analysisstrategy}. 
The selection is applied on top of the common preselection as defined in Section~\ref{sec:objdef},
and the SRs are optimised to the wino/bino (+) scenario.

\begin{DIFnomarkup}
\vskip 1em\subsection{Search regions} \label{ssec:offshell:reg}
\end{DIFnomarkup}
The \SRoffWZ selection targets the \ofs \WZ region 
by requiring $\mllmin, \mllmax < 75~\GeV$.
The \mllmax is the largest SFOS lepton pair invariant mass in the event,
and the double requirement helps to maximally suppress combinatorial backgrounds with an \ons $Z$ boson.
Further variables used in the \ofs \WZ selection assume \minmll-based lepton assignment to the $Z^{*}$- and $W^{*}$-boson candidates unless otherwise indicated.
The event preselection vetoes events with a \bjet to reduce contamination from \ttbar,
\begin{DIFnomarkup}
requires the three leptons to be well separated in \smash{$\mindRtl = \min[\Delta R(\ell_i,\ell_j); \text{for all lepton pairs}~\scalebox{0.9}{$(\ell_i,\ell_j)$}]$},
\end{DIFnomarkup}
and requires a lower bound on \minmll of 1~\GeV\ to remove  events with collimated leptons for which FNP lepton background estimation is challenging.
Finally, \mllmin mass ranges of $[3.0,3.2]$ and $[9,12]$~\GeV\ are vetoed to avoid contributions
from $J/\psi$ and $\Upsilon$ resonance backgrounds associated with a FNP lepton,
except in the jet-inclusive high \met regions ($\met > 200~\GeV$) where the contribution is negligible.
 
Preselected events are further divided into four categories based on the multiplicity of jets with $\pt>30~\GeV$ (\njetsth), and on \met.
Jet-veto categories \SRlowzj and \SRhighzj reject events containing jets and select low and high \met, respectively.
Jet-inclusive categories \SRlownj and \SRhighnj require at least one jet and also separate the events with low and high \met.
As the \met is harder in the jet-inclusive categories, due to the recoil between the $\textchinoonepmninotwo$ system and the jets,
the boundary between the low and high \met bins is set at 50~\GeV~for the jet-veto categories and at 200~\GeV~for the jet-inclusive categories.
The \SRlowzj, \SRlownj and \SRhighzj primarily target signals with moderate mass splitting ($\dm \sim [40,90]~\GeV$),
and rely mostly on moderate kinematics and lepton triggers.
The \SRhighnj also target signals with highly compressed mass spectra ($\dm \lesssim 40~\GeV$) -- resulting in events with very soft leptons --
by exploiting events with large \met recoiling against hard hadronic activity.
Initial lepton \pT requirements are kept as loose as possible, $\pT>10~\GeV$~for \SRlowzj, \SRlownj, and \SRhighzj, and $\pT>4.5\,(3.0)~\GeV$ for $e$ ($\mu$) in \nopagebreak\smash{\SRhighnj};
however, the selection is restricted by the trigger requirements (Section~\ref{sec:objdef}) and some further requirements are applied in the bin-by-bin SR optimisation as discussed in the following.
}
 
% The next lines are included from the .//tables/ofs_preselection.tex input file
\begin{table}[t!]
\begin{center}
\caption{
Summary of the preselection criteria applied in the SRs of the \ofs \WZ selection.
In rows where only one value is given it applies to all regions.
`-' indicates no requirement is applied for a given variable/region.
}
\label{tab:ofs:presel}
\adjustbox{max width=0.99\textwidth}{
\begin{tabular}{l@{\hskip -0.5ex}P{2.8cm}@{\hskip 2ex}P{2.8cm}@{\hskip 5.0ex}P{2.8cm}@{\hskip 2ex}P{3.2cm}@{\hskip 1.5ex}}
\hline\hline 
& \multicolumn{4}{c}{Preselection requirements} \\
\cline{2-3}\cline{4-5} 
\rule{0pt}{\dimexpr.7\normalbaselineskip+1mm}
Variable                        & \SRlowzj & \SRlownj & \SRhighzj & \SRhighnj \\[0.1cm]
\hline 
\nlbl, \nlsig                   & \multicolumn{4}{c}{= 3}                    \\
\nSFOS                          & \multicolumn{4}{c}{$\geq 1$}               \\
\mllmax [\GeV]                  & \multicolumn{4}{c}{$<75$}                  \\
\mllmin [\GeV]                  & \multicolumn{4}{c}{$\in [1,75]$}           \\
\nbjets                         & \multicolumn{4}{c}{$= 0$}                  \\
\mindRtl                        & \multicolumn{4}{c}{$>0.4$}                 \\
\hline\tabvspace 
Resonance veto \mllmin [\GeV]   & \multicolumn{3}{c|}{$\notin [3,3.2]$, $\notin [9,12]$} & - \\
Trigger                         & \multicolumn{2}{c|}{(multi-)lepton\hspace{5ex}\ } & \multicolumn{2}{c}{((multi-)lepton\ ||\ \met)} \\
\njetsth                        & $= 0$ & $\geq 1$ & $= 0$ & $\geq 1$                                                  \\
\met [\GeV]                     & $<50$  & $<200$ & $>50$ & $>200$                                                     \\
\metsig                         & $>1.5$ & $>3.0$ & $>3.0$ & $>3.0$                                                    \\
\ptl{1}, \ptl{2}, \ptl{3} [GeV] & \multicolumn{3}{c|}{$>10$} & \ $>4.5(3.0)$ \ for $e(\mu)$ \\
$|\mtl-\mZ|$ [\GeV]             & \multicolumn{2}{c|}{$>20$ {($\lW=e$ only)}\hspace{5ex}\ } & \multicolumn{2}{c}{-}       \\
\mindRossf                      & \multicolumn{2}{c|}{$[0.6,2.4]$ {($\lW=e$ only)}\hspace{5ex}\ } & \multicolumn{2}{c}{-} \\
\hline\hline 
\end{tabular}}
\end{center}
\end{table}
% End of text imported from the .//tables/ofs_preselection.tex input file
 
\pagebreak
Further preselection criteria are applied to reduce the contamination from \ZjetZgam.
First, a lower bound is set to ensure $\metsig>1.5$ or $3.0$, depending on the SR category.
For \SRlow, events are then treated separately for different flavours of the lepton from the $W$-boson decay (\lW),
selected using \mllmZ-based lepton assignment to best capture the SM background topology for rejection.
To suppress the contribution from  $Z(+\gamma) \rightarrow \ell\ell ee$ caused by bremsstrahlung from prompt electrons and subsequent photon conversions,
if \lW is an electron,
the trilepton invariant mass $\mtl$ is required to be off the $Z$-boson peak ($|\mtl-\mZ|>20~\GeV$),
and the minimum angular distance between all SFOS lepton pairs must be within $\mindRossf \in [0.6,2.4]$,
with \mindRossf defined as \begin{DIFnomarkup}$\min[\Delta R(\ell_{i},\ell_{j}); \text{for all SFOS lepton pairs}~\scalebox{0.9}{$(\ell_i,\ell_j)$}]$\end{DIFnomarkup}.
The preselection criteria and categorisation are \nolinebreak summarised in Table~\ref{tab:ofs:presel}.
 
The primary discriminant in \SRoffWZ is \mllmin.
This variable serves as a proxy for the mass splitting of the targeted signals,
and displays a characteristic kinematic edge at their mass-splitting value: $\mllmin = \dm$,\linebreak
as demonstrated in Figure~\ref{fig:analysisstrategy:ofs:mll_and_mt2}.
A shape fit over the \mllmin spectrum is performed in each SR category.\linebreak
Seven \mllmin bins are defined with boundaries at 1, 12, 15, 20, 30, 40, 60 and 75~\GeV, and labelled \SRj{a} to \SRj{g};
the \mllmin bin labels are added to the region names as defined above.
Signal regions \SRj{a}~are dropped everywhere except in \SRhighnj,
to avoid low-mass resonance backgrounds.
 
\newcommand{\ptvec}{\mathbf{p}_\mathrm{T}}
\newcommand{\qtvec}{\mathbf{q}_\mathrm{T}}
A second, similar kinematic edge is present in stransverse mass \mttwo~\cite{Lester:1999tx,Barr:2003rg},
reflecting the kinematic constraint originating from the $\textchinoonepm \ra \Wstar \textninoone$ decay chain.
In this selection, \mttwo is constructed by assigning the dilepton system providing \mllmin ($\ell_1\ell_2$) to one visible particle leg,
and the remaining lepton ($\ell_3$) to the other leg:
\begin{linenomath*}
\begin{equation*}
\mttwo^{m_\chi} \left(\ptvec^{\ell_1\ell_2},\ptvec^{\ell_3},\ptvec^\mathrm{miss}\right)
= \underset{\qtvec}{\min}
\left(\max\left[
\mt\left(\ptvec^{\ell_1\ell_2},\qtvec,m_\chi\right),
\mt\left(\ptvec^{\ell_3},\ptmissvec-\qtvec,m_\chi\right)
\right]\right),
\end{equation*}
\end{linenomath*}
where the transverse mass \mT in this \mttwo formula is defined by
\begin{linenomath*}
\begin{equation*}
\mt\left(\ptvec^{\ell},\qtvec,m_\chi\right)
= \sqrt{ m_\ell^2 + m_\chi^2
+ 2\left(\sqrt{(\pt^{\ell})^2 + m_\ell^2} \sqrt{q_\mathrm{T}^2 + m_\chi^2} -
\ptvec^{\ell} \cdot \qtvec
\right)
}.
\end{equation*}
\end{linenomath*}
A hypothesised mass $m_\chi$ is assigned to each invisible particle leg, corresponding to the $\textninoone$ mass;
$m_\chi$ is fixed to 100~\GeV\ in this selection.\footnote{The dependency of the performance on hypothetical invisible particle mass \smash{$m_\chi$} is generally small
except when assuming \smash{$m_\chi \sim 0~\GeV$} for signals with finite \textninoone mass, where the signal kinematic edges become significantly smeared.}
The kinematic edge for signals appears at $\mttwoh = \dm(\textninotwo,\textninoone) + 100~\GeV$ as illustrated in Figure~\ref{fig:analysisstrategy:ofs:mll_and_mt2}.
To take advantage of this feature, a sliding cut is applied per \mllmin bin, requiring \mttwoh to be smaller than the upper \mllmin bin edge\ +\ 100~\GeV.
SM backgrounds can exceed the boundary and are suppressed, while a large fraction of the signal contribution targeted by a given bin is retained.
The cut is particularly effective in the lowest \mllmin bins, targeting the smallest mass splittings:
e.g.\ in \SRhighnji{a} ($\mllmin \in [1,12]~\GeV$) the total background is reduced by a factor of three following $\mttwoh<112~\GeV$,
while the efficiency for $\dm=10~\GeV$~signals is $>95\%$.
 
\begin{figure}[t!]
\centering
\includegraphics[width=0.48\columnwidth]{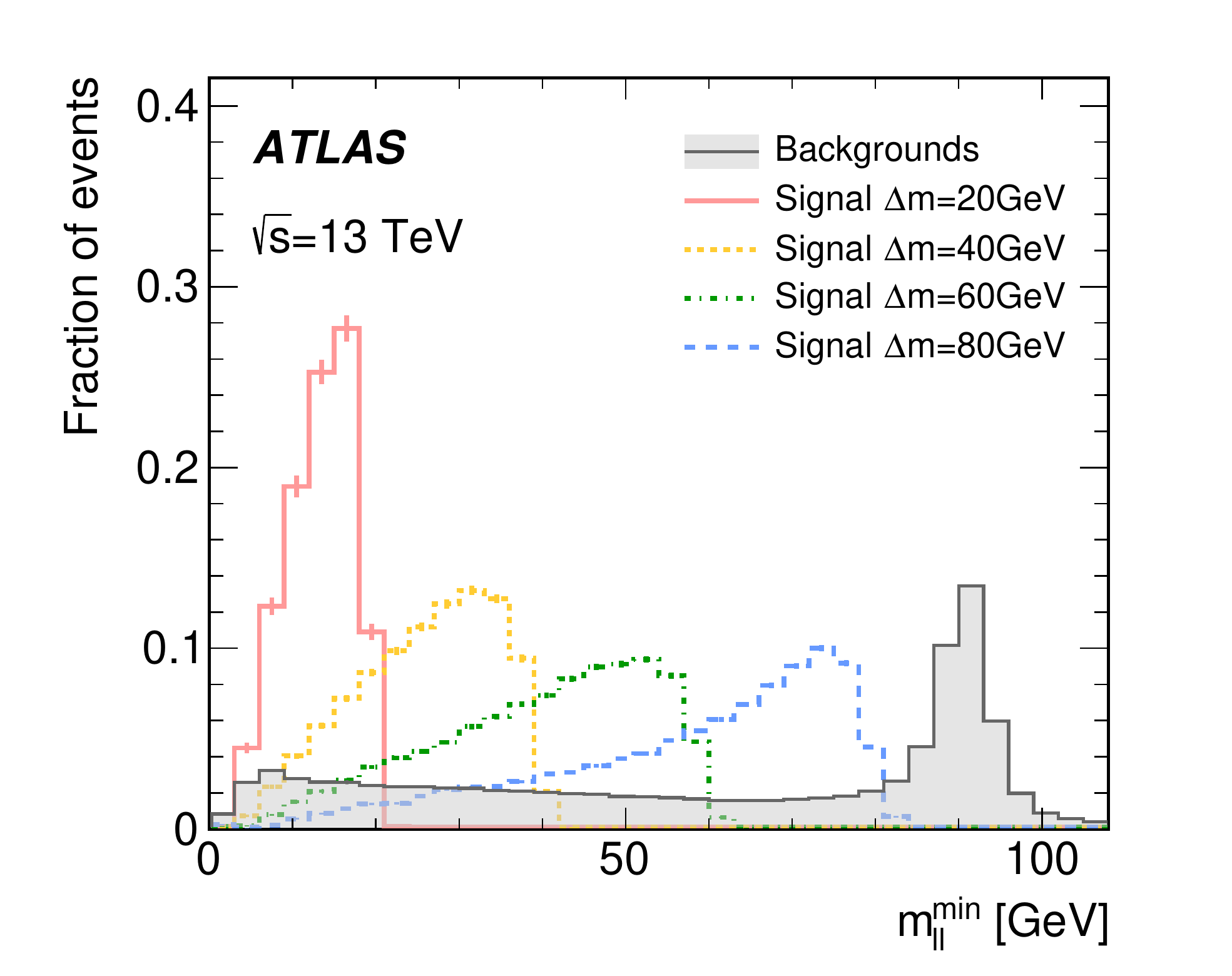}
\includegraphics[width=0.48\columnwidth]{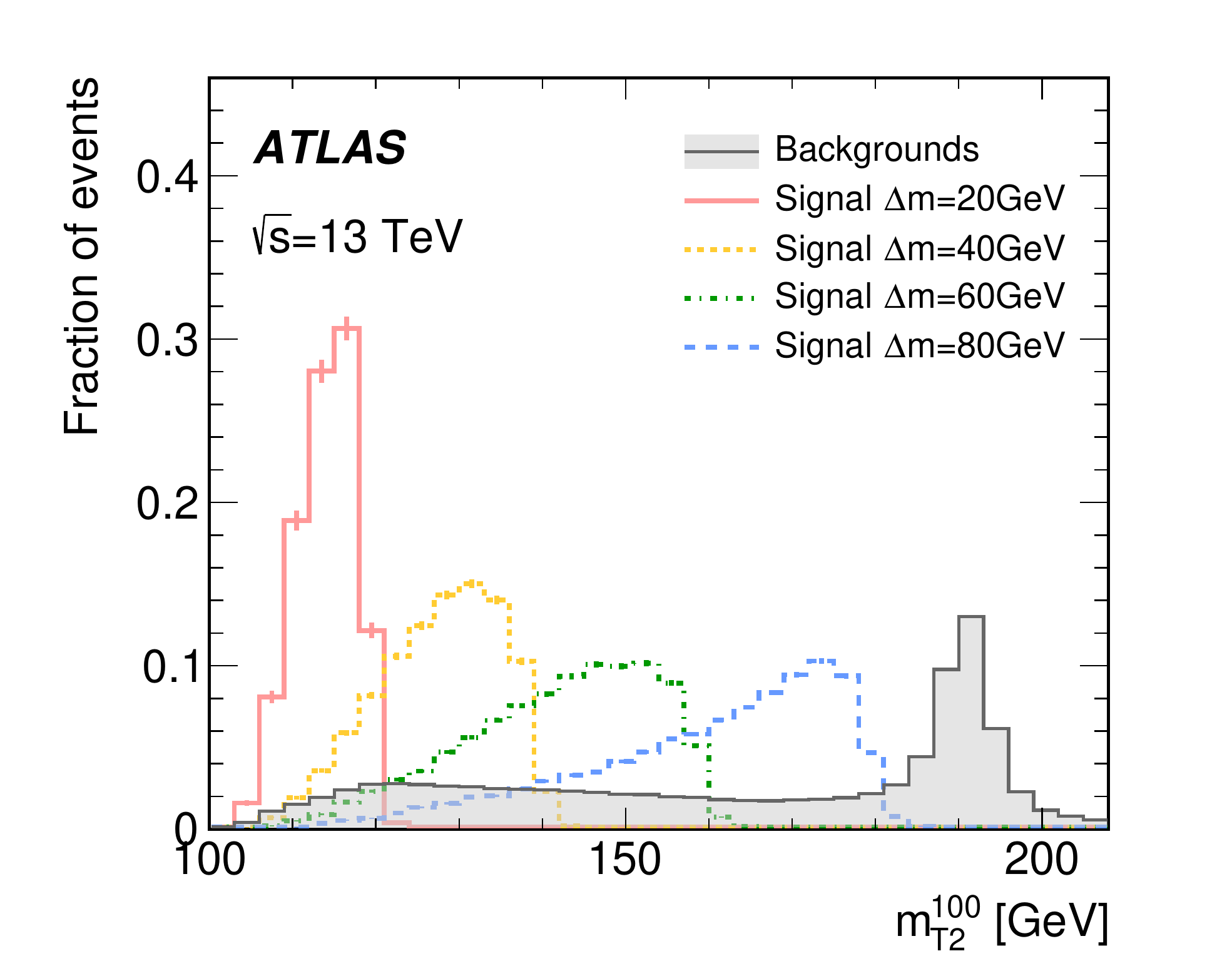}
\caption{
Distributions of (left) \mllmin and (right) \mttwoh showing the expected SM background as well as signals with various mass splittings \dmNN
(\smash{$m(\hspace{1pt}\raisebox{-2pt}{\chinoonepm}\hspace{-2pt})=m(\hspace{1pt}\raisebox{-2pt}{\ninotwo}\hspace{-2pt})=200~\GeV$}),
for a selection of exactly three baseline and signal leptons.
The distributions are normalised to unity.
Signals demonstrate a cut-off in both variables matching the mass splitting, while backgrounds do not.
The dominant background in this selection is \WZ, with the $Z$-boson mass peak visible in both distributions.
}
\label{fig:analysisstrategy:ofs:mll_and_mt2}
\end{figure}
 
Event selection is tightened further by employing various background rejection criteria,
optimised separately for each \SRoffWZ category and each \mllmin bin.
The discriminating variables used and the detailed bin-by-bin cut values are summarised in Table~\ref{tab:ofs:sel1}.
 
% The next lines are included from the .//tables/ofs_selectionSR.tex input file
\begin{table}[t!]
\begin{center}
\caption{
Summary of the selection criteria for SRs for the \ofs \WZ selection.
\SRoffWZ preselection criteria are applied (Table~\ref{tab:ofs:presel}).
`-' indicates no requirement is applied for a given variable/region,
while $\times$ is marked for regions that aren't considered.
}
\label{tab:ofs:sel1}
\adjustbox{max width=0.99\textwidth}{
\begin{tabular}{l*{9}{P{1.3cm}}}
\hline\hline 
& \multicolumn{9}{c}{Selection requirements} \\
\cline{2-10} 
Variable        & \texttt{a} & \texttt{b} & \texttt{c} & \texttt{d} & \texttt{e} & \texttt{f1} & \texttt{f2} & \texttt{g1} & \texttt{g2} \\
\hline 
\mllmin [\GeV]  & [1, 12] & [12, 15] & [15, 20] & [20, 30] & [30, 40] & \multicolumn{2}{c}{[40, 60]} & \multicolumn{2}{c}{[60, 75]} \\
\hline\hline 
\rule{0pt}{\dimexpr.7\normalbaselineskip+1mm}
& \multicolumn{9}{c}{\rSR{\rWZof}{\rlo\rmet}{}~common} \\[0.1cm]
\cline{2-10} 
\mllmax [\GeV]     & $\times$ & $<60$ & $<60$ & $<60$ & $<60$ & - & - & - & - \\
\mtmllmin [\GeV]   & $\times$ & $<50$ & $<50$ & $<50$ & $<60$ & $<60$ & $>90$ & $<60$ & $>90$ \\
\mttwoh [\GeV]     & $\times$ & $<115$ & $<120$ & $<130$ & - & - & - & - & - \\
\mindRossf         & $\times$ & $<1.6$ & $<1.6$ & $<1.6$ & - & - & - & - & - \\
\ptl{1},\ptl{2},\ptl{3} [\GeV] & $\times$ & $>10$ & $>10$ & $>10$ & $>10$ & $>15$ & $>15$ & $>15$ & $>15$ \\
\cline{2-10} 
\rule{0pt}{\dimexpr.7\normalbaselineskip+1mm}
& \multicolumn{9}{c}{\rSR{\rWZof}{\rlo\rmet}{-\rzj}} \\[0.1cm]
\cline{2-10} 
\ptlepovermet      & $\times$ & $<1.1$& $<1.1$  & $<1.1$ & $<1.3$ & $<1.4$ & $<1.4$ & $<1.4$ & $<1.4$ \\
\mtl [\GeV] & $\times$ & - & - & - & - & $>100$ & $>100$ & $>100$ & $>100$ \\
\cline{2-10} 
\rule{0pt}{\dimexpr.7\normalbaselineskip+1mm}
& \multicolumn{9}{c}{\rSR{\rWZof}{\rlo\rmet}{-\rnj}} \\[0.1cm]
\cline{2-10} 
\ptlepovermet      & $\times$ & $<1.0$& $<1.0$  & $<1.0$ & $<1.0$ & $<1.2$ & $<1.2$ & $<1.2$ & $<1.2$ \\
\hline\hline 
\rule{0pt}{\dimexpr.7\normalbaselineskip+1mm}
& \multicolumn{9}{c}{\rSR{\rWZof}{\rhi\rmet}{}~common} \\[0.1cm]
\cline{2-10} 
\mttwoh [\GeV]     & $<112$ & $<115$ & $<120$ & $<130$ & $<140$ & $<160$ & $<160$ & $<175$ & $<175$ \\
\cline{2-10} 
\rule{0pt}{\dimexpr.7\normalbaselineskip+1mm}
& \multicolumn{9}{c}{\rSR{\rWZof}{\rhi\rmet}{-\rzj}} \\[0.1cm]
\cline{2-10} 
\ptl{1},\ptl{2},\ptl{3} [\GeV] & $\times$ & \multicolumn{7}{c}{$>25$, $>15$, $>10$} & \\
\mtmllmin [\GeV]   & $\times$ & $<50$ & $<50$ & $<60$ & $<60$ & $<70$ & $>90$ & $<70$ & $>90$ \\
\cline{2-10} 
\rule{0pt}{\dimexpr.7\normalbaselineskip+1mm}
& \multicolumn{9}{c}{\rSR{\rWZof}{\rhi\rmet}{-\rnj}} \\[0.1cm]
&   &   &   &   &   & \multicolumn{2}{c}{f} & \multicolumn{2}{c}{g} \\
\cline{2-10} 
\ptl{1},\ptl{2},\ptl{3} [\GeV] & & \multicolumn{7}{c}{$>4.5\,(3.0)$ \ for $e\,(\mu)$} & \\
\ptlepovermet      & $<0.2$ & $<0.2$ & $<0.3$ & $<0.3$ & $<0.3$ & \multicolumn{2}{c}{$<1.0$} & \multicolumn{2}{c}{$<1.0$} \\
\hline\hline 
\end{tabular}}
\end{center}
\end{table}
% End of text imported from the .//tables/ofs_selectionSR.tex input file
 
In order to reduce the FNP lepton background contributions from \ZjetZgam and \ttbar,
lepton \pt thresholds are raised in \SRlowzj, \SRlownj and \SRhighzj.
In these same three categories, the transverse mass \mtmllmin is used to suppress the SM \WZ contribution;
the \mtmllmin variable is constructed using the $W$ lepton after \mllmin-based lepton assignment
and marked with `$\scalebox{0.8}{\text{mllmin}}$' to distinguish it from the \mT variable in the \ons \WZ selection.
The SRs target phase space either below or above the SM $W$-boson peak present at $\mtmllmin \sim m_W$.
An upper bound of $\mtmllmin <$ 50--70~\GeV\ is applied in low \mllmin bins,
while the \SRj{f} and \SRj{g} bins are split into two parts below (\SRj{f1}, \SRj{g1}) and above (\SRj{f2}, \SRj{g2}) the Jacobian peak of SM \WZ.
 
In \SRlow, the selection on \mindRossf is tightened in the low \mllmin bins, exploiting the topology with a relatively boosted $Z^{*}$ in the target signatures,
and a lower bound on \mtl is applied for the high \mllmin bins to reject the SM $Z\rightarrow 4\ell$ background peaking at $\mtl \sim m_Z$.
The ratio of the magnitude of a vectorial \pT sum of the three leptons, $\ptlep$, to \met, is labelled $\ptlepovermet$ and
represents the extent to which the transverse momentum of the hard-scatter $\textchinoonepm\textninotwo$ system, recoiling against ISR jets, is converted into leptons as opposed to \met.
Due to the presence of a massive $\textninoone$, contributing to the \met, signals tend to populate lower parts of the \ptlepovermet spectrum than SM backgrounds,
particularly for the compressed signals in the high \met regions where the \met is almost fully generated by the ISR jets.
A tight upper bound $\ptlepovermet$ is therefore imposed in the low \mllmin bins of \SRhighnj.
 
After applying the selection criteria,
for the wino/bino (+) model \begin{DIFnomarkup}\raisebox{0.1em}{\textchinoonepmninotwo}\end{DIFnomarkup}~signal sample with NLSP masses
of 200~\GeV\ and a mass splitting of $\dm=20$~\GeV,
the \SRlowzj, \SRlownj, \SRhighzj, and \SRhighnj regions (taking the union of the bins inside each region)
have acceptance times efficiency values of
$2.2\times 10^{-5}$, $1.1\times 10^{-5}$, $3.4\times 10^{-6}$, and $6.0\times 10^{-5}$, respectively.
Similarly, for a mass splitting of $\dm=60$~\GeV, values of
$1.6\times 10^{-4}$, $1.7\times 10^{-4}$, $2.8\times 10^{-4}$, and $7.9\times 10^{-5}$ are found.
The acceptance times efficiency values for the wino/bino ($-$) and higgsino model signal samples
are typically 15\%--55\% and 20\%--60\% lower, depending on the region.
 
\begin{DIFnomarkup}
\FloatBarrier
\pagebreak
\vskip 1.0em\subsection{Background estimation} \label{ssec:offshell:bkg}
\end{DIFnomarkup}

{
The selection criteria for the CRs and the VRs for \WZ estimation are summarised in Table~\ref{tab:bg:offShellCRVR}.
An \ons $Z$ boson ($\mll \in [81,101]~\GeV$) is required to ensure orthogonality to the \SRoffWZ,
and an upper bound on \met ensures orthogonality to the \SRonWZ.
A lower bound on $\mT$ is applied to suppress the \ZjetZgam background.
The CRs are further split into two bins (\CRofWZzj and \CRofWZnj), based on the absence or presence of jets,
to constrain \WZ events without or with hard ISR jets separately with individual normalisation factors.
 
Three validation regions are defined in the region with $\mllmin,\mll<75~\GeV$, similar to \SRoffWZ.
First, \VRofWZzj and \VRofWZnj are designed to validate the \WZ estimation in the \SRlow phase space.
A window in \mt around the Jacobian peak ($\mT \in [60,90]~\GeV$) is selected to enhance \WZ,
as well as to ensure the orthogonality with respect to the SRs.
Further kinematic selection criteria similar to those in \SRlow are applied.
Two additional variables are employed in the \VRofWZzj to suppress the signal contamination in the region.
The $W$-boson mass, \mWPZB, is reconstructed assuming the \WZ topology and balanced longitudinal momenta of the $W$ and $Z$ bosons,
and \DRlWMET is defined by $\sqrt{\scalebox{0.85}{$\eta_{\ell_{\W}}^2 + \Delta \phi(\ell_{\W}, \met)^2$}}$
where leptons are assigned according to the \minmll approach, and $\ell_{\W}$ is the lepton associated with the $W$ boson.
Since \mWPZB peaks around \mW with a long tail to higher masses for \WZ background, while signals tend to have a flatter distribution, $\mWPZB > 75~\GeV$ is found to effectively reduce signal contamination.
}
 
{
In the very low \mllmin region, \VRofWZnjmll is used to validate the \WZ estimation in the \SRhigh phase space.
This region has the low-mass resonance veto applied and a lower bound on \ptlepovermet to ensure orthogonality with the SRs.
Other kinematic cuts are loosened relative to \SRhigh, or removed entirely, to increase the number of data events in the region.
The \WZ purity is 85\%--90\% in the CRs and 70\%--75\% in the VRs.
The contamination from the benchmark signals is negligible in the CRs and below 15\% in the VRs.

The \VRoftt selection criteria are summarised in Table~\ref{tab:bg:offShellCRVR}.
At least one $b$-jet is required to maintain orthogonality with the SRs,
$\met>50~\GeV$ is required to suppress the \ZjetZgam contribution,
and the low-mass resonance background veto is applied. 
The \ttbar purity in this region is approximately 65\%.

The \ZjetZgam background is estimated using the FF method as described in Section~\ref{ssec:analysisstrategy:bkgs}.
The FF measurement region for the \offShell selection, \CRofFF, is summarised in Table~\ref{tab:bg:offShellCRVRFF}.
The $Z$-boson candidate is selected by requiring $|\mll-\mZ|<15~\GeV$, and
$\met<40~\GeV$~and $\mt<30~\GeV$~are applied to reject contamination from \WZ.
Additionally $\mtl>105~\GeV$~is applied to suppress $Z\rightarrow 4\ell$.
To increase the number of FNP lepton candidates at high $\pt$,
the overlap removal procedure described in Section~\ref{sec:objdef} is modified for this FF measurement so that muons overlapping with jets are always kept.
Finally, a jet veto is applied except for events where the FNP lepton candidate is a muon with $\pt>30~\GeV$,
in which case $\njetsth \le 1$ is required in order to account for the special muon-vs-jet overlap-removal treatment applied to this region.
 
The FFs are derived separately per lepton flavour of FNP lepton candidates and per signal lepton criterion, i.e.\ with or without applying the non-prompt BDT,
and are parameterised as a function of lepton \pt and \met in the event.
Typical FF values are 0.2--0.4 (0.2--0.6) without the BDT applied,
and 0.05--0.2 (0.07--0.2) when applying the BDT,
for electrons (muons) in a \pt range of 4.5--30 (3.0--30)~\GeV.
The parameterisation in \met is used to reflect the variation of FNP lepton source with \met, which is required in order to model the shape of fake \met correctly.
Typically the fraction of FNP leptons originating from heavy-flavour decays varies with \met,
in the range 20\%--30\% (60\%--70\%) for electrons (muons),
because of the neutrinos from the leptonic $b$-/$c$-decays.
 
% The next lines are included from the .//tables/ofs_CRVRWZ.tex input file
\begin{table}[p!]
\begin{center}
\caption{Summary of the selection criteria for the CRs and VRs for \WZ and \ttbar, for the \ofs \WZ selection.
In rows where only one value is given it applies to all regions.
`-' indicates no requirement is applied for a given variable/region.
}
\label{tab:bg:offShellCRVR}
\adjustbox{max width=\textwidth}{
\begin{tabular}{l@{\hspace{-0.5ex}}*{2}{P{1.6cm}}|*{2}{P{1.6cm}}c|c}
\hline\hline 
\tabvspace
Variable                        & \CRofWZzj & \CRofWZnj              & \VRofWZzj    & \VRofWZnj         & \VRofWZnjmll              & \VRoftt \\[0.1cm]
\hline 
\nlbl, \nlsig                   & \multicolumn{6}{c}{= 3}                    \\
\nSFOS                          & \multicolumn{6}{c}{$\geq 1$} \\
Trigger                         & \multicolumn{6}{c}{((multi-)lepton\ ||\ \met)} \\
\mindRtl                        & \multicolumn{6}{c}{$>0.4$}                 \\
\nbjets                         & \multicolumn{2}{c|}{$= 0$}         & \multicolumn{3}{c|}{$= 0$}                                   & $\geq 1$ \\
\mll    [\GeV]                  & \multicolumn{2}{c|}{$\in [81,101]$}& \multicolumn{3}{c|}{$<75$}                                   & $<75$ \\
\njetsth                        & $= 0$      & $\geq 1$              & $ =0$         & $\geq 1$          & $\geq 1$                 & - \\
\met [\GeV]                     & $<50$     & $<50$                  & $<50$        & $<80$             & $>80$                     & $>50$ \\
\metsig                         & \multicolumn{2}{c|}{-}             & $>1.5$       & $>1.5$            & $>1.5$                           & - \\
\mt  [\GeV]                     & $>50$     & $>50$                  & \multicolumn{2}{c}{$\in[60,90]$} & $>30$                     & - \\
\mllmin    [\GeV]               & \multicolumn{2}{c|}{-}             & \multicolumn{2}{c}{$\in[12,75]$} & $\in[1,12]$               & $\in [1,75]$ \\
Resonance veto \mllmin [\GeV]   & \multicolumn{2}{c|}{-}             & -            & -                 & $\notin [3,3.2]$, $\notin [9,12]$ &  $\notin [3,3.2], \notin [9,12]$ \\
\ptl{1}, \ptl{2}, \ptl{3} [\GeV]& \multicolumn{2}{c|}{$>10$}         & $>10$        & $>10$             & -                         & - \\
\mindR                          & \multicolumn{2}{c|}{-}             & \multicolumn{2}{c}{$[0.6,2.4]$ {($\ell_{\W}=e$ only)}} & - & -\\
$|\mtl-\mZ|$ [\GeV]             & \multicolumn{2}{c|}{-}             & \multicolumn{2}{c}{$>20$ {($\ell_{\W}=e$ only)}} & -       & - \\
\mWPZB       [\GeV]             & \multicolumn{2}{c|}{-}             & $>75$        & -                 & -                         & - \\
\DRlWMET                        & \multicolumn{2}{c|}{-}             & $>2.6$       & -                 & -                         & - \\
\ptlepovermet                   & \multicolumn{2}{c|}{-}             & -            & -                 & $>0.3$                    & - \\
\hline\hline 
\end{tabular}}
\end{center}
\end{table}
% End of text imported from the .//tables/ofs_CRVRWZ.tex input file
 
% The next lines are included from the .//tables/ofs_CRVRFF.tex input file
\begin{table}[p!]
\begin{center}
\caption{Summary of the selection criteria for the CRs and VRs for \ZjetZgam, for the \ofs \WZ selection.
The corresponding anti-ID regions used for the \ZjetZgam prediction follow the same selection criteria,
except that at least one of the leptons is anti-ID instead of signal.
`-' indicates no requirement is applied for a given variable/region.
}
\label{tab:bg:offShellCRVRFF}
\adjustbox{max width=\textwidth}{
\begin{tabular}{lP{3.6cm}P{2.4cm}|*{3}{P{2.0cm}}}
\hline\hline 
\tabvspace
Variable                        & \CRofFF         & \CRoftt                  & \VRofFFzj       & \VRofFFnj      & \VRofFFnjpt \\[0.1cm]
\hline
\nlbl, \nlsig                   & \multicolumn{5}{c}{= 3}                    \\
\mindRtl                        & \multicolumn{5}{c}{$>0.4$}                 \\
Trigger                         & \multicolumn{1}{c|}{dilepton} &  \multicolumn{4}{c}{((multi-)lepton\ ||\ \met)} \\
\nSFOS                          & $\geq 1$        & $= 0$                     & \multicolumn{3}{c}{$\geq 1$}   \\
\nbjets                         & $= 0$           & $= 0$ or $\geq 1$         & \multicolumn{3}{c}{$= 0$}       \\
\mll    [\GeV]                  & $\in[\mZ-15,\mZ+15]$   & -                        & \multicolumn{3}{c}{$<75$}      \\
\njetsth                        & $\leq 1$ if $\pt^{\ell_{\W}=\mu}>30~\GeV$ & - & $= 0$       & $\geq 1$       & $\geq 1$ \\
& $= 0$ otherwise  &                          &                 &                & \\
\met [\GeV]                     & $<40$           & $>50$                    & $<50$           & $<200$         & $\in[50,200]$ \\
\metsig                         & -               & -                        &  $\in[0.5,1.5]$ & $\in[0.5,3.0]$ & $\in[0.5,3.0]$ \\
\ptl{1}, \ptl{2}, \ptl{3} [\GeV]& -               & $>10$                    &  $>10$          & $>10$          & $<10$ \\
\mllmin [\GeV]                  & -               & -                        &  $\in[12,75]$   & $\in[12,75]$   & $\in[1,75]$ \\
\mt  [\GeV]                     & $<30$           & -                        &  \multicolumn{3}{c}{$<50$} \\
\mindR                          & -               & -                        &  \multicolumn{3}{c}{$[0.6,2.4]$ {($\ell_{\W}=e$ only)}} \\
\mtl [\GeV]                     & $>105$          & -                        &  \multicolumn{3}{c}{$[81.2,101.2]$ {($\ell_{\W}=e$ only)}} \\
\hline\hline 
\end{tabular}}
\end{center}
\end{table}
% End of text imported from the .//tables/ofs_CRVRFF.tex input file
 
\begin{DIFnomarkup}\FloatBarrier\pagebreak\end{DIFnomarkup}
The contribution of non-\ZjetZgam processes is subtracted using MC simulated samples.
A small normalisation correction is applied to the \ttbar events in the simulated anti-ID region
to account for the different \antiID lepton efficiencies in data and MC simulation.
Normalisation factors are derived separately depending on the \lW flavour and the $b$-jet multiplicity in the event.
They are measured using the data events in a \ttbar-enriched control region, \CRoftt, and are found to be between 0.88 and 0.95.
The \CRoftt selection requires there to be no SFOS lepton pair in the event, as well as $\ptl{3}>10~\GeV$ and $\met>50~\GeV$ to enhance the \ttbar purity.

Two sources of uncertainty specific to the estimation in \SRoffWZ are considered in addition to those described in Section~\ref{ssec:analysisstrategy:systematics}.
The FF parameterisation uncertainty is evaluated from the effect of using a different \met binning (\met$<50$~\GeV, 50\% larger bin size),
or a 3D parameterisation in lepton \pT, \met and lepton $\eta$, additionally taking into account the dependency on lepton $\eta$.
The impact on the \ZjetZgam background yields in the CRs is $\sim 5\%$, and 1\%--7\% in the SRs/VRs.
The uncertainty from disabling the muon-vs-jet overlap removal procedure in the FF measurement region
is assessed by comparing those FFs with alternative FFs measured with muon-vs-jet overlap removal applied for events with a FNP muon candidate of $\pt<30~\GeV$.
The variation in the estimated \ZjetZgam yields in the SRs/CRs/VRs is found to be 5\%--15\%.
 
The yields predicted by the FF method are cross-checked in dedicated VRs enriched in FNP lepton backgrounds, as summarised in Table~\ref{tab:bg:offShellCRVRFF}.
The \metsig selection is inverted with respect to the SRs to ensure orthogonality.
First, \VRofFFzj and \VRofFFnj are designed to validate the yields in \SRlowzj and \SRlownj, respectively,
while \VRofFFnjpt aims to cross-check the modelling of FNP leptons with $\pt<10~\GeV$ specifically.
The \ZjetZgam purity is in the VRs is 50\%--80\%, while the contamination from signals is negligible.
}

\begin{DIFnomarkup}\FloatBarrier\end{DIFnomarkup}
{
Performing the background-only fit, \WZ normalisation factors of
$1.06\pm0.03$ (\CRofWZzj)
and
$0.93\pm0.03$ (\CRofWZnj)
are determined.
Examples of kinematic distributions in the CRs and VRs, demonstrating good agreement,
are presented in Figure~\ref{fig:offShell:CRVR}.
Observed and expected yields for all CRs and VRs are summarised in Figure~\ref{fig:results:ofWZ_pull}. 
 
The systematic uncertainties considered in the \offShell selection 
are summarised in Figure~\ref{fig:offShell:summary_syst},
grouped as discussed in Section~\ref{ssec:analysisstrategy:systematics}.
As the expected yields can vary by an order of magnitude throughout the regions, bin-to-bin fluctuations are expected in both the statistical and experimental uncertainty;
these uncertainties are often dominant in bins with limited MC statistics in the phase space of the selection.
The FNP lepton uncertainty is naturally more important in bins with larger FNP lepton background contributions,
and can fluctuate in bins with few events in the corresponding anti-ID sample, such as \SRhighzji{b} and \SRhighnji{b}.
The modelling uncertainty is larger in the presence of ISR jets and at higher values of \met; the fluctuation in \SRlownji{g2} originates from the effect of the QCD scale uncertainty on the \WZ background.
\begin{DIFnomarkup}\vskip 0.5em\end{DIFnomarkup}

\begin{figure}[bh!]
\centering
\includegraphics[width=0.49\columnwidth]{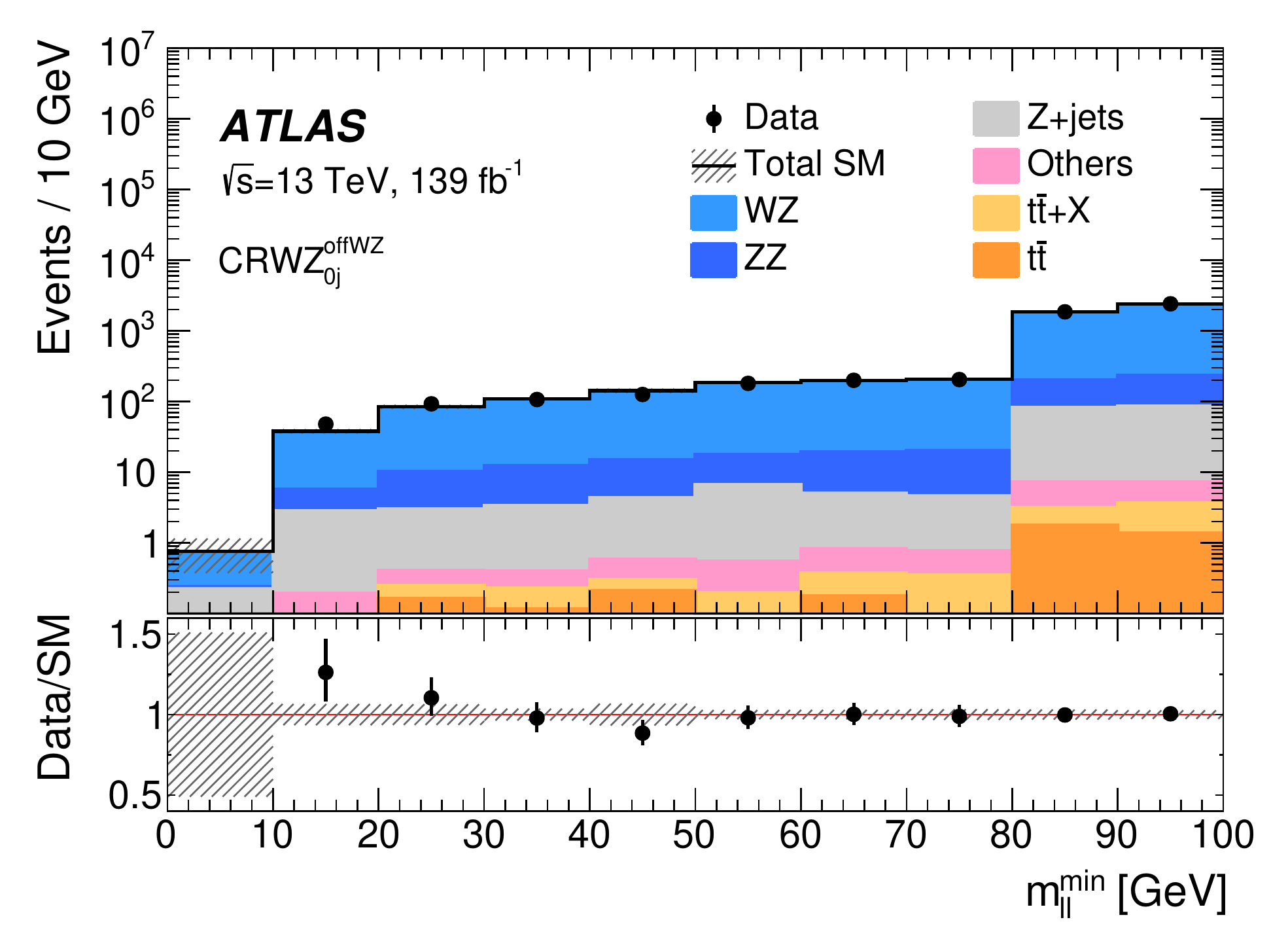}
\includegraphics[width=0.49\columnwidth]{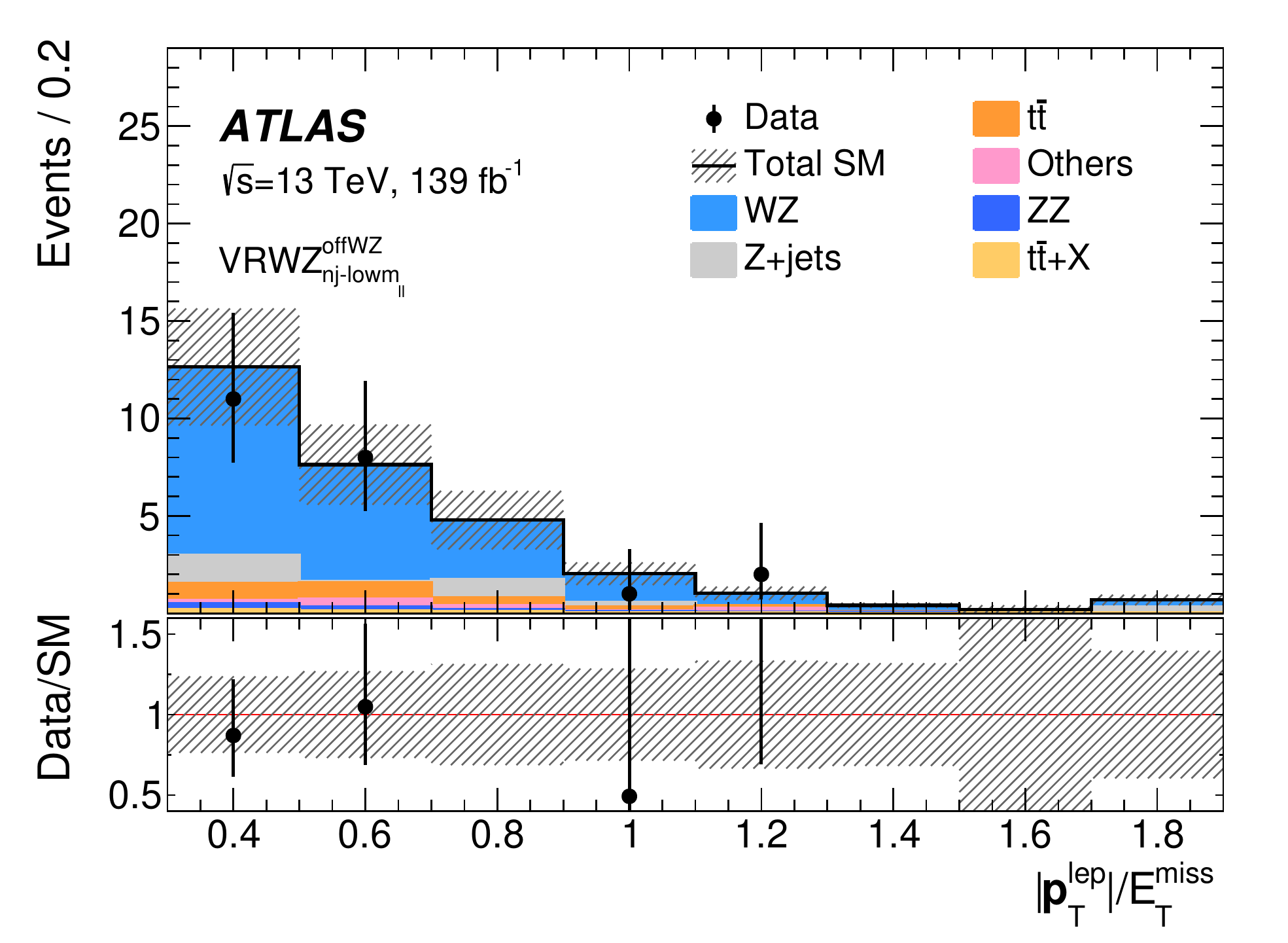}
\includegraphics[width=0.49\columnwidth]{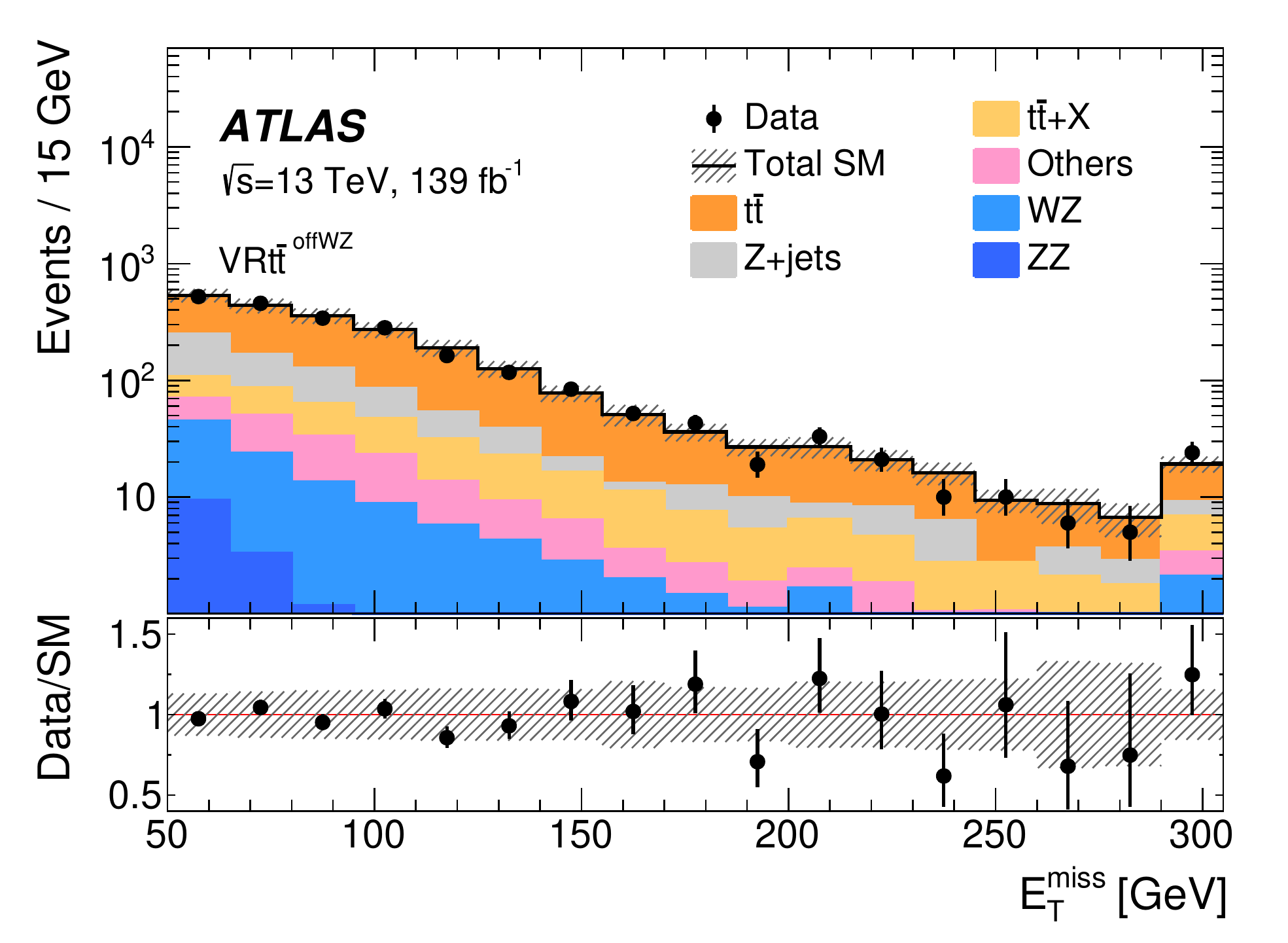}
\includegraphics[width=0.49\columnwidth]{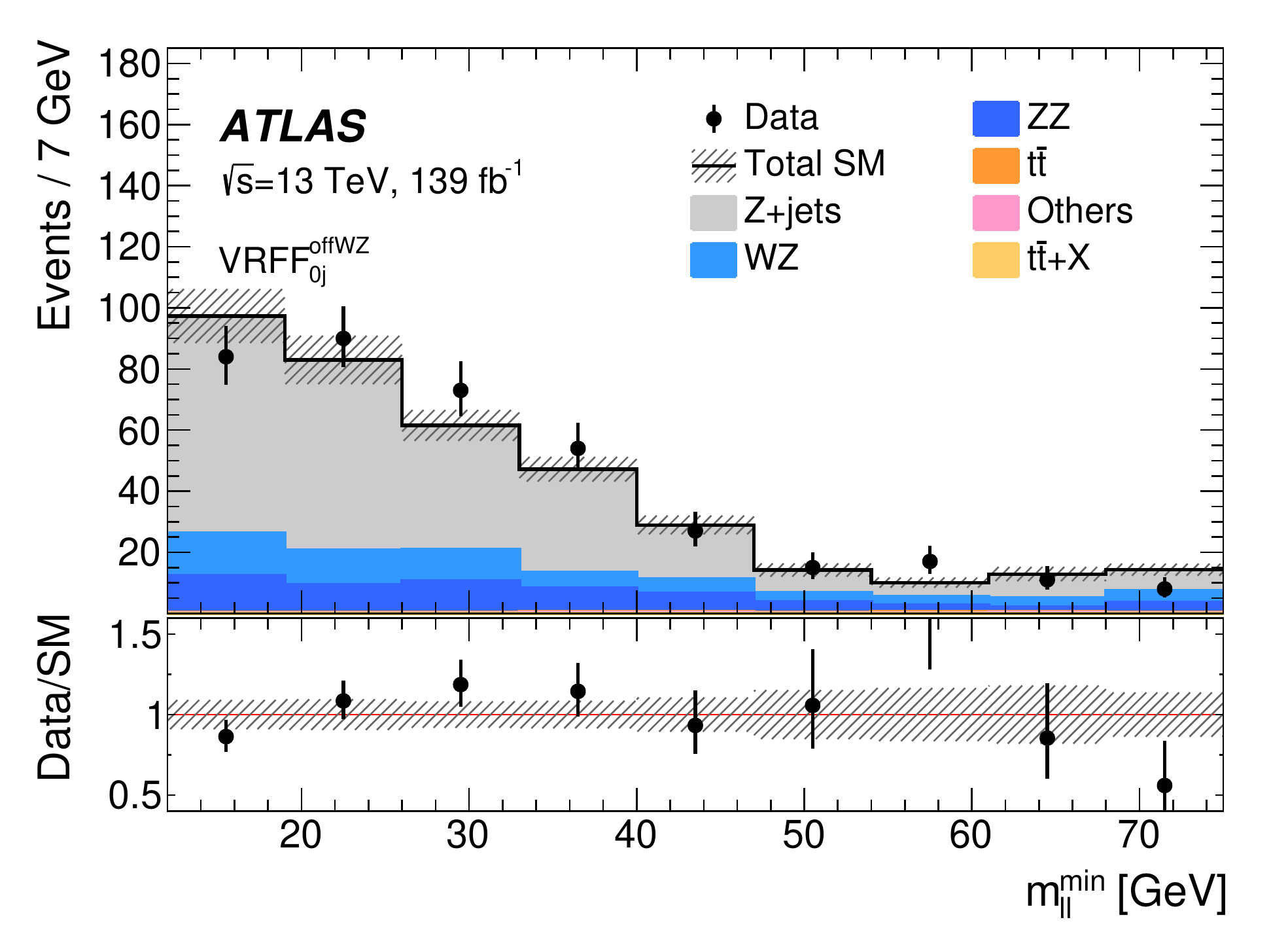}
\caption{
Example kinematic distributions after the background-only fit, showing the data and the post-fit expected background,
in regions of the \ofs \WZ selection.
The figure shows
(top left) the \mllmin distribution in \CRofWZzj,
(top right) the \ptlepovermet distribution in \VRofWZnjmll,
(bottom left) the \met distribution in \VRoftt, and
(bottom right) the \mllmin distribution in \VRofFFzj.
The last bin includes overflow.
The `Others' category contains backgrounds from single-top, \WW, triboson, Higgs and rare top processes.
The bottom panel shows the ratio of the observed data to the predicted yields.
The hatched bands indicate the combined theoretical, experimental, and MC statistical uncertainties.
The slope change in the bottom left \met distribution illustrates the selection extension with \met triggered events,
which start contributing at $\met \gtrsim 200~\GeV$.
}
\label{fig:offShell:CRVR}
\end{figure}
}
 
\begin{figure}[tbp]
\centering
{\includegraphics[width=0.9\textwidth]{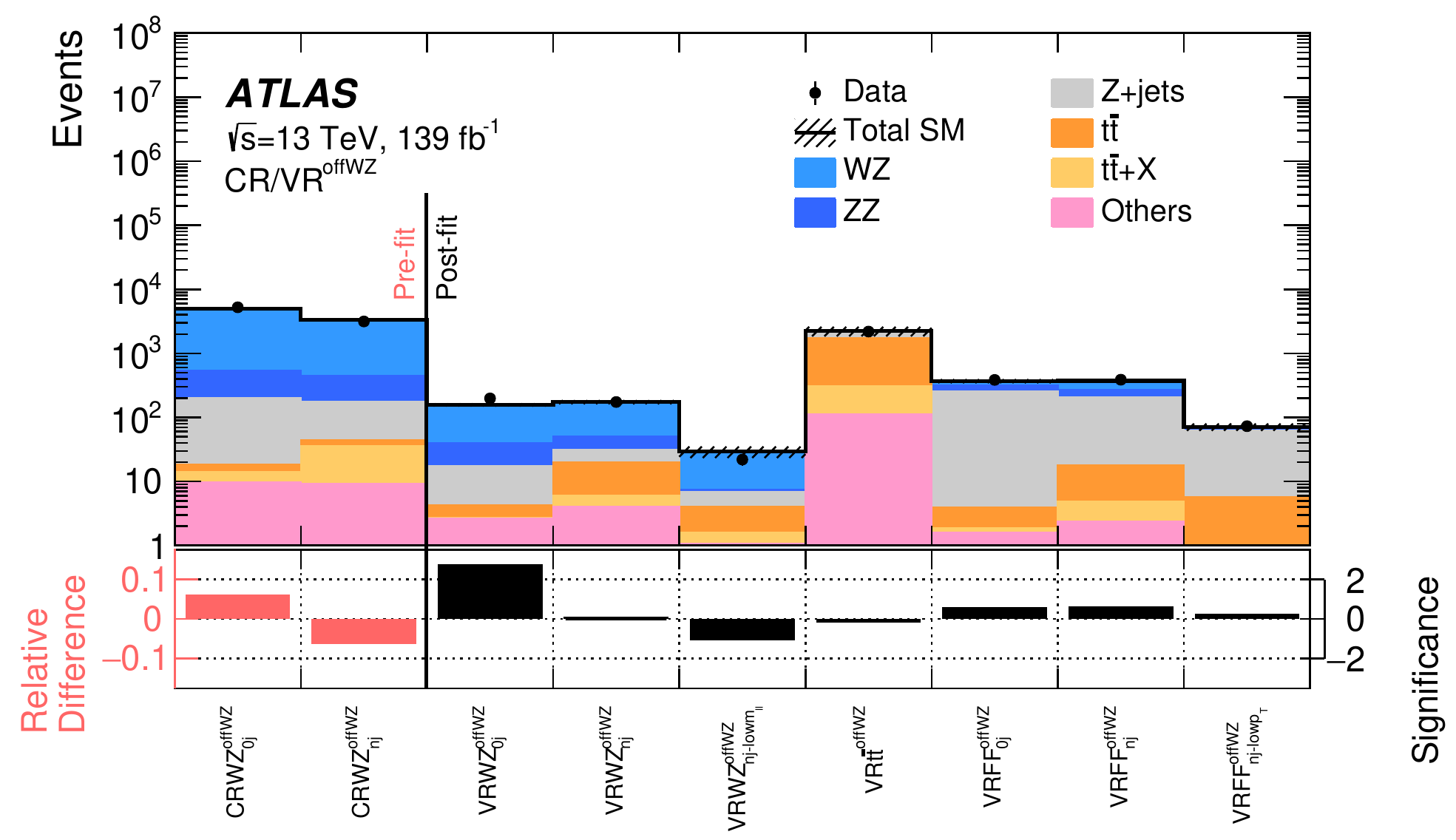}}
\caption{
Comparison of the observed data and expected SM background yields in the CRs and VRs of the \ofs \WZ selection.
The SM prediction is taken from the background-only fit.
The `Others' category contains the single-top, \WW, triboson, Higgs and rare top processes.
The hatched band indicates the combined theoretical, experimental, and MC statistical uncertainties.
The bottom panel shows the significance of the difference between the observed and expected yields,
calculated with the profile likelihood method from Ref.~\cite{Cousins:2007bmb}, adding a minus sign if the yield is below the prediction.
}
\label{fig:results:ofWZ_pull}
\end{figure}
 
\begin{figure}[b]
\centering
\includegraphics[width=0.70\columnwidth]{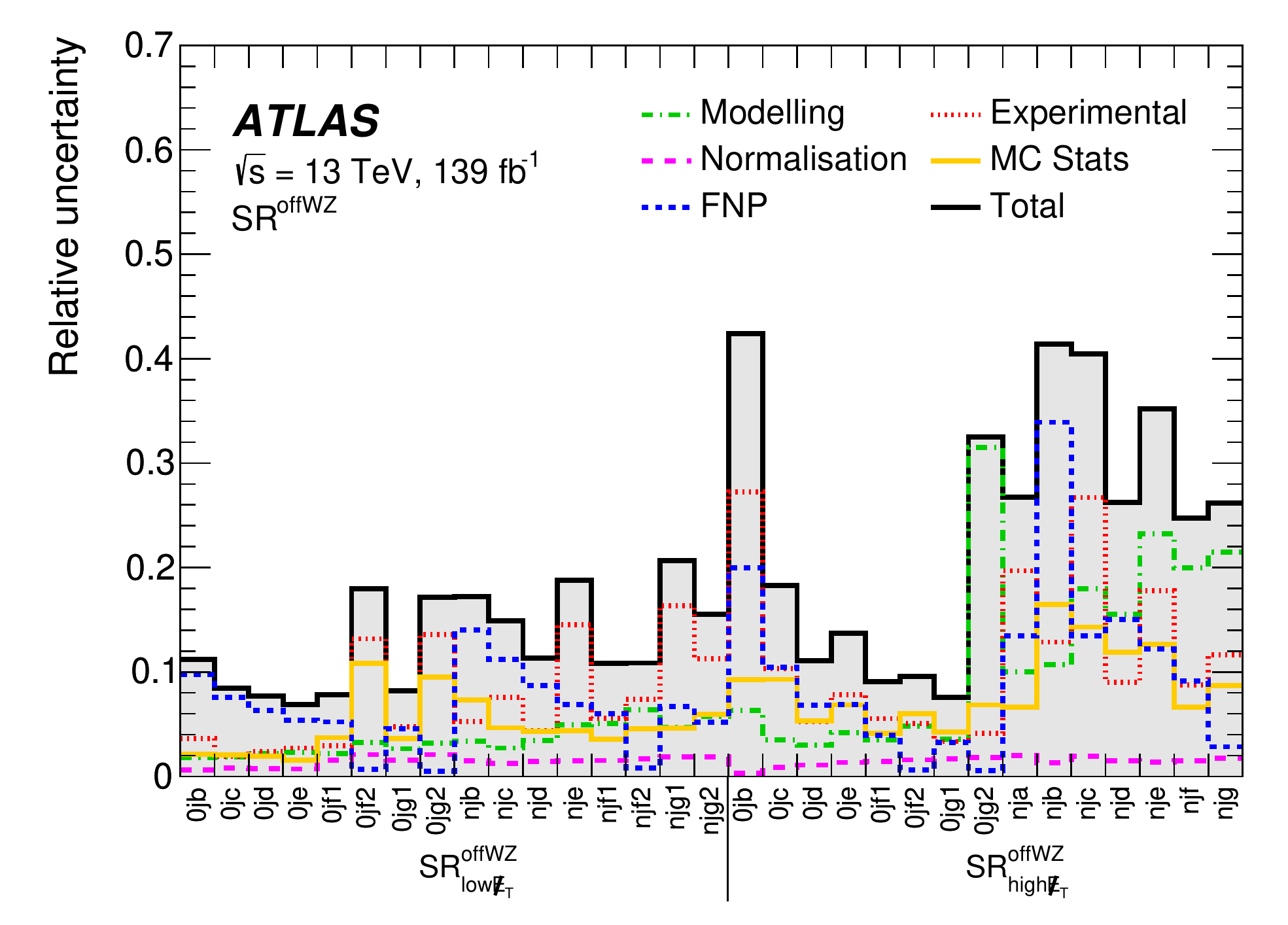}
\caption{
Breakdown of the total systematic uncertainties in the background prediction for the SRs of the \ofs \WZ selection.
}
\label{fig:offShell:summary_syst}
\end{figure}
 
% End of text imported from the .//sections/offshell.tex input file
% The next lines are included from the .//sections/results.tex input file
\FloatBarrier
\begin{DIFnomarkup}
\vskip 1em\section{Results}
\label{sec:results}
\end{DIFnomarkup}
\FloatBarrier
 
The observed data in the \ons \WZ, \ofs \WZ, and \Wh SRs are compared with the background expectation obtained from the background-only fits described in Section~\ref{ssec:analysisstrategy:fit}. 
The results are summarised
in Tables~\ref{table:WZ_SR}--\ref{table:Wh_SR} as well as visualised in Figures~\ref{fig:results:onShellWZPull}--\ref{fig:results:onShellWhPull} for the \SRonWZ and \SRWh regions,
and in Tables~\ref{tab:results:offShell_SRlow}--\ref{tab:results:offShell_SRhigh} and Figure~\ref{fig:results:offShellPull} for the \SRoffWZ.
Post-fit distributions of key kinematic observables are shown for the \SRonWZ and \SRWh regions in Figure~\ref{fig:results:ons:distSR}
and for the \SRoffWZ regions in Figure~\ref{fig:results:ofs:distSR}.
 
To illustrate the sensitivity to various \textchinoonepmninotwo signals throughout the regions, representative signal MC predictions are overlaid on the figures.
The sensitivity to \WZ-mediated models, when the mass difference between the \chinoonepm/\ninotwo and \ninoone is large, is driven by the \SRWZ with large \mt and \met values.
On the other hand, when the mass splitting is close to the $Z$-boson mass,
the sensitivity is dominated by the high \HT region and moderate \mt and \met bins of the $\nJ = 0$ and low \HT regions.
For the \Wh-mediated scenarios the sensitivity is driven by \SRWhSF and \SRWhDF regions, with \SRWhDFi{1} contributing the most.
% The next lines are included from the .//sections/BkgOnly_WZ_SRs.tex input file

\begin{table}[h!]
\begin{center}
\caption{
Observed and expected yields after the background-only fit in the SRs for the \ons \WZ selection.
The normalisation factors of the \WZ sample are extracted separately for the \rzj, \rlHT and \rhHT regions, and are treated separately in the combined fit.
The `Others' category contains the single-top, \WW, triboson, Higgs and rare top processes.
Combined statistical and systematic uncertainties are presented.
}
\label{table:WZ_SR}
\maxsizebox*{0.98\textwidth}{\textheight}{
\begin{tabular}{lccccccc}
\toprule
\toprule
{Regions}           & \SRWZi{1}            & \SRWZi{2}            & \SRWZi{3}            & \SRWZi{4}            & \SRWZi{5}            & \SRWZi{6}            & \SRWZi{7}              \\[0.1cm]
\cline{1-8}
Observed           & $331$              & $31$              & $3$              & $2$              & $42$              & $7$              & $3$                    \\
\cline{1-8}
Fitted  SM          & $314 \pm 33$~~          & $35 \pm 6$~~          & $4.1 \pm 1.0$          & $1.2 \pm 0.5$          & $58 \pm 5$~~          & $8.0 \pm 0.9$          & $5.8 \pm 1.0$              \\
\cline{1-8}
\WZ          & $294 \pm 31$~~          & $32 \pm 5$~~         & $3.7 \pm 0.9$          & $0.9 \pm 0.5$          & $48 \pm 4$~~          & $7.1 \pm 0.8$          & $5.0 \pm 0.9$              \\
\ZZ          & $12.1 \pm 3.1$~~          & $0.66 \pm 0.35$          & $0.08 \pm 0.04$          & $0.04 \pm 0.02$          & $2.3 \pm 0.6$          & $0.12 \pm 0.04$          & $0.08 \pm 0.03$              \\
\ttbar          & $2.8 \pm 0.8$          & $0.36 \pm 0.26$          & $0.04 \pm 0.01$          & $0.00\pm^{0.01}_{0.00}$~\,          & $1.4 \pm 0.4$          & $0.00\pm^{0.01}_{0.00}$~\,          &  $0.04 \pm 0.02$             \\
\ZjetZgam    & $0.01 \pm 0.01$          & $0.14 \pm 0.14$          & $0.05 \pm 0.06$          & $0.06 \pm 0.04$          & $2.8 \pm 2.3$          & $0.3 \pm 0.4$          & $0.26 \pm 0.17$              \\  
\ttbarX          & $0.16 \pm 0.06$          & $0.13 \pm 0.05$          & $0.03 \pm 0.04$          & $0.01 \pm 0.01$          & $0.10 \pm 0.06$          & $0.05 \pm 0.03$          & $0.01 \pm 0.01$              \\
Others          & $5.1 \pm 0.8$          & $1.1 \pm 0.4$          & $0.21 \pm 0.06$          & $0.17 \pm 0.06$          & $3.2 \pm 0.5$          & $0.38 \pm 0.11$          & $0.34 \pm 0.10$              \\
 
\bottomrule 
\toprule
 
{Regions}           & \SRWZi{8}            & \SRWZi{9}            & \SRWZi{10}            & \SRWZi{11}            & \SRWZi{12}            & \SRWZi{13}            & \SRWZi{14}              \\[0.1cm]
\cline{1-8}
Observed           & $1$              & $77$              & $11$              & $0$              & $0$              & $111$              & $19$                    \\
\cline{1-8}
Fitted SM          & $0.8 \pm 0.4$          & $90 \pm 20$          & $13.4 \pm 2.4$~~          & $0.5 \pm 0.4$          & $0.49 \pm 0.24$          & $89 \pm 11$          & $16.0 \pm 1.4$ ~~             \\
\cline{1-8}
\WZ          & $0.44 \pm 0.32$          & $77 \pm 19$          & $11.3 \pm 2.4$~~          & $0.37 \pm 0.31$          & $0.38 \pm 0.22$          & $72 \pm 9$~~          & $13.4 \pm 1.3$~~              \\
\ZZ          & $0.01 \pm 0.01$          & $1.9 \pm 0.9$          & $0.24 \pm 0.13$          & $0.01 \pm 0.01$          & $0.01 \pm 0.01$          & $5.8 \pm 2.8$          & $0.39 \pm 0.18$              \\
\ttbar          & $0.00\pm^{0.01}_{0.00}$~\,          & $3.3 \pm 0.9$          & $0.45 \pm 0.28$          & $0.00\pm^{0.01}_{0.00}$~\,          & $0.00\pm^{0.01}_{0.00}$~\,          & $6.0 \pm 1.4$          & $0.24 \pm 0.17$              \\
\ZjetZgam          & $0.28 \pm 0.20$          & $4 \pm 5$          & $0.2 \pm 0.4$          & $0.02 \pm 0.03$          & $0.02 \pm 0.03$          & $0.02 \pm 0.03$          & $0.02 \pm 0.03$              \\
\ttbarX          & $0 \pm 0$          & $1.3 \pm 0.4$          & $0.40 \pm 0.14$          & $0.05 \pm 0.04$          & $0.02 \pm 0.01$          & $1.6 \pm 0.5$          & $0.56 \pm 0.16$              \\
Others          & $0.08 \pm 0.06$          & $2.3 \pm 0.5$          & $0.79 \pm 0.22$          & $0.08 \pm 0.05$          & $0.08 \pm 0.03$          & $3.5 \pm 0.7$          & $1.37 \pm 0.33$  \\
\bottomrule
\toprule
{Regions }           & \SRWZi{15}            & \SRWZi{16}            & \SRWZi{17}            & \SRWZi{18}            & \SRWZi{19}            & \SRWZi{20}              \\[0.1cm]
\cline{1-7}
Observed           & $5$              & $1$              & $13$              & $9$              & $3$              & $1$                    \\
\cline{1-7}
Fitted SM          & $2.8 \pm 0.6$          & $1.30 \pm 0.27$          & $13.7 \pm 2.6$~~          & $9.2 \pm 1.3$          & $2.3 \pm 0.4$          & $1.09 \pm 0.13$                  \\
\cline{1-7}
\WZ          & $2.3 \pm 0.6$          & $1.07 \pm 0.24$                   &  $10.2 \pm 1.9$~~          & $6.7 \pm 0.8$          & $1.58 \pm 0.24$          & $0.87 \pm 0.12$                \\
\ZZ          & $0.07 \pm 0.04$          & $0.04 \pm 0.03$          & $0.13 \pm 0.06$          & $0.10 \pm 0.04$          & $0.02 \pm 0.01$          & $0.02 \pm 0.01$              \\
\ttbar          & $0.00\pm^{0.01}_{0.00}$~\,          & $0.00\pm^{0.01}_{0.00}$~\,          & $0.77 \pm 0.32$          & $0.45 \pm 0.26$          & $0.00\pm^{0.01}_{0.00}$~\,          & $0.00\pm^{0.01}_{0.00}$~\,              \\
\ZjetZgam         & $0.02 \pm 0.02$          & $0.07 \pm 0.08$          & $1 \pm 1$          & $0.7 \pm 1.0$          & $0.25 \pm 0.34$          & $0.02 \pm 0.02$              \\
\ttbarX          & $0.07 \pm 0.03$          & $0.00\pm^{0.03}_{0.00}$~\,          & $0.53 \pm 0.17$          & $0.33 \pm 0.10$          & $0.07 \pm 0.04$          & $0.03 \pm 0.02$              \\
Others          & $0.37 \pm 0.11$          & $0.12 \pm 0.04$          & $1.1 \pm 0.8$          & $0.9 \pm 0.7$          & $0.27 \pm 0.07$          & $0.18 \pm 0.05$              \\
\cline{1-7}
\end{tabular}
}\begin{DIFnomarkup}\vspace{-1.0em}\mbox{\ }\end{DIFnomarkup}
\end{center}
\end{table}
 
% End of text imported from the .//sections/BkgOnly_WZ_SRs.tex input file

\begin{DIFnomarkup}\newpage\end{DIFnomarkup}
For the \WZ-mediated models targeted with the \SRoffWZ, with mass differences between the \chinoonepm/\ninotwo and \ninoone smaller than the $Z$-boson mass,
the sensitivity to signals with different $\Dm$ depends on the \mllmin range of the bins.
The bins with larger (smaller) \mllmin values are sensitive to signals with larger (smaller) mass splittings;
for the lowest mass-splitting signals, only \SRhighnji{a} has sensitivity.
 
No significant deviation from the SM background prediction is found in any of the SRs,
and none of the deviations agree with any of the benchmark signal hypotheses.
The maximum deviation of the data from the background expectation is in \SRlowzj{d} with a $2.3\sigma$ data excess,
followed by a $2.1\sigma$ deficit in \SRhighzji{f2},
a $2.0\sigma$ excess in \SRWhDFi{1},
and a $2.0\sigma$ deficit in \SRWZi{5};
the significances are computed following the profile likelihood method in Ref.~\cite{Cousins:2007bmb}.
 
% The next lines are included from the .//sections/BkgOnly_Wh_SRs.tex input file

\begin{table}[h!]
\begin{center}
\caption{
Observed and expected yields after the background-only fit in the SRs for the \Wh selection.
The normalisation factors of the \WZ sample are extracted separately for the \rzj, \rlHT and \rhHT regions, and are treated separately in the combined fit.
The `Others' category contains the single-top, \WW, \ttbarX and rare top processes.
Combined statistical and systematic uncertainties are presented.
}
\label{table:Wh_SR}
 
\maxsizebox*{\textwidth}{\textheight}{
\begin{tabular}{lccccccc}
\toprule
\toprule
{Regions}           & \SRWhSFi{1}            & \SRWhSFi{2}            & \SRWhSFi{3}            & \SRWhSFi{4}            & \SRWhSFi{5}            & \SRWhSFi{6}            & \SRWhSFi{7}              \\[0.1cm]
\cline{1-8}
Observed           & $152$              & $14$              & $8$              & $47$              & $6$              & $15$              & $19$                    \\
\cline{1-8}
Fitted SM          & $136 \pm 13$~~        & $13.5 \pm 1.7$~~        & $4.3 \pm 0.9$          & $50 \pm 5$~~        & $4.3 \pm 0.7$          & $20.2 \pm 2.1$~~        & $16.0 \pm 2.1$~~              \\
\cline{1-8}
\WZ          & $107 \pm 12$~~         & $10.2 \pm 1.7$~~        & $3.8 \pm 0.8$          & $32 \pm 4$~~          & $2.7 \pm 0.6$          & $12.3 \pm 1.6$~~        & $10.8 \pm 1.7$~~            \\
\ttbar          & $10.3 \pm 2.5$~~        & $1.6 \pm 0.6$          & $0.13 \pm 0.12$          & $7.7 \pm 1.9$          & $0.74 \pm 0.34$          & $3.5 \pm 1.0$          & $2.5 \pm 0.7$              \\
\ZjetZgam          & $2.5 \pm 2.9$          & $0.00\pm^{0.02}_{0.00}$~\,        & $0.00\pm^{0.02}_{0.00}$~\,        & $2.0 \pm 1.6$          & $0.00\pm^{0.04}_{0.00}$~\,        & $0.00\pm^{0.04}_{0.00}$~\,        & $0.00\pm^{0.02}_{0.00}$~\,            \\
Higgs          & $5.7 \pm 0.6$          & $0.69 \pm 0.07$          & $0.20 \pm 0.03$          & $3.12 \pm 0.31$          & $0.26 \pm 0.05$          & $1.29 \pm 0.14$          & $0.81 \pm 0.09$              \\
 
Triboson     & $1.9 \pm 0.5$          & $0.22 \pm 0.07$          & $0.07 \pm 0.02$          & $1.4 \pm 0.4$          & $0.28 \pm 0.09$          & $0.61 \pm 0.18$          & $0.83 \pm 0.24$        \\
Others          & $8.6 \pm 1.9$          & $0.84 \pm 0.11$          & $0.08 \pm 0.05$          & $4.0 \pm 0.5$          & $0.23 \pm 0.24$          & $2.54 \pm 0.22$          & $1.11 \pm 0.15$   \\
\bottomrule 
\toprule
{Regions}           & \SRWhSFi{8}            & \SRWhSFi{9}            & \SRWhSFi{10}            & \SRWhSFi{11}            & \SRWhSFi{12}            & \SRWhSFi{13}            & \SRWhSFi{14}              \\[0.1cm]
\cline{1-8}
Observed           & $113$              & $184$              & $28$              & $5$              & $82$              & $16$              & $4$                    \\ \cline{1-8}
Fitted SM         & $108 \pm 13$~~        & $180 \pm 17$~~        & $31 \pm 4$~~        & $6.6 \pm 0.9$          & $90 \pm 11$          & $18.7 \pm 2.6$~~        & $2.5 \pm 0.7$              \\ \cline{1-8}
\WZ          & $54 \pm 6$~~        & $127 \pm 13$~~        & $19.3 \pm 2.3$~~          & $5.3 \pm 0.8$          & $47 \pm 6$~~        & $6.8 \pm 1.7$          & $1.26 \pm 0.26$              \\
\ttbar          & $21 \pm 6$~~        & $33 \pm 10$          & $8.2 \pm 2.3$          & $0.7 \pm 0.5$          & $28 \pm 8$~~        & $8.0 \pm 2.2$          & $0.9 \pm 0.5$              \\
\ZjetZgam          & $19 \pm 10$          & $2.3 \pm 1.9$          & $1.0 \pm 1.3$          & $0.10 \pm 0.21$          & $2.1 \pm 3.1$          & $1.2 \pm 0.7$          & $0.00\pm^{0.12}_{0.00}$~\,              \\
Higgs          & $1.91 \pm 0.19$          & $3.63 \pm 0.35$          & $0.67 \pm 0.06$          & $0.15 \pm 0.02$          & $2.98 \pm 0.25$          & $0.61 \pm 0.07$          & $0.07 \pm 0.07$              \\
Triboson     & $0.79 \pm 0.24$          & $1.4 \pm 0.4$          & $0.41 \pm 0.13$          & $0.12 \pm 0.05$          & $1.6 \pm 0.5$          & $0.56 \pm 0.18$          & $0.13 \pm 0.05$         \\
Others          & $11.1 \pm 2.2$~~        & $12.2 \pm 2.2$~~        & $1.8 \pm 0.4$          & $0.22 \pm 0.05$          & $9.0 \pm 1.1$          & $1.6 \pm 0.7$          & $0.10 \pm 0.05$        \\
\bottomrule
\toprule
 
{Regions }           & \SRWhSFi{15}            & \SRWhSFi{16}            & \SRWhSFi{17}            & \SRWhSFi{18}            & \SRWhSFi{19}     & \SRWhDFi{1}            & \SRWhDFi{2}                        \\[0.1cm]
\cline{1-8}
Observed           & $51$              & $5$              & $37$              & $7$              & $4$  & $10$              & $10$                   \\
\cline{1-8}
Fitted SM          & $46 \pm 7$~~        & $9.8 \pm 1.6$          & $43 \pm 7$~~        & $12.6 \pm 1.7$~~        & $1.8 \pm 0.4$  & $4.5 \pm 0.8$          & $7.0 \pm 2.3$              \\
\cline{1-8}
\WZ          & $18.9 \pm 2.2$~~        & $3.9 \pm 0.8$          & $35 \pm 6$~~        & $9.8 \pm 1.6$          & $1.44 \pm 0.32$       & $0.44 \pm 0.14$          & $1.05 \pm 0.20$         \\
\ttbar          & $18 \pm 6$~~        & $3.2 \pm 1.3$          & $1.00 \pm 0.34$          & $0.33 \pm 0.17$          & $0.00 \pm^{0.01}_{0.00}$~\,  & $1.0 \pm 0.6$        & $1.7 \pm 1.1$               \\
\ZjetZgam          & $ 0.00\pm^{0.12}_{0.00}$~\,        & $0.00\pm^{0.12}_{0.00}$~\,        & $0.00\pm^{0.12}_{0.00}$~\,        & $0.00\pm^{0.12}_{0.00}$~\,        & $0.00\pm^{0.12}_{0.00}$~\, & $0.00\pm^{0.20}_{0.00}$~\,        & $2.5 \pm 2.0$                 \\
Higgs          & $2.06 \pm 0.23$          & $0.36 \pm 0.05$          & $1.02 \pm 0.12$          & $0.44 \pm 0.05$          & $0.05 \pm 0.05$    & $1.59 \pm 0.22$          & $0.96 \pm 0.11$             \\
Triboson     & $1.5 \pm 0.4$          & $0.53 \pm 0.17$          & $2.5 \pm 0.7$          & $1.3 \pm 0.4$          & $0.2 \pm 0.1$    & $0.66 \pm 0.15$          & $0.64 \pm 0.16$             \\
 
Others          & $5.0 \pm 0.6$          & $1.8 \pm 0.5$          & $3.0 \pm 0.7$          & $0.73 \pm 0.15$          & $0.14 \pm 0.05$    & $0.81 \pm 0.09$          & $0.21 \pm 0.07$           \\
 
\cline{1-8}
\cline{1-8}
\end{tabular}
}\begin{DIFnomarkup}\vspace{-3.5em}\mbox{\ }\end{DIFnomarkup}
\end{center}
\end{table}
% End of text imported from the .//sections/BkgOnly_Wh_SRs.tex input file

\FloatBarrier

\begin{figure}[tbp]
\centering
{\includegraphics[width=0.83\textwidth]{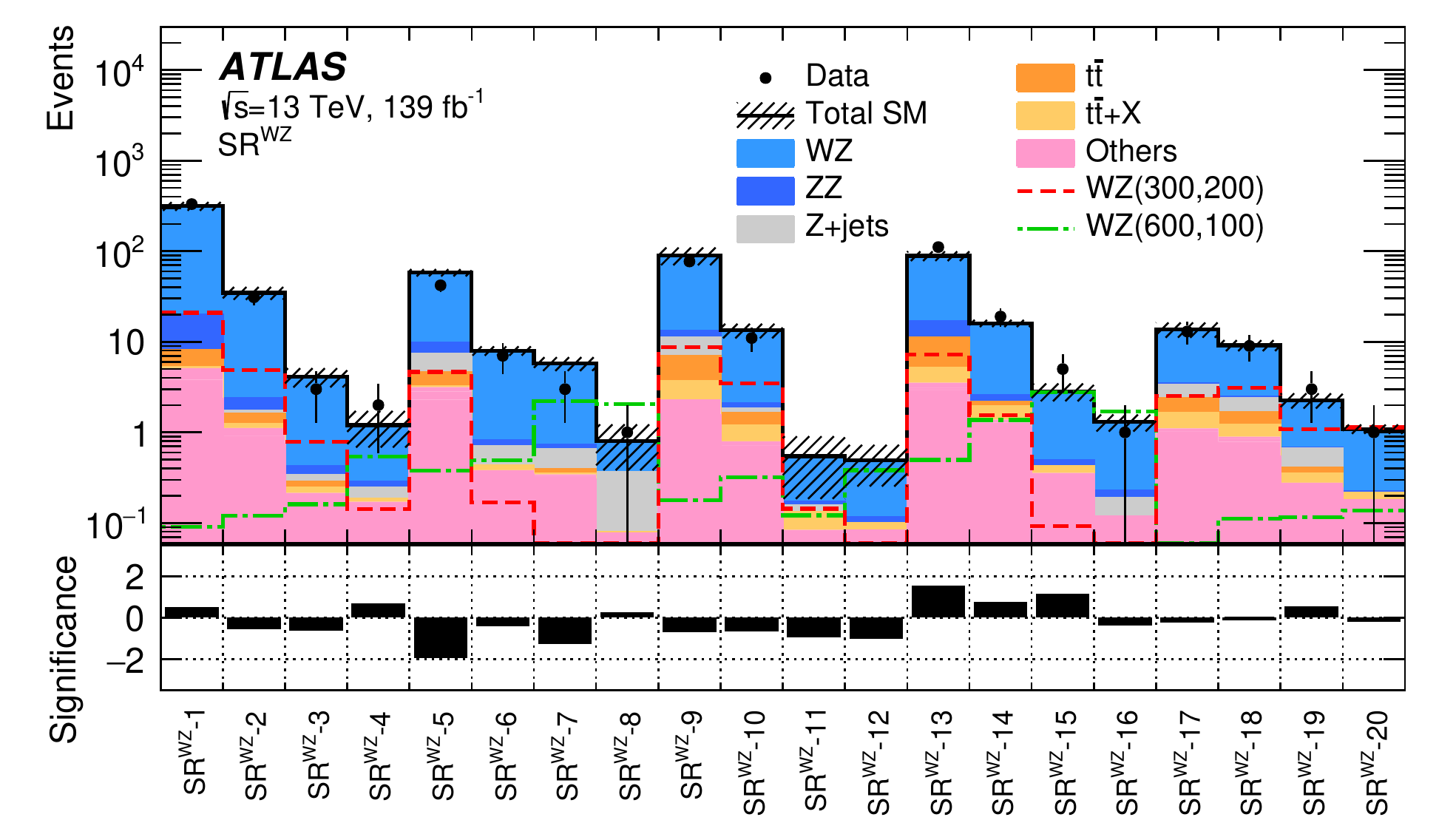}}
\caption{
Comparison of the observed data and expected SM background yields in the SRs of the \ons \WZ selection.
The SM prediction is taken from the background-only fit.
The `Others' category contains the single-top, \WW, triboson, Higgs and rare top processes.
The hatched band indicates the combined theoretical, experimental, and MC statistical uncertainties.
Distributions for wino/bino (+) \textchinoonepmninotwo \ra\ \WZ~signals are overlaid, with mass values given as \CNmasspair~\GeV.
The bottom panel shows the significance of the difference between the observed and expected yields,
calculated with the profile likelihood method from Ref.~\cite{Cousins:2007bmb}, adding a minus sign if the yield is below the prediction.
}
\label{fig:results:onShellWZPull}
\end{figure}

\begin{figure}[tbp]
\centering
{\includegraphics[width=0.83\textwidth]{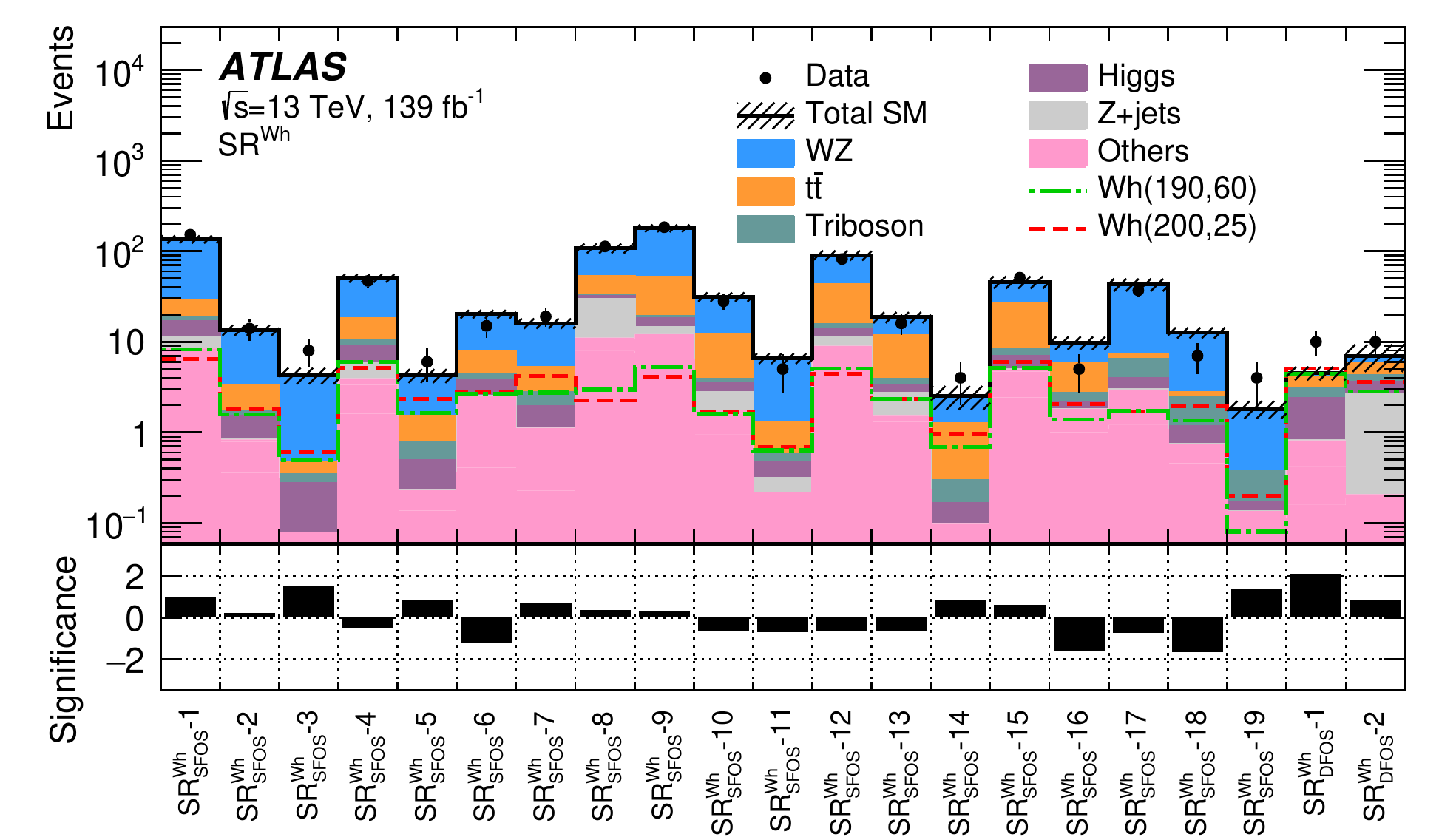}}
\caption{
Comparison of the observed data and expected SM background yields in the SRs of the \Wh selection.
The SM prediction is taken from the background-only fit.
The `Others' category contains the single-top, \WW, \ttbarX and rare top processes.
The hatched band indicates the combined theoretical, experimental, and MC statistical uncertainties.
Distributions for wino/bino (+) \textchinoonepmninotwo \ra\ \Wh~signals are overlaid, with mass values given as \CNmasspair~\GeV.
The bottom panel shows the significance of the difference between the observed and expected yields,
calculated with the profile likelihood method from Ref.~\cite{Cousins:2007bmb}, adding a minus sign if the yield is below the prediction.
}
\label{fig:results:onShellWhPull}
\end{figure}
 
\FloatBarrier
% The next lines are included from the .//sections/yields_SR_offShell_1.tex input file
\begin{table}[tbp]
\caption{
Observed and expected yields after the background-only fit in \SRlow.
The normalisation factors of the \WZ sample are extracted separately for 0j and nj, and are treated separately in the combined fit.
The `Others' category contains the single-top, \WW, triboson, Higgs and rare top processes.
Combined statistical and systematic uncertainties are presented.
}
\label{tab:results:offShell_SRlow}
\maxsizebox*{\textwidth}{\textheight}{
\begin{tabular}{l r@{$\ \pm\ $}lr@{$\ \pm\ $}lr@{$\ \pm\ $}lr@{$\ \pm\ $}lr@{$\ \pm\ $}lr@{$\ \pm\ $}l}
\toprule
\toprule
Region          & \tcc{\SRlowzji{b}}           & \tcc{\SRlowzji{c}}           & \tcc{\SRlowzji{d}}           & \tcc{\SRlowzji{e}}           & \tcc{\SRlowzji{f1}}          & \tcc{\SRlowzji{f2}}          \\[-0.05cm]
\midrule
Observed        & \tcc{$25$}                   & \tcc{$42$}                   & \tcc{$77$}                   & \tcc{$101$}                  & \tcc{$33$}                   & \tcc{$7$}                    \\
\midrule
Fitted SM events & $32$    & $4$                & $44$    & $4$                & $54$    & $4$                & $91$    & $6$                & $32.2$  & $2.5$              & $5.9$   & $1.1$              \\
\midrule
\WZ              & $7.6$   & $0.9$              & $13.8$  & $1.3$              & $16.3$  & $1.9$              & $25.6$  & $1.8$              & $20.1$  & $1.5$              & $4.9$   & $1.0$              \\
\ZZ              & $5.5$   & $1.3$              & $7.4$   & $1.2$              & $9.6$   & $1.6$              & $21.8$  & $3.2$              & $2.7$   & $1.1$              & $0.43$  & $0.14$             \\
\Zjet            & $19.1$  & $3.2$              & $22.7$  & $3.4$              & $26.5$  & $3.5$              & $40$    & $5$                & $7.2$   & $1.7$              & $0.00$  & $_{0.00}^{0.04}$   \\
\ttbar           & $0.05$  & $_{0.05}^{0.18}$   & $0.11$  & $_{0.11}^{0.17}$   & $0.38$  & $0.22$             & $1.1$   & $0.4$              & $0.78$  & $0.29$             & $0.08$  & $_{0.08}^{0.10}$   \\
\ttbarX          & $0.007$ & $_{0.007}^{0.019}$ & $0.002$ & $_{0.002}^{0.008}$ & $0.009$ & $_{0.009}^{0.019}$ & $0.019$ & $_{0.019}^{0.026}$ & $0.026$ & $0.026$            & $0.010$ & $_{0.010}^{0.015}$ \\
Others           & $0.045$ & $0.031$            & $0.30$  & $0.12$             & $1.3$   & $0.6$              & $1.9$   & $0.6$              & $1.4$   & $0.4$              & $0.51$  & $0.18$             \\
\bottomrule
\toprule
Region          & \tcc{\SRlowzji{g1}} & \tcc{\SRlowzji{g2}}           & \tcc{\SRlownji{b}} & \tcc{\SRlownji{c}}         & \tcc{\SRlownji{d}} & \tcc{\SRlownji{e}}      \\[-0.05cm]
\midrule
Observed        & \tcc{$34$}          & \tcc{$9$}                     & \tcc{$6$}          & \tcc{$13$}                 & \tcc{$17$}         & \tcc{$14$}              \\
\midrule
Fitted SM events & $34.7$  & $2.8$     & $6.3$   & $1.1$               & $3.5$   & $0.6$    & $8.0$   & $1.2$            & $13.5$ & $1.5$     & $18.2$ & $3.4$          \\
\midrule
\WZ              & $21.4$  & $2.1$     & $5.2$   & $1.0$               & $1.62$  & $0.30$   & $3.2$   & $0.6$            & $6.0$  & $0.8$     & $8.6$  & $1.3$          \\
\ZZ              & $4.7$   & $1.4$     & $0.45$  & $0.14$              & $0.45$  & $0.13$   & $0.72$  & $0.22$           & $1.00$ & $0.28$    & $1.4$  & $0.9$          \\
\Zjet            & $6.6$   & $1.6$     & $0.001$ & $_{0.001}^{0.029}$  & $1.2$   & $0.5$    & $3.7$   & $0.9$            & $4.5$  & $1.2$     & $3.3$  & $1.3$          \\
\ttbar           & $0.8$   & $0.4$     & $0.36$  & $0.21$              & $0.15$  & $0.13$   & $0.28$  & $0.14$           & $1.5$  & $0.4$     & $3.3$  & $0.9$          \\
\ttbarX          & $0.039$ & $0.025$   & $0.003$ & $_{0.003}^{0.008}$  & $0.030$ & $0.013$  & $0.052$ & $0.019$          & $0.24$ & $0.06$    & $0.33$ & $0.07$         \\
Others           & $1.16$  & $0.27$    & $0.27$  & $0.09$              & $0.006$ & $0.004$  & $0.14$  & $_{0.14}^{0.34}$ & $0.21$ & $0.06$    & $1.3$  & $_{1.3}^{1.8}$ \\
\bottomrule
\addlinespace[0.25\aboverulesep]
\cmidrule[\heavyrulewidth]{1-9}
Region          & \tcc{\SRlownji{f1}} & \tcc{\SRlownji{f2}}       & \tcc{\SRlownji{g1}}     & \tcc{\SRlownji{g2}}       & \tcc{} & \tcc{} \\[-0.05cm]
\cmidrule[0.6pt]{1-9}
Observed        & \tcc{$25$}          & \tcc{$20$}                & \tcc{$22$}              & \tcc{$12$}                & \tcc{} & \tcc{} \\
\cmidrule[0.6pt]{1-9}
Fitted SM events & $23.4$ & $2.5$      & $17.9$ & $1.9$            & $17.0$ & $3.5$          & $12.4$ & $1.9$            & \tcc{} & \tcc{} \\
\cmidrule[0.6pt]{1-9}
\WZ              & $11.1$ & $1.2$      & $9.4$  & $1.1$            & $10.0$ & $1.2$          & $7.3$  & $1.3$            & \tcc{} & \tcc{} \\
\ZZ              & $4.0$  & $1.6$      & $0.66$ & $0.25$           & $1.1$  & $_{1.1}^{2.6}$ & $0.34$ & $0.11$           & \tcc{} & \tcc{} \\
\Zjet            & $2.2$  & $1.4$      & $0.00$ & $_{0.00}^{0.14}$ & $1.8$  & $1.1$          & $0.0$  & $_{0.0}^{0.6}$   & \tcc{} & \tcc{} \\
\ttbar           & $4.6$  & $1.1$      & $5.7$  & $1.2$            & $3.0$  & $0.8$          & $2.9$  & $0.7$            & \tcc{} & \tcc{} \\
\ttbarX          & $0.44$ & $0.09$     & $0.72$ & $0.11$           & $0.36$ & $0.08$         & $0.44$ & $0.09$           & \tcc{} & \tcc{} \\
Others           & $1.0$  & $0.4$      & $1.4$  & $0.9$            & $0.71$ & $0.21$         & $1.4$  & $0.6$            & \tcc{} & \tcc{} \\
\cmidrule[\heavyrulewidth]{1-9}
\addlinespace[-1.\belowrulesep]
\cmidrule[\heavyrulewidth]{1-9}
\end{tabular}
}
\end{table}
 
% End of text imported from the .//sections/yields_SR_offShell_1.tex input file
% The next lines are included from the .//sections/yields_SR_offShell_2.tex input file
\begin{table}[tbp]
\caption{
Observed and expected yields after the background-only fit in \SRhigh.
The normalisation factors of the \WZ sample are extracted separately for 0j and nj, and are treated separately in the combined fit.
The `Others' category contains the single-top, \WW, triboson, Higgs and rare top processes.
Combined statistical and systematic uncertainties are presented.
}
\label{tab:results:offShell_SRhigh}
\maxsizebox*{\textwidth}{\textheight}{
\begin{tabular}{l r@{$\ \pm\ $}lr@{$\ \pm\ $}lr@{$\ \pm\ $}lr@{$\ \pm\ $}lr@{$\ \pm\ $}l}
\toprule
\toprule
Region          & \tcc{\SRhighzji{b}}          & \tcc{\SRhighzji{c}}          & \tcc{\SRhighzji{d}} & \tcc{\SRhighzji{e}} & \tcc{\SRhighzji{f1}} \\[-0.05cm]
\midrule
Observed        & \tcc{$1$}                    & \tcc{$4$}                    & \tcc{$11$}          & \tcc{$13$}          & \tcc{$37$}           \\
\midrule
Fitted SM events & $1.5$   & $0.7$              & $4.3$   & $0.8$              & $14.0$  & $1.6$     & $11.5$ & $1.6$      & $35.7$ & $3.2$       \\
\midrule
\WZ              & $0.20$  & $_{0.20}^{0.27}$   & $1.5$   & $0.5$              & $6.0$   & $0.9$     & $6.1$  & $1.1$      & $20.5$ & $2.1$       \\
\ZZ              & $0.5$   & $0.5$              & $0.31$  & $0.12$             & $1.8$   & $0.8$     & $0.89$ & $0.24$     & $3.1$  & $1.0$       \\
\Zjet            & $0.81$  & $0.31$             & $1.7$   & $0.4$              & $4.4$   & $1.0$     & $1.1$  & $0.8$      & $4.3$  & $1.4$       \\
\ttbar           & $0.05$  & $0.05$             & $0.45$  & $0.17$             & $0.64$  & $0.28$    & $1.8$  & $0.6$      & $4.4$  & $1.0$       \\
\ttbarX          & $0.003$ & $_{0.003}^{0.014}$ & $0.009$ & $_{0.009}^{0.013}$ & $0.029$ & $0.015$   & $0.08$ & $0.04$     & $0.11$ & $0.05$      \\
Others           & $0.014$ & $_{0.014}^{0.018}$ & $0.3$   & $_{0.3}^{0.4}$     & $1.1$   & $0.4$     & $1.6$  & $0.4$      & $3.3$  & $0.8$       \\
\bottomrule
\toprule
Region          & \tcc{\SRhighzji{f2}}       & \tcc{\SRhighzji{g1}} & \tcc{\SRhighzji{g2}}      & \tcc{\SRhighnji{a}} & \tcc{\SRhighnji{b}}          \\[-0.05cm]
\midrule
Observed        & \tcc{$14$}                 & \tcc{$43$}           & \tcc{$17$}                & \tcc{$3$}           & \tcc{$2$}                    \\
\midrule
Fitted SM events & $25.5$ & $2.4$             & $39.5$ & $3.0$       & $21$   & $7$              & $6.0$   & $1.6$     & $1.4$   & $0.6$              \\
\midrule
\WZ              & $16.0$  & $2.3$            & $26.4$ & $2.2$       & $15$   & $7$              & $3.8$   & $1.2$     & $0.57$  & $0.18$             \\
\ZZ              & $0.95$  & $0.35$           & $3.0$  & $0.9$       & $0.58$ & $0.17$           & $0.044$ & $0.023$   & $0.009$ & $0.005$            \\
\Zjet            & $0.00$  & $_{0.00}^{0.15}$ & $3.4$  & $1.3$       & $0.00$ & $_{0.00}^{0.11}$ & $1.5$   & $0.8$     & $0.5$   & $0.5$              \\
\ttbar           & $4.4$   & $1.0$            & $4.3$  & $0.9$       & $3.1$  & $0.7$            & $0.6$   & $0.5$     & $0.14$  & $_{0.14}^{0.15}$   \\
\ttbarX          & $0.109$ & $0.030$          & $0.16$ & $0.05$      & $0.09$ & $0.04$           & $0.16$  & $0.06$    & $0.014$ & $_{0.014}^{0.025}$ \\
Others           & $4.0$   & $1.0$            & $2.3$  & $0.8$       & $2.0$  & $0.5$            & $0.038$ & $0.030$   & $0.22$  & $0.22$             \\
\bottomrule
\toprule
Region          & \tcc{\SRhighnji{c}}        & \tcc{\SRhighnji{d}}        & \tcc{\SRhighnji{e}}      & \tcc{\SRhighnji{f}}      & \tcc{\SRhighnji{g}}        \\[-0.05cm]
\midrule
Observed        & \tcc{$2$}                  & \tcc{$2$}                  & \tcc{$2$}                & \tcc{$11$}               & \tcc{$4$}                  \\
\midrule
Fitted SM events & $2.1$   & $0.8$            & $5.4$   & $1.4$            & $3.0$   & $1.1$          & $9.9$   & $2.5$          & $6.8$   & $1.8$            \\
\midrule
\WZ              & $1.25$  & $0.25$           & $2.5$   & $0.4$            & $1.31$  & $0.25$         & $4.5$   & $0.7$          & $3.7$   & $0.6$            \\
\ZZ              & $0.020$ & $0.011$          & $0.014$ & $0.013$          & $0.029$ & $0.014$        & $0.081$ & $0.033$        & $0.050$ & $0.020$          \\
\Zjet            & $0.04$  & $_{0.04}^{0.28}$ & $0.7$   & $_{0.7}^{0.8}$   & $0.0$   & $_{0.0}^{0.4}$ & $0.6$   & $_{0.6}^{0.9}$ & $0.00$  & $_{0.00}^{0.19}$ \\
\ttbar           & $0.6$   & $0.5$            & $1.3$   & $0.8$            & $1.2$   & $1.0$          & $3.4$   & $2.0$          & $2.5$   & $1.6$            \\
\ttbarX          & $0.027$ & $0.023$          & $0.08$  & $0.08$           & $0.09$  & $0.04$         & $0.31$  & $0.08$         & $0.21$  & $0.07$           \\
Others           & $0.14$  & $_{0.14}^{0.36}$ & $0.8$   & $0.6$            & $0.33$  & $0.21$         & $1.0$   & $0.4$          & $0.3$   & $_{0.3}^{0.4}$   \\
\bottomrule
\bottomrule
\end{tabular}
}
\end{table}
% End of text imported from the .//sections/yields_SR_offShell_2.tex input file
 
\FloatBarrier
\begin{figure}[t!]
\begin{DIFnomarkup}\vskip -1em\end{DIFnomarkup}
\centering
\includegraphics[width=0.85\columnwidth]{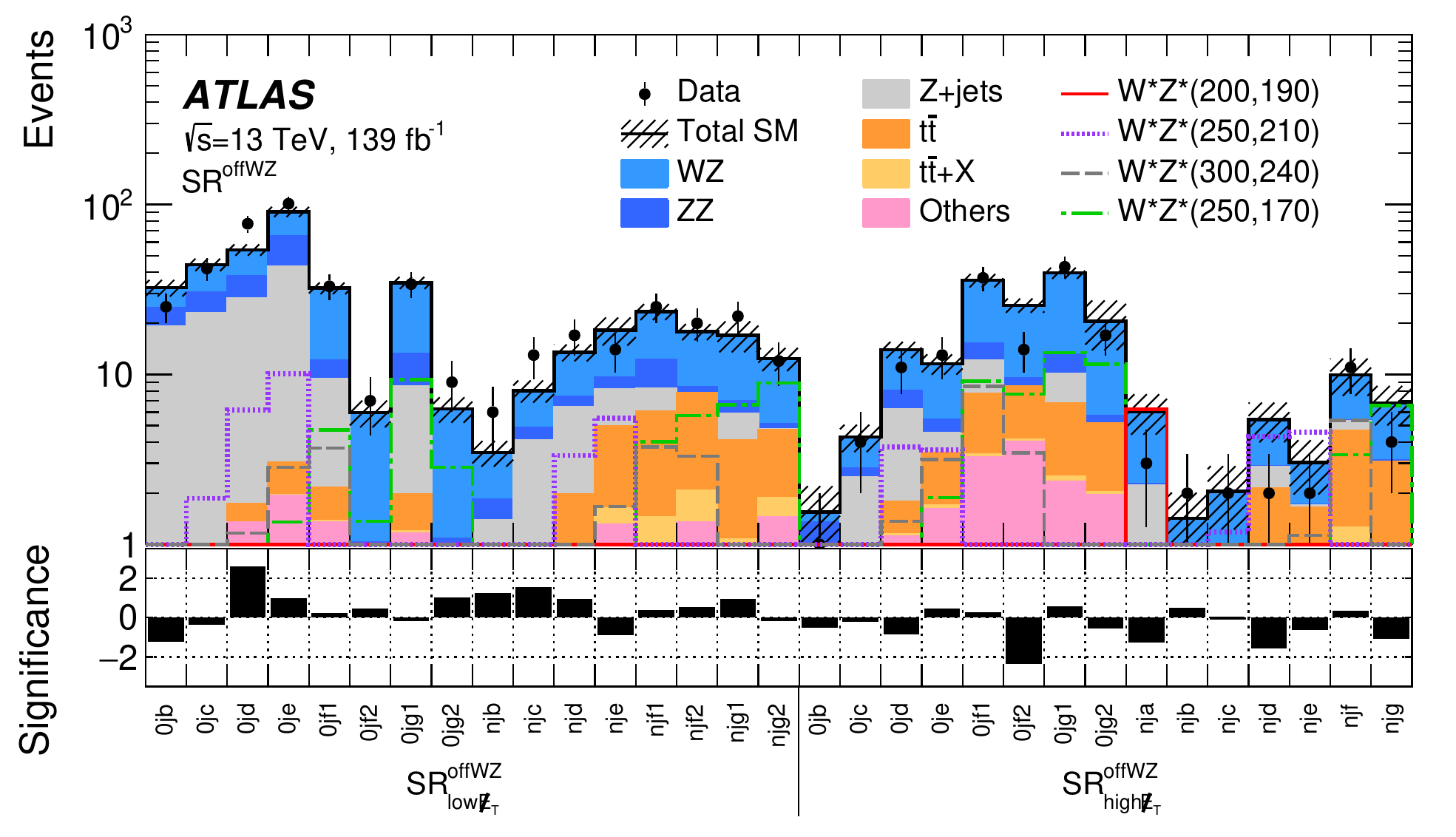}\begin{DIFnomarkup}\vskip -0.5em\end{DIFnomarkup}
\caption{
Comparison of the observed data and expected SM background yields in the SRs of the \ofs \WZ selection.
The SM prediction is taken from the background-only fit.
The `Others' category contains the single-top, \WW, triboson, Higgs and rare top processes.
The hatched band indicates the combined theoretical, experimental, and MC statistical uncertainties.
Distributions for wino/bino (+) \textchinoonepmninotwo \ra\ \WZStar~signals are overlaid, with mass values given as \CNmasspair~\GeV.
The bottom panel shows the significance of the difference between the observed and expected yields,
calculated with the profile likelihood method from Ref.~\cite{Cousins:2007bmb}, adding a minus sign if the yield is below the prediction.
}
\label{fig:results:offShellPull}
\end{figure}
 
\begin{figure}[p!]
\centering
\includegraphics[width=0.48\columnwidth]{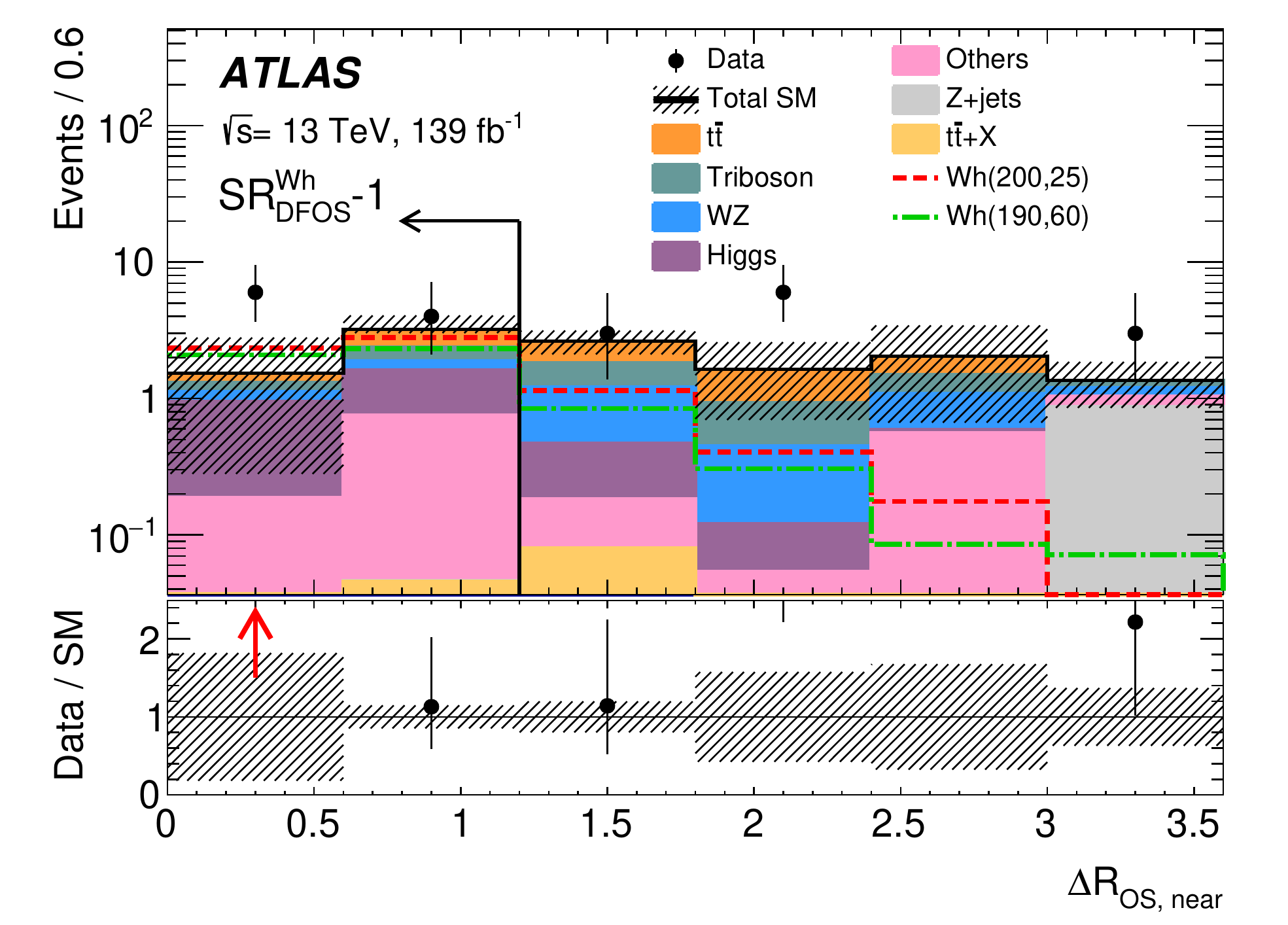}
\includegraphics[width=0.48\columnwidth]{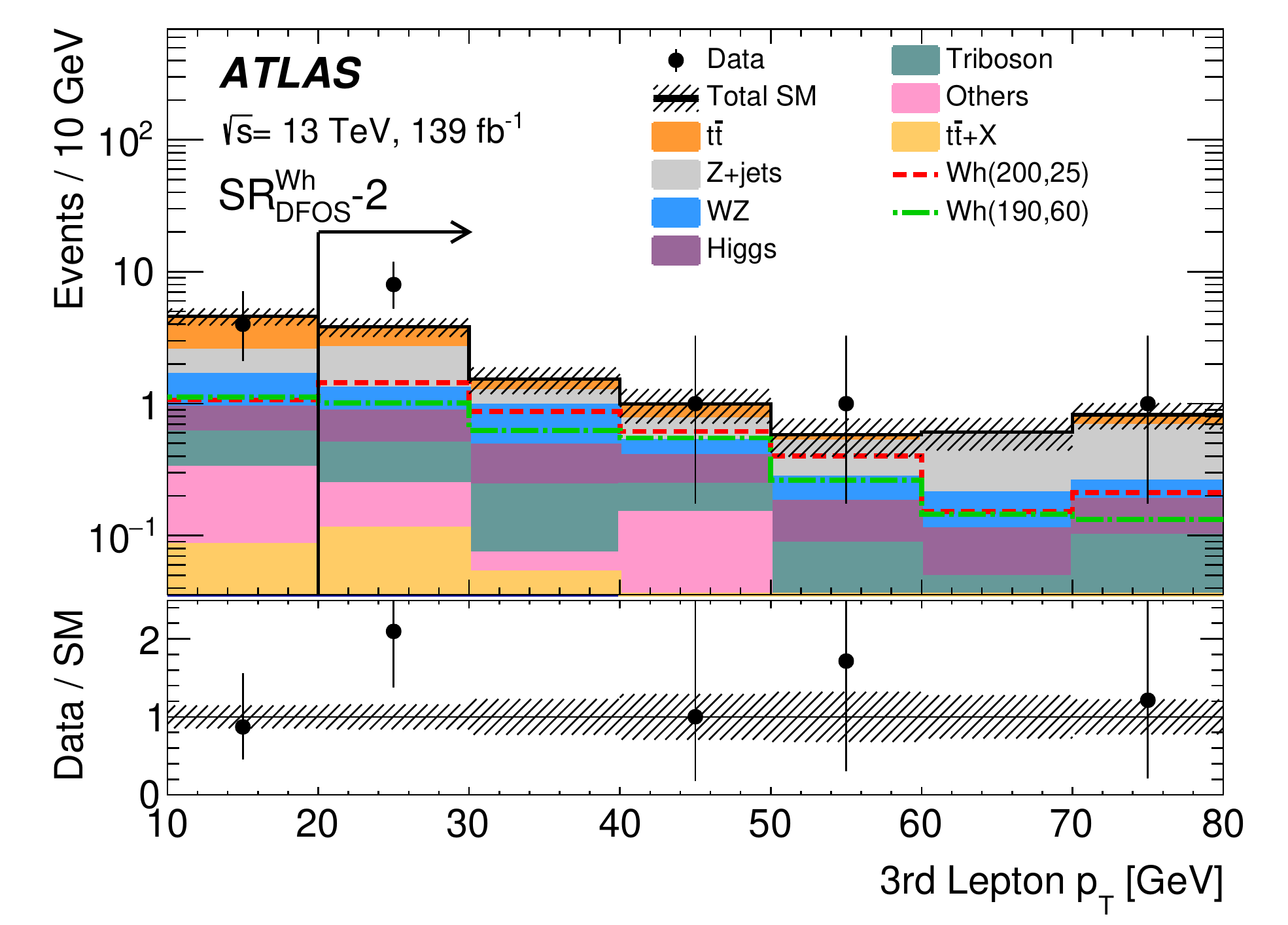}
\includegraphics[width=0.48\columnwidth]{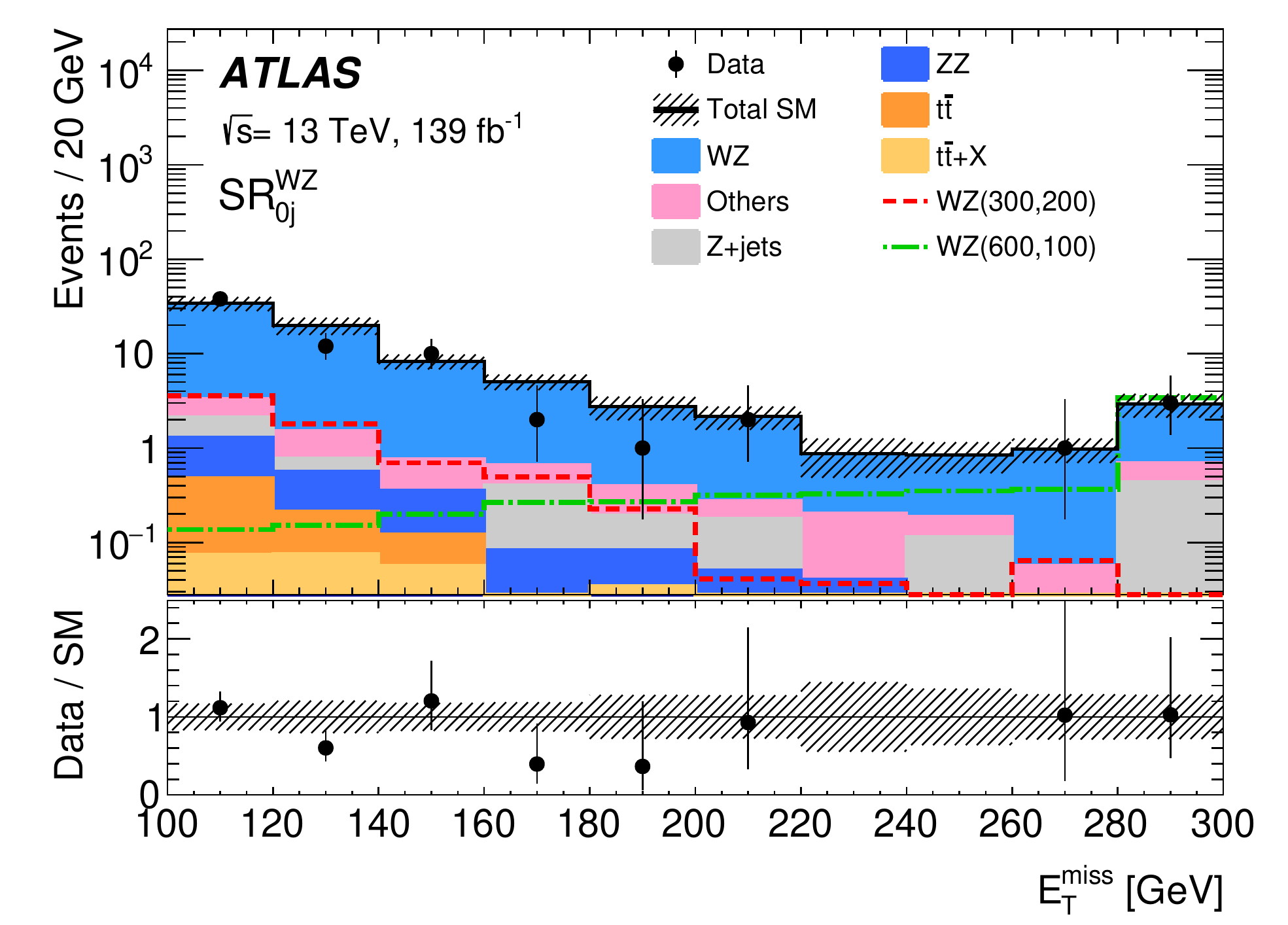}
\includegraphics[width=0.48\columnwidth]{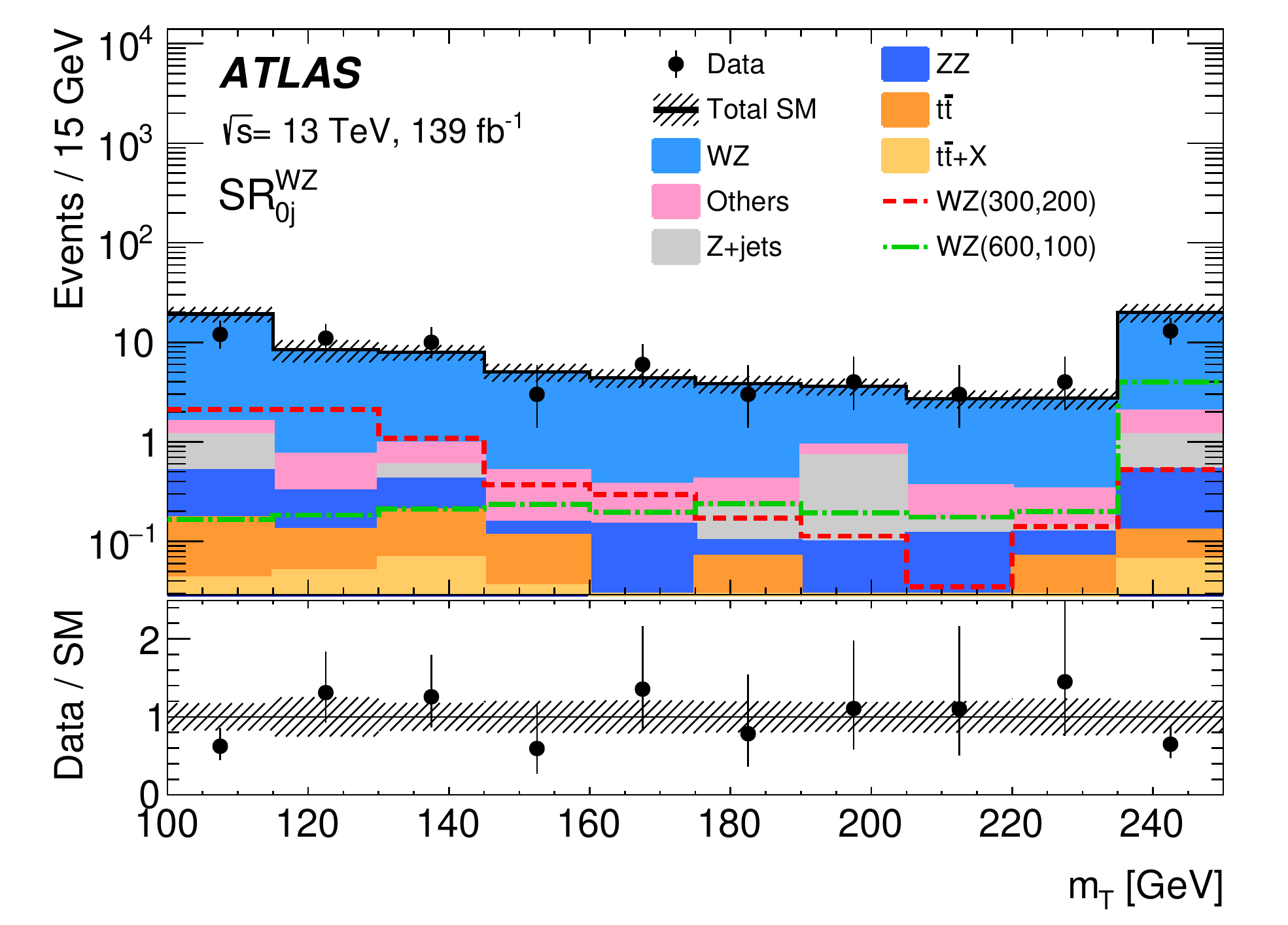}
\caption{
Kinematic distributions after the background-only fit showing the data and the post-fit expected background, in SRs of the \ons \WZ and \Wh selections.
The figure shows
(top left) the \dRnear distribution in \SRWhDFi{1},
(top right) the 3rd leading lepton's \pT in \SRWhDFi{2},
and the (bottom left) \met and (bottom right) \mT distributions in \SRonWZzj (with all SR-i bins of \SRonWZzj summed).
The SR selections are applied for each distribution, except for the variable shown, for which the selection is indicated by a black arrow.
The last bin includes overflow.
The `Others' category contains backgrounds from single-top, \WW, triboson, Higgs and rare top processes, except in the top panels, where triboson and Higgs production contributions are shown separately, and \ttbarX is merged into Others.
Distributions for wino/bino (+) \textchinoonepmninotwo \ra\ \WZ/\Wh~signals are overlaid, with mass values given as \CNmasspair~\GeV.
The bottom panel shows the ratio of the observed data to the predicted yields.
Ratio values outside the graph range are indicated by a red arrow.
The hatched bands indicate the combined theoretical, experimental, and MC statistical uncertainties.
}
\label{fig:results:ons:distSR}
\end{figure}
 
\begin{DIFnomarkup}
\begin{figure}[p!]
\centering
\includegraphics[width=0.48\columnwidth]{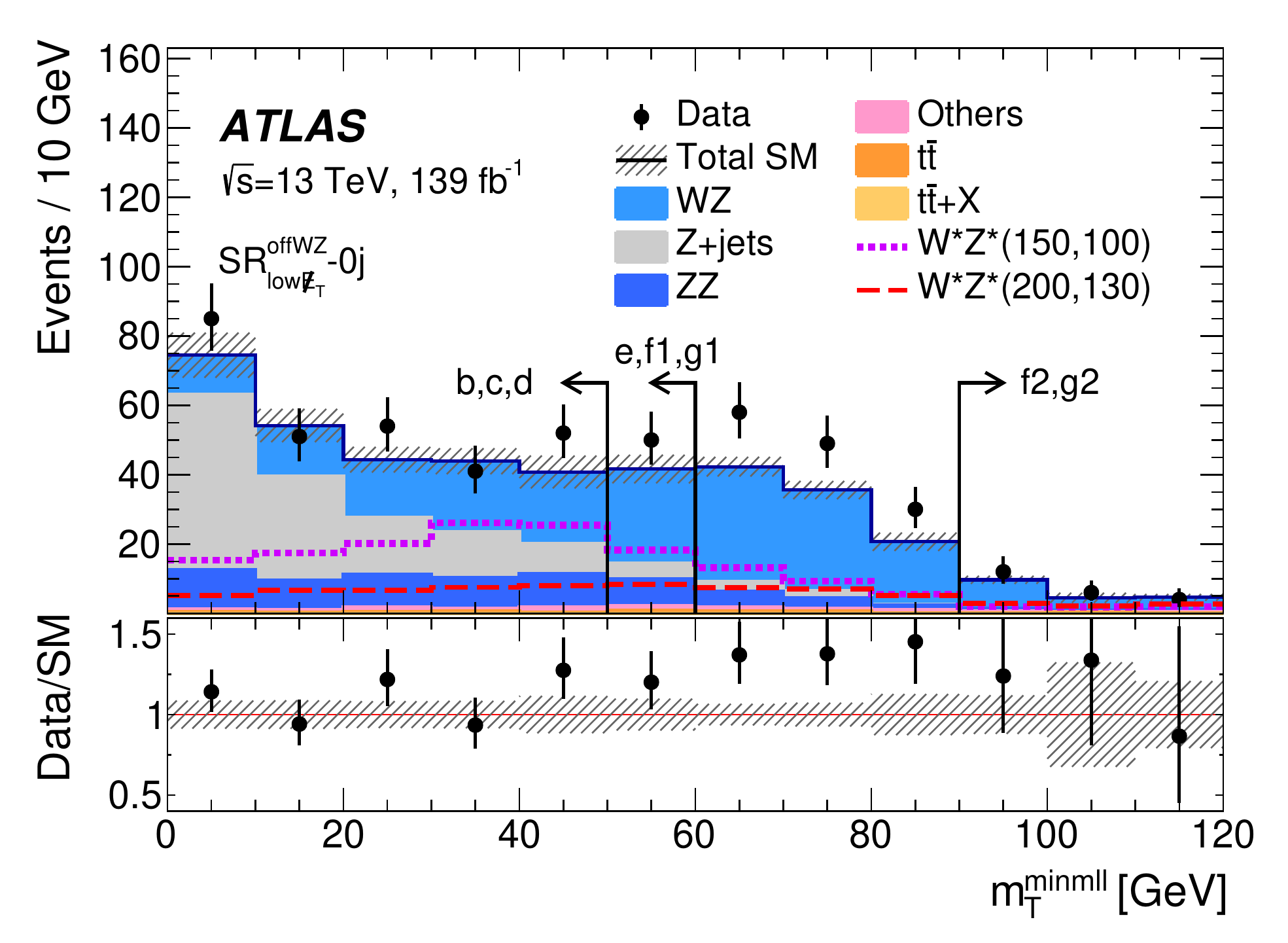}
\includegraphics[width=0.48\columnwidth]{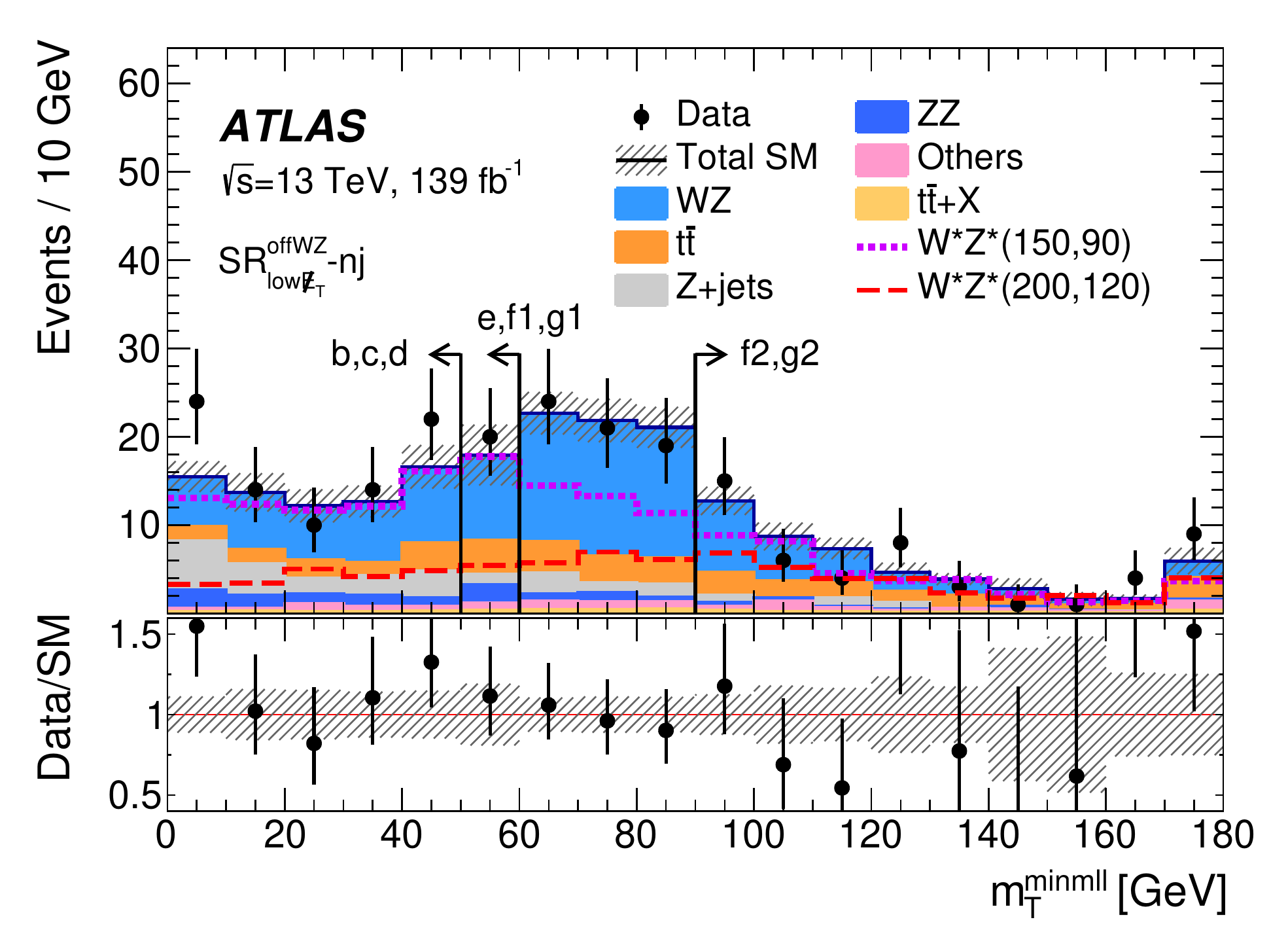}
\includegraphics[width=0.48\columnwidth]{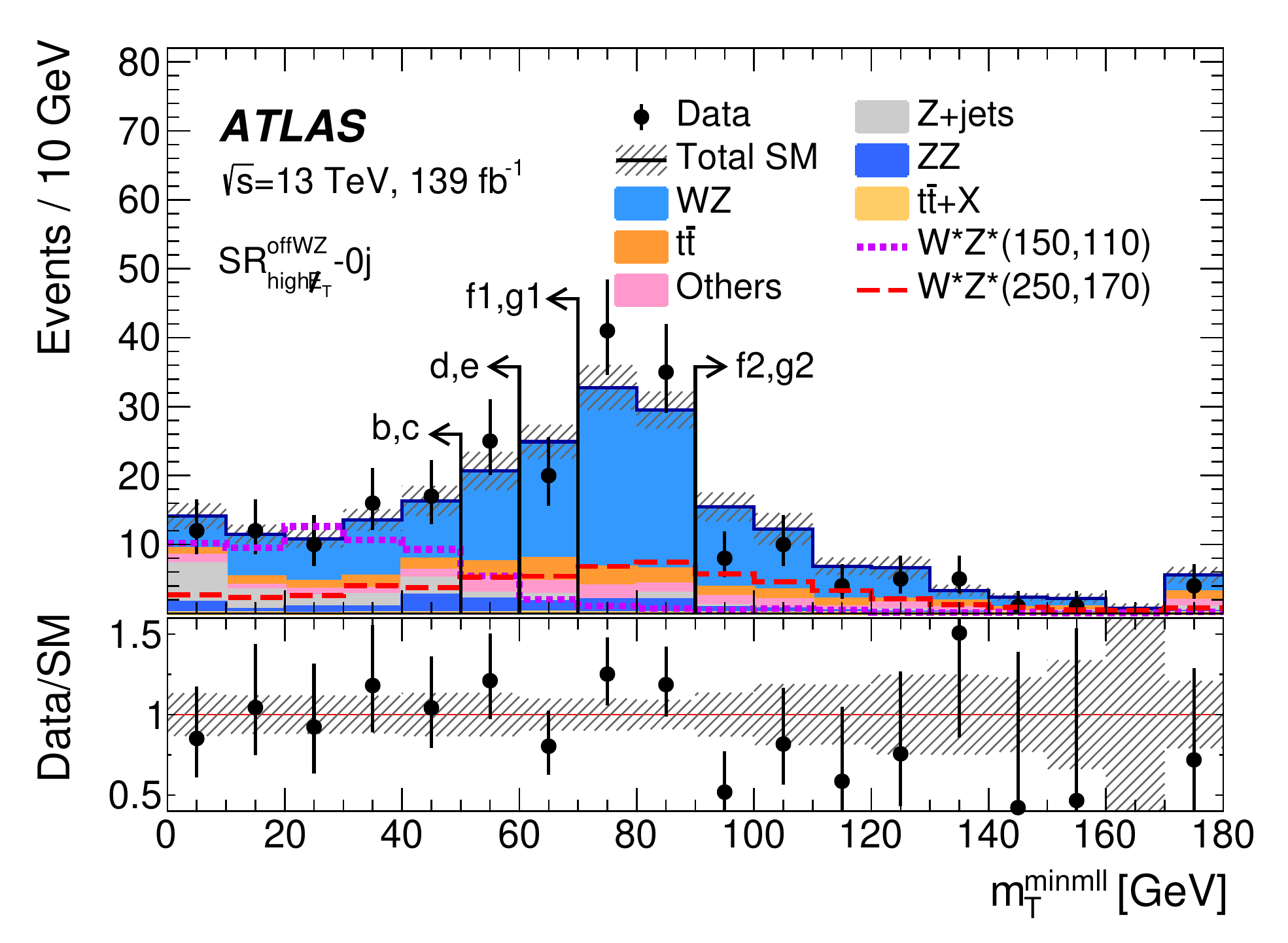}
\includegraphics[width=0.48\columnwidth]{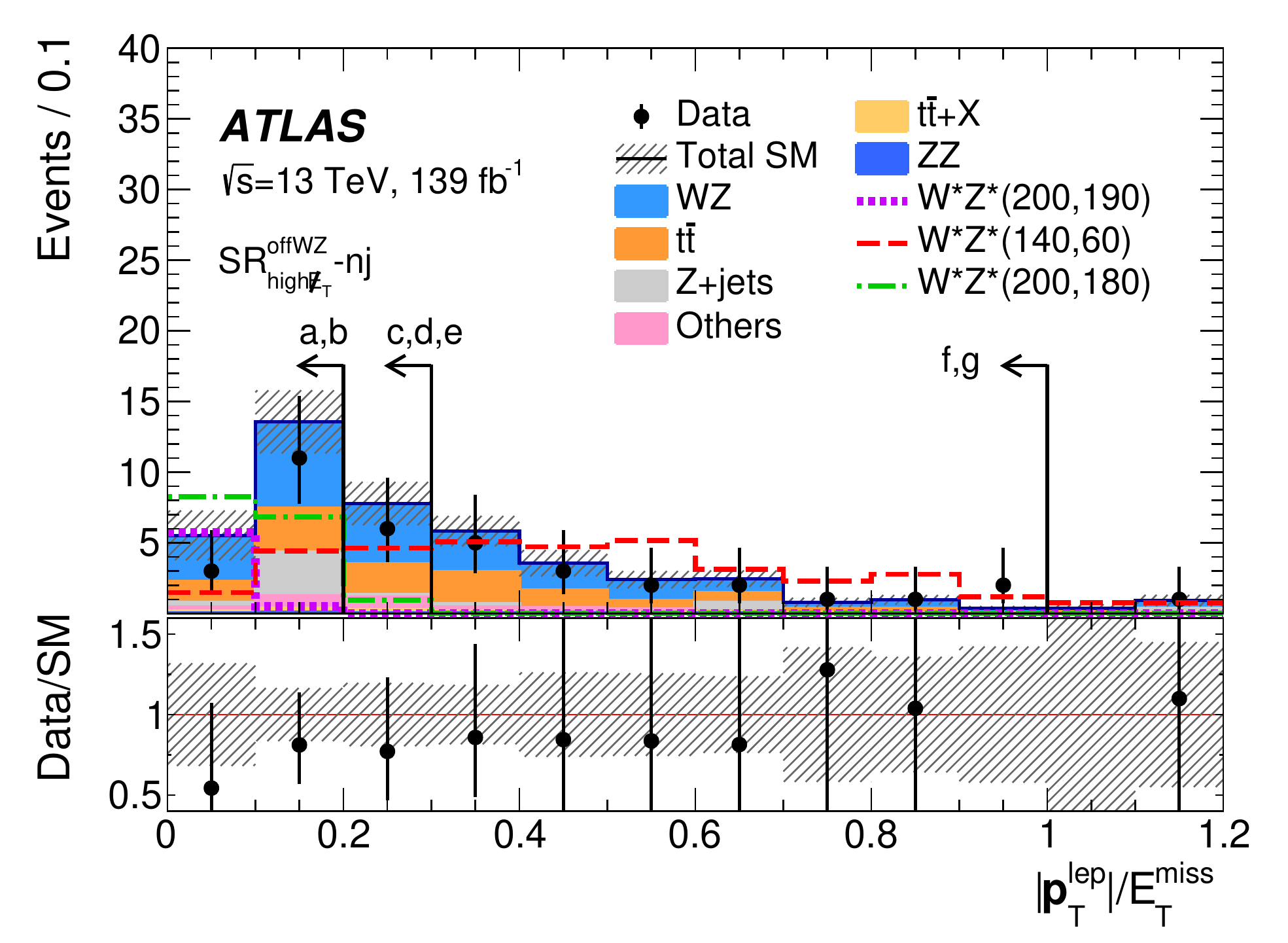}
\caption{
Kinematic distributions after the background-only fit showing the data and the post-fit expected background, in SRs of the \ofs \WZ selection.
The figure shows the \mtmllmin distribution in
(top left) \SRlowzj, (top right) \SRlownj and (bottom left) \SRhighzj, and the \ptlepovermet distribution in (bottom right) \SRhighnj.
The contributing \mllmin mass bins within each \SRoffWZ category are summed.
The SR selections are applied for each distribution, except for the variable shown, for which the selection is indicated by a black arrow.
The last bin includes overflow.
The `Others' category contains backgrounds from single-top, \WW, triboson, Higgs and rare top processes.
Distributions for wino/bino (+) \textchinoonepmninotwo \ra\ \WZ~signals are overlaid, with mass values given as \CNmasspair~\GeV.
The bottom panel shows the ratio of the observed data to the predicted yields.
The hatched bands indicate the combined theoretical, experimental, and MC statistical uncertainties.
}
\label{fig:results:ofs:distSR}
\end{figure}
\end{DIFnomarkup}
 
\begin{DIFnomarkup}
\subsection{Model-independent limits on new physics in inclusive regions}\label{ssec:results:incl}
\end{DIFnomarkup}
 
Model-independent upper limits and discovery $p$-values in the SRs are derived by performing the discovery fits as described in Section~\ref{ssec:analysisstrategy:fit}.
The set of single-bin signal regions used in the fits, referred to as `inclusive SRs',
is constructed by logically grouping adjoining, disjoint, nominal SRs of the \onShell, \Wh and \offShell selections.
Multiple, sometimes overlapping, regions are defined to capture signatures
with different unknown \mllmin shapes and jet multiplicities inclusively.
Based on the best expected discovery sensitivity and using a number of signal points covering both the \WZ- and \Wh-mediated scenarios and different mass splittings,
12 inclusive SRs are formed by merging \SRonWZ and \SRWh regions, creating \incSRWZ and \incSRWh, respectively.
They are summarised in Table~\ref{tab:ons:disc}.
Similarly, 17 inclusive SRs are formed by merging \SRoffWZ regions, creating \incSRlowhigh; their definitions are summarised in Table~\ref{tab:ofs:disc}.
For \incSRlowhigh, contiguous jet-veto regions are merged with jet-inclusive regions, as the \mllmin shape of a signal is assumed to be insensitive to jet multiplicity.
The \SRlow and \SRhigh regions are kept separate, while the \SRhighnj regions are considered separately for \begin{DIFnomarkup}\smash{$\minmll < 20~\GeV$}\end{DIFnomarkup},
as this selection provides the best sensitivity to low-mass-splitting models.
 
% The next lines are included from the .//tables/ons_selectionSRinc.tex input file
\begin{table}[p]
\begin{center}
\caption{Summary of the selection criteria for the inclusive SRs in the \ons \WZ and \Wh selections.}
\label{tab:ons:disc}
\adjustbox{max width=\textwidth}{
\begin{tabular}{l*{4}{P{3.5cm}}}
\hline\hline\tabvthinspace
& \multicolumn{4}{c}{\SRonWZ (\mll $\in [75,105]~\GeV)$} \\\cline{2-5}
& \multicolumn{2}{c}{$\nJ = 0$} & \multicolumn{2}{c}{$\nJ>0$} \\\cline{2-5}
\mt [\GeV] & \multicolumn{4}{c}{\MET [\GeV]} \\ \hline\tabvthinspace
$[100,160]$  & \incSRWZi{1}: $[100,200]$ & \incSRWZi{2}: $>200$ & \incSRWZi{3}: $[150,250]$ & \incSRWZi{4}: $>250$ \\\tabvthinspace
$>160$     & \multicolumn{2}{c}{\incSRWZi{5}: $>200$} & \multicolumn{2}{c}{\incSRWZi{6}: $>200$} \\\hline\hline
& \multicolumn{4}{c}{\SRWhSF (\mll $\leq 75~\GeV$)} \\\cline{2-5}\tabvthinspace
& \multicolumn{2}{c}{$\nJ = 0$} & \multicolumn{2}{c}{$\nJ>0$} \\\cline{2-5}
\mt [\GeV] & \multicolumn{4}{c}{\MET [\GeV]} \\ \hline\tabvthinspace
$[0,100]$    & \multicolumn{2}{c}{\incSRWhSFi{7}: $>50$} & \multicolumn{2}{c}{-} \\\tabvthinspace
$[100,160]$& \multicolumn{2}{c}{\incSRWhSFi{8}: $>50$} & \multicolumn{2}{c}{\incSRWhSFi{9}: $>75$} \\\tabvthinspace
$>160$     & \multicolumn{2}{c}{\incSRWhSFi{10}: $>50$} & \multicolumn{2}{c}{\incSRWhSFi{11}: $>75$} \\\hline\hline\tabvthinspace
& \multicolumn{4}{c}{\SRWhDF}  \\\cline{2-5}\tabvthinspace
& \multicolumn{4}{c}{\incSRWhDFi{12}: $\nJ \in [0,2]$, \dRnear $<$ 1.2, 3rd lepton \pt $>$20 \GeV} \\
\hline \hline
\end{tabular}
}
\end{center}
\end{table}
% End of text imported from the .//tables/ons_selectionSRinc.tex input file
% The next lines are included from the .//tables/ofs_selectionSRinc.tex input file
\begin{table}[p]
\begin{center}
\caption{Summary of the selection criteria for the inclusive SRs in the \ofs \WZ selection.}
\label{tab:ofs:disc}
\adjustbox{max width=\textwidth}{
\begin{tabular}{l*{6}{P{3.3cm}}}
\hline\hline
\rule{0pt}{\dimexpr.7\normalbaselineskip+1mm}
& \multicolumn{4}{c}{\incSRhighnj} & \\[0.1cm]
& \texttt{a}      & \texttt{b}       & \texttt{c1}     & \texttt{c2} & \\\cline{2-5}
\mllmin [\GeV] & $[1,12]$ & $[12,15]$ & $[1,20]$ & $[15,20]$ & \\\cline{2-5}
\rule{0pt}{\dimexpr.7\normalbaselineskip+1mm}
& \SRhighnjd{a} & \SRhighnjd{b} & \SRhighnjd{a-c} & \SRhighnjd{c} & \\[0.1cm]\hline\hline
\rule{0pt}{\dimexpr.7\normalbaselineskip+1mm}
& \multicolumn{2}{c}{\incSRlow} & \multicolumn{2}{c}{\incSRhigh} & \\[0.1cm]
\rule{0pt}{\dimexpr.7\normalbaselineskip+1mm}
& \texttt{b}       & \texttt{c}                  & \texttt{b}       & \texttt{c}       &   \\\cline{2-5}
\mllmin [\GeV] & $[12,15]$ & $[12,20]$            & $[12,15]$ & $[12,20]$ & \\\cline{2-5}
\rule{0pt}{\dimexpr.7\normalbaselineskip+1mm}
& \SRlowzjd{b}, \SRlownjd{b} & \SRlowzjd{b-c}, \SRlownjd{b-c} & \SRhighzjd{b}, \SRhighnjd{b} & \SRhighzjd{b-c}, \SRhighnjd{b-c} \\[0.1cm]\hline\hline
& \multicolumn{5}{c}{\incSRlowhigh} \\
& \texttt{d} & \texttt{e1} & \texttt{e2} & \texttt{f1} & \texttt{f2} \\\cline{2-6}
\mllmin [\GeV] & $[12,30]$ & $[12,40]$ & $[20,40]$ & $[12,60]$ & $[30,60]$ \\\cline{2-6}
\rule{0pt}{\dimexpr.7\normalbaselineskip+1mm}
& \SRlowzjd{b-d}, \SRlownjd{b-d}, \SRhighzjd{b-d}, \SRhighnjd{b-d} & \SRlowzjd{b-e}, \SRlownjd{b-e}, \SRhighzjd{b-e}, \SRhighnjd{b-e} & \SRlowzjd{c-e}, \SRlownjd{c-e}, \SRhighzjd{c-e}, \SRhighnjd{c-e} & \SRlowzjd{c-f2}, \SRlownjd{c-f2}, \SRhighzjd{c-f2}, \SRhighnjd{c-f} &  \SRlowzjd{e-f2}, \SRlownjd{e-f2}, \SRhighzjd{e-f2}, \SRhighnjd{e-f} \\[0.1cm] \hline\hline
& \multicolumn{4}{c}{\incSRlowhigh} & \\
& \texttt{g1} & \texttt{g2} & \texttt{g3} & \texttt{g4} & \\\cline{2-5}
\mllmin [\GeV] & $[12,75]$ & $[30,75]$ & $[40,75]$ & $[60,75]$ & \\\cline{2-5}
\rule{0pt}{\dimexpr.7\normalbaselineskip+1mm}
& \SRlowzjd{b-g2}, \SRlownjd{b-g2}, \SRhighzjd{b-g2}, \SRhighnjd{b-g} & \SRlowzjd{e-g2}, \SRlownjd{e-g2}, \SRhighzjd{e-g2}, \SRhighnjd{e-g} & \SRlowzjd{f1-g2}, \SRlownjd{f1-g2}, \SRhighzjd{f1-g2}, \SRhighnjd{f1-g} & \SRlowzjd{g1-g2}, \SRlownjd{g1-g2}, \SRhighzjd{g1-g2}, \SRhighnjd{g} & \\
\hline\hline
\end{tabular}
}
\end{center}
\end{table}
% End of text imported from the .//tables/ofs_selectionSRinc.tex input file
 
The 95\% CL upper limits on the generic BSM cross section are calculated by performing a discovery fit for each target SR and its associated CRs, using pseudo-experiments.
Results are reported in Table~\ref{table.results.exclxsec.pval.upperlimit.Disc1} (\ref{tab:results:offShell_discSRs})
for the \onShell and \WhAna selections (\offShell selection).
The tables list
the observed ($N_{\text{obs}}$) and expected ($N_{\text{exp}}$) yields in the inclusive SRs,
the upper limits on the observed ($S^{95}_{\mathrm{obs}}$) and expected ($S^{95}_{\mathrm{exp}}$) number of BSM events,
and the visible cross section ($\sigma^{95}_{\mathrm{vis}}$) reflecting the product of the production cross section, the acceptance,
and the selection efficiency for a BSM process;
the $p$-value and significance ($Z$) for the background-only hypothesis are also presented.
\begin{DIFnomarkup}\vspace{-2em}\mbox{\ }\end{DIFnomarkup}
 
% The next lines are included from the .//sections/log_Onshell_FullSyst_Discovery.tex input file
 
\begin{table}[p!]
\centering
\caption{
Observed ($N_{\text{obs}}$) yields after the discovery fit and expected ($N_{\text{exp}}$) after the background-only fit, for the inclusive SRs of the \ons \WZ and \Wh selections.
The third and fourth columns list the 95\% CL upper limits on the visible cross section ($\sigma_{\text{vis}}^{95}$) and on the number of signal events ($S_\text{obs}^{95}$).
The fifth column ($S_\text{exp}^{95}$) shows the 95\% CL upper limit on the number of signal events, given the expected number (and $\pm 1\sigma$ excursions of the expectation) of background events.
The last two columns indicate the \CLb value, i.e.\ the confidence level observed for the background-only hypothesis, and the discovery $p$-value ($p(s = 0)$).
If the observed yield is below the expected yield, the $p$-value is capped at 0.5.}
\label{table.results.exclxsec.pval.upperlimit.Disc1}
\setlength{\tabcolsep}{0.8pc}
\adjustbox{max width=0.85\textwidth}{
\begin{tabular*}{\textwidth}{L{2.8cm}r@{\extracolsep{1.9pc}}r@{\extracolsep{0.16pc}}L{0.85cm}rrr@{\extracolsep{0pc}}lrr}\cline{1-10}\tabvspacei{1.5}
SR              & $N_{\text{obs}}$ & \multicolumn{2}{c}{$N_{\text{exp}}\mbox{\ \ \,}$}         & $\sigma_{\text{vis}}^{95}$ [fb]  &  $S_\text{obs}^{95}$  & \multicolumn{2}{c}{\ \,$S_\text{exp}^{95}$} & \CLb & $p(s=0)$ ($Z$)  \\\cline{1-10}\tabvspacei{1.5}
\incSRWZi{1}	   		& 34			&  $38$ 		& $\pm\,5$			& $0.10$ 		&  $14  $ 		& $ { 16   }$ & ${}^{ +7   }_{ -4   }$ 		& $0.32$ 		& $ 0.50$~$(0.00)$ \\[0.04cm]
\incSRWZi{2}	   		& 2			& $1.2$ 		& $\pm\,0.5$	   & $0.04$ 		&  $5.0$ 		& $ { 4.0  }$ & ${}^{ +1.6 }_{ -0.7 }$ 		& $0.76$ 		& $ 0.23$~$(0.73)$ \\[0.04cm]
\incSRWZi{3}	   		& 4			& $6.5$ 		& $\pm\,1.1$		& $0.03$ 		& $4.8$ 		   & $ { 6.5  }$ & ${}^{ +2.6 }_{ -1.8 }$ 		& $0.19$ 		& $ 0.50$~$(0.00)$  \\[0.04cm]
\incSRWZi{4}	   		& 25			&  $31$ 		& $\pm\,6$			& $0.09$ 		&  $12  $ 		& $ { 15   }$ & ${}^{ +6   }_{ -4   }$ 		& $0.25$ 		& $ 0.50$~$(0.00)$ \\[0.04cm]
\incSRWZi{5}	   		& 1			& $5.2$ 		& $\pm\,1.1$		& $0.03$ 		&  $3.9$ 		& $ { 5.8  }$ & ${}^{ +2.2 }_{ -1.4 }$ 		& $0.03$ 		& $ 0.50$~$(0.00)$ \\[0.04cm]
\incSRWZi{6}	   		& 23			& $16.4 $ 	& $\pm\,1.4$		& $0.12$ 		&  $17.0$ 		& $ { 10.3 }$ & ${}^{ +3.9 }_{ -3.0 }$ 		& $0.93$ 		& $ 0.07$~$(1.48)$ \\[0.04cm]
\incSRWhSFi{7 }			& 174			& $150$ 		& $\pm\,14$  		& $0.41$ 		&  $58  $ 		& $ { 38   }$ & ${}^{ +15   }_{ -11   }$ 	& $0.90$ 		& $ 0.10$~$(1.27)$ \\[0.04cm]
\incSRWhSFi{8 }			& 53			&  $55$ 		& $\pm\,5$   		& $0.12$ 		&  $17  $ 		& $ { 18   }$ & ${}^{ +7   }_{ -5   }$ 		& $0.42$ 		& $ 0.50$~$(0.00)$  \\[0.04cm]
\incSRWhSFi{9 }			& 34  		& $36 $ 		& $\pm\,4$	    	& $0.10$ 		&  $14  $ 		& $ { 15   }$ & ${}^{ +6   }_{ -4   }$ 		& $0.40$ 		& $ 0.50$~$(0.00)$  \\[0.04cm]
\incSRWhSFi{10}			& 56			&  $55$ 		& $\pm\,7$ 		   & $0.16$ 		&  $22  $ 		& $ { 21   }$ & ${}^{ +8   }_{ -6   }$ 		& $0.55$ 		& $ 0.41$~$(0.22)$ \\[0.04cm]
\incSRWhSFi{11}			& 41			&  $45$ 		& $\pm\,6$ 		   & $0.11$ 		&  $16  $ 		& $ { 18   }$ & ${}^{ +7   }_{ -5   }$ 		& $0.34$ 		& $ 0.50$~$(0.00)$ \\[0.04cm]
\incSRWhDFi{12}			& 18			&  $11.5$ 	& $\pm\,4.1$ 		& $0.12$ 		&  $17.0$ 		& $ { 10.5 }$ & ${}^{ +4.2 }_{ -2.7 }$ 		& $0.92$ 		& $ 0.07$~$(1.48)$  \\[0.04cm]
\cline{1-10}

\end{tabular*}
}
\end{table}
 
% End of text imported from the .//sections/log_Onshell_FullSyst_Discovery.tex input file
% The next lines are included from the .//sections/yields_discSR_offShell.tex input file
\begin{table}[p!]
\centering
\caption{
Observed ($N_{\text{obs}}$) yields after the discovery fit and expected ($N_{\text{exp}}$) after the background-only fit, for the inclusive SRs of the \ofs \WZ selection.
The third and fourth columns list the 95\% CL upper limits on the visible cross section ($\sigma_{\text{vis}}^{95}$) and on the number of signal events ($S_\text{obs}^{95}$).
The fifth column ($S_\text{exp}^{95}$) shows the 95\% CL upper limit on the number of signal events, given the expected number (and $\pm 1\sigma$ excursions of the expectation) of background events.
The last two columns indicate the \CLb value, i.e.\ the confidence level observed for the background-only hypothesis, and the discovery $p$-value (p$(s = 0)$).
If the observed yield is below the expected yield, the $p$-value is capped at 0.5.
}
\label{tab:results:offShell_discSRs}
\setlength{\tabcolsep}{0.8pc}
\adjustbox{max width=0.85\textwidth}{
\begin{tabular*}{\textwidth}{L{2.8cm}r@{\extracolsep{1.9pc}}r@{\extracolsep{0.16pc}}L{0.85cm}rrr@{\extracolsep{0pc}}lrr}\cline{1-10}\tabvspacei{1.5}
SR & $N_{\text{obs}}$ & \multicolumn{2}{c}{$N_{\text{exp}}\mbox{\ \ \ \,}$} & $\sigma_{\text{vis}}^{95}$ [fb] &
$S_\text{obs}^{95}$  & \multicolumn{2}{c}{$S_\text{exp}^{95}$} & \CLb & $p(s=0)$ ($Z$)  \\\cline{1-10}\tabvspacei{1.5}
\incSRhighji{nja}  & $3$   &  $6.0$ & $\pm\,1.6$ & $0.03$ & $4.6$ & $6.3$ & ${}^{ +2.4 }_{ -2.0 }$  & $0.16$ & $ 0.50$~$(0.00)$ \\[0.04cm]
\incSRhighji{njb}  & $2$   &  $1.4$ & $\pm\,0.6$ & $0.03$ & $4.8$ & $4.0$ & ${}^{ +1.6 }_{ -0.7 }$  & $0.71$ & $ 0.30$~$(0.53)$ \\[0.04cm]
\incSRhighji{njc1} & $7$   &  $9.5$ & $\pm\,2.2$ & $0.05$ & $7.0$ & $8.4$ & ${}^{ +2.9 }_{ -2.2 }$  & $0.28$ & $ 0.50$~$(0.00)$ \\[0.04cm]
\incSRhighji{njc2} & $2$   &  $2.1$ & $\pm\,0.8$ & $0.03$ & $4.7$ & $4.6$ & ${}^{ +1.8 }_{ -1.1 }$  & $0.52$ & $ 0.50$~$(0.00)$ \\[0.04cm]
\incSRlowji{b}     & $31$  &  $36$  & $\pm\,4$   & $0.09$ & $12$  & $15 $ & ${}^{ +6 }_{ -4 }$       & $0.25$ & $ 0.50$~$(0.00)$ \\[0.04cm]
\incSRhighji{b}    & $3$   &  $3.0$ & $\pm\,0.9$ & $0.04$ & $5.4$ & $5.2$ & ${}^{ +2.0 }_{ -1.3 }$  & $0.53$ & $ 0.50$~$(0.00)$ \\[0.04cm]
\incSRlowji{c}     & $86$  &  $88$  & $\pm\,7$   & $0.17$ & $23$  & $24 $ & ${}^{ +9 }_{ -7 }$       & $0.44$ & $ 0.50$~$(0.00)$ \\[0.04cm]
\incSRhighji{c}    & $9$   &  $9.3$ & $\pm\,1.5$ & $0.06$ & $7.7$ & $7.7$ & ${}^{ +3.4 }_{ -1.8 }$  & $0.50$ & $ 0.50$~$(0.00)$ \\[0.04cm]
\incSRj{d}         & $202$ &  $184$ & $\pm\,12$  & $0.37$ & $51$  & $37 $ & ${}^{ +14 }_{ -11 }$     & $0.84$ & $ 0.16$~$(0.99)$ \\[0.04cm]
\incSRj{e1}        & $332$ &  $308$ & $\pm\,17$  & $0.49$ & $68$  & $49 $ & ${}^{ +19 }_{ -15 }$     & $0.84$ & $ 0.16$~$(1.00)$ \\[0.04cm]
\incSRj{e2}        & $298$ &  $269$ & $\pm\,15$  & $0.50$ & $69$  & $46 $ & ${}^{ +17 }_{ -14 }$     & $0.90$ & $ 0.10$~$(1.29)$ \\[0.04cm]
\incSRj{f1}        & $479$ &  $457$ & $\pm\,22$  & $0.56$ & $78$  & $63 $ & ${}^{ +22 }_{ -20 }$     & $0.77$ & $ 0.23$~$(0.75)$ \\[0.04cm]
\incSRj{f2}        & $277$ &  $272$ & $\pm\,13$  & $0.33$ & $46$  & $42 $ & ${}^{ +17 }_{ -12 }$     & $0.60$ & $ 0.37$~$(0.34)$ \\[0.04cm]
\incSRj{g1}        & $620$ &  $593$ & $\pm\,28$  & $0.69$ & $96$  & $74 $ & ${}^{ +29 }_{ -22 }$     & $0.77$ & $ 0.21$~$(0.79)$ \\[0.04cm]
\incSRj{g2}        & $418$ &  $408$ & $\pm\,20$  & $0.46$ & $64$  & $57 $ & ${}^{ +23 }_{ -15 }$     & $0.65$ & $ 0.32$~$(0.47)$ \\[0.04cm]
\incSRj{g3}        & $288$ &  $285$ & $\pm\,16$  & $0.35$ & $48$  & $47 $ & ${}^{ +19 }_{ -12 }$     & $0.55$ & $ 0.38$~$(0.30)$ \\[0.04cm]
\incSRj{g4}        & $141$ &  $136$ & $\pm\,10$  & $0.25$ & $35$  & $31 $ & ${}^{ +13 }_{ -8 }$      & $0.64$ & $ 0.35$~$(0.39)$ \\[0.04cm]
\cline{1-10}
\end{tabular*}
}
\end{table}

% End of text imported from the .//sections/yields_discSR_offShell.tex input file

\FloatBarrier
\begin{DIFnomarkup}
\subsection{Constraints on \WZ- and \Wh-mediated models}\label{ssec:results:excl}
\end{DIFnomarkup}
 
Constraints on the target simplified models are derived using the nominal SRs discussed in Sections~\ref{ssec:onshellAndWh:reg} and \ref{ssec:offshell:reg}.
The results are statistically combined with the previous results for the electroweakino regions (\SRE) of the two-lepton search targeting compressed mass spectra~\cite{SUSY-2018-16}, referred to as the compressed selection.
Model-dependent 95\% CL exclusion limits are calculated by performing the exclusion fits as described in Section~\ref{ssec:analysisstrategy:fit}.
When performing the combination, common experimental uncertainties are treated as correlated between regions and processes.
Theoretical uncertainties of the background and signal are treated as correlated between regions only,
while statistical uncertainties are considered uncorrelated between regions and processes.

All regions of the \ons \WZ, \ofs \WZ, and compressed selections were explicitly designed to be orthogonal,
allowing a statistical combination of the results.
The \ons and \ofs \WZ selections are orthogonal due to the \mll and \met requirements,
while the \ofs \WZ and compressed selections are orthogonal by lepton multiplicity.
Results are combined where greater exclusion power is expected over the individual results,
ignoring contributions from search regions that do not add sensitivity in a given region of phase space.
This approach results in multiple pairwise combinations of
the \ons and \ofs \WZ selections,
and the \ofs \WZ and compressed selections,
in bands of the $(\dm,\,\mNtwo)$ plane.

Four separate fits are performed to obtain constraints for the following simplified models:
\begin{itemize}
\begin{DIFnomarkup}
\item \mbox{the wino/bino (+) \WZmed} combining the \ons \WZ, \ofs \WZ, and compressed selections,
\item \mbox{the wino/bino (+) \Whmed} using the \Wh selection only,
\item \mbox{the wino/bino ($-$) \WZmed} combining the \ofs \WZ and compressed selections,
\end{DIFnomarkup}
\item \mbox{the higgsino \WZmed} combining the \ofs \WZ and compressed selections.
\end{itemize}\begin{DIFnomarkup}\vskip 1.0em\end{DIFnomarkup}

For the \WZmed in the wino/bino (+) scenario, only the \SRonWZ are sensitive for mass splittings $\dm$ above $100~\GeV$.
Conversely, the \SRoffWZ dominate the intermediate mass-splitting region, with sensitivity in the $\dm = [5,100]~\GeV$ range.
In the most compressed region, the \SRE are important, driving the result for $\dm$ below $10~\GeV$ and adding sensitivity up to $\dm = 50~\GeV$.
Given these contributions, the $\dm$ range is split into five bands: [$<$8, 8--28, 28--78, 78--108, $>$108]~\GeV.
In each of these bands the combination considers respectively the \SRE only, the \SRE and \SRoffWZ, the \SRoffWZ only, the \SRoffWZ and \SRonWZ, and the \SRonWZ only.
In the wino/bino ($-$) scenario, the \ons \WZ selection is not used, and three bands are defined for the combination as [$<$8, 8--28, $>$28]~\GeV.
In the higgsino scenario, the mass-splitting range is restricted to $\dm \leq 60~\GeV$ for model consistency, and the \ons \WZ selection is not used.
The combination is defined using three bands: [$<$8, 8--42, $>$42]~\GeV,
respectively considering the \SRE only, the \SRE and \SRoffWZ, and the \SRoffWZ only.
The ranges used are illustrated for the different scenarios in Figure~\ref{fig:combiranges}.
 
\begin{figure}[htp]
\centering
\includegraphics[width=0.68\textwidth]{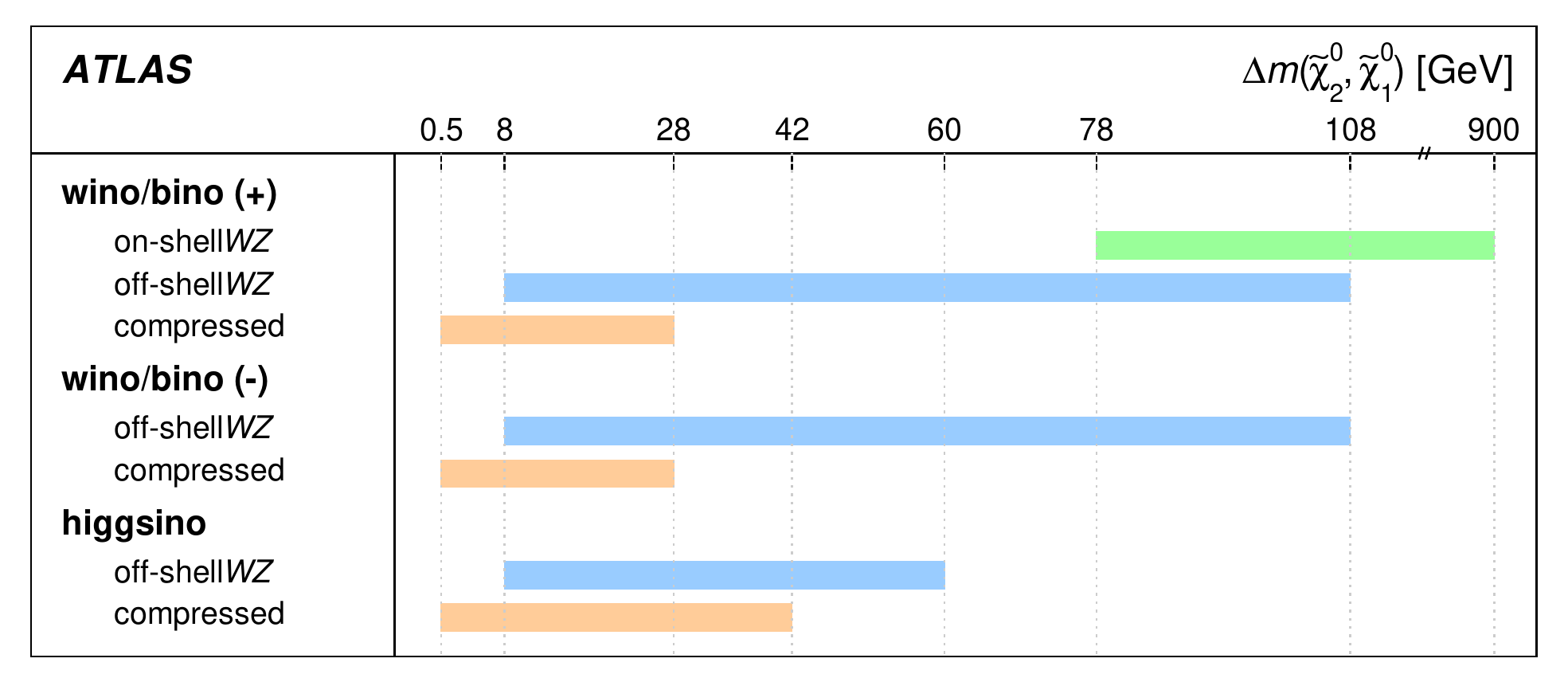}
\caption{Illustration of the selections considered for the combined result for each scenario, dependent on \dm.}
\label{fig:combiranges}
\end{figure}
 
Expected and observed exclusion contours are reported as a function of the \textninoone and \textchinoonepm/\textninotwo masses,
and shown in Figure~\ref{fig:results:limit_WZ} (\WZmed) and Figure~\ref{fig:results:limit_Wh} (\Whmed).
The combined results are shown together with the individual contributions.
For each mass point, a \CLs value is derived to assess the probability of compatibility between the observed data and the signal-plus-background prediction obtained by the exclusion fit.
For the \WZmed, the results are obtained by statistically combining the \SRonWZ, \SRoffWZ and \SRE contributions, following the prescription outlined above.
For the \Whmed, the results are taken from a simultaneous fit of the 19 bins of \SRWh.

{
\begin{DIFnomarkup}\pagebreak\enlargethispage*{2\baselineskip}\end{DIFnomarkup}
For the wino/bino (+) \WZmed, shown in Figure~\ref{fig:results:limit_WZ} (top panels), observed (expected) lower limits for equal-mass \textchinoonepm/\textninotwo are set at 640~(660)~\GeV\ for massless \textninoone,
and up to 300~(300)~\GeV\ for scenarios with mass splittings \dm near \mZ, driven by the \ons \WZ selection.
The exclusion for the scenarios with $\dm < m_Z$ is driven by the \ofs \WZ selection. 
For \textchinoonepm and \textninotwo decaying via off-shell \WZ bosons,
observed and expected limits are set at values up to 300~\GeV\ for $\dm > 35~\GeV$,
and up to 210--300~\GeV\ for $\dm = 20$--35~\GeV.
Below $\dm = 15~\GeV$ the observed and expected limits are extended by the combination with the compressed selection,
up to 240~\GeV\ for $\dm = 10~\GeV$, and down to as low as $\dm = 2~\GeV$ for a \textchinoonepm/\textninotwo mass of 100~\GeV.
Furthermore, constraints are calculated in the bino--wino co-annihilation dark-matter scenario
by determining the area in the two-dimensional mass plane that yields a thermal dark-matter relic density equal to the observed value~\cite{Aghanim:2015xee}.
Figure~\ref{fig:results:limit_WZ} (top right) shows this area in blue,  
with the over- and under-abundant regions marked above and below;
\textchinoonepm/\textninotwo (\textninoone) masses are excluded in this dark-matter scenario up to 210~(195)~\GeV.
 
The obtained wino/bino (+) exclusion limits are greatly improved compared to
the previous equivalent search presented by the ATLAS experiment using the Run\,1, 8~\TeV~dataset~\cite{SUSY-2013-12}
(shown as a light grey shaded area in Figure~\ref{fig:results:limit_WZ}, top panels),
due to a combination of increased production cross section at the increased collision centre-of-mass energy,
larger data sample, and improved analysis techniques.

Expected and observed exclusion contours are also derived for the \WZmed in the wino/bino ($-$) and higgsino scenarios,
shown in Figure~\ref{fig:results:limit_WZ} (bottom panels) as a function of the \textninoone and \textninotwo masses.
The results are obtained by statistically combining the \SRoffWZ and \SRE contributions, following the prescription outlined above.
 
In the wino/bino ($-$) scenario, shown in Figure~\ref{fig:results:limit_WZ} (bottom left), observed (expected) lower limits for equal-mass \textchinoonepm/\textninotwo are set
at values up to 310 (300)~\GeV\ for mass splittings \dm around 80~\GeV, and up to 250 (250)~\GeV\ for \dm around 40~\GeV.
For \dm of 10--20~\GeV, the impact of the combination of the \ofs \WZ and compressed results is the largest,
and raises the expected limit to \textchinoonepm/\textninotwo masses of 270~\GeV,
with the observed limit still showing a mild deficit similar to that visible in the compressed contribution.
At a \textchinoonepm/\textninotwo mass of 100~\GeV, the observed (expected) exclusion extends down to $\dm = 1~(1.5)$~\GeV.
 
In the higgsino scenario,
shown in Figure~\ref{fig:results:limit_WZ} (bottom right),
with the \textchinoonepm mass between that of the \textninoone and \textninotwo,
limits are set for mass splittings \dm up to 60~\GeV.
For \dm between 30 and 60~\GeV, observed (expected) limits extend to around 150--210 (160--215)~\GeV.
The impact of the combination of the \ofs \WZ and compressed results is largest in the $\dm = 15$--30~\GeV\ range,
improving on the individual result by up to 15~\GeV.
Below $\dm = 20~\GeV$, the result is dominated by the compressed contribution,
and limits extend down to $\dm = 2~\GeV$.
}
 
\begin{DIFnomarkup}\pagebreak\end{DIFnomarkup}
\begin{figure}[tb]
\centering
\includegraphics[width=0.485\textwidth]{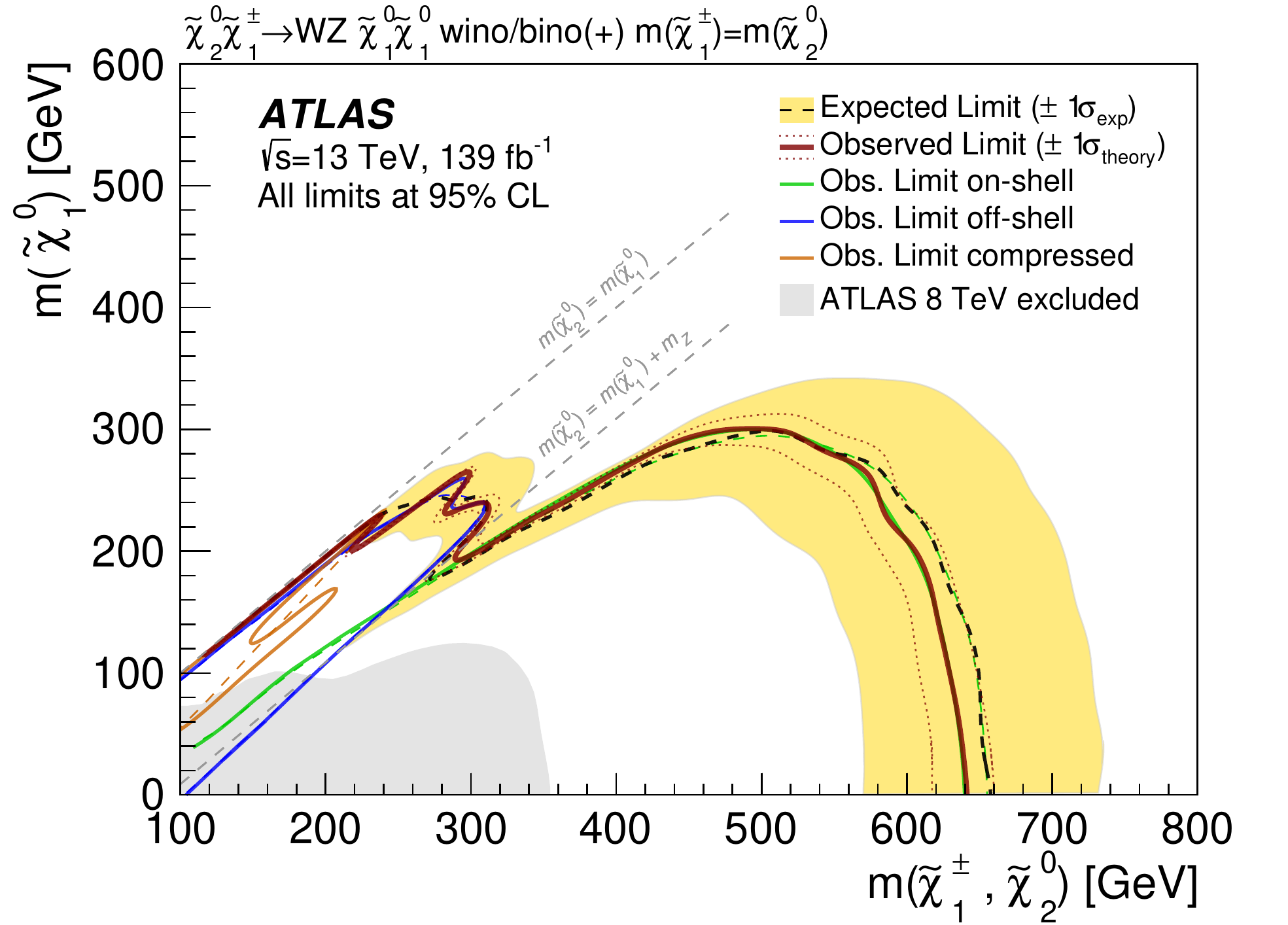}
\includegraphics[width=0.485\textwidth]{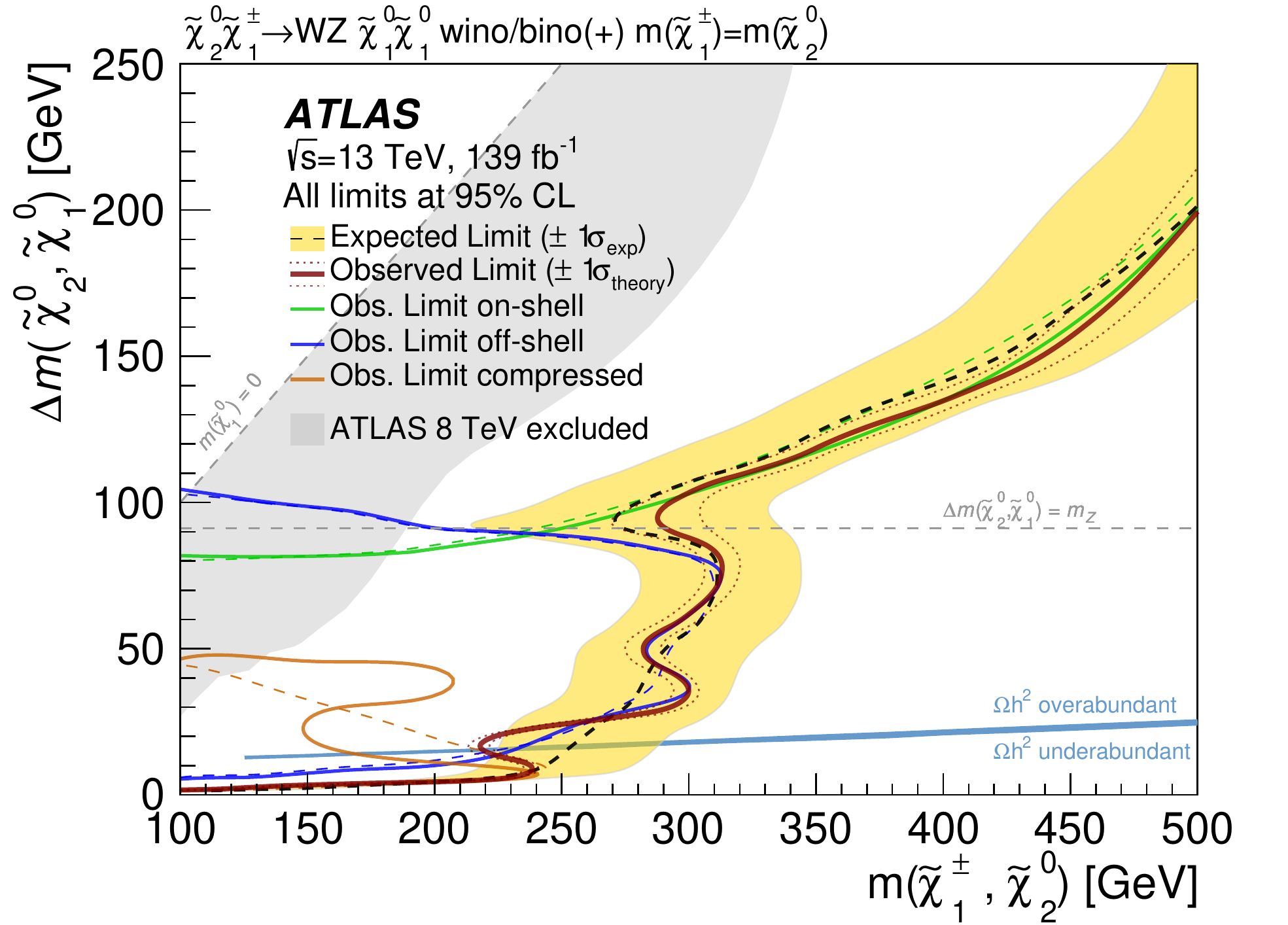}
\includegraphics[width=0.485\textwidth]{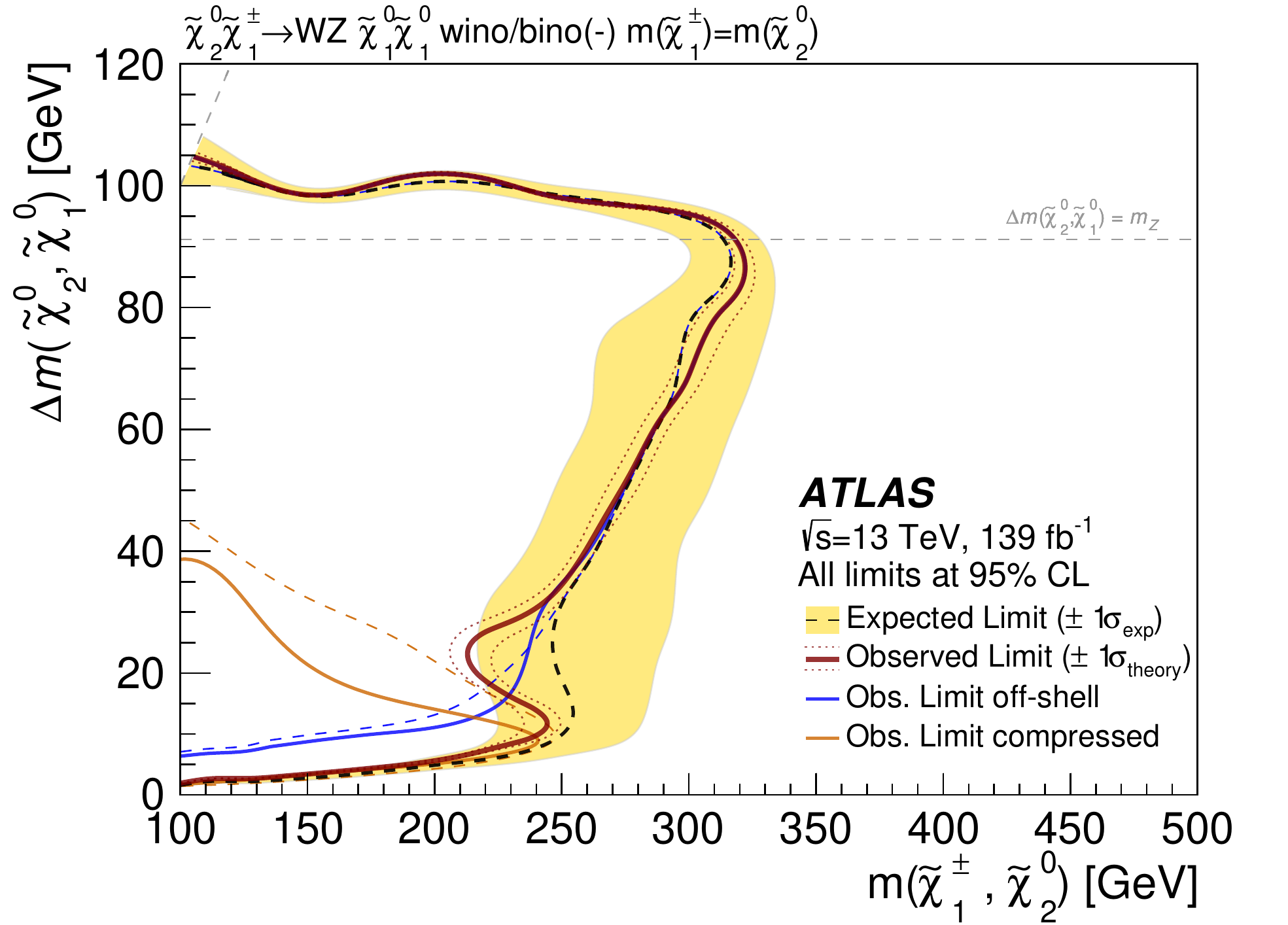}
\includegraphics[width=0.485\textwidth]{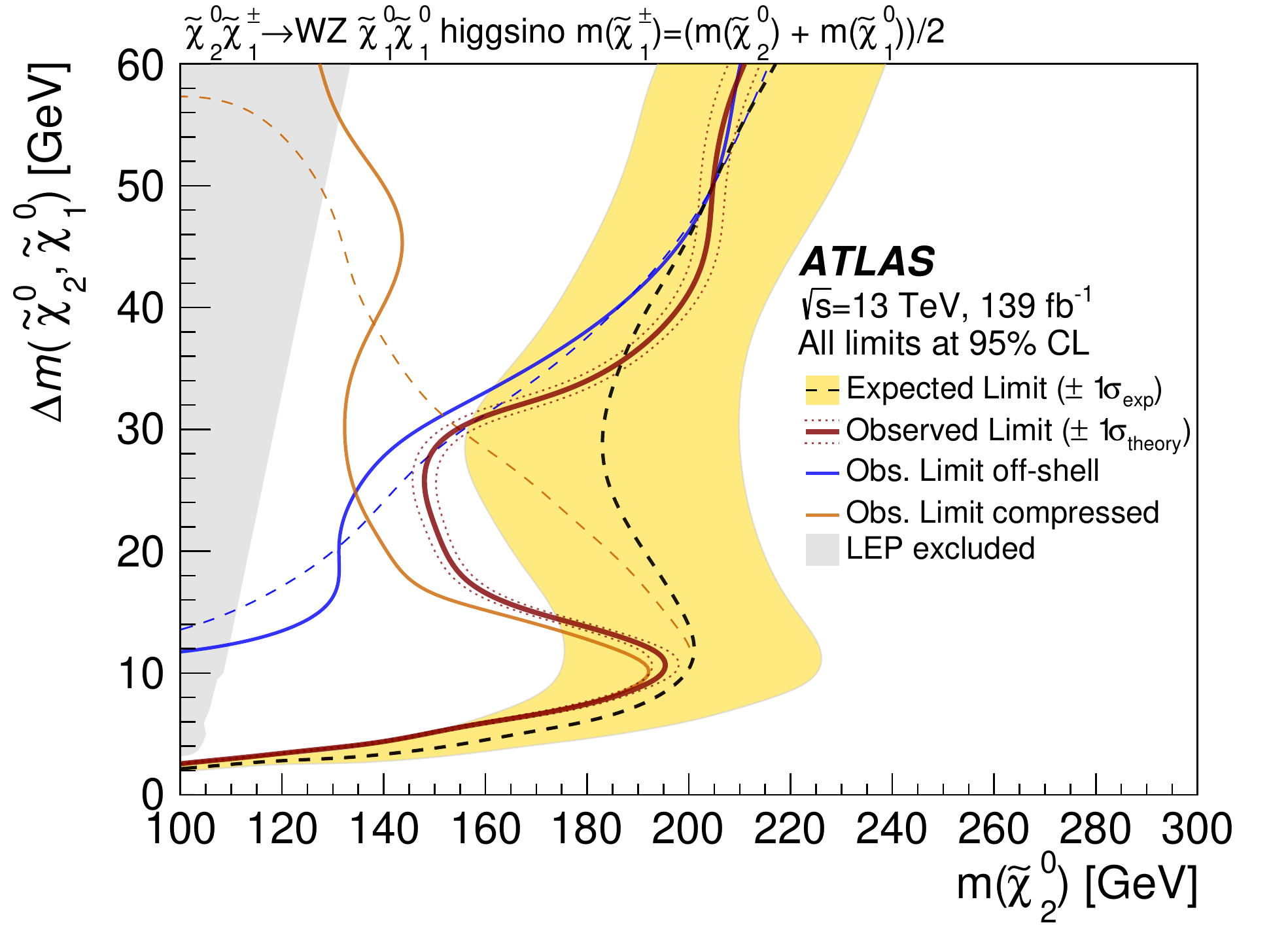}
\caption{
Exclusion limits obtained for the \WZ-mediated models in the
(top left and right) wino/bino (+) scenario,
(bottom left) the wino/bino ($-$) scenario,
and (bottom right) the higgsino scenario.
The expected 95\% CL sensitivity (dashed black line) is shown with $\pm1\sigma_{\text{exp}}$ (yellow band)
from experimental systematic uncertainties and statistical uncertainties in the data yields,
and the observed limit (red solid line) is shown with $\pm1\sigma_{\text{theory}}$ (dotted red lines) from signal cross-section uncertainties.
The statistical combination of the \ons \WZ, \ofs \WZ, and compressed results is shown as the main contour,
while the observed (expected) limits for each individual selection are overlaid in green, blue, and orange solid (dashed) lines, respectively.
The exclusion is shown projected (top left) onto the $m(\textchinoonepm,\,\textninotwo)$ vs $m(\textninoone)$ plane or (top right and bottom) onto the $m(\textninotwo)$ vs $\dm$ plane.
The light grey area denotes (top) the constraints obtained by the previous equivalent analysis in ATLAS using the 8~\TeV~20.3~\ifb\ dataset~\cite{SUSY-2013-12}, and (bottom right) the LEP lower \textchinoonepm mass limit~\cite{LEPlimits}.
The pale blue line in the top right panel represents the mass-splitting range that yields a dark-matter relic density equal to the observed relic density, $\Omega h^2=0.1186\pm0.0020$~\cite{Aghanim:2015xee}, when the mass parameters of all the decoupled SUSY partners are set to $5~\TeV$ and $\tan\beta$ is chosen
such that the lightest Higgs boson's mass is consistent with the observed value of the SM Higgs~\cite{duan2018probing}.
The area above (below) the blue line represents a dark-matter relic density larger (smaller) than the observed.
}
\label{fig:results:limit_WZ}
\end{figure}
 
{
\begin{DIFnomarkup}\enlargethispage*{4\baselineskip}\end{DIFnomarkup}
The obtained results for the wino/bino ($-$) and higgsino scenarios
complement the previous compressed result using two-lepton final states as well.
These results from the \ofs \WZ selection in three-lepton final states
make full use of the larger data sample and target a novel phase space in the intermediately compressed \dmNN region.
The new results extend the exclusion by up to 100~\GeV\ in \textninotwo mass.
 
For the wino/bino (+) \Whmed, observed (expected) lower limits for equal-mass \textchinoonepm/\textninotwo are set at values up to 190~(240)~\GeV\ for \textninoone masses below 20~\GeV,
as shown in Figure~\ref{fig:results:limit_Wh}.
The observed exclusion is weaker than the expected exclusion,
which is explained by the mild excess found in \SRWhDF;
the limits are, however, compatible within $2\sigma$.
The obtained observed (expected) limits show an improvement of up to 40 (80)~\GeV~compared to the previous Run\,1, 8~\TeV, ATLAS search~\cite{SUSY-2013-12}.
}
 
\begin{figure}[tbp]
\centering
\includegraphics[width=0.495\textwidth]{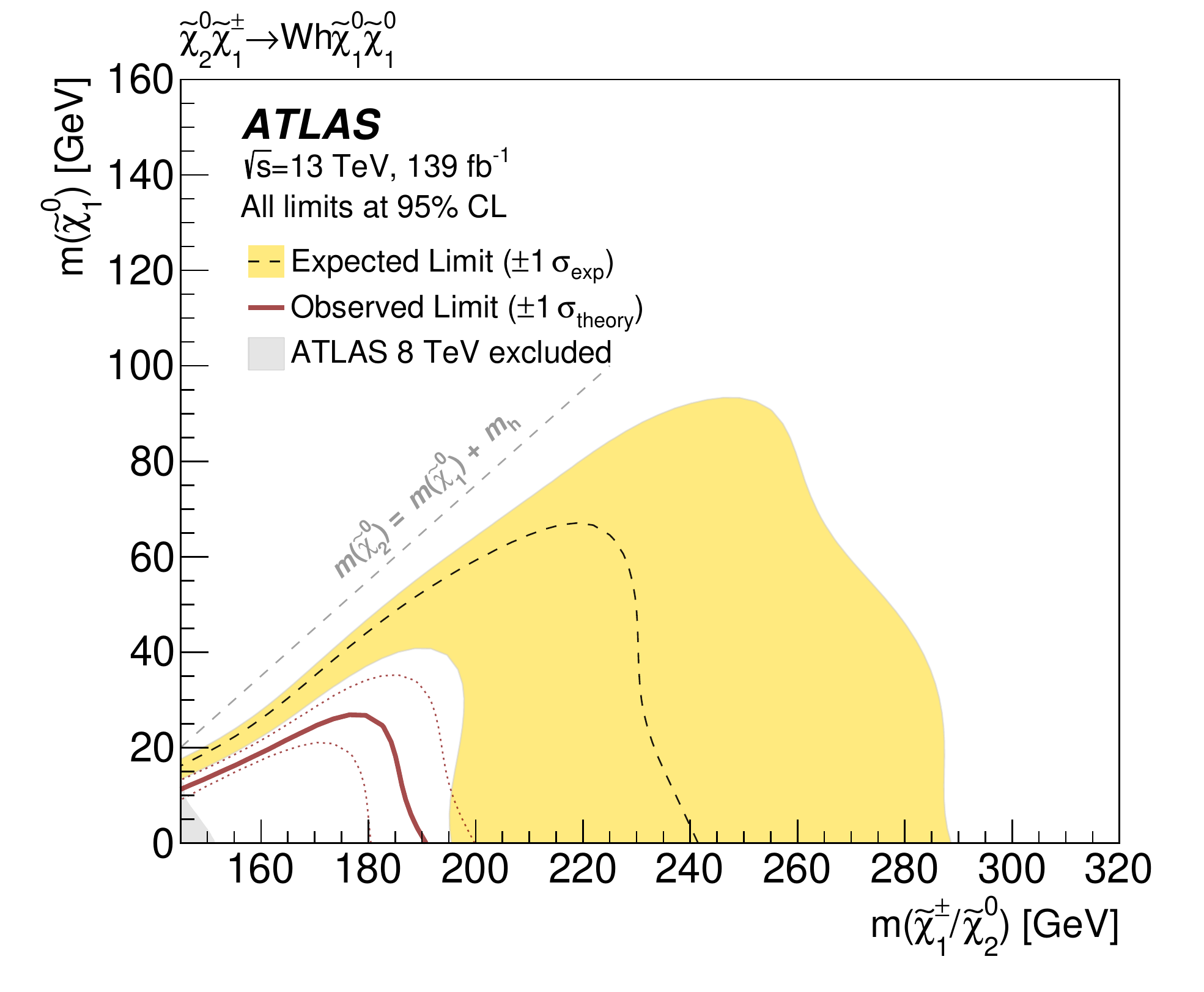}
\caption{
Exclusion limits obtained for the \Whmed in the wino/bino (+) scenario,
calculated using the \Wh SRs
and projected onto the $m(\textchinoonepm,\,\textninotwo)$ vs $m(\textninoone)$ plane.
The expected 95\% CL sensitivity (dashed black line) is shown with $\pm1\sigma_{\text{exp}}$ (yellow band)
from experimental systematic uncertainties and statistical uncertainties in the data yields,
and the observed limit (red solid line) is shown with $\pm1\sigma_{\text{theory}}$ (dotted red lines) from signal cross-section uncertainties.
The light grey area denotes the constraints obtained by the previous equivalent analysis in ATLAS using the 8~\TeV~20.3~\ifb\ dataset~\cite{SUSY-2013-12}.
}
\label{fig:results:limit_Wh}
\end{figure}

% End of text imported from the .//sections/results.tex input file
% The next lines are included from the .//sections/rj.tex input file
 
\FloatBarrier
\begin{DIFnomarkup}
\vskip 1em\section{\RJR selection and results}\label{sec:rj}
\end{DIFnomarkup}
 
To follow up on an earlier ATLAS search performed
using the \RJR~(RJR) technique
with the 2015--2016, 36~\ifb\ dataset~\cite{SUSY-2017-03},
the search in this paper includes two signal regions in which the original search observed excesses of three-lepton events.
The original search in the two regions is repeated following the same methods, updated to use the full \RunTwo dataset.
The \RJSRlow region targets low-mass wino/bino (+) \textchinoonepmninotwo production,
while the \RJSRISR region targets wino/bino (+) \textchinoonepmninotwo production in association with ISR and mass differences \dm near the $Z$-boson mass.
The excesses in \RJSRlow and \RJSRISR observed in the 36~\ifb\ result
correspond to local significances of 2.1$\sigma$ and 3.0$\sigma$, respectively.
 
The RJR technique endeavours to resolve the ambiguities inherent in reconstructing original particles
for event decays including invisible particles, e.g.\ SUSY particles.
By analysing the event starting from the laboratory frame and boosting back to the parent particle's rest frame,
assuming given decay chains, the technique can resolve the \textchinoonepm and \textninotwo particles.
For this search, both the standard decay tree applied to a three-lepton final state (representing the decay of pair-produced sparticles into a final state with two invisible objects and three leptons, in the laboratory frame)
and the ISR decay tree (representing the decay of an intermediate sparticle into a visible and an invisible component, recoiling from ISR activity, in the centre-of-mass frame) are considered.
Using the reconstructed leptons, jets, and missing transverse momentum as inputs,
the algorithm assigns each particle to a parent sparticle.
ISR jets are selected by minimising the invariant mass of the system formed by the candidate jets and the sparticle system, in the centre-of-mass frame.
The algorithm then determines the smallest Lorentz-invariant configuration of the particles' four-momenta guaranteeing a non-negative mass parameter for the invisible particles.
Finally, object or frame momenta and derived variables can be considered in each of the different frames of each decay tree.
 
The search in the \rjtl selection regions follows a similar strategy
for background estimation, systematic uncertainty treatment, and statistical interpretation
to that outlined for the \ons \WZ, \ofs \WZ, and \Wh selections in Section~\ref{sec:analysisstrategy}.
For the search in \RJSRlow (\RJSRISR), the SM diboson background is taken from MC simulation samples and normalised in a dedicated control region \RJCRVV (\RJCRVVISR) and validated in a validation region \RJVRVV (\RJVRVVISR).
The selection criteria for each of the regions follow the original search~\cite{SUSY-2017-03},
except for an additional jet-veto ($n_{\text{jets}}=0$) in \RJCRVV and \RJVRVV
which guarantees the orthogonality between the low-mass and ISR regions.
The FNP lepton background component, including \ttbar, $tW$, $WW$ and \Zjet SM background contributions, is estimated in a data-driven way using the matrix method~\cite{TOPQ-2010-01}.
The method derives the number of events with one or two FNP leptons by relating the yields for tighter (signal tagged) and looser (baseline tagged) lepton identification criteria.
The result is a function of the real-lepton identification efficiencies and the FNP lepton misidentification probabilities.
The remaining SM backgrounds, including multiboson and Higgs boson production, and top-pair production in association with a boson, are estimated from MC simulation in all analysis regions.
Beyond the treatment of experimental and theoretical systematical uncertainties following the general strategy in Section~\ref{ssec:analysisstrategy:systematics},
uncertainties are assigned to the matrix-method FNP lepton background estimation,
accounting for limited numbers of events in the measurement region,
potentially different compositions (heavy flavour, light flavour, or conversions) between SRs and CRs,
and the uncertainty from the subtraction of prompt-lepton contributions using MC simulation samples.

Performing the background-only fit, diboson normalisation factors of 0.92$\pm$0.07 (\RJCRVV) and 0.92$\pm$0.05 (\RJCRVVISR) are determined.
Observed and expected yields for all CRs and VRs are summarised in Figure~\ref{fig:RJ:CRVRsum}
and a summary of the considered systematic uncertainties is presented in
Figure~\ref{newfig:rjsystematics}, grouped as discussed in Section~\ref{ssec:analysisstrategy:systematics}.

\begin{figure}[b!]
\centering
{\includegraphics[width=0.85\textwidth]{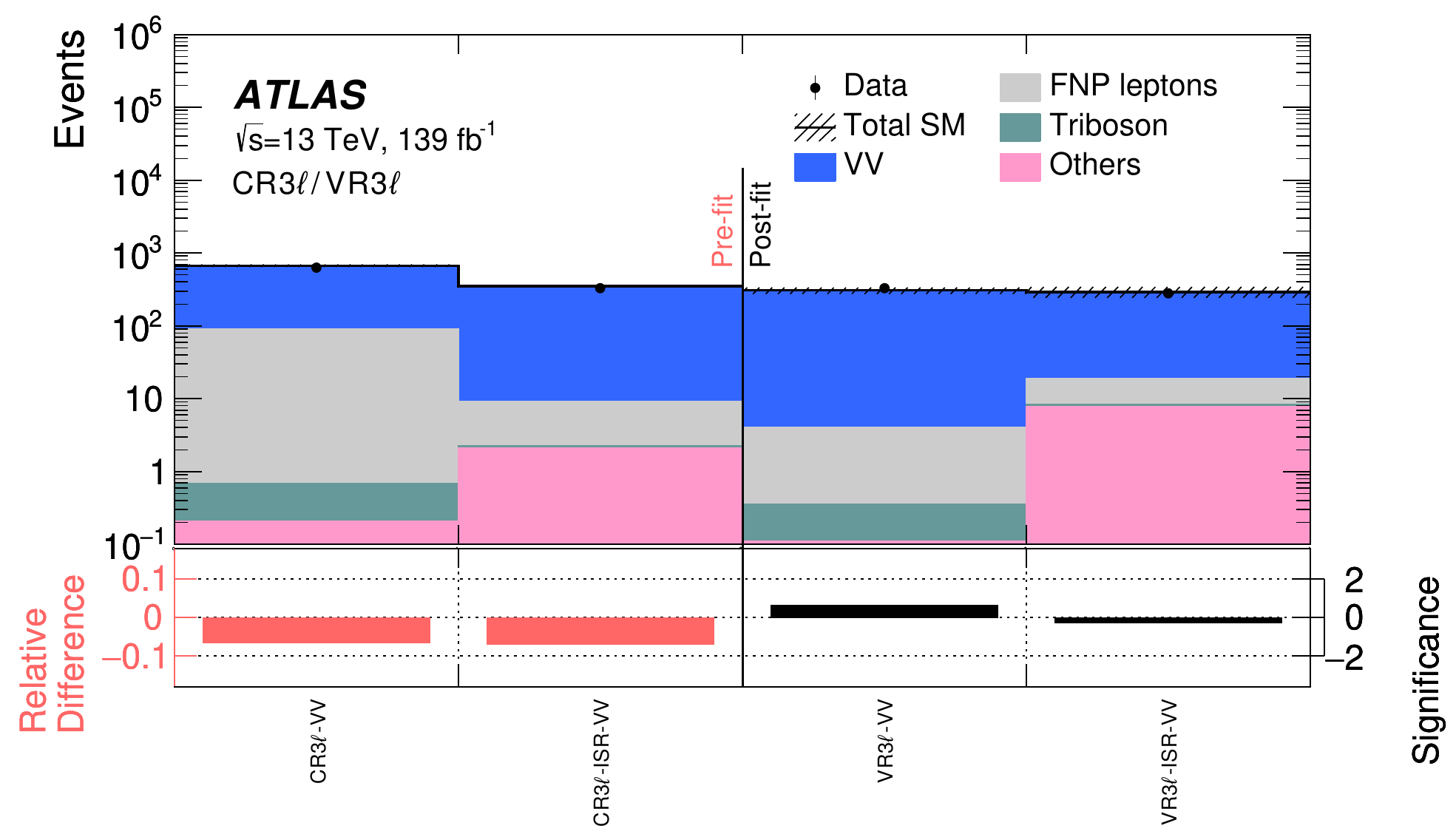}}
\caption{
Comparison of the observed data and expected SM background yields in the CRs and VRs of the \rjtl selection.
The SM prediction is taken from the background-only fit.
The `FNP leptons' category contains backgrounds from \ttbar, $tW$, $WW$ and \Zjet~processes.
The `Others' category contains backgrounds from Higgs and rare top processes.
The hatched band indicates the combined theoretical, experimental, and MC statistical uncertainties.
The bottom panel shows the significance of the difference between the observed and expected yields,
calculated with the profile likelihood method from Ref.~\cite{Cousins:2007bmb}, adding a minus sign if the yield is below the prediction.
}
\label{fig:RJ:CRVRsum}
\end{figure}
 
\begin{figure}[tp!]
\centering
\includegraphics[width=0.57\textwidth]{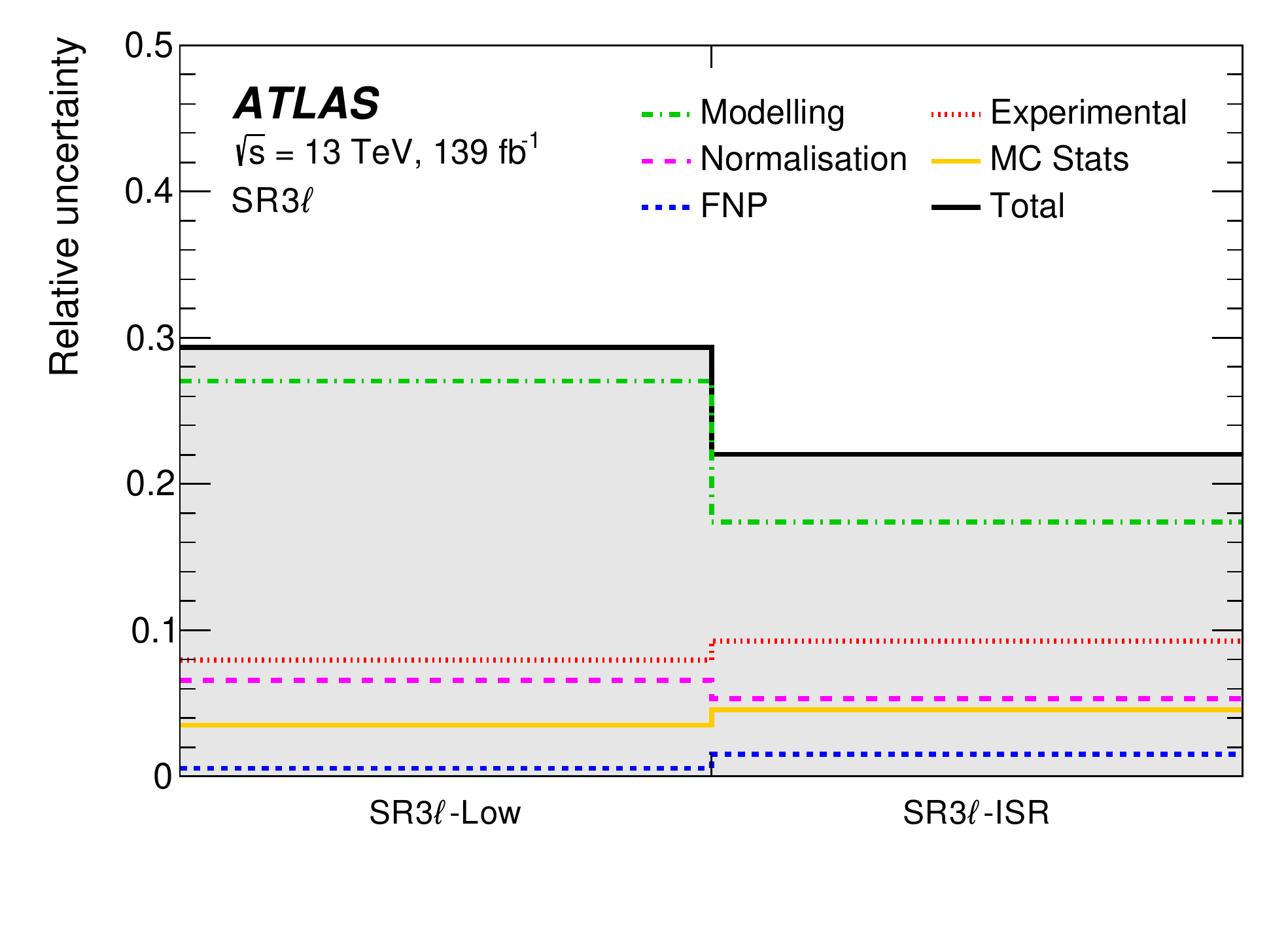}\begin{DIFnomarkup}\vskip -1.6em\end{DIFnomarkup}
\caption{
Breakdown of the total systematic uncertainties in the background prediction for the SRs of the \rjtl selection.
}
\label{newfig:rjsystematics}
\end{figure}
 
\begin{DIFnomarkup}\FloatBarrier\end{DIFnomarkup}
 
The observed data in \RJSRlow and \RJSRISR are compared with the background expectation obtained by the background-only fit.
The results are reported in Table~\ref{tab:RJ:yieldsSR} and
post-fit distributions of key observables for the SRs are shown in Figure~\ref{fig:RJ:SR}.
For the low-mass \rjtl selection, Figure~\ref{fig:RJ:SR} shows
the leading lepton's transverse momentum, \smash{\ptl{1}},
and the scalar momentum sum, \smash{\rjhpp}, of the three visible particles (the leptons) and the invisible particles (the LSPs and the neutrino),
in the pair-produced parent sparticle--sparticle (PP) frame and assuming the standard decay tree.
For the ISR \rjtl selection, Figure~\ref{fig:RJ:SR} shows
the vector sum of the transverse momenta of all objects, \rjptcm,
and the fraction of the total momentum of the sparticle system carried by the invisible system, \rjrisr,
in the centre-of-mass (CM) frame and assuming the ISR decay tree.
Good agreement with the background-only hypothesis is observed in both SRs.
The deviations from the SM expectation as found in the 36~\ifb\ result are reduced and no longer significant when including the additional 103~\ifb~of data from the 2017--2018 datasets.
 
Model-independent results for \RJSRlow and \RJSRISR are shown in Table~\ref{tab:results:RJ_model_indep}.
The 95\% CL upper limits on the generic BSM cross section are calculated by performing a discovery fit for each target SR and its associated CR, using pseudo-experiments.
The table lists
the upper limits on the observed ($S^{95}_{\mathrm{obs}}$) and expected ($S^{95}_{\mathrm{exp}}$) number of BSM events in the inclusive SRs,
and the visible cross section ($\sigma^{95}_{\mathrm{vis}}$) reflecting the product of the production cross section, the acceptance,
and the selection efficiency for a BSM process;
the $p$-value and significance ($Z$) for the background-only hypothesis are also presented.

% The next lines are included from the .//tables/rj_yields_SR.tex input file
\begin{table}[bh!]
\centering
\caption{
Observed and expected yields after the background-only fit in the SRs for the \rjtl selection.
The `FNP leptons' category contains backgrounds from \ttbar, $tW$, $WW$ and \Zjet~processes.
The `Others' category contains backgrounds from Higgs and rare top processes.
Combined statistical and systematic uncertainties are presented.
}
\label{tab:RJ:yieldsSR}
\adjustbox{max width=0.56\textwidth}{
\setlength{\tabcolsep}{0cm}
\begin{tabular}{l@{\hspace{0.5cm}} *{2}{R{1.65cm}P{0.4cm}L{1.6cm}}}
\midrule
Region           & \multicolumn{3}{c}{\RJSRlow} & \multicolumn{3}{c}{\RJSRISR}  \\
\midrule
Observed           & \multicolumn{3}{c}{$53$}              & \multicolumn{3}{c}{$25$}      \\
\midrule
Fitted SM  & $49$ & $\pm$ & $14$          & $17$ & $\pm$ & $4$      \\
\midrule
Diboson              & $47 $ & $\pm$ & $14$      & $16$ & $\pm$ & $4$       \\
FNP leptons          & $1.36  $ & $\pm$ & $0.29$       & $0.83 $ & $\pm$ & $0.27$      \\
Triboson           & $0.40  $ & $\pm$ & $0.14$       & $0.14 $ & $\pm$ & $0.06$       \\
Others             & $0.052  $ & $\pm$ & $0.029$       & $0.41 $ & $\pm$ & $0.21$       \\
\midrule
\end{tabular}
}
\end{table}
% End of text imported from the .//tables/rj_yields_SR.tex input file
\FloatBarrier
 
\begin{figure}[p!]
\centering
\includegraphics[width=0.485\columnwidth]{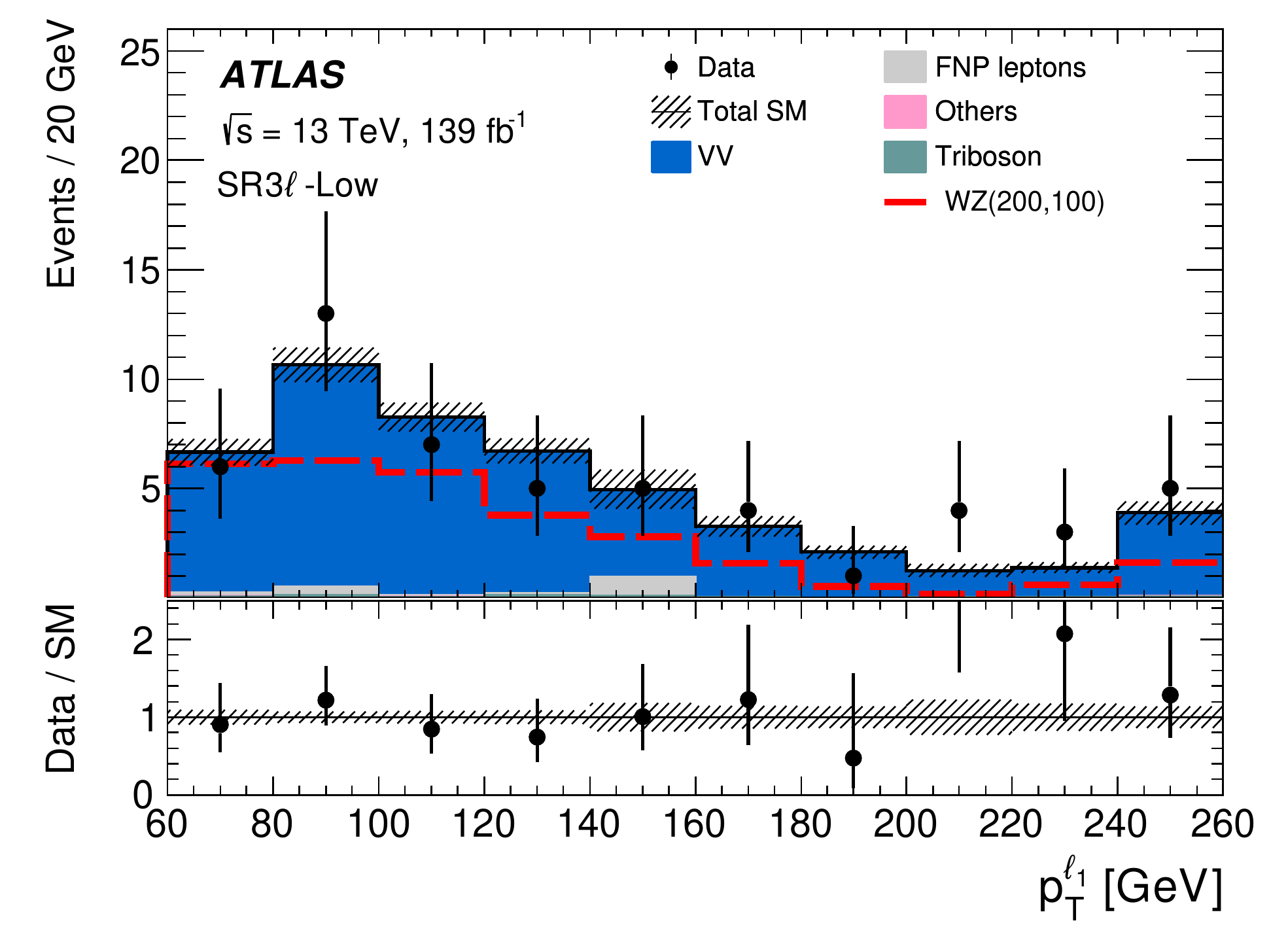}
\includegraphics[width=0.485\columnwidth]{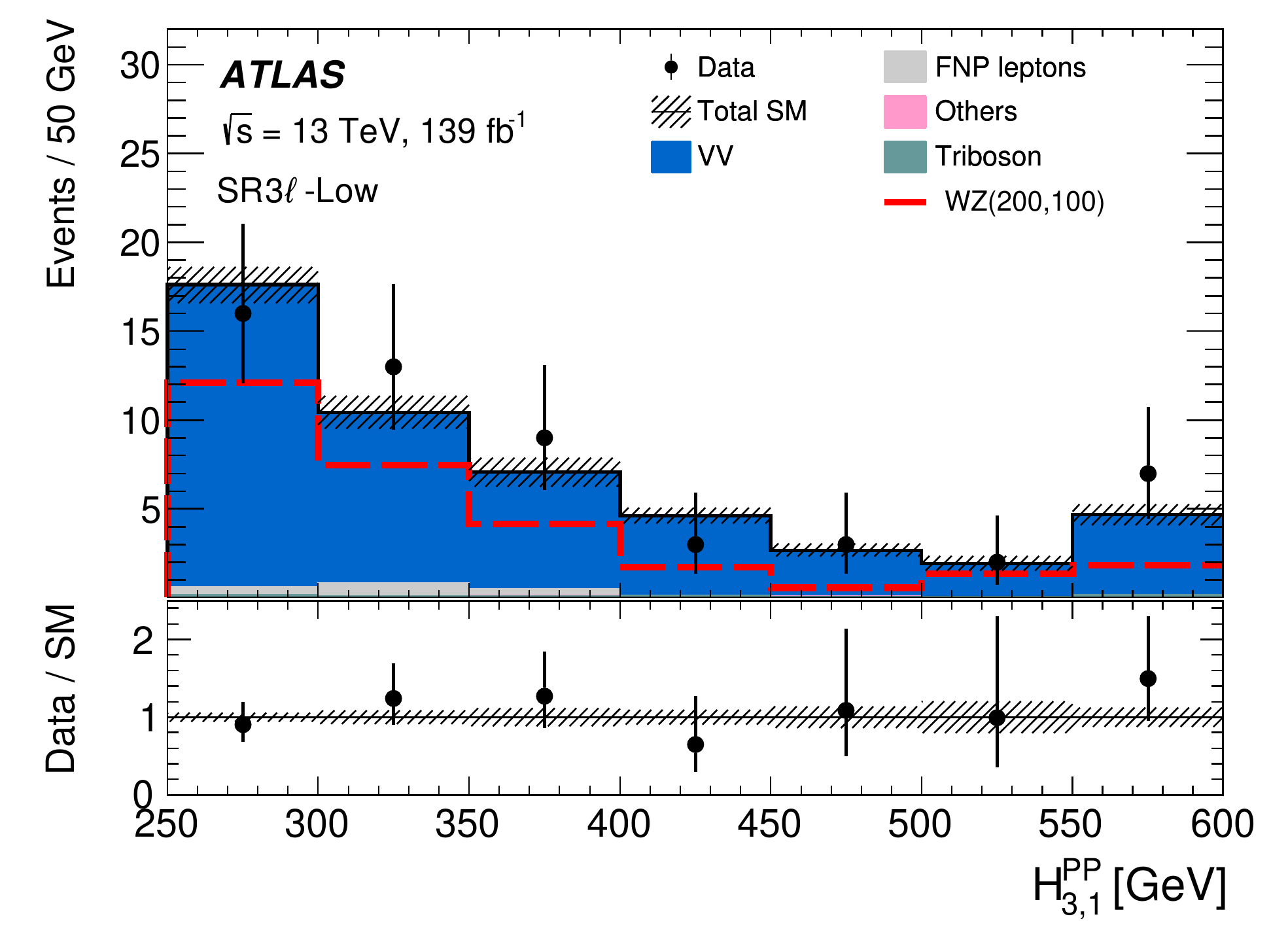}
\includegraphics[width=0.485\columnwidth]{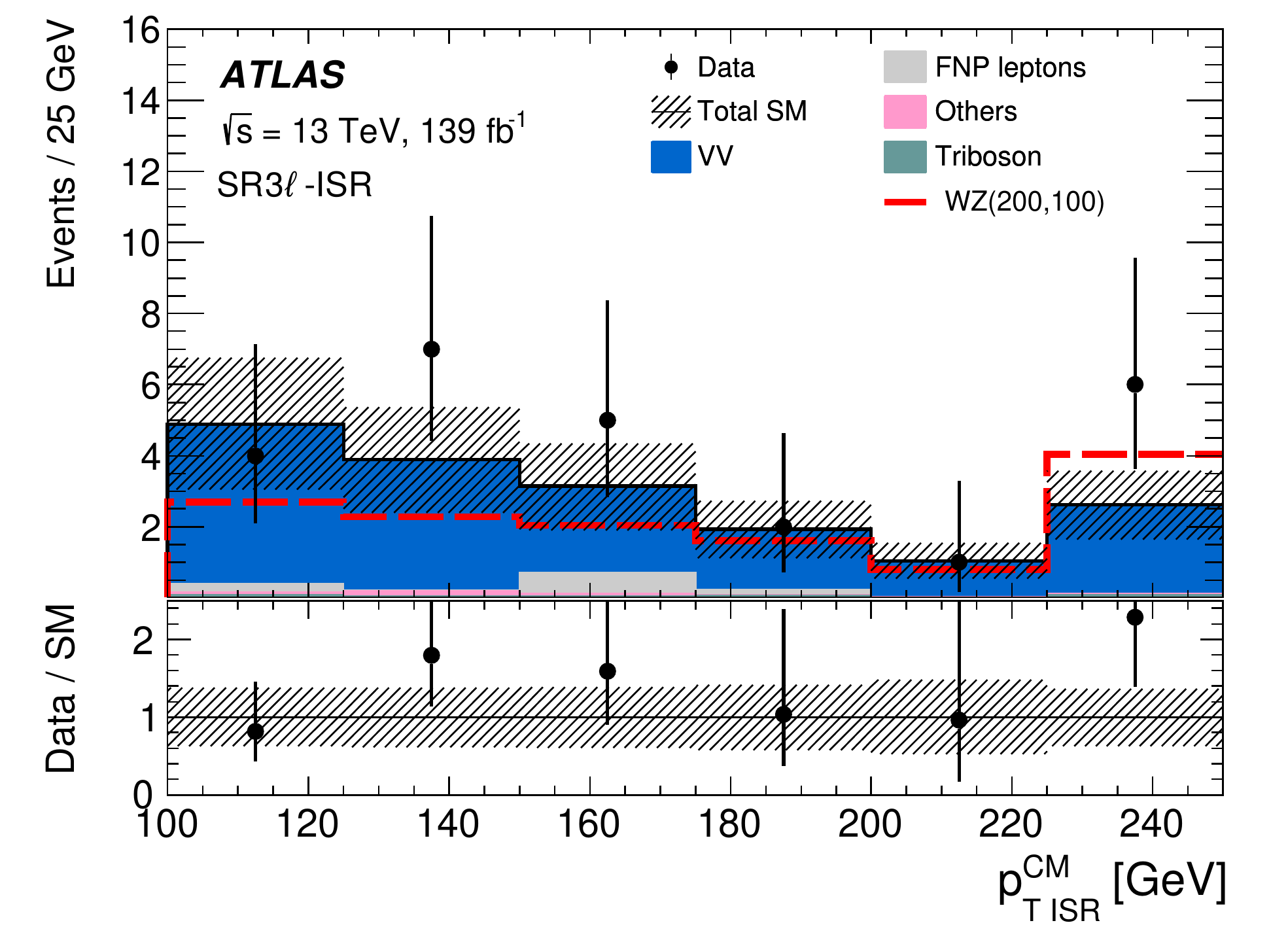}
\includegraphics[width=0.485\columnwidth]{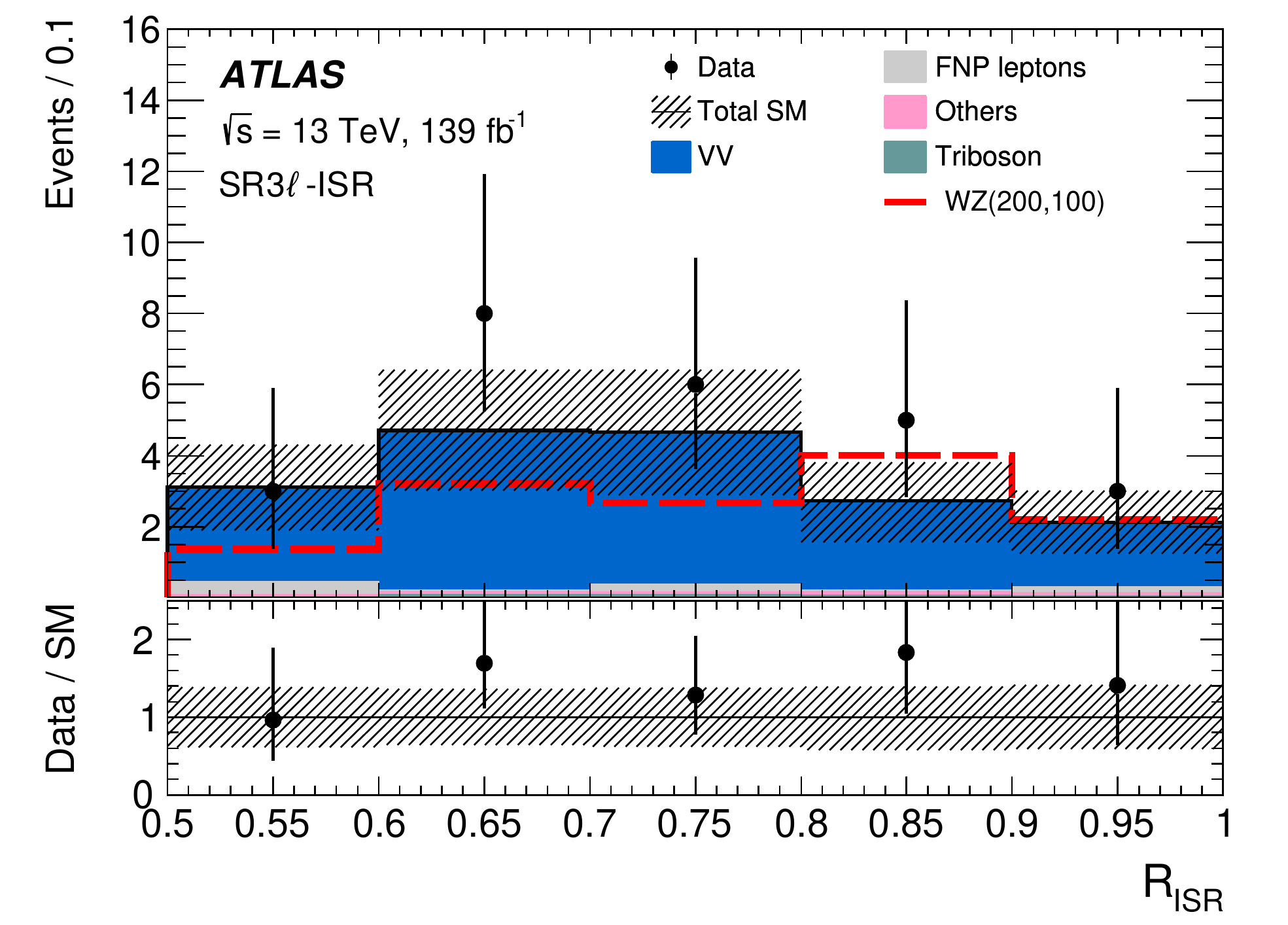}
\caption{
Example of kinematic distributions after the background-only fit,
showing the data and the post-fit expected background,
in regions of the \rjtl selection.
The figure shows the
(top left) \ptl{1}~
and (top right) \rjhpp~
distributions in \RJSRlow,
and the (bottom left) \rjptisr~
and (bottom right) \rjrisr~
distributions in \RJSRISR.
The last bin includes overflow.
The `FNP leptons' category contains backgrounds from \ttbar, $tW$, $WW$ and \Zjet~processes.
The `Others' category contains backgrounds from Higgs and rare top processes.
Distributions for wino/bino (+) \textchinoonepmninotwo \ra\ \WZ~signals are overlaid, with mass values given as \CNmasspair~\GeV.
The bottom panel shows the ratio of the observed data to the predicted yields.
The hatched bands indicate the combined theoretical, experimental, and MC statistical uncertainties.
}
\label{fig:RJ:SR}
\end{figure}
 
% The next lines are included from the .//tables/rj_discSR.tex input file
\begin{table}[p]
\centering
\setlength{\tabcolsep}{0.8pc}
\caption{
Results of the discovery fit for the SRs of the \rjtl selection, calculated using pseudo-experiments.
The first and second column list the 95\% CL upper limits on the visible cross section ($\sigma_{\text{vis}}^{95}$) and on the number of signal events ($S_\text{obs}^{95}$).
The third column ($S_\text{exp}^{95}$) shows the 95\% CL upper limit on the number of signal events, given the expected number (and $\pm 1\sigma$ excursions of the expectation) of background events.
The last two columns indicate the \CLb value, i.e.\ the confidence level observed for the background-only hypothesis, and the discovery $p$-value ($p(s = 0)$).
If the observed yield is below the expected yield, the $p$-value is capped at 0.5.
\vspace{0.5em}
}
\label{tab:results:RJ_model_indep}
\adjustbox{max width=0.85\textwidth}{
\centering
\begin{tabular}{L{2.8cm}llllll}\cline{1-6}\tabvlargespace
SR &
$\sigma_{\text{vis}}^{95}$ [fb] &
$S_\text{obs}^{95}$  & $S_\text{exp}^{95}$ & \CLb & $p(s=0)$ ($Z$)  \\\cline{1-6}\tabvspace
\RJSRlow  & $0.24$ &  $33$ & $ { 30 }^{ +10 }_{ -8 }$ & $0.61$ & $ 0.39$~$(0.28)$ \\[0.04cm]
\RJSRISR  & $0.14$ &  $19$ & $ { 12 }^{ +5 }_{ -4 }$ & $0.89$ & $ 0.09$~$(1.32)$ \\[0.04cm]
\cline{1-6}
\end{tabular}
}
\end{table}
 
% End of text imported from the .//tables/rj_discSR.tex input file
% End of text imported from the .//sections/rj.tex input file
% The next lines are included from the .//sections/conclusion.tex input file
\FloatBarrier
\begin{DIFnomarkup}
\vskip 1em\section{Conclusion}
\label{sec:conclusion}
\end{DIFnomarkup}
Results of a search for chargino--neutralino pair production decaying via \WZ, $W^{*}Z^{*}$ or \Wh into three-lepton final states are presented.
A dataset of $\sqrt{s}=13~\TeV$ proton--proton collisions corresponding to an integrated luminosity of 139~\ifb, collected by the ATLAS experiment at the CERN LHC, is used.
Events with three light-flavour charged leptons and missing transverse momentum are preselected,
and three selections are developed with a signal region strategy optimised for chargino--neutralino signals decaying via \WZ, $W^{*}Z^{*}$ and \Wh, respectively.
A fourth selection targeting the chargino--neutralino signals decaying via \WZ using the Recursive Jigsaw Reconstruction technique is also studied,
to follow up on the excesses observed in the previous ATLAS result using the same method and event selection.
In all the selections the data are found to be consistent with predictions of the Standard Model.
The results are interpreted for simplified models with wino or higgsino production.
A statistical combination is performed to include the result of an ATLAS search probing the final state with two soft leptons using the same dataset.
 
Assuming a simplified model with wino production decaying to a bino LSP, exclusion limits at 95\% confidence level are placed
on the minimum \textchinoonepm/\textninotwo mass,
extending the reach of previous searches~\cite{
SUSY-2013-12,
SUSY-2016-24,
SUSY-2017-03,
CMS-SUS-17-004,
SUSY-2018-06,
SUSY-2018-16,
}.
Limits are set
at 640~\GeV~for the \WZmed signals
in the limit of massless \textninoone,
improving by about 140~\GeV;
and at 300~\GeV~for mass-splittings between \textchinoonepm/\textninotwo and \textninoone close to \mZ,
improving by about 100~\GeV.
In the case of a mass splitting of 5--90~\GeV, \textchinoonepm/\textninotwo masses up to 200--300~\GeV~for the \WZmed are excluded.
The limit extends down to a smallest mass splitting of 2~\GeV~for a \textchinoonepm mass of 100~\GeV.
The dependency on a model parameter -- the sign of the \mNN product -- is also tested,
and comparable limits are found for the two scenarios.
For the \Whmed signals, the limit
on the minimum \textchinoonepm/\textninotwo mass is set at 190~\GeV, for \textninoone masses below 20~\GeV.
 
Limits are also set for simplified models with a higgsino LSP triplet,
for the first time including results from three-lepton final states,
which increases sensitivity to scenarios with moderate mass splittings.
Combined with the two-lepton analysis targeting compressed mass spectra,
the exclusion limits at 95\% confidence level are placed on the minimum \textninotwo mass
up to 210~\GeV\ for \WZmed signals with a mass splitting of 2--60~\GeV.
In these models, searches in the three-lepton final state enhance the sensitivity in the experimentally challenging region with mass splitting greater than 30~\GeV.

\vspace{1em}
 
% End of text imported from the .//sections/conclusion.tex input file
 
\begin{DIFnomarkup}
\section*{Acknowledgements}
\end{DIFnomarkup}
% The next lines are included from the .//acknowledgements/Acknowledgements.tex input file

We thank CERN for the very successful operation of the LHC, as well as the
support staff from our institutions without whom ATLAS could not be
operated efficiently.
 
We acknowledge the support of ANPCyT, Argentina; YerPhI, Armenia; ARC, Australia; BMWFW and FWF, Austria; ANAS, Azerbaijan; SSTC, Belarus; CNPq and FAPESP, Brazil; NSERC, NRC and CFI, Canada; CERN; ANID, Chile; CAS, MOST and NSFC, China; Minciencias, Colombia; MSMT CR, MPO CR and VSC CR, Czech Republic; DNRF and DNSRC, Denmark; IN2P3-CNRS and CEA-DRF/IRFU, France; SRNSFG, Georgia; BMBF, HGF and MPG, Germany; GSRT, Greece; RGC and Hong Kong SAR, China; ISF and Benoziyo Center, Israel; INFN, Italy; MEXT and JSPS, Japan; CNRST, Morocco; NWO, Netherlands; RCN, Norway; MNiSW and NCN, Poland; FCT, Portugal; MNE/IFA, Romania; JINR; MES of Russia and NRC KI, Russian Federation; MESTD, Serbia; MSSR, Slovakia; ARRS and MIZ\v{S}, Slovenia; DST/NRF, South Africa; MICINN, Spain; SRC and Wallenberg Foundation, Sweden; SERI, SNSF and Cantons of Bern and Geneva, Switzerland; MOST, Taiwan; TAEK, Turkey; STFC, United Kingdom; DOE and NSF, United States of America. In addition, individual groups and members have received support from BCKDF, CANARIE, Compute Canada, CRC and IVADO, Canada; Beijing Municipal Science \& Technology Commission, China; COST, ERC, ERDF, Horizon 2020 and Marie Sk{\l}odowska-Curie Actions, European Union; Investissements d'Avenir Labex, Investissements d'Avenir Idex and ANR, France; DFG and AvH Foundation, Germany; Herakleitos, Thales and Aristeia programmes co-financed by EU-ESF and the Greek NSRF, Greece; BSF-NSF and GIF, Israel; La Caixa Banking Foundation, CERCA Programme Generalitat de Catalunya and PROMETEO and GenT Programmes Generalitat Valenciana, Spain; G\"{o}ran Gustafssons Stiftelse, Sweden; The Royal Society and Leverhulme Trust, United Kingdom.
 
The crucial computing support from all WLCG partners is acknowledged gratefully, in particular from CERN, the ATLAS Tier-1 facilities at TRIUMF (Canada), NDGF (Denmark, Norway, Sweden), CC-IN2P3 (France), KIT/GridKA (Germany), INFN-CNAF (Italy), NL-T1 (Netherlands), PIC (Spain), ASGC (Taiwan), RAL (UK) and BNL (USA), the Tier-2 facilities worldwide and large non-WLCG resource providers. Major contributors of computing resources are listed in Ref.~\cite{ATL-SOFT-PUB-2020-001}.
 
% End of text imported from the .//acknowledgements/Acknowledgements.tex input file

\printbibliography

@article{Fuks:2012qx,
      author         = "Fuks, Benjamin and Klasen, Michael and Lamprea, David R.
                        and Rothering, Marcel",
      title          = "{Gaugino production in proton-proton collisions at a
                        center-of-mass energy of 8 TeV}",
      journal        = "JHEP",
      volume         = "10",
      pages          = "081",
      doi            = "10.1007/JHEP10(2012)081",
      year           = "2012",
      eprint         = "1207.2159",
      archivePrefix  = "arXiv",
      primaryClass   = "hep-ph",
      reportNumber   = "IPHC-PHENO-12-07, MS-TP-12-05",
      SLACcitation   = "%%CITATION = ARXIV:1207.2159;%%",
}

@article{Fuks:2013vua,
      author         = "Fuks, Benjamin and Klasen, Michael and Lamprea, David R.
                        and Rothering, Marcel",
      title          = "{Precision predictions for electroweak superpartner
                        production at hadron colliders with {\textsc{Resummino}}}",
      journal        = "Eur. Phys. J. C",
      volume         = "73",
      pages          = "2480",
      doi            = "10.1140/epjc/s10052-013-2480-0",
      year           = "2013",
      eprint         = "1304.0790",
      archivePrefix  = "arXiv",
      primaryClass   = "hep-ph",
      reportNumber   = "CERN-PH-TH-2013-064, IPHC-PHENO-13-02, MS-TP-13-06",
      SLACcitation   = "%%CITATION = ARXIV:1304.0790;%%",
}

@article{LUCID2,
  author={G. Avoni and others},
  title={The new LUCID-2 detector for luminosity measurement and monitoring in ATLAS},
  journal={JINST},
  volume={13},
  number={07},
  pages={P07017},
  doi="10.1088/1748-0221/13/07/P07017",
  year={2018},
}

@article{Albahri:2021ixb,
    author = "Albahri, T. and others",
    collaboration = "Muon g-2",
    title = "{Measurement of the anomalous precession frequency of the muon in the Fermilab Muon $g-2$ Experiment}",
    eprint = "2104.03247",
    archivePrefix = "arXiv",
    primaryClass = "hep-ex",
    reportNumber = "FERMILAB-PUB-21-183-E",
    doi = "10.1103/PhysRevD.103.072002",
    journal = "Phys. Rev. D",
    volume = "103",
    number = "7",
    pages = "072002",
    year = "2021"
}

@article{Aoyama:2020ynm,
    author = "Aoyama, T. and others",
    title = "{The anomalous magnetic moment of the muon in the Standard Model}",
    eprint = "2006.04822",
    archivePrefix = "arXiv",
    primaryClass = "hep-ph",
    reportNumber = "FERMILAB-PUB-20-207-T, INT-PUB-20-021, KEK Preprint 2020-5,
  MITP/20-028, KEK Preprint 2020-5, MITP/20-028, CERN-TH-2020-075, IFT-UAM/CSIC-20-74, LMU-ASC 18/20, LTH 1234,
  LU TP 20-20, LTH 1234, LU TP 20-20, MAN/HEP/2020/003, PSI-PR-20-06, UWThPh 2020-14, ZU-TH 18/20",
    doi = "10.1016/j.physrep.2020.07.006",
    journal = "Phys. Rept.",
    volume = "887",
    pages = "1--166",
    year = "2020"
}

@article{moroi1996muon,
    author = "Moroi, Takeo",
    archivePrefix = "arXiv",
    doi = "10.1103/PhysRevD.53.6565",
    eprint = "hep-ph/9512396",
    journal = "Phys.\ Rev.\ D",
    related        = "moroi1996-err",
    relatedstring  = "Erratum:",
    pages = "6565--6575",
    reportNumber = "LBL-38099",
    title = "{Muon anomalous magnetic dipole moment in the minimal supersymmetric standard model}",
    volume = "53",
    year = "1996"
}

@article{feng2000supernatural,
    author = "Feng, Jonathan L. and Moroi, Takeo",
    archivePrefix = "arXiv",
    doi = "10.1103/PhysRevD.61.095004",
    eprint = "hep-ph/9907319",
    journal = "Phys.\ Rev.\ D",
    pages = "095004",
    reportNumber = "IASSNS-HEP-99-65",
    title = "{Supernatural supersymmetry: Phenomenological implications of anomaly-mediated supersymmetry breaking}",
    volume = "61",
    year = "2000"
}

@article{endo2020muon,    
    author = "Endo, Motoi and Hamaguchi, Koichi and Iwamoto, Sho and Kitahara, Teppei",
    title = "{Muon $g - 2$ vs LHC Run 2 in supersymmetric models}",
    eprint = "2001.11025",
    archivePrefix = "arXiv",
    primaryClass = "hep-ph",
    reportNumber = "UT-20-01, IPMU20-0009, KEK-TH-2188",
    doi = "10.1007/JHEP04(2020)165",
    journal = "JHEP",
    volume = "04",
    pages = "165",
    year = "2020"
}

@article{DeSanctis:2007yoa,
      author         = "De Sanctis, U. and Lari, T. and Montesano, S. and
                        Troncon, C.",
      title          = "{Perspectives for the detection and measurement of
                        supersymmetry in the focus point region of mSUGRA models
                        with the ATLAS detector at LHC}",
      journal        = "Eur. Phys. J. C",
      volume         = "52",
      year           = "2007",
      pages          = "743-758",
      doi            = "10.1140/epjc/s10052-007-0415-3",
      eprint         = "0704.2515",
      archivePrefix  = "arXiv",
      primaryClass   = "hep-ex",
      reportNumber   = "ATLAS-SCIENTIFIC-NOTE-SN-ATLAS-2007-062",
      SLACcitation   = "%%CITATION = ARXIV:0704.2515;%%"
}

@article{Profumo:2017ntc,
      author         = "Profumo, Stefano and Stefaniak, Tim and Stephenson
                        Haskins, Laurel",
      title          = "{Not-so-well-tempered neutralino}",
      journal        = "Phys. Rev. D",
      volume         = "96",
      year           = "2017",
      number         = "5",
      pages          = "055018",
      doi            = "10.1103/PhysRevD.96.055018",
      eprint         = "1706.08537",
      archivePrefix  = "arXiv",
      primaryClass   = "hep-ph",
      reportNumber   = "SCIPP-17-08",
      SLACcitation   = "%%CITATION = ARXIV:1706.08537;%%"
}

@article{Barbieri:2009ev,
    author = "Barbieri, Riccardo and Pappadopulo, Duccio",
    archivePrefix = "arXiv",
    doi = "10.1088/1126-6708/2009/10/061",
    eprint = "0906.4546",
    journal = "JHEP",
    pages = "061",
    primaryClass = "hep-ph",
    title = "{S-particles at their naturalness limits}",
    volume = "10",
    year = "2009"
}

@article{Papucci:2011wy,
    author = "Papucci, Michele and Ruderman, Joshua T. and Weiler, Andreas",
    archivePrefix = "arXiv",
    doi = "10.1007/JHEP09(2012)035",
    eprint = "1110.6926",
    journal = "JHEP",
    pages = "035",
    primaryClass = "hep-ph",
    reportNumber = "DESY-11-193, CERN-PH-TH-265",
    title = "{Natural SUSY endures}",
    volume = "09",
    year = "2012"
}

@article{Baer:2011ec,
    author = "Baer, Howard and Barger, Vernon and Huang, Peisi",
    archivePrefix = "arXiv",
    doi = "10.1007/JHEP11(2011)031",
    eprint = "1107.5581",
    journal = "JHEP",
    pages = "031",
    primaryClass = "hep-ph",
    title = "{Hidden SUSY at the LHC: the light higgsino-world scenario and the role of a lepton collider}",
    volume = "11",
    year = "2011"
}

@article{Baer:2012up,
    author = "Baer, Howard and Barger, Vernon and Huang, Peisi and Mustafayev, Azar and Tata, Xerxes",
    archivePrefix = "arXiv",
    doi = "10.1103/PhysRevLett.109.161802",
    eprint = "1207.3343",
    journal = "Phys.\ Rev.\ Lett.",
    pages = "161802",
    primaryClass = "hep-ph",
    title = "{Radiative Natural Supersymmetry with a 125 GeV Higgs Boson}",
    volume = "109",
    year = "2012"
}

@article{Griest:1990kh,
    author = "Griest, Kim and Seckel, David",
    doi = "10.1103/PhysRevD.43.3191",
    journal = "Phys.\ Rev.\ D",
    pages = "3191--3203",
    reportNumber = "CFPA-TH-90-001A, BA-90-79",
    title = "{Three exceptions in the calculation of relic abundances}",
    volume = "43",
    year = "1991"
}

@article{Edsjo:1997bg,
    author = "Edsjo, Joakim and Gondolo, Paolo",
    archivePrefix = "arXiv",
    doi = "10.1103/PhysRevD.56.1879",
    eprint = "hep-ph/9704361",
    journal = "Phys.\ Rev.\ D",
    pages = "1879--1894",
    reportNumber = "UUITP-11-97, MPI-PHT-97-27",
    title = "{Neutralino relic density including coannihilations}",
    volume = "56",
    year = "1997"
}

@article{duan2018probing,
    author = {Duan, Guang Hua and Hikasa, Ken-ichi and Ren, Jie and Wu, Lei and Yang, Jin Min},
    title = "{Probing bino-wino coannihilation dark matter below the neutrino floor at the LHC}",
    eprint = "1804.05238",
    archivePrefix = "arXiv",
    primaryClass = "hep-ph",
    doi = "10.1103/PhysRevD.98.015010",
    journal = "Phys. Rev. D",
    volume = "98",
    number = "1",
    pages = "015010",
    year = "2018"
}

@article{Aghanim:2015xee,
      author         = "Aghanim, N. and others",
      title          = "{Planck 2015 results. XI. CMB power spectra, likelihoods,
                        and robustness of parameters}",
      collaboration  = "Planck",
      journal        = "Astron. Astrophys.",
      volume         = "594",
      year           = "2016",
      pages          = "A11",
      doi            = "10.1051/0004-6361/201526926",
      eprint         = "1507.02704",
      archivePrefix  = "arXiv",
      primaryClass   = "astro-ph.CO",
      SLACcitation   = "%%CITATION = ARXIV:1507.02704;%%"
}

@article{Lester:1999tx,
      author         = "Lester, C. G. and Summers, D. J.",
      title          = "{Measuring masses of semi-invisibly decaying particles
                        pair produced at hadron colliders}",
      journal        = "Phys. Lett. B",
      volume         = "463",
      year           = "1999",
      pages          = "99-103",
      doi            = "10.1016/S0370-2693(99)00945-4",
      eprint         = "hep-ph/9906349",
      archivePrefix  = "arXiv",
      reportNumber   = "CAVENDISH-HEP-99-07",
      SLACcitation   = "%%CITATION = HEP-PH/9906349;%%"
}

@article{Barr:2003rg,
      author         = "Barr, Alan and Lester, Christopher and Stephens, P.",
      title          = "{A variable for measuring masses at hadron colliders when missing energy is expected; $m_\text{T2}$: the truth behind the glamour}",
      journal        = "J. Phys. G",
      volume         = "29",
      year           = "2003",
      pages          = "2343-2363",
      doi            = "10.1088/0954-3899/29/10/304",
      eprint         = "hep-ph/0304226",
      archivePrefix  = "arXiv",
      reportNumber   = "CAVENDISH-HEP-2002-02-14",
      SLACcitation   = "%%CITATION = HEP-PH/0304226;%%"
}

@Booklet{LEPlimits,
      author        = "{ALEPH, DELPHI, L3, OPAL Experiments}",
      title         = "{Combined LEP Chargino Results, up to 208 GeV for low DM}",
      howpublished  = "{LEPSUSYWG/02-04.1}",
      url           = "http://lepsusy.web.cern.ch/lepsusy/www/inoslowdmsummer02/charginolowdm_pub.html",
      year          = "2002"
}

@article{Heister:2001nk,
      author         = "{ALEPH Collaboration}",
      title          = "{Search for scalar leptons in $e^+e^-$ collisions at
                        center-of-mass energies up to 209 GeV}",
      collaboration  = "ALEPH",
      journal        = "Phys. Lett. B",
      volume         = "526",
      year           = "2002",
      pages          = "206-220",
      doi            = "10.1016/S0370-2693(01)01494-0",
      eprint         = "hep-ex/0112011",
      archivePrefix  = "arXiv",
      reportNumber   = "CERN-EP-2001-086",
      SLACcitation   = "%%CITATION = HEP-EX/0112011;%%"
}

@article{Heister:2003zk,
      author         = "{ALEPH Collaboration}",
      title          = "{Absolute mass lower limit for the lightest neutralino of
                        the MSSM from $e^+e^-$ data at $\sqrt{s}$ up to 209 GeV}",
      collaboration  = "ALEPH",
      journal        = "Phys. Lett. B",
      volume         = "583",
      year           = "2004",
      pages          = "247-263",
      doi            = "10.1016/j.physletb.2003.12.066",
      reportNumber   = "CERN-EP-2003-077",
      SLACcitation   = "%%CITATION = PHLTA,B583,247;%%"
}

@article{Abdallah:2003xe,
      author         = "{DELPHI Collaboration}",
      title          = "{Searches for supersymmetric particles in $e^+e^-$
                        collisions up to 208 GeV and interpretation of the results
                        within the MSSM}",
      collaboration  = "DELPHI",
      journal        = "Eur. Phys. J. C",
      volume         = "31",
      year           = "2003",
      pages          = "421-479",
      doi            = "10.1140/epjc/s2003-01355-5",
      eprint         = "hep-ex/0311019",
      archivePrefix  = "arXiv",
      reportNumber   = "CERN-EP-2003-007",
      SLACcitation   = "%%CITATION = HEP-EX/0311019;%%"
}

@article{Achard:2003ge,
      author         = "{L3 Collaboration}",
      title          = "{Search for scalar leptons and scalar quarks at LEP}",
      collaboration  = "L3",
      journal        = "Phys. Lett. B",
      volume         = "580",
      year           = "2004",
      pages          = "37-49",
      doi            = "10.1016/j.physletb.2003.10.010",
      eprint         = "hep-ex/0310007",
      archivePrefix  = "arXiv",
      reportNumber   = "CERN-EP-2003-059",
      SLACcitation   = "%%CITATION = HEP-EX/0310007;%%"
}

@article{Abbiendi:2003ji,
      author         = "{OPAL Collaboration}",
      title          = "{Search for anomalous production of di-lepton events with
                        missing transverse momentum in $e^+e^-$ collisions at
                        $\sqrt{s}$ = 183--209 GeV}",
      collaboration  = "OPAL",
      journal        = "Eur. Phys. J. C",
      volume         = "32",
      year           = "2004",
      pages          = "453-473",
      doi            = "10.1140/epjc/s2003-01466-y",
      eprint         = "hep-ex/0309014",
      archivePrefix  = "arXiv",
      reportNumber   = "CERN-EP-2003-040",
      SLACcitation   = "%%CITATION = HEP-EX/0309014;%%"
}

@article{Heister:2002jca,
      author         = "{ALEPH Collaboration}",
      title          = "{Absolute lower limits on the masses of selectrons and
                        sneutrinos in the MSSM}",
      collaboration  = "ALEPH",
      journal        = "Phys. Lett. B",
      volume         = "544",
      year           = "2002",
      pages          = "73-88",
      doi            = "10.1016/S0370-2693(02)02471-1",
      eprint         = "hep-ex/0207056",
      archivePrefix  = "arXiv",
      reportNumber   = "CERN-EP-2002-055",
      SLACcitation   = "%%CITATION = HEP-EX/0207056;%%"
}

@article{Heister:2002mn,
      author         = "{ALEPH Collaboration}",
      title          = "{Search for charginos nearly mass degenerate with the
                        lightest neutralino in $e^+e^-$ collisions at center-of-mass
                        energies up to 209 GeV}",
      collaboration  = "ALEPH",
      journal        = "Phys. Lett. B",
      volume         = "533",
      year           = "2002",
      pages          = "223-236",
      doi            = "10.1016/S0370-2693(02)01584-8",
      eprint         = "hep-ex/0203020",
      archivePrefix  = "arXiv",
      reportNumber   = "CERN-EP-2002-020",
      SLACcitation   = "%%CITATION = HEP-EX/0203020;%%"
}

@article{Acciarri:2000wy,
      author         = "{L3 Collaboration}",
      title          = "{Search for charginos with a small mass difference with
                        the lightest supersymmetric particle at $\sqrt{s}$ =
                        189 GeV}",
      collaboration  = "L3",
      journal        = "Phys. Lett. B",
      volume         = "482",
      year           = "2000",
      pages          = "31-42",
      doi            = "10.1016/S0370-2693(00)00488-3",
      eprint         = "hep-ex/0002043",
      archivePrefix  = "arXiv",
      reportNumber   = "CERN-EP-2000-018",
      SLACcitation   = "%%CITATION = HEP-EX/0002043;%%"
}

@article{Abbiendi:2002vz,
      author         = "{OPAL Collaboration}",
      title          = "{Search for nearly mass-degenerate charginos and
                        neutralinos at LEP}",
      collaboration  = "OPAL",
      journal        = "Eur. Phys. J. C",
      volume         = "29",
      year           = "2003",
      pages          = "479-489",
      doi            = "10.1140/epjc/s2003-01237-x",
      eprint         = "hep-ex/0210043",
      archivePrefix  = "arXiv",
      reportNumber   = "CERN-EP-2002-063",
      SLACcitation   = "%%CITATION = HEP-EX/0210043;%%"
}

@article{Jackson:2017gcy,
      author         = "Jackson, Paul and Rogan, Christopher",
      title          = "{Recursive jigsaw reconstruction: HEP event analysis in the presence of kinematic and combinatoric ambiguities}",
      journal        = "Phys. Rev. D",
      volume         = "96",
      year           = "2017",
      number         = "11",
      pages          = "112007",
      doi            = "10.1103/PhysRevD.96.112007",
      eprint         = "1705.10733",
      archivePrefix  = "arXiv",
      primaryClass   = "hep-ph",
      SLACcitation   = "%%CITATION = ARXIV:1705.10733;%%"
}

@article{Jackson:2016mfb,
    author = "Jackson, Paul and Rogan, Christopher and Santoni, Marco",
    title = "{Sparticles in motion: Analyzing compressed SUSY scenarios with a new method of event reconstruction}",
    eprint = "1607.08307",
    archivePrefix = "arXiv",
    primaryClass = "hep-ph",
    doi = "10.1103/PhysRevD.95.035031",
    journal = "Phys. Rev. D",
    volume = "95",
    number = "3",
    pages = "035031",
    year = "2017"
}

@article{Heinrich:2021gyp,
    author = "Heinrich, Lukas and Feickert, Matthew and Stark, Giordon and Cranmer, Kyle",
    title = "{pyhf: pure-Python implementation of HistFactory statistical models}",
    doi = "10.21105/joss.02823",
    journal = "J. Open Source Softw.",
    volume = "6",
    number = "58",
    pages = "2823",
    year = "2021"
}

@software{pyhfzenodo,
  author       = {Lukas Heinrich and Matthew Feickert and Giordon Stark},
  title        = {scikit-hep/pyhf: v0.5.2},
  month        = jul,
  year         = 2020,
  publisher    = {Zenodo},
  doi          = {10.5281/zenodo.4018115},
  url          = {https://doi.org/10.5281/zenodo.4018115}
}

@article{Fuks:2017rio,
      author         = "Fuks, Benjamin and Klasen, Michael and Schmiemann, Saskia and Sunder, Marthijn",
      title          = "{Realistic simplified gaugino-higgsino models in the MSSM}",
      journal        = "Eur. Phys. J. C",
      volume         = "78",
      year           = "2018",
      number         = "3",
      pages          = "209",
      doi            = "10.1140/epjc/s10052-018-5695-2",
      eprint         = "1710.09941",
      archivePrefix  = "arXiv",
      primaryClass   = "hep-ph",
      reportNumber   = "MS-TP-17-14",
      SLACcitation   = "%%CITATION = ARXIV:1710.09941;%%"
}

@article{Gunion:1984yn,
    author = "Gunion, J. F. and Haber, Howard E.",
    title = "{Higgs Bosons in Supersymmetric Models (I)}",
    reportNumber = "SLAC-PUB-3404",
    doi = "10.1016/0550-3213(86)90340-8",
    journal = "Nucl. Phys. B",
    volume = "272",
    pages = "76",
    year = "1986",
    related        = "Gunion:1984yn-err",
    relatedstring  = "Erratum:",
}

@article{Quintero:2014lqa,
    author = "Quintero, Nestor and Diaz-Cruz, J. Lorenzo and Lopez Castro, Gabriel",
    title = "{Lepton pair emission in the top quark decay $t \to bW^+\ell^-\ell^+$}",
    eprint = "1403.3044",
    archivePrefix = "arXiv",
    primaryClass = "hep-ph",
    doi = "10.1103/PhysRevD.89.093014",
    journal = "Phys. Rev. D",
    volume = "89",
    number = "9",
    pages = "093014",
    year = "2014"
}

@article{Borschensky:2014cia,
      author         = "Borschensky, Christoph and Kr{\"a}mer, Michael and Kulesza,
                        Anna and Mangano, Michelangelo and Padhi, Sanjay and
                        Plehn, Tilman and Portell, Xavier",
      title          = "{Squark and gluino production cross sections in pp
                        collisions at \(\sqrt{s} = 13, 14, 33\) and \(100\,\text{TeV}\)}",
      journal        = "Eur. Phys. J. C",
      volume         = "74",
      year           = "2014",
      pages          = "3174",
      doi            = "10.1140/epjc/s10052-014-3174-y",
      eprint         = "1407.5066",
      archivePrefix  = "arXiv",
      primaryClass   = "hep-ph",
      reportNumber   = "MS-TP-14-25, CERN-PH-TH-2014-137, TTK-14-13",
      SLACcitation   = "%%CITATION = ARXIV:1407.5066;%%"
}

@article{Golfand:1971iw,
      author         = {Golfand, Y.A.  and Likhtman, E.P.},
      title          = "{Extension of the Algebra of Poincare Group Generators
                        and Violation of P Invariance}",
      journal        = "JETP Lett.",
      volume         = "13",
      year           = "1971",
      pages          = "323",
      note           = "[Pisma Zh. Eksp. Teor. Fiz. \textbf{13} (1971) 452]",
      SLACcitation   = "%%CITATION = JTPLA,13,323;%%"
}

@article{Volkov:1973ix,
      author         = "Volkov, D.V. and Akulov, V.P.",
      title          = "{Is the neutrino a goldstone particle?}",
      journal        = "Phys. Lett. B",
      volume         = "46",
      year           = "1973",
      pages          = "109",
      doi            = "10.1016/0370-2693(73)90490-5",
      SLACcitation   = "%%CITATION = PHLTA,B46,109;%%"
}

@article{Wess:1974tw,
      author         = "Wess, J. and Zumino, B.",
      title          = "{Supergauge transformations in four dimensions}",
      journal        = "Nucl. Phys. B",
      volume         = "70",
      year           = "1974",
      pages          = "39",
      doi            = "10.1016/0550-3213(74)90355-1",
      SLACcitation   = "%%CITATION = NUPHA,B70,39;%%"
}

@article{Salam:1974ig,
      author         = "Salam, Abdus and Strathdee, J.",
      title          = "{Super-symmetry and non-Abelian gauges}",
      journal        = "Phys. Lett. B",
      volume         = "51",
      year           = "1974",
      pages          = "353",
      doi            = "10.1016/0370-2693(74)90226-3",
      reportNumber   = "IC/74/36",
      SLACcitation   = "%%CITATION = PHLTA,B51,353;%%"
}

@article{Wess:1974jb,
      author         = "Wess, J. and Zumino, B.",
      title          = "{Supergauge invariant extension of quantum electrodynamics}",
      journal        = "Nucl. Phys. B",
      volume         = "78",
      year           = "1974",
      pages          = "1",
      doi            = "10.1016/0550-3213(74)90112-6",
      reportNumber   = "CERN-TH-1857",
      SLACcitation   = "%%CITATION = NUPHA,B78,1;%%"
}

@article{Ferrara:1974pu,
      author         = "Ferrara, S. and Zumino, B.",
      title          = "{Supergauge invariant Yang-Mills theories}",
      journal        = "Nucl. Phys. B",
      volume         = "79",
      year           = "1974",
      pages          = "413",
      doi            = "10.1016/0550-3213(74)90559-8",
      reportNumber   = "CERN-TH-1866",
      SLACcitation   = "%%CITATION = NUPHA,B79,413;%%"
}

@article{Fayet:1976et,
      author         = "Fayet, Pierre",
      title          = "{Supersymmetry and weak, electromagnetic and strong
                        interactions}",
      journal        = "Phys. Lett. B",
      volume         = "64",
      year           = "1976",
      pages          = "159",
      doi            = "10.1016/0370-2693(76)90319-1",
      reportNumber   = "PTENS 76/14",
      SLACcitation   = "%%CITATION = PHLTA,B64,159;%%"
}

@article{Fayet:1977yc,
      author         = "Fayet, Pierre",
      title          = "{Spontaneously broken supersymmetric theories of weak,
                        electromagnetic and strong interactions}",
      journal        = "Phys. Lett. B",
      volume         = "69",
      year           = "1977",
      pages          = "489",
      doi            = "10.1016/0370-2693(77)90852-8",
      reportNumber   = "LPTENS 77/11",
      SLACcitation   = "%%CITATION = PHLTA,B69,489;%%"
}

@article{Farrar:1978xj,
      author         = "Farrar, Glennys R. and Fayet, Pierre",
      title          = "{Phenomenology of the production, decay, and detection of
                        new hadronic states associated with supersymmetry}",
      journal        = "Phys. Lett. B",
      volume         = "76",
      year           = "1978",
      pages          = "575",
      doi            = "10.1016/0370-2693(78)90858-4",
      reportNumber   = "CALT-68-648",
      SLACcitation   = "%%CITATION = PHLTA,B76,575;%%"
}

@article{Goldberg:1983nd,
      author         = "Goldberg, H.",
      title          = "{Constraint on the Photino Mass from Cosmology}",
      journal        = "Phys. Rev. Lett.",
      volume         = "50",
      year           = "1983",
      pages          = "1419",
      doi            = "10.1103/PhysRevLett.50.1419",
      related        = "Goldberg:1983nd-err",
      relatedstring  = "Erratum:",
      reportNumber   = "NUB-2592",
      SLACcitation   = "%%CITATION = PRLTA,50,1419;%%"
}

@article{Ellis:1983ew,
      author         = "Ellis, John and Hagelin, J.S. and Nanopoulos, Dimitri V. and Olive, Keith A. and Srednicki, M.",
      title          = "{Supersymmetric relics from the big bang}",
      booktitle      = "{IN *BATAVIA 1984, PROCEEDINGS, INNER SPACE/OUTER SPACE*,
                        458-459., In *Srednicki, M.A. (ed.): Particle physics and
                        cosmology* 223-246}",
      journal        = "Nucl. Phys. B",
      volume         = "238",
      year           = "1984",
      pages          = "453",
      doi            = "10.1016/0550-3213(84)90461-9",
      reportNumber   = "SLAC-PUB-3171",
      SLACcitation   = "%%CITATION = NUPHA,B238,453;%%"
}

@article{Chamseddine:1982jx,
      author         = "Chamseddine, Ali H. and Arnowitt, Richard L. and Nath, Pran",
      title          = "{Locally Supersymmetric Grand Unification}",
      journal        = "Phys. Rev. Lett.",
      volume         = "49",
      year           = "1982",
      pages          = "970",
      doi            = "10.1103/PhysRevLett.49.970",
      reportNumber   = "NUB-2559",
      SLACcitation   = "%%CITATION = PRLTA,49,970;%%"
}

@article{Barbieri:1982eh,
      author         = "Barbieri, Riccardo and Ferrara, S. and Savoy, Carlos A.",
      title          = "{Gauge Models with Spontaneously Broken Local Supersymmetry}",
      journal        = "Phys. Lett. B",
      volume         = "119",
      year           = "1982",
      pages          = "343",
      doi            = "10.1016/0370-2693(82)90685-2",
      reportNumber   = "CERN-TH-3365",
      SLACcitation   = "%%CITATION = PHLTA,B119,343;%%"
}

@article{Kane:1993td,
      author         = "Kane, Gordon L. and Kolda, Christopher F. and Roszkowski, Leszek and Wells, James D.",
      title          = "{Study of constrained minimal supersymmetry}",
      journal        = "Phys. Rev. D",
      volume         = "49",
      year           = "1994",
      pages          = "6173",
      doi            = "10.1103/PhysRevD.49.6173",
      eprint         = "hep-ph/9312272",
      archivePrefix  = "arXiv",
      reportNumber   = "UM-TH-93-24",
      SLACcitation   = "%%CITATION = HEP-PH/9312272;%%"
}

@article{Barbieri:1987fn,
      author         = "Barbieri, Riccardo and Giudice, G.F.",
      title          = "{Upper bounds on supersymmetric particle masses}",
      journal        = "Nucl. Phys. B",
      volume         = "306",
      year           = "1988",
      pages          = "63",
      doi            = "10.1016/0550-3213(88)90171-X",
      reportNumber   = "CERN-TH-4825/87",
      SLACcitation   = "%%CITATION = NUPHA,B306,63;%%"
}

@article{deCarlos:1993yy,
      author         = "de Carlos, B. and Casas, J.A.",
      title          = "{One-loop analysis of the electroweak breaking in supersymmetric models and the fine-tuning problem}",
      journal        = "Phys. Lett. B",
      volume         = "309",
      year           = "1993",
      pages          = "320",
      doi            = "10.1016/0370-2693(93)90940-J",
      eprint         = "hep-ph/9303291",
      archivePrefix  = "arXiv",
      reportNumber   = "CERN-TH-6835-93, IEM-FT-70-93",
      SLACcitation   = "%%CITATION = HEP-PH/9303291;%%"
}

@article{Girardello:1981wz,
      author         = "Girardello, L. and Grisaru, Marcus T.",
      title          = "{Soft Breaking of Supersymmetry}",
      journal        = "Nucl. Phys. B",
      volume         = "194",
      year           = "1982",
      pages          = "65",
      doi            = "10.1016/0550-3213(82)90512-0",
      reportNumber   = "Print-81-0592 (HARVARD)",
      SLACcitation   = "%%CITATION = NUPHA,B194,65;%%"
}

@article{Dimopoulos:1981zb,
      author         = "Dimopoulos, Savas and Georgi, Howard",
      title          = "{Softly broken supersymmetry and SU(5)}",
      journal        = "Nucl. Phys. B",
      volume         = "193",
      year           = "1981",
      pages          = "150",
      doi            = "10.1016/0550-3213(81)90522-8",
      reportNumber   = "HUTP-81/A022",
      SLACcitation   = "%%CITATION = NUPHA,B193,150;%%"
}

@article{Sakai:1981gr,
      author         = "Sakai, N.",
      title          = "{Naturalness in supersymmetric GUTS}",
      journal        = "Z. Phys. C",
      volume         = "11",
      year           = "1981",
      pages          = "153",
      doi            = "10.1007/BF01573998",
      reportNumber   = "TU/81/225",
      SLACcitation   = "%%CITATION = ZEPYA,C11,153;%%"
}

@article{Dimopoulos:1981yj,
      author         = "Dimopoulos, S. and Raby, S. and Wilczek, Frank",
      title          = "{Supersymmetry and the scale of unification}",
      journal        = "Phys. Rev. D",
      volume         = "24",
      year           = "1981",
      pages          = "1681",
      doi            = "10.1103/PhysRevD.24.1681",
      reportNumber   = "NSF-ITP-81-31",
      SLACcitation   = "%%CITATION = PHRVA,D24,1681;%%"
}

@article{Ibanez:1981yh,
      author         = "Ib{\'a}{\~n}ez, Luis E. and Ross, Graham G.",
      title          = "{Low-energy predictions in supersymmetric grand unified
                        theories}",
      journal        = "Phys. Lett. B",
      volume         = "105",
      year           = "1981",
      pages          = "439",
      doi            = "10.1016/0370-2693(81)91200-4",
      reportNumber   = "OXFORD-TP 65/81",
      SLACcitation   = "%%CITATION = PHLTA,B105,439;%%"
}

@article{Alwall:2008ve,
      author         = "Alwall, Johan and Le, My-Phuong and Lisanti, Mariangela and Wacker, Jay G.",
      title          = "{Searching for directly decaying gluinos at the Tevatron}",
      journal        = "Phys. Lett. B",
      volume         = "666",
      year           = "2008",
      pages          = "34",
      doi            = "10.1016/j.physletb.2008.06.065",
      eprint         = "0803.0019",
      archivePrefix  = "arXiv",
      primaryClass   = "hep-ph",
      reportNumber   = "SLAC-PUB-13149",
      SLACcitation   = "%%CITATION = ARXIV:0803.0019;%%"
}

@article{Alwall:2008ag,
      author         = "Alwall, Johan and Schuster, Philip and Toro, Natalia",
      title          = "{Simplified models for a first characterization of new physics at the LHC}",
      journal        = "Phys. Rev. D",
      volume         = "79",
      year           = "2009",
      pages          = "075020",
      doi            = "10.1103/PhysRevD.79.075020",
      eprint         = "0810.3921",
      archivePrefix  = "arXiv",
      primaryClass   = "hep-ph",
      reportNumber   = "SLAC-PUB-13425, SU-ITP-08-24",
      SLACcitation   = "%%CITATION = ARXIV:0810.3921;%%"
}

@article{Alves:2011wf,
      author         = "Alves, Daniele and others",
      title          = "{Simplified models for LHC new physics searches}",
      journal        = "J. Phys. G",
      volume         = "39",
      year           = "2012",
      pages          = "105005",
      doi            = "10.1088/0954-3899/39/10/105005",
      eprint         = "1105.2838",
      archivePrefix  = "arXiv",
      primaryClass   = "hep-ph",
      reportNumber   = "SLAC-PUB-15045, FERMILAB-PUB-11-842-A-PPD",
      SLACcitation   = "%%CITATION = ARXIV:1105.2838;%%"
}

@article{Beenakker:1999xh,
      author         = "Beenakker, W. and others",
      title          = "{Production of Charginos, Neutralinos, and Sleptons at Hadron Colliders}",
      journal        = "Phys. Rev. Lett.",
      volume         = "83",
      pages          = "3780",
      doi            = "10.1103/PhysRevLett.83.3780",
      year           = "1999",
      related        = "Beenakker:1999xh-err",
      relatedstring  = "Erratum:",
      eprint         = "hep-ph/9906298",
      archivePrefix  = "arXiv",
      reportNumber   = "CERN-TH-99-159, DESY-99-055, DTP-99-44, MADPH-99-1114,
                        ANL-HEP-PR-99-71",
      SLACcitation   = "%%CITATION = HEP-PH/9906298;%%",
}

@article{Debove:2010kf,
      author         = "Debove, Jonathan and Fuks, Benjamin and Klasen, Michael",
      title          = "{Threshold resummation for gaugino pair production at hadron colliders}",
      journal        = "Nucl. Phys. B",
      volume         = "842",
      year           = "2011",
      pages          = "51",
      doi            = "10.1016/j.nuclphysb.2010.08.016",
      eprint         = "1005.2909",
      archivePrefix  = "arXiv",
      primaryClass   = "hep-ph",
      reportNumber   = "IPHC-PHENO-10-02, LPSC-10-050",
      SLACcitation   = "%%CITATION = ARXIV:1005.2909;%%"
}

@article{Fiaschi:2018hgm,
      author         = "Fiaschi, Juri and Klasen, Michael",
      title          = "{Neutralino-chargino pair production at NLO+NLL with resummation-improved parton density functions for LHC Run~II}",
      journal        = "Phys. Rev. D",
      volume         = "98",
      year           = "2018",
      number         = "5",
      pages          = "055014",
      doi            = "10.1103/PhysRevD.98.055014",
      eprint         = "1805.11322",
      archivePrefix  = "arXiv",
      primaryClass   = "hep-ph",
      reportNumber   = "MS-TP-18-19",
      SLACcitation   = "%%CITATION = ARXIV:1805.11322;%%"
}

@article{Baak:2014wma,
      author         = "Baak, M. and Besjes, G.J. and C{\^o}t{\'e}, D. and Koutsman, A. and Lorenz, J. and Short, D.",
      title          = "{HistFitter software framework for statistical data analysis}",
      journal        = "Eur. Phys. J. C",
      volume         = "75",
      year           = "2015",
      pages          = "153",
      doi            = "10.1140/epjc/s10052-015-3327-7",
      eprint         = "1410.1280",
      archivePrefix  = "arXiv",
      primaryClass   = "hep-ex",
      SLACcitation   = "%%CITATION = ARXIV:1410.1280;%%",
}

@Article{Cacciari:2008gp,
     author    = "Cacciari, Matteo and Salam, Gavin P. and Soyez, Gregory",
     title     = "{The anti-\(k_{t}\) jet clustering algorithm}",
     journal   = "JHEP",
     volume    = "04",
     year      = "2008",
     pages     = "063",
     eprint    = "0802.1189",
     archivePrefix = "arXiv",
     primaryClass  =  "hep-ph",
     doi       = "10.1088/1126-6708/2008/04/063",
     SLACcitation  = "%%CITATION = 0802.1189;%%"
}

@Article{Fastjet,
      author         = "Cacciari, Matteo and Salam, Gavin P. and Soyez, Gregory",
      title          = "{FastJet user manual}",
      journal        = "Eur. Phys. J. C",
      volume         = "72",
      year           = "2012",
      pages          = "1896",
      doi            = "10.1140/epjc/s10052-012-1896-2",
      eprint         = "1111.6097",
      archivePrefix  = "arXiv",
      primaryClass   = "hep-ph",
      reportNumber   = "CERN-PH-TH-2011-297",
      SLACcitation   = "%%CITATION = ARXIV:1111.6097;%%"
}

@Article{Pumplin:2002vw,
     author    = "Pumplin, J. and others",
     title     = "{New Generation of Parton Distributions with Uncertainties from Global QCD Analysis}",
     journal   = "JHEP",
     volume    = "07",
     year      = "2002",
     pages     = "012",
     doi       = "10.1088/1126-6708/2002/07/012",
     eprint    = "hep-ph/0201195",
     archivePrefix = "arXiv",
     SLACcitation  = "%%CITATION = HEP-PH/0201195;%%"
}

@article{Ball:2012cx,
      author         = "Ball, Richard D. and others",
      title          = "{Parton distributions with LHC data}",
      journal        = "Nucl. Phys. B",
      volume         = "867",
      year           = "2013",
      pages          = "244",
      doi            = "10.1016/j.nuclphysb.2012.10.003",
      eprint         = "1207.1303",
      archivePrefix  = "arXiv",
      primaryClass   = "hep-ph",
      reportNumber   = "EDINBURGH-2012-08, IFUM-FT-997, FR-PHENO-2012-014,
                        RWTH-TTK-12-25, CERN-PH-TH-2012-037, SFB-CPP-12-47\,
                        --CERN-PH-TH-2012-037",
      SLACcitation   = "%%CITATION = ARXIV:1207.1303;%%"
}

@article{Ball:2014uwa,
      author         = "Ball, Richard D. and others",
      title          = "{Parton distributions for the LHC run II}",
      collaboration  = "NNPDF",
      journal        = "JHEP",
      volume         = "04",
      year           = "2015",
      pages          = "040",
      doi            = "10.1007/JHEP04(2015)040",
      eprint         = "1410.8849",
      archivePrefix  = "arXiv",
      primaryClass   = "hep-ph",
      reportNumber   = "EDINBURGH-2014-15, IFUM-1034-FT, CERN-PH-TH-2013-253,
                        OUTP-14-11P, CAVENDISH-HEP-14-11",
      SLACcitation   = "%%CITATION = ARXIV:1410.8849;%%"
}

@article{Czakon:2012pz,
      author         = "Czakon, Michal and Mitov, Alexander",
      title          = "{NNLO corrections to top pair production at hadron
                        colliders: the quark-gluon reaction}",
      journal        = "JHEP",
      volume         = "01",
      pages          = "080",
      doi            = "10.1007/JHEP01(2013)080",
      year           = "2013",
      eprint         = "1210.6832",
      archivePrefix  = "arXiv",
      primaryClass   = "hep-ph",
      SLACcitation   = "%%CITATION = ARXIV:1210.6832;%%",
}

@article{Sjostrand:2014zea,
      author         = "Sj{\"o}strand, Torbj{\"o}rn and Ask, Stefan and Christiansen,
                        Jesper R. and Corke, Richard and Desai, Nishita and Ilten,
                        Philip and Mrenna, Stephen and Prestel, Stefan and
                        Rasmussen, Christine O. and Skands, Peter Z.",
      title          = "{An introduction to PYTHIA 8.2}",
      journal        = "Comput. Phys. Commun.",
      volume         = "191",
      year           = "2015",
      pages          = "159",
      doi            = "10.1016/j.cpc.2015.01.024",
      eprint         = "1410.3012",
      archivePrefix  = "arXiv",
      primaryClass   = "hep-ph",
      reportNumber   = "LU-TP-14-36, MCNET-14-22, CERN-PH-TH-2014-190,
                        FERMILAB-PUB-14-316-CD, DESY-14-178, SLAC-PUB-16122,
                        --FERMILAB-PUB-14-316-CD",
      SLACcitation   = "%%CITATION = ARXIV:1410.3012;%%"
}

@article{Lonnblad:2001iq,
      author         = "L{\"o}nnblad, Leif",
      title          = "{Correcting the Colour-Dipole Cascade Model
                  with Fixed Order Matrix Elements}",
      journal        = "JHEP",
      volume         = "05",
      year           = "2002",
      pages          = "046",
      doi            = "10.1088/1126-6708/2002/05/046",
      eprint         = "hep-ph/0112284",
      archivePrefix  = "arXiv",
      XprimaryClass   = "hep-ph",
      reportNumber   = "LU-TP-01-38",
      SLACcitation   = "%%CITATION = HEP-PH/0112284;%%"
}

@article{Lonnblad:2011xx,
      author         = "L{\"o}nnblad, Leif and Prestel, Stefan",
      title          = "{Matching tree-level matrix elements with interleaved
                        showers}",
      journal        = "JHEP",
      volume         = "03",
      year           = "2012",
      pages          = "019",
      doi            = "10.1007/JHEP03(2012)019",
      eprint         = "1109.4829",
      archivePrefix  = "arXiv",
      primaryClass   = "hep-ph",
      SLACcitation   = "%%CITATION = ARXIV:1109.4829;%%"
}

@Article{Frixione:2008yi,
     author    = "Frixione, Stefano and Laenen, Eric and Motylinski, Patrick and White, Chris and Webber, Bryan R.",
     title     = "{Single-top hadroproduction in association with a \(W\) boson}",
     journal   = "JHEP",
     volume    = "07",
     year      = "2008",
     pages     = "029",
     eprint    = "0805.3067",
     archivePrefix = "arXiv",
     primaryClass  =  "hep-ph",
     doi       = "10.1088/1126-6708/2008/07/029"
}

@Article{Alwall:2014hca,
      author         = "Alwall, J. and Frederix, R. and Frixione, S. and Hirschi,
                        V. and Maltoni, F. and Mattelaer, O. and Shao, H. -S. and
                        Stelzer, T. and Torrielli, P. and Zaro, M.",
      title          = "{The automated computation of tree-level and
                        next-to-leading order differential cross sections, and
                        their matching to parton shower simulations}",
      journal        = "JHEP",
      volume         = "07",
      year           = "2014",
      pages          = "079",
      doi            = "10.1007/JHEP07(2014)079",
      eprint         = "1405.0301",
      archivePrefix  = "arXiv",
      primaryClass   = "hep-ph",
      reportNumber   = "CERN-PH-TH-2014-064, CP3-14-18, LPN14-066, MCNET-14-09,
                        ZU-TH-14-14",
      SLACcitation   = "%%CITATION = ARXIV:1405.0301;%%"
}

@article{Artoisenet:2012st,
      author         = "Artoisenet, Pierre and Frederix, Rikkert and Mattelaer,
                        Olivier and Rietkerk, Robbert",
      title          = "{Automatic spin-entangled decays of heavy resonances in
                        Monte Carlo simulations}",
      journal        = "JHEP",
      volume         = "03",
      pages          = "015",
      doi            = "10.1007/JHEP03(2013)015",
      year           = "2013",
      eprint         = "1212.3460",
      archivePrefix  = "arXiv",
      primaryClass   = "hep-ph",
      reportNumber   = "NIKHEF-2012-021, CERN-PH-TH-2012-329",
      SLACcitation   = "%%CITATION = ARXIV:1212.3460;%%",
}

@Article{Bahr:2008pv,
      author         = "B{\"a}hr, M. and others",
      title          = "{Herwig++ physics and manual}",
      journal        = "Eur. Phys. J. C",
      volume         = "58",
      year           = "2008",
      pages          = "639",
      doi            = "10.1140/epjc/s10052-008-0798-9",
      eprint         = "0803.0883",
      archivePrefix  = "arXiv",
      primaryClass   = "hep-ph",
      reportNumber   = "CERN-PH-TH-2008-038, CAVENDISH-HEP-08-03, KA-TP-05-2008,
                        DCPT-08-22, IPPP-08-11, CP3-08-05",
      SLACcitation   = "%%CITATION = ARXIV:0803.0883;%%"
}

@Article{Bellm:2015jjp,
      author         = "Bellm, Johannes and others",
      title          = "{Herwig 7.0/Herwig++ 3.0 release note}",
      journal        = "Eur. Phys. J. C",
      volume         = "76",
      year           = "2016",
      number         = "4",
      pages          = "196",
      doi            = "10.1140/epjc/s10052-016-4018-8",
      eprint         = "1512.01178",
      archivePrefix  = "arXiv",
      primaryClass   = "hep-ph",
      reportNumber   = "CERN-PH-TH-2015-289, MAN-HEP-2015-15, IFJPAN-IV-2015-13,
                        HERWIG-2015-01, KA-TP-18-2015, DCPT-15-142, MCNET-15-28,
                        IPPP-15-71, --HERWIG-2015-01",
      SLACcitation   = "%%CITATION = ARXIV:1512.01178;%%"
}

@Article{Nason:2004rx,
      author         = "Nason, Paolo",
      title          = "{A new method for combining NLO QCD with shower Monte Carlo algorithms}",
      journal        = "JHEP",
      volume         = "11",
      pages          = "040",
      doi            = "10.1088/1126-6708/2004/11/040",
      year           = "2004",
      eprint         = "hep-ph/0409146",
      archivePrefix  = "arXiv",
}

@Article{Frixione:2007vw,
      author         = "Frixione, Stefano and Nason, Paolo and Oleari, Carlo",
      title          = "{Matching NLO QCD computations with parton shower
                        simulations: the POWHEG method}",
      journal        = "JHEP",
      volume         = "11",
      pages          = "070",
      doi            = "10.1088/1126-6708/2007/11/070",
      year           = "2007",
      eprint         = "0709.2092",
      archivePrefix  = "arXiv",
      primaryClass   = "hep-ph",
}

@Article{Alioli:2010xd,
      author         = "Alioli, Simone and Nason, Paolo and Oleari, Carlo and Re,
                        Emanuele",
      title          = "{A general framework for implementing NLO calculations in
                        shower Monte Carlo programs: the POWHEG BOX}",
      journal        = "JHEP",
      volume         = "06",
      pages          = "043",
      doi            = "10.1007/JHEP06(2010)043",
      year           = "2010",
      eprint         = "1002.2581",
      archivePrefix  = "arXiv",
      primaryClass   = "hep-ph",
}

@article{Hartanto:2015uka,
      author         = "Hartanto, Heribertus B. and J{\"a}ger, Barbara and Reina,
                        Laura and Wackeroth, Doreen",
      title          = "{Higgs boson production in association with top quarks in
                        the POWHEG BOX}",
      journal        = "Phys. Rev. D",
      volume         = "91",
      year           = "2015",
      number         = "9",
      pages          = "094003",
      doi            = "10.1103/PhysRevD.91.094003",
      eprint         = "1501.04498",
      archivePrefix  = "arXiv",
      primaryClass   = "hep-ph",
      SLACcitation   = "%%CITATION = ARXIV:1501.04498;%%"
}

@article{Alioli:2009je,
      author         = "Alioli, Simone and Nason, Paolo and Oleari, Carlo and Re,
                        Emanuele",
      title          = "{NLO single-top production matched with shower in POWHEG:
                        \(s\)- and \(t\)-channel contributions}",
      journal        = "JHEP",
      volume         = "09",
      year           = "2009",
      pages          = "111",
      doi            = "10.1088/1126-6708/2009/09/111",
      eprint         = "0907.4076",
      archivePrefix  = "arXiv",
      primaryClass   = "hep-ph",
      SLACcitation   = "%%CITATION = ARXIV:0907.4076;%%",
      related        = "Alioli:2009je-err",
      relatedstring  = "Erratum:",
}

@article{Frederix:2012dh,
      author         = "Frederix, Rikkert and Re, Emanuele and Torrielli, Paolo",
      title          = "{Single-top \(t\)-channel hadroproduction in the four-flavour
                        scheme with POWHEG and aMC@NLO}",
      journal        = "JHEP",
      volume         = "09",
      year           = "2012",
      pages          = "130",
      doi            = "10.1007/JHEP09(2012)130",
      eprint         = "1207.5391",
      archivePrefix  = "arXiv",
      primaryClass   = "hep-ph",
      reportNumber   = "ZU-TH-14-12, IPPP-12-54, DCPT-12-108,
                        CERN-PH-TH-2012-206, MCNET-12-10, LPN12-084",
      SLACcitation   = "%%CITATION = ARXIV:1207.5391;%%"
}

@article{Aliev:2010zk,
      author         = "Aliev, M. and Lacker, H. and Langenfeld, U. and Moch, S.
                        and Uwer, P. and Wiedermann, M.",
      title          = "{HATHOR -- HAdronic Top and Heavy quarks crOss section
                        calculatoR}",
      journal        = "Comput. Phys. Commun.",
      volume         = "182",
      year           = "2011",
      pages          = "1034-1046",
      doi            = "10.1016/j.cpc.2010.12.040",
      eprint         = "1007.1327",
      archivePrefix  = "arXiv",
      primaryClass   = "hep-ph",
      reportNumber   = "DESY-10-091, HU-EP-10-33, SFB-CPP-10-60",
      SLACcitation   = "%%CITATION = ARXIV:1007.1327;%%"
}

@article{Re:2010bp,
      author         = "Re, Emanuele",
      title          = "{Single-top \(Wt\)-channel production matched with parton
                        showers using the POWHEG method}",
      journal        = "Eur. Phys. J. C",
      volume         = "71",
      year           = "2011",
      pages          = "1547",
      doi            = "10.1140/epjc/s10052-011-1547-z",
      eprint         = "1009.2450",
      archivePrefix  = "arXiv",
      primaryClass   = "hep-ph",
      reportNumber   = "IPPP-10-74, DCPT-10-148",
      SLACcitation   = "%%CITATION = ARXIV:1009.2450;%%"
}

@article{Kant:2014oha,
      author         = "Kant, P. and Kind, O. M. and Kintscher, T. and Lohse, T.
                        and Martini, T. and Mölbitz, S. and Rieck, P. and Uwer, P.",
      title          = "{HatHor for single top-quark production: Updated
                        predictions and uncertainty estimates for single top-quark
                        production in hadronic collisions}",
      journal        = "Comput. Phys. Commun.",
      volume         = "191",
      year           = "2015",
      pages          = "74-89",
      doi            = "10.1016/j.cpc.2015.02.001",
      eprint         = "1406.4403",
      archivePrefix  = "arXiv",
      primaryClass   = "hep-ph",
      reportNumber   = "HU-EP-14-22",
      SLACcitation   = "%%CITATION = ARXIV:1406.4403;%%"
}

@article{Kidonakis:2010ux,
      author         = "Kidonakis, Nikolaos",
      title          = "{Two-loop soft anomalous dimensions for single top quark
                        associated production with a \(W^{-}\) or \(H^{-}\)}",
      journal        = "Phys. Rev. D",
      volume         = "82",
      year           = "2010",
      pages          = "054018",
      doi            = "10.1103/PhysRevD.82.054018",
      eprint         = "1005.4451",
      archivePrefix  = "arXiv",
      primaryClass   = "hep-ph",
      SLACcitation   = "%%CITATION = ARXIV:1005.4451;%%"
}

@inproceedings{Kidonakis:2013zqa,
      author         = "Kidonakis, Nikolaos",
      title          = "{Top Quark Production}",
      booktitle      = "{Proceedings, Helmholtz International Summer School on
                        Physics of Heavy Quarks and Hadrons (HQ 2013)}",
      eventdate      = "2013-07-15/2013-07-28",
      venue          = "JINR, Dubna, Russia",
      pages          = "139-168",
      doi            = "10.3204/DESY-PROC-2013-03/Kidonakis",
      eprint         = "1311.0283",
      archivePrefix  = "arXiv",
      primaryClass   = "hep-ph",
      reportNumber   = "KSU-HEP-110113",
      SLACcitation   = "%%CITATION = ARXIV:1311.0283;%%"
}

@article{Beneke:2011mq,
      author         = "Beneke, M. and Falgari, P. and Klein, S. and Schwinn, C.",
      title          = "{Hadronic top-quark pair production with NNLL threshold
                        resummation}",
      journal        = "Nucl. Phys. B",
      volume         = "855",
      year           = "2012",
      pages          = "695-741",
      doi            = "10.1016/j.nuclphysb.2011.10.021",
      eprint         = "1109.1536",
      archivePrefix  = "arXiv",
      primaryClass   = "hep-ph",
      reportNumber   = "TTK-11-38, ITP-UU-11-26, SPIN-11-19, FR-PHENO-2011-015,
                        SFB-CPP-11-49",
      SLACcitation   = "%%CITATION = ARXIV:1109.1536;%%"
}

@article{Anastasiou:2003ds,
      author         = "Anastasiou, Charalampos and Dixon, Lance J. and Melnikov,
                        Kirill and Petriello, Frank",
      title          = "{High precision QCD at hadron colliders: Electroweak
                        gauge boson rapidity distributions at next-to-next-to leading order}",
      journal        = "Phys. Rev. D",
      volume         = "69",
      year           = "2004",
      pages          = "094008",
      doi            = "10.1103/PhysRevD.69.094008",
      eprint         = "hep-ph/0312266",
      archivePrefix  = "arXiv",
      reportNumber   = "SLAC-PUB-10288, UH-511-1042-03",
      SLACcitation   = "%%CITATION = HEP-PH/0312266;%%"
}

@article{Cacciari:2011hy,
      author         = "Cacciari, Matteo and Czakon, Michal and Mangano,
                        Michelangelo and Mitov, Alexander and Nason, Paolo",
      title          = "{Top-pair production at hadron colliders with
                        next-to-next-to-leading logarithmic soft-gluon
                        resummation}",
      journal        = "Phys. Lett. B",
      volume         = "710",
      pages          = "612-622",
      doi            = "10.1016/j.physletb.2012.03.013",
      year           = "2012",
      eprint         = "1111.5869",
      primaryClass   = "hep-ph",
      archivePrefix  = "arXiv",
      reportNumber   = "CERN-PH-TH-2011-277, TTK-11-54",
      SLACcitation   = "%%CITATION = ARXIV:1111.5869;%%",
}

@article{Czakon:2012zr,
      author         = "Czakon, Michal and Mitov, Alexander",
      title          = "{NNLO corrections to top-pair production at hadron
                        colliders: the all-fermionic scattering channels}",
      journal        = "JHEP",
      volume         = "12",
      pages          = "054",
      doi            = "10.1007/JHEP12(2012)054",
      year           = "2012",
      eprint         = "1207.0236",
      archivePrefix  = "arXiv",
      primaryClass   = "hep-ph",
      SLACcitation   = "%%CITATION = ARXIV:1207.0236;%%",
}

@article{Czakon:2013goa,
      author         = "Czakon, Michal and Fiedler, Paul and Mitov, Alexander",
      title          = "{Total Top-Quark Pair-Production Cross Section at
                        Hadron Colliders Through \(O(\alpha_S^4)\)}",
      journal        = "Phys. Rev. Lett.",
      volume         = "110",
      pages          = "252004",
      doi            = "10.1103/PhysRevLett.110.252004",
      year           = "2013",
      eprint         = "1303.6254",
      archivePrefix  = "arXiv",
      primaryClass   = "hep-ph",
      reportNumber   = "CERN-PH-TH-2013-056, TTK-13-08",
      SLACcitation   = "%%CITATION = ARXIV:1303.6254;%%",
}

@article{Baernreuther:2012ws,
      author         = "B{\"a}rnreuther, Peter and Czakon, Michal and Mitov,
                        Alexander",
      title          = "{Percent-Level-Precision Physics at the
                  Tevatron: Next-to-Next-to-Leading Order QCD
                  Corrections to \(q \bar{q} \to t \bar{t} + X\)}",
      journal        = "Phys. Rev. Lett.",
      volume         = "109",
      pages          = "132001",
      doi            = "10.1103/PhysRevLett.109.132001",
      year           = "2012",
      eprint         = "1204.5201",
      archivePrefix  = "arXiv",
      primaryClass   = "hep-ph",
      SLACcitation   = "%%CITATION = ARXIV:1204.5201;%%",
}

@article{Frixione:2007nw,
      author         = "Frixione, Stefano and Nason, Paolo and Ridolfi, Giovanni",
      title          = "{A positive-weight next-to-leading-order Monte Carlo for
                        heavy flavour hadroproduction}",
      journal        = "JHEP",
      volume         = "09",
      pages          = "126",
      doi            = "10.1088/1126-6708/2007/09/126",
      year           = "2007",
      eprint         = "0707.3088",
      archivePrefix  = "arXiv",
      primaryClass   = "hep-ph",
      SLACcitation   = "%%CITATION = ARXIV:0707.3088;%%",
}

@article{deFlorian:2016spz,
      author         = "de Florian, D. and others",
      title          = "{Handbook of LHC Higgs Cross Sections: 4. Deciphering the
                        Nature of the Higgs Sector}",
      collaboration  = "LHC Higgs Cross Section Working Group",
      doi            = "10.23731/CYRM-2017-002",
      year           = "2016",
      eprint         = "1610.07922",
      archivePrefix  = "arXiv",
      primaryClass   = "hep-ph",
      reportNumber   = "FERMILAB-FN-1025-T, CERN-2017-002-M",
      SLACcitation   = "%%CITATION = ARXIV:1610.07922;%%"
}

@Article{Hoeche:2009rj,
      author         = "H{\"o}che, Stefan and Krauss, Frank and Schumann, Steffen and Siegert, Frank",
      title          = "{QCD matrix elements and truncated showers}",
      journal        = "JHEP",
      volume         = "05",
      pages          = "053",
      doi            = "10.1088/1126-6708/2009/05/053",
      year           = "2009",
      eprint         = "0903.1219",
      archivePrefix  = "arXiv",
      primaryClass   = "hep-ph",
}

@Article{Gleisberg:2008fv,
      author         = "Gleisberg, Tanju and H{\"o}che, Stefan",
      title          = "{Comix, a new matrix element generator}",
      journal        = "JHEP",
      volume         = "12",
      pages          = "039",
      doi            = "10.1088/1126-6708/2008/12/039",
      year           = "2008",
      eprint         = "0808.3674",
      archivePrefix  = "arXiv",
      primaryClass   = "hep-ph",
}

@Article{Schumann:2007mg,
      author         = "Schumann, Steffen and Krauss, Frank",
      title          = "{A parton shower algorithm based on Catani--Seymour dipole factorisation}",
      journal        = "JHEP",
      volume         = "03",
      pages          = "038",
      doi            = "10.1088/1126-6708/2008/03/038",
      year           = "2008",
      eprint         = "0709.1027",
      archivePrefix  = "arXiv",
      primaryClass   = "hep-ph",
}

@article{Hoeche:2012yf,
      author         = "H{\"o}che, Stefan and Krauss, Frank and Sch{\"o}nherr, Marek and
                        Siegert, Frank",
      title          = "{QCD matrix elements + parton showers. The NLO case}",
      journal        = "JHEP",
      volume         = "04",
      year           = "2013",
      pages          = "027",
      doi            = "10.1007/JHEP04(2013)027",
      eprint         = "1207.5030",
      archivePrefix  = "arXiv",
      primaryClass   = "hep-ph",
      reportNumber   = "SLAC-PUB-15191, IPPP-12-52, DCPT-12-104, LPN12-081,
                        FR-PHENO-2012-017, MCNET-12-09, --FR-PHENO-2012-017",
      SLACcitation   = "%%CITATION = ARXIV:1207.5030;%%"
}

@article{Bothmann:2019yzt,
      author = "Bothmann, Enrico and others",
      title = "{Event generation with Sherpa 2.2}",
      journal = "SciPost Phys.",
      volume = "7",
      year = "2019",
      number = "3",
      pages = "034",
      doi = "10.21468/SciPostPhys.7.3.034",
      reportNumber = "FERMILAB-PUB-19-218-T, SLAC-PUB-17433, IPPP/19/42, MCNET-19-11",
      eprint = "1905.09127",
      archivePrefix = "arXiv",
      primaryClass = "hep-ph",
}

@article{Hoeche:2011fd,
      author         = "H{\"o}che, Stefan and Krauss, Frank and Sch{\"o}nherr, Marek and
                        Siegert, Frank",
      title          = "{A critical appraisal of NLO+PS matching methods}",
      journal        = "JHEP",
      volume         = "09",
      year           = "2012",
      pages          = "049",
      doi            = "10.1007/JHEP09(2012)049",
      eprint         = "1111.1220",
      archivePrefix  = "arXiv",
      primaryClass   = "hep-ph",
      reportNumber   = "SLAC-PUB-14661, IPPP-11-67, DCPT-11-134, LPN11-58,
                        FR-PHENO-2011-019, MCNET-11-24",
      SLACcitation   = "%%CITATION = ARXIV:1111.1220;%%"
}

@article{Catani:2001cc,
      author         = "Catani, S. and Krauss, F. and Kuhn, R. and Webber, B. R.",
      title          = "{QCD Matrix Elements + Parton Showers}",
      journal        = "JHEP",
      volume         = "11",
      year           = "2001",
      pages          = "063",
      doi            = "10.1088/1126-6708/2001/11/063",
      eprint         = "hep-ph/0109231",
      archivePrefix  = "arXiv",
      reportNumber   = "CERN-TH-2000-367, CAVENDISH-HEP-00-03",
      SLACcitation   = "%%CITATION = HEP-PH/0109231;%%"
}

@article{Cascioli:2011va,
      author         = "Cascioli, Fabio and Maierh{\"o}fer, Philipp and Pozzorini, Stefano",
      title          = "{Scattering Amplitudes with Open Loops}",
      journal        = "Phys. Rev. Lett.",
      volume         = "108",
      year           = "2012",
      pages          = "111601",
      doi            = "10.1103/PhysRevLett.108.111601",
      eprint         = "1111.5206",
      archivePrefix  = "arXiv",
      primaryClass   = "hep-ph",
      reportNumber   = "ZU-TH-23-11, LPN11-66",
      SLACcitation   = "%%CITATION = ARXIV:1111.5206;%%"
}

@article{Denner:2016kdg,
      author         = "Denner, Ansgar and Dittmaier, Stefan and Hofer, Lars",
      title          = "{\textsc{Collier}: A fortran-based complex one-loop library in
                        extended regularizations}",
      journal        = "Comput. Phys. Commun.",
      volume         = "212",
      year           = "2017",
      pages          = "220-238",
      doi            = "10.1016/j.cpc.2016.10.013",
      eprint         = "1604.06792",
      archivePrefix  = "arXiv",
      primaryClass   = "hep-ph",
      reportNumber   = "FR-PHENO-2016-003, ICCUB-16-016",
      SLACcitation   = "%%CITATION = ARXIV:1604.06792;%%"
}

@article{Czakon:2011xx,
  author         = "Czakon, Michal and Mitov, Alexander",
  title          = "{Top++: A program for the calculation of the top-pair cross-section at hadron colliders}",
  journal        = "Comput. Phys. Commun.",
  volume         = "185",
  year           = "2014",
  pages          = "2930",
  doi            = "10.1016/j.cpc.2014.06.021",
  eprint         = "1112.5675",
  archivePrefix  = "arXiv",
  primaryClass   = "hep-ph",
  reportNumber   = "CERN-PH-TH-2011-303, TTK-11-58",
}

@article{Anastasiou:2016cez,
      author         = "Anastasiou, Charalampos and Duhr, Claude and Dulat, Falko
                        and Furlan, Elisabetta and Gehrmann, Thomas and Herzog,
                        Franz and Lazopoulos, Achilleas and Mistlberger, Bernhard",
      title          = "{High precision determination of the gluon fusion Higgs
                        boson cross-section at the LHC}",
      journal        = "JHEP",
      volume         = "05",
      year           = "2016",
      pages          = "058",
      doi            = "10.1007/JHEP05(2016)058",
      eprint         = "1602.00695",
      archivePrefix  = "arXiv",
      primaryClass   = "hep-ph",
      reportNumber   = "CP3-16-01, ZU-TH-27-15, NIKHEF-2016-004,
                        CERN-TH-2016-006",
      SLACcitation   = "%%CITATION = ARXIV:1602.00695;%%"
}

@article{Anastasiou:2015ema,
      author         = "Anastasiou, Charalampos and Duhr, Claude and Dulat, Falko
                        and Herzog, Franz and Mistlberger, Bernhard",
      title          = "{Higgs Boson Gluon-Fusion Production in QCD at Three
                        Loops}",
      journal        = "Phys. Rev. Lett.",
      volume         = "114",
      year           = "2015",
      pages          = "212001",
      doi            = "10.1103/PhysRevLett.114.212001",
      eprint         = "1503.06056",
      archivePrefix  = "arXiv",
      primaryClass   = "hep-ph",
      reportNumber   = "CERN-PH-TH-2015-055, CP3-15-07",
      SLACcitation   = "%%CITATION = ARXIV:1503.06056;%%"
}

@article{Dulat:2018rbf,
      author         = "Dulat, Falko and Lazopoulos, Achilleas and Mistlberger,
                        Bernhard",
      title          = "{iHixs 2 -- Inclusive Higgs cross sections}",
      journal        = "Comput. Phys. Commun.",
      volume         = "233",
      year           = "2018",
      pages          = "243-260",
      doi            = "10.1016/j.cpc.2018.06.025",
      eprint         = "1802.00827",
      archivePrefix  = "arXiv",
      primaryClass   = "hep-ph",
      reportNumber   = "CERN-TH-2018-019, SLAC-PUB-17222",
      SLACcitation   = "%%CITATION = ARXIV:1802.00827;%%"
}

@article{Actis:2008ug,
      author         = "Actis, Stefano and Passarino, Giampiero and Sturm,
                        Christian and Uccirati, Sandro",
      title          = "{NLO electroweak corrections to Higgs boson production at
                        hadron colliders}",
      journal        = "Phys. Lett. B",
      volume         = "670",
      year           = "2008",
      pages          = "12-17",
      doi            = "10.1016/j.physletb.2008.10.018",
      eprint         = "0809.1301",
      archivePrefix  = "arXiv",
      primaryClass   = "hep-ph",
      reportNumber   = "PITHA-08-20, SFB-CPP-08-62, TTP08-38",
      SLACcitation   = "%%CITATION = ARXIV:0809.1301;%%"
}

@article{Bonetti:2018ukf,
      author         = "Bonetti, Marco and Melnikov, Kirill and Tancredi,
                        Lorenzo",
      title          = "{Higher order corrections to mixed QCD-EW contributions
                        to Higgs boson production in gluon fusion}",
      journal        = "Phys. Rev. D",
      volume         = "97",
      year           = "2018",
      number         = "5",
      pages          = "056017",
      doi            = "10.1103/PhysRevD.97.056017",
      eprint         = "1801.10403",
      archivePrefix  = "arXiv",
      primaryClass   = "hep-ph",
      reportNumber   = "CERN-TH-2018-011, TTP18-004",
      SLACcitation   = "%%CITATION = ARXIV:1801.10403;%%",
      related = "Bonetti:2018ukf-err",
      relatedstring  = "Erratum:",
}

@article{Aglietti:2004nj,
      author         = "Aglietti, U. and Bonciani, R. and Degrassi, G. and
                        Vicini, A.",
      title          = "{Two-loop light fermion contribution to Higgs production
                        and decays}",
      journal        = "Phys. Lett. B",
      volume         = "595",
      year           = "2004",
      pages          = "432-441",
      doi            = "10.1016/j.physletb.2004.06.063",
      eprint         = "hep-ph/0404071",
      archivePrefix  = "arXiv",
      xprimaryClass   = "hep-ph",
      reportNumber   = "ROME1-1373-04, FREIBURG-THEP-04-05, RM3-TH-04-06,
                        IFUM-788-FT",
      SLACcitation   = "%%CITATION = HEP-PH/0404071;%%"
}

@article{Ciccolini:2007jr,
     author         = "Ciccolini, M. and Denner, Ansgar and Dittmaier, S.",
     title          = "{Strong and Electroweak Corrections to the Production of Higgs + 2 Jets via Weak Interactions at the Large Hadron Collider}",
     journal        = "Phys. Rev. Lett.",
     volume         = "99",
     pages          = "161803",
     doi            = "10.1103/PhysRevLett.99.161803",
     year           = "2007",
     eprint         = "0707.0381",
     archivePrefix  = "arXiv",
     primaryClass   = "hep-ph",
}

@Article{Ciccolini:2007ec,
    author    = "Ciccolini, Mariano and Denner, Ansgar and Dittmaier, Stefan",
    title     = "{Electroweak and QCD corrections to Higgs production via
                 vector-boson fusion at the CERN LHC}",
    journal   = "Phys. Rev. D",
    volume    = "77",
    year      = "2008",
    pages     = "013002",
    eprint    = "0710.4749",
    archivePrefix = "arXiv",
    primaryClass  =  "hep-ph",
    doi       = "10.1103/PhysRevD.77.013002",
    SLACcitation  = "%%CITATION = 0710.4749;%%"
}

@Article{Bolzoni:2010xr,
    author    = "Bolzoni, Paolo and Maltoni, Fabio and Moch, Sven-Olaf and
                 Zaro, Marco",
    title     = "{Higgs Boson Production via Vector-Boson Fusion at Next-to-Next-to-Leading Order in QCD}",
    journal   = "Phys. Rev. Lett.",
    volume    = "105",
    year      = "2010",
    pages     = "011801",
    eprint    = "1003.4451",
    archivePrefix = "arXiv",
    primaryClass  =  "hep-ph",
    doi       = "10.1103/PhysRevLett.105.011801",
    SLACcitation  = "%%CITATION = 1003.4451;%%"
}

@article{Agostinelli:2002hh,
      author         = "{GEANT4 Collaboration} and Agostinelli, S. and others",
      title          = "{\textsc{Geant4} -- a simulation toolkit}",
      journal        = "Nucl. Instrum. Meth. A",
      volume         = "506",
      year           = "2003",
      pages          = "250",
      doi            = "10.1016/S0168-9002(03)01368-8",
      reportNumber   = "SLAC-PUB-9350, FERMILAB-PUB-03-339",
      SLACcitation   = "%%CITATION = NUIMA,A506,250;%%"
}

@Article{Read:2002hq,
    Author = {Read, Alexander L.},
    Journal = {J. Phys. G},
    Pages = {2693},
    doi = "10.1088/0954-3899/28/10/313",
    Title = "{Presentation of search results: the \(CL_S\) technique}",
    Volume = {28},
    Year = {2002}
}

@Article{Cowan:2010js,
   Author = {Cowan, Glen and Cranmer, Kyle and Gross, Eilam and Vitells, Ofer},
   Title = {{Asymptotic formulae for likelihood-based tests of new physics}},
   Journal = {Eur. Phys. J. C},
   Volume = {71},
   Year = {2011},
   Pages = {1554},
   doi = "10.1140/epjc/s10052-011-1554-0",
   Eprint = "1007.1727",
   Archiveprefix = {arXiv},
   Primaryclass = {physics.data-an},
   related = "Cowan:2010js-err",
   relatedstring  = "Erratum:",
}

@Article{Cousins:2007bmb,
      author         = "Cousins, Robert D. and Linnemann, James T. and Tucker,
                        Jordan",
      title          = "{Evaluation of three methods for calculating statistical
                        significance when incorporating a systematic uncertainty
                        into a test of the background-only hypothesis for a
                        Poisson process}",
      journal        = "Nucl. Instrum. Meth. A",
      volume         = "595",
      year           = "2008",
      number         = "2",
      pages          = "480",
      doi            = "10.1016/j.nima.2008.07.086",
      eprint         = "physics/0702156",
      archivePrefix  = "arXiv",
      primaryClass   = "physics.data-an",
      SLACcitation   = "%%CITATION = PHYSICS/0702156;%%"
}

@Article{PERF-2007-01,
    author         = "{ATLAS Collaboration}",
    title          = "{The ATLAS Experiment at the CERN Large Hadron Collider}",
    journal        = "JINST",
    volume         = "3",
    year           = "2008",
    pages          = "S08003",
    doi            = "10.1088/1748-0221/3/08/S08003",
    primaryClass   = "hep-ex",
}

@Article{SOFT-2010-01,
    author         = "{ATLAS Collaboration}",
    title          = "{The ATLAS Simulation Infrastructure}",
    journal        = "Eur. Phys. J. C",
    volume         = "70",
    year           = "2010",
    pages          = "823",
    doi            = "10.1140/epjc/s10052-010-1429-9",
    eprint         = "1005.4568",
    archivePrefix  = "arXiv",
    primaryClass   = "physics.ins-det",
}

@Article{TOPQ-2010-01,
    author         = "{ATLAS Collaboration}",
    title          = "{Measurement of the top quark-pair production cross section with ATLAS in \(pp\) collisions at \(\sqrt{s} = 7\,\text{TeV}\)}",
    journal        = "Eur. Phys. J. C",
    volume         = "71",
    year           = "2011",
    pages          = "1577",
    doi            = "10.1140/epjc/s10052-011-1577-6",
    reportNumber   = "CERN-PH-EP-2010-064",
    eprint         = "1012.1792",
    archivePrefix  = "arXiv",
    primaryClass   = "hep-ex",
}

@Article{STDM-2011-24,
    author         = "{ATLAS Collaboration}",
    title          = "{Measurement of the \(W W\) cross section in \(\sqrt{s} = 7\,\text{TeV}\) \(pp\) collisions with the ATLAS detector and limits on anomalous gauge couplings}",
    journal        = "Phys. Lett. B",
    volume         = "712",
    year           = "2012",
    pages          = "289",
    doi            = "10.1016/j.physletb.2012.05.003",
    reportNumber   = "CERN-PH-EP-2012-060",
    eprint         = "1203.6232",
    archivePrefix  = "arXiv",
    primaryClass   = "hep-ex",
}

@Article{STDM-2012-23,
    author         = "{ATLAS Collaboration}",
    title          = "{Measurement of the \(Z/\gamma^*\) boson transverse momentum distribution in \(pp\) collisions at \(\sqrt{s} = 7\,\text{TeV}\) with the ATLAS detector}",
    journal        = "JHEP",
    volume         = "09",
    year           = "2014",
    pages          = "145",
    doi            = "10.1007/JHEP09(2014)145",
    reportNumber   = "CERN-PH-EP-2014-075",
    eprint         = "1406.3660",
    archivePrefix  = "arXiv",
    primaryClass   = "hep-ex",
}

@Article{SUSY-2013-12,
    author         = "{ATLAS Collaboration}",
    title          = "{Search for direct production of charginos and neutralinos in events with three leptons and missing transverse momentum in \(\sqrt{s} = 8\,\text{TeV}\) \(pp\) collisions with the ATLAS detector}",
    journal        = "JHEP",
    volume         = "04",
    year           = "2014",
    pages          = "169",
    doi            = "10.1007/JHEP04(2014)169",
    reportNumber   = "CERN-PH-EP-2014-019",
    eprint         = "1402.7029",
    archivePrefix  = "arXiv",
    primaryClass   = "hep-ex",
}

@Article{PERF-2014-02,
    author         = "{ATLAS Collaboration}",
    title          = "{Determination of jet calibration and energy resolution in proton--proton collisions at \(\sqrt{s} = 8\,\text{TeV}\) using the ATLAS detector}",
    journal        = "Eur. Phys. J. C",
    volume         = "80",
    year           = "2020",
    pages          = "1104",
    doi            = "10.1140/epjc/s10052-020-08477-8",
    reportNumber   = "CERN-EP-2019-057",
    eprint         = "1910.04482",
    archivePrefix  = "arXiv",
    primaryClass   = "hep-ex",
}

@Article{PERF-2014-03,
    author         = "{ATLAS Collaboration}",
    title          = "{Performance of pile-up mitigation techniques for jets in \(pp\) collisions at \(\sqrt{s} = 8\,\text{TeV}\) using the ATLAS detector}",
    journal        = "Eur. Phys. J. C",
    volume         = "76",
    year           = "2016",
    pages          = "581",
    doi            = "10.1140/epjc/s10052-016-4395-z",
    reportNumber   = "CERN-PH-EP-2015-206",
    eprint         = "1510.03823",
    archivePrefix  = "arXiv",
    primaryClass   = "hep-ex",
}

@Article{PERF-2014-07,
    author         = "{ATLAS Collaboration}",
    title          = "{Topological cell clustering in the ATLAS calorimeters and its performance in LHC Run~1}",
    journal        = "Eur. Phys. J. C",
    volume         = "77",
    year           = "2017",
    pages          = "490",
    doi            = "10.1140/epjc/s10052-017-5004-5",
    reportNumber   = "CERN-PH-EP-2015-304",
    eprint         = "1603.02934",
    archivePrefix  = "arXiv",
    primaryClass   = "hep-ex",
}

@Article{PERF-2015-01,
    author         = "{ATLAS Collaboration}",
    title          = "{Reconstruction of primary vertices at the ATLAS experiment in Run~1 proton--proton collisions at the LHC}",
    journal        = "Eur. Phys. J. C",
    volume         = "77",
    year           = "2017",
    pages          = "332",
    doi            = "10.1140/epjc/s10052-017-4887-5",
    reportNumber   = "CERN-EP-2016-150",
    eprint         = "1611.10235",
    archivePrefix  = "arXiv",
    primaryClass   = "hep-ex",
}

@Article{PERF-2016-04,
    author         = "{ATLAS Collaboration}",
    title          = "{Jet energy scale measurements and their systematic uncertainties in proton--proton collisions at \(\sqrt{s} = 13\,\text{TeV}\) with the ATLAS detector}",
    journal        = "Phys. Rev. D",
    volume         = "96",
    year           = "2017",
    pages          = "072002",
    doi            = "10.1103/PhysRevD.96.072002",
    reportNumber   = "CERN-EP-2017-038",
    eprint         = "1703.09665",
    archivePrefix  = "arXiv",
    primaryClass   = "hep-ex",
}

@Article{PERF-2016-07,
    author         = "{ATLAS Collaboration}",
    title          = "{Performance of missing transverse momentum reconstruction with the ATLAS detector using proton--proton collisions at \(\sqrt{s} = 13\,\text{TeV}\)}",
    journal        = "Eur. Phys. J. C",
    volume         = "78",
    year           = "2018",
    pages          = "903",
    doi            = "10.1140/epjc/s10052-018-6288-9",
    reportNumber   = "CERN-EP-2017-274",
    eprint         = "1802.08168",
    archivePrefix  = "arXiv",
    primaryClass   = "hep-ex",
}

@Article{SUSY-2016-24,
    author         = "{ATLAS Collaboration}",
    title          = "{Search for electroweak production of supersymmetric particles in final states with two or three leptons at \(\sqrt{s} = 13\,\text{TeV}\) with the ATLAS detector}",
    journal        = "Eur. Phys. J. C",
    volume         = "78",
    year           = "2018",
    pages          = "995",
    doi            = "10.1140/epjc/s10052-018-6423-7",
    reportNumber   = "CERN-EP-2017-303",
    eprint         = "1803.02762",
    archivePrefix  = "arXiv",
    primaryClass   = "hep-ex",
}

@Article{TRIG-2016-01,
    author         = "{ATLAS Collaboration}",
    title          = "{Performance of the ATLAS trigger system in 2015}",
    journal        = "Eur. Phys. J. C",
    volume         = "77",
    year           = "2017",
    pages          = "317",
    doi            = "10.1140/epjc/s10052-017-4852-3",
    reportNumber   = "CERN-EP-2016-241",
    eprint         = "1611.09661",
    archivePrefix  = "arXiv",
    primaryClass   = "hep-ex",
}

@Article{HIGG-2017-02,
    author         = "{ATLAS Collaboration}",
    title          = "{Evidence for the associated production of the Higgs boson and a top quark pair with the ATLAS detector}",
    journal        = "Phys. Rev. D",
    volume         = "97",
    year           = "2018",
    pages          = "072003",
    doi            = "10.1103/PhysRevD.97.072003",
    reportNumber   = "CERN-EP-2017-281",
    eprint         = "1712.08891",
    archivePrefix  = "arXiv",
    primaryClass   = "hep-ex",
}

@Article{SUSY-2017-01,
    author         = "{ATLAS Collaboration}",
    title          = "{Search for chargino and neutralino production in final states with a Higgs boson and missing transverse momentum at \(\sqrt{s} = 13\,\text{TeV}\) with the ATLAS detector}",
    journal        = "Phys. Rev. D",
    volume         = "100",
    year           = "2019",
    pages          = "012006",
    doi            = "10.1103/PhysRevD.100.012006",
    reportNumber   = "CERN-EP-2018-306",
    eprint         = "1812.09432",
    archivePrefix  = "arXiv",
    primaryClass   = "hep-ex",
}

@Article{SUSY-2017-03,
    author         = "{ATLAS Collaboration}",
    title          = "{Search for chargino--neutralino production using recursive jigsaw reconstruction in final states with two or three charged leptons in proton--proton collisions at \(\sqrt{s} = 13\,\text{TeV}\) with the ATLAS detector}",
    journal        = "Phys. Rev. D",
    volume         = "98",
    year           = "2018",
    pages          = "092012",
    doi            = "10.1103/PhysRevD.98.092012",
    reportNumber   = "CERN-EP-2018-113",
    eprint         = "1806.02293",
    archivePrefix  = "arXiv",
    primaryClass   = "hep-ex",
}

@Article{DAPR-2018-01,
    author         = "{ATLAS Collaboration}",
    title          = "{ATLAS data quality operations and performance for 2015--2018 data-taking}",
    journal        = "JINST",
    volume         = "15",
    year           = "2020",
    pages          = "P04003",
    doi            = "10.1088/1748-0221/15/04/P04003",
    reportNumber   = "CERN-EP-2019-207",
    eprint         = "1911.04632",
    archivePrefix  = "arXiv",
    primaryClass   = "physics.ins-det",
}

@Article{EGAM-2018-01,
    author         = "{ATLAS Collaboration}",
    title          = "{Electron and photon performance measurements with the ATLAS detector using the 2015--2017 LHC proton--proton collision data}",
    journal        = "JINST",
    volume         = "14",
    year           = "2019",
    pages          = "P12006",
    doi            = "10.1088/1748-0221/14/12/P12006",
    reportNumber   = "CERN-EP-2019-145",
    eprint         = "1908.00005",
    archivePrefix  = "arXiv",
    primaryClass   = "hep-ex",
}

@Article{FTAG-2018-01,
    author         = "{ATLAS Collaboration}",
    title          = "{ATLAS \(b\)-jet identification performance and efficiency measurement with \(t\bar{t}\) events in \(pp\) collisions at \(\sqrt{s} = 13\,\text{TeV}\)}",
    journal        = "Eur. Phys. J. C",
    volume         = "79",
    year           = "2019",
    pages          = "970",
    doi            = "10.1140/epjc/s10052-019-7450-8",
    reportNumber   = "CERN-EP-2019-132",
    eprint         = "1907.05120",
    archivePrefix  = "arXiv",
    primaryClass   = "hep-ex",
}

@Article{JETM-2018-05,
    author         = "{ATLAS Collaboration}",
    title          = "{Jet energy scale and resolution measured in proton--proton collisions at \(\sqrt{s} = 13\,\text{TeV}\) with the ATLAS detector}",
    year           = "2020",
    reportNumber   = "CERN-EP-2020-083",
    eprint         = "2007.02645",
    archivePrefix  = "arXiv",
    primaryClass   = "hep-ex",
}

@Article{MUON-2018-03,
    author         = "{ATLAS Collaboration}",
    title          = "{Muon reconstruction and identification efficiency in ATLAS using the full Run~2 \(pp\) collision data set at \(\sqrt{s} = 13\,\text{TeV}\)}",
    year           = "2020",
    reportNumber   = "CERN Preprint ID: CERN-EP-2020-199",
    eprint         = "2012.00578",
    archivePrefix  = "arXiv",
    primaryClass   = "hep-ex",
}

@Article{PIX-2018-001,
    author         = "Abbott, B. and others",
    title          = "{Production and integration of the ATLAS Insertable B-Layer}",
    journal        = "JINST",
    volume         = "13",
    year           = "2018",
    pages          = "T05008",
    doi            = "10.1088/1748-0221/13/05/T05008",
    eprint         = "1803.00844",
    archivePrefix  = "arXiv",
    primaryClass   = "physics.ins-det",
}

@Article{SUSY-2018-06,
    author         = "{ATLAS Collaboration}",
    title          = "{Search for chargino--neutralino production with mass splittings near the electroweak scale in three-lepton final states in \(\sqrt{s} = 13\,\text{TeV}\) \(pp\) collisions with the ATLAS detector}",
    journal        = "Phys. Rev. D",
    volume         = "101",
    year           = "2020",
    pages          = "072001",
    doi            = "10.1103/PhysRevD.101.072001",
    reportNumber   = "CERN-EP-2019-263",
    eprint         = "1912.08479",
    archivePrefix  = "arXiv",
    primaryClass   = "hep-ex",
}

@Article{SUSY-2018-16,
    author         = "{ATLAS Collaboration}",
    title          = "{Searches for electroweak production of supersymmetric particles with compressed mass spectra in \(\sqrt{s} = 13\,\text{TeV}\) \(pp\) collisions with the ATLAS detector}",
    journal        = "Phys. Rev. D",
    volume         = "101",
    year           = "2020",
    pages          = "052005",
    doi            = "10.1103/PhysRevD.101.052005",
    reportNumber   = "CERN-EP-2019-242",
    eprint         = "1911.12606",
    archivePrefix  = "arXiv",
    primaryClass   = "hep-ex",
}

@Article{SUSY-2018-23,
    author         = "{ATLAS Collaboration}",
    title          = "{Search for direct production of electroweakinos in final states with missing transverse momentum and a Higgs boson decaying into photons in \(pp\) collisions at \(\sqrt{s} = 13\,\text{TeV}\) with the ATLAS detector}",
    journal        = "JHEP",
    volume         = "10",
    year           = "2020",
    pages          = "005",
    doi            = "10.1007/JHEP10(2020)005",
    reportNumber   = "CERN-EP-2019-204",
    eprint         = "2004.10894",
    archivePrefix  = "arXiv",
    primaryClass   = "hep-ex",
}

@Article{SUSY-2018-32,
    author         = "{ATLAS Collaboration}",
    title          = "{Search for electroweak production of charginos and sleptons decaying into final states with two leptons and missing transverse momentum in \(\sqrt{s} = 13\,\text{TeV}\) \(pp\) collisions using the ATLAS detector}",
    journal        = "Eur. Phys. J. C",
    volume         = "80",
    year           = "2020",
    pages          = "123",
    doi            = "10.1140/epjc/s10052-019-7594-6",
    reportNumber   = "CERN-EP-2019-106",
    eprint         = "1908.08215",
    archivePrefix  = "arXiv",
    primaryClass   = "hep-ex",
}

@Article{TRIG-2018-01,
    author         = "{ATLAS Collaboration}",
    title          = "{Performance of the ATLAS muon triggers in Run~2}",
    journal        = "JINST",
    volume         = "15",
    year           = "2020",
    pages          = "P09015",
    doi            = "10.1088/1748-0221/15/09/p09015",
    reportNumber   = "CERN-EP-2020-031",
    eprint         = "2004.13447",
    archivePrefix  = "arXiv",
    primaryClass   = "hep-ex",
}

@Article{TRIG-2018-05,
    author         = "{ATLAS Collaboration}",
    title          = "{Performance of electron and photon triggers in ATLAS during LHC Run~2}",
    journal        = "Eur. Phys. J. C",
    volume         = "80",
    year           = "2020",
    pages          = "47",
    doi            = "10.1140/epjc/s10052-019-7500-2",
    reportNumber   = "CERN-EP-2019-169",
    eprint         = "1909.00761",
    archivePrefix  = "arXiv",
    primaryClass   = "hep-ex",
}

@Article{SUSY-2019-08,
    author         = "{ATLAS Collaboration}",
    title          = "{Search for direct production of electroweakinos in final states with one lepton, missing transverse momentum and a Higgs boson decaying into two \(b\)-jets in \(pp\) collisions at \(\sqrt{s} = 13\,\text{TeV}\) with the ATLAS detector}",
    journal        = "Eur. Phys. J. C",
    volume         = "80",
    year           = "2020",
    pages          = "691",
    doi            = "10.1140/epjc/s10052-020-8050-3",
    reportNumber   = "CERN-EP-2019-188",
    eprint         = "1909.09226",
    archivePrefix  = "arXiv",
    primaryClass   = "hep-ex",
}

@Article{TRIG-2019-01,
    author         = "{ATLAS Collaboration}",
    title          = "{Performance of the missing transverse momentum triggers for the ATLAS detector during Run-2 data taking}",
    journal        = "JHEP",
    volume         = "08",
    year           = "2020",
    pages          = "080",
    doi            = "10.1007/JHEP08(2020)080",
    reportNumber   = "CERN-EP-2020-050",
    eprint         = "2005.09554",
    archivePrefix  = "arXiv",
    primaryClass   = "hep-ex",
}

@Booklet{ATL-SOFT-PUB-2020-001,
    author         = "{ATLAS Collaboration}",
    title          = "{ATLAS Computing Acknowledgements}",
    howpublished   = "{ATL-SOFT-PUB-2020-001}",
    url            = "https://cds.cern.ch/record/2717821",
}

@Report{ATLAS-TDR-2010-19,
    author         = "{ATLAS Collaboration}",
    title          = "{ATLAS Insertable B-Layer Technical Design Report}",
    type           = "ATLAS-TDR-19; CERN-LHCC-2010-013",
    year           = "2010",
    url            = "https://cds.cern.ch/record/1291633",
    related        = "ATLAS-TDR-2010-19-addm",
    relatedstring  = "Addendum:",
}

@Article{CMS-SUS-16-034,
    author         = "{CMS Collaboration}",
    title          = "{Search for new phenomena in final states with two opposite-charge, same-flavor leptons, jets, and missing transverse momentum in \(pp\) collisions at \(\sqrt{s} = 13\,\text{TeV}\)}",
    journal        = "JHEP",
    volume         = "03",
    year           = "2018",
    pages          = "076",
    doi            = "10.1007/s13130-018-7845-2",
    reportNumber   = "CERN-EP-2017-170",
    eprint         = "1709.08908",
    archivePrefix  = "arXiv",
    primaryClass   = "hep-ex",
}

@Article{CMS-SUS-16-039,
    author         = "{CMS Collaboration}",
    title          = "{Search for electroweak production of charginos and neutralinos in multilepton final states in proton--proton collisions at \(\sqrt{s} = 13\,\text{TeV}\)}",
    journal        = "JHEP",
    volume         = "03",
    year           = "2018",
    pages          = "166",
    doi            = "10.1007/JHEP03(2018)166",
    reportNumber   = "CERN-EP-2017-121",
    eprint         = "1709.05406",
    archivePrefix  = "arXiv",
    primaryClass   = "hep-ex",
}

@Article{CMS-SUS-16-043,
    author         = "{CMS Collaboration}",
    title          = "{Search for electroweak production of charginos and neutralinos in \(WH\) events in proton--proton collisions at \(\sqrt{s} = 13\,\text{TeV}\)}",
    journal        = "JHEP",
    volume         = "11",
    year           = "2017",
    pages          = "029",
    doi            = "10.1007/JHEP11(2017)029",
    reportNumber   = "CERN-EP-2017-113",
    eprint         = "1706.09933",
    archivePrefix  = "arXiv",
    primaryClass   = "hep-ex",
}

@Article{CMS-SUS-16-045,
    author         = "{CMS Collaboration}",
    title          = "{Search for supersymmetry with Higgs boson to diphoton decays using the razor variables at \(\sqrt{s} = 13\,\text{TeV}\)}",
    journal        = "Phys. Lett. B",
    volume         = "779",
    year           = "2018",
    pages          = "166",
    doi            = "10.1016/j.physletb.2017.12.069",
    reportNumber   = "CERN-EP-2017-158",
    eprint         = "1709.00384",
    archivePrefix  = "arXiv",
    primaryClass   = "hep-ex",
}

@Article{CMS-SUS-16-048,
    author         = "{CMS Collaboration}",
    title          = "{Search for new physics in events with two soft oppositely charged leptons and missing transverse momentum in proton--proton collisions at \(\sqrt{s} = 13\,\text{TeV}\)}",
    journal        = "Phys. Lett. B",
    volume         = "782",
    year           = "2018",
    pages          = "440",
    doi            = "10.1016/j.physletb.2018.05.062",
    reportNumber   = "CERN-EP-2017-336",
    eprint         = "1801.01846",
    archivePrefix  = "arXiv",
    primaryClass   = "hep-ex",
}

@Article{CMS-SUS-17-004,
    author         = "{CMS Collaboration}",
    title          = "{Combined search for electroweak production of charginos and neutralinos in proton--proton collisions at \(\sqrt{s} = 13\,\text{TeV}\)}",
    journal        = "JHEP",
    volume         = "03",
    year           = "2018",
    pages          = "160",
    doi            = "10.1007/JHEP03(2018)160",
    reportNumber   = "CERN-EP-2017-283",
    eprint         = "1801.03957",
    archivePrefix  = "arXiv",
    primaryClass   = "hep-ex",
}

@Article{CMS-SUS-17-007,
    author         = "{CMS Collaboration}",
    title          = "{Search for supersymmetry with a compressed mass spectrum in the vector boson fusion topology with 1-lepton and 0-lepton final states in proton--proton collisions at \(\sqrt{s} = 13\,\text{TeV}\)}",
    journal        = "JHEP",
    volume         = "08",
    year           = "2019",
    pages          = "150",
    doi            = "10.1007/JHEP08(2019)150",
    reportNumber   = "CERN-EP-2019-093",
    eprint         = "1905.13059",
    archivePrefix  = "arXiv",
    primaryClass   = "hep-ex",
}

@Booklet{ATLAS-CONF-2014-018,
    author         = "{ATLAS Collaboration}",
    title          = "{Tagging and suppression of pileup jets with the ATLAS detector}",
    howpublished   = "{ATLAS-CONF-2014-018}",
    url            = "https://cds.cern.ch/record/1700870",
    year           = "2014",
}

@Booklet{ATLAS-CONF-2015-029,
    author         = "{ATLAS Collaboration}",
    title          = "{Selection of jets produced in \(13~\text{TeV}\) proton--proton collisions with the ATLAS detector}",
    howpublished   = "{ATLAS-CONF-2015-029}",
    url            = "https://cds.cern.ch/record/2037702",
    year           = "2015",
}

@Booklet{ATLAS-CONF-2018-023,
    author         = "{ATLAS Collaboration}",
    title          = "{\(E_{\text{T}}^{\text{miss}}\) performance in the ATLAS detector using 2015--2016 LHC \(pp\) collisions}",
    howpublished   = "{ATLAS-CONF-2018-023}",
    url            = "https://cds.cern.ch/record/2625233",
    year           = "2018",
}

@Booklet{ATLAS-CONF-2018-038,
    author         = "{ATLAS Collaboration}",
    title          = "{Object-based missing transverse momentum significance in the ATLAS Detector}",
    howpublished   = "{ATLAS-CONF-2018-038}",
    url            = "https://cds.cern.ch/record/2630948",
    year           = "2018",
}

@Booklet{ATLAS-CONF-2019-021,
    author         = "{ATLAS Collaboration}",
    title          = "{Luminosity determination in \(pp\) collisions at \(\sqrt{s} = 13\,\text{TeV}\) using the ATLAS detector at the LHC}",
    howpublished   = "{ATLAS-CONF-2019-021}",
    url            = "https://cds.cern.ch/record/2677054",
    year           = "2019",
}

@Article{EvtGen,
    author         = "{D. J. Lange}",
    title          = "{The EvtGen particle decay simulation package}",
    journal        = "Nucl. Instrum. Meth. A",
    volume         = "462",
    year           = "2001",
    pages          = "152",
    doi            = "10.1016/S0168-9002(01)00089-4",
    primaryClass   = "hep-ex",
}

@Booklet{ATL-PHYS-PUB-2010-005,
    author         = "{ATLAS Collaboration}",
    title          = "{Prospects for Higgs boson searches using the \(H \rightarrow WW^{(*)} \rightarrow \ell\nu\ell\nu\) decay mode with the ATLAS detector at \(10~\text{TeV}\)}",
    howpublished   = "{ATL-PHYS-PUB-2010-005}",
    url            = "https://cds.cern.ch/record/1270568",
    year           = "2010",
}

@Booklet{ATL-PHYS-PUB-2010-013,
    author         = "{ATLAS Collaboration}",
    title          = "{The simulation principle and performance of the ATLAS fast calorimeter simulation FastCaloSim}",
    howpublished   = "{ATL-PHYS-PUB-2010-013}",
    url            = "https://cds.cern.ch/record/1300517",
    year           = "2010",
}

@Booklet{ATL-PHYS-PUB-2014-021,
    author         = "{ATLAS Collaboration}",
    title          = "{ATLAS Pythia~8 tunes to \(7~\text{TeV}\) data}",
    howpublished   = "{ATL-PHYS-PUB-2014-021}",
    url            = "https://cds.cern.ch/record/1966419",
    year           = "2014",
}

@Booklet{ATL-PHYS-PUB-2015-026,
    author         = "{ATLAS Collaboration}",
    title          = "{Vertex Reconstruction Performance of the ATLAS Detector at \(\sqrt{s} = 13~\text{TeV}\)}",
    howpublished   = "{ATL-PHYS-PUB-2015-026}",
    url            = "https://cds.cern.ch/record/2037717",
    year           = "2015",
}

@Booklet{ATL-PHYS-PUB-2016-017,
    author         = "{ATLAS Collaboration}",
    title          = "{The Pythia~8 A3 tune description of ATLAS minimum bias and inelastic measurements incorporating the Donnachie--Landshoff diffractive model}",
    howpublished   = "{ATL-PHYS-PUB-2016-017}",
    url            = "https://cds.cern.ch/record/2206965",
    year           = "2016",
}

@Booklet{ATL-PHYS-PUB-2016-020,
    author         = "{ATLAS Collaboration}",
    title          = "{Studies on top-quark Monte Carlo modelling for Top2016}",
    howpublished   = "{ATL-PHYS-PUB-2016-020}",
    url            = "https://cds.cern.ch/record/2216168",
    year           = "2016",
}

@Booklet{ATL-PHYS-PUB-2017-005,
    author         = "{ATLAS Collaboration}",
    title          = "{Multi-Boson Simulation for \(13~\text{TeV}\) ATLAS Analyses}",
    howpublished   = "{ATL-PHYS-PUB-2017-005}",
    url            = "https://cds.cern.ch/record/2261933",
    year           = "2017",
}

@Booklet{ATL-PHYS-PUB-2017-006,
    author         = "{ATLAS Collaboration}",
    title          = "{ATLAS simulation of boson plus jets processes in Run~2}",
    howpublished   = "{ATL-PHYS-PUB-2017-006}",
    url            = "https://cds.cern.ch/record/2261937",
    year           = "2017",
}

@Booklet{ATL-PHYS-PUB-2017-013,
    author         = "{ATLAS Collaboration}",
    title          = "{Optimisation and performance studies of the ATLAS \(b\)-tagging algorithms for the 2017-18 LHC run}",
    howpublished   = "{ATL-PHYS-PUB-2017-013}",
    url            = "https://cds.cern.ch/record/2273281",
    year           = "2017",
}

@Booklet{ATL-PHYS-PUB-2019-029,
    author         = "{ATLAS Collaboration}",
    title          = "{Reproduction searches for new physics with the ATLAS experiment through publication of full statistical likelihoods}",
    howpublished   = "{ATL-PHYS-PUB-2019-029}",
    url            = "https://cds.cern.ch/record/2684863",
    year           = "2019",
}
 
\begin{DIFnomarkup}
\end{DIFnomarkup}

\clearpage
% ATLAS Collaboration author list
% Reference date of SUSY-2019-09 is 2021-02-13
% Author list last updated on date 03-JUN-21
% Data extracted on 03-Jun-2021 for paper reference SUSY-2019-09
% at 10:05am
 
\begin{flushleft}
\hypersetup{urlcolor=black}
{\Large The ATLAS Collaboration}

\bigskip

\AtlasOrcid[0000-0002-6665-4934]{G.~Aad}$^\textrm{\scriptsize 99}$,    
\AtlasOrcid[0000-0002-5888-2734]{B.~Abbott}$^\textrm{\scriptsize 125}$,    
\AtlasOrcid[0000-0002-7248-3203]{D.C.~Abbott}$^\textrm{\scriptsize 100}$,    
\AtlasOrcid[0000-0002-2788-3822]{A.~Abed~Abud}$^\textrm{\scriptsize 34}$,    
\AtlasOrcid[0000-0002-1002-1652]{K.~Abeling}$^\textrm{\scriptsize 51}$,    
\AtlasOrcid[0000-0002-2987-4006]{D.K.~Abhayasinghe}$^\textrm{\scriptsize 91}$,    
\AtlasOrcid[0000-0002-8496-9294]{S.H.~Abidi}$^\textrm{\scriptsize 27}$,    
\AtlasOrcid[0000-0002-8279-9324]{O.S.~AbouZeid}$^\textrm{\scriptsize 38}$,    
\AtlasOrcid[0000-0001-5329-6640]{H.~Abramowicz}$^\textrm{\scriptsize 158}$,    
\AtlasOrcid[0000-0002-1599-2896]{H.~Abreu}$^\textrm{\scriptsize 157}$
% ...
\bigskip
\\

% ...
$^{27}$Physics Department, Brookhaven National Laboratory, Upton NY; United States of America.\\
% ...
$^{34}$CERN, Geneva; Switzerland.\\
% ...
$^{38}$Niels Bohr Institute, University of Copenhagen, Copenhagen; Denmark.\\
% ...
$^{51}$II. Physikalisches Institut, Georg-August-Universit\"{a}t G\"ottingen, G\"ottingen; Germany.\\
% ...
$^{91}$Department of Physics, Royal Holloway University of London, Egham; United Kingdom.\\
% ...
$^{99}$CPPM, Aix-Marseille Universit\'e, CNRS/IN2P3, Marseille; France.\\
$^{100}$Department of Physics, University of Massachusetts, Amherst MA; United States of America.\\
% ...
$^{125}$Homer L. Dodge Department of Physics and Astronomy, University of Oklahoma, Norman OK; United States of America.\\
% ...
$^{157}$Department of Physics, Technion, Israel Institute of Technology, Haifa; Israel.\\
$^{158}$Raymond and Beverly Sackler School of Physics and Astronomy, Tel Aviv University, Tel Aviv; Israel.\\
% ...

\end{flushleft}

% Created with Glance <Atlas.Glance@cern.ch>
 
\end{document}